\documentclass[a4paper,twocolumn,11pt,unpublished,allowtoday]{quantumarticle}
\pdfoutput=1
\usepackage[T1]{fontenc}
\usepackage{amsmath,amssymb,booktabs,graphicx,microtype,tikz,xurl,adjustbox,placeins}
\usetikzlibrary{arrows.meta,positioning,calc,fit,backgrounds,shapes.geometric}
\usepackage[numbers]{natbib}
\usepackage{hyperref}
\usepackage{float}
\usepackage[compat=0.8]{yquant}

\tikzset{
 wire/.style={line width=.55pt},
 flow/.style={-{Stealth[length=2mm]},line width=.6pt},
 box/.style={rectangle,draw=black,fill=white,align=center,inner sep=2mm,font=\small,line width=.5pt},
 note/.style={font=\scriptsize,align=center},
 secondary/.style={note,text=black!65},
 var/.style={circle,draw,fill=white,minimum size=6mm,inner sep=.6mm,font=\small},
 site/.style={circle,draw,fill=white,minimum size=6mm,inner sep=0pt,font=\scriptsize},
 pauli/.style={circle,draw,fill=black!5,minimum size=7mm,inner sep=.8mm,font=\small},
 initial pauli/.style={pauli,fill=black!12},
 pauli edge/.style={flow},
 keep/.style={box,inner sep=1.5mm}
}

\definecolor{pppurple}{RGB}{198,81,205}
\definecolor{ppcyan}{RGB}{42,185,209}
\tikzset{
 pauli/.append style={
   minimum size=10mm,inner sep=0pt,font=\scriptsize,
   left color=pppurple,right color=pppurple!15,shading angle=135},
 initial pauli/.append style={
   left color=ppcyan,right color=ppcyan!15,shading angle=135}
}
\definecolor{highlight}{RGB}{128,128,0}

\pgfdeclarelayer{edgelayer}
\pgfdeclarelayer{nodelayer}
\newenvironment{semiwebaddition}{%
  \par\begingroup\color{black}\hypersetup{allcolors=black}%
}{\par\endgroup}

\newcommand{\semiwebcaption}[1]{%
  \captionsetup{font+={color=black},labelfont+={color=black}}%
  \caption{#1}}

\definecolor{semicapZ}{RGB}{216,248,216}
\definecolor{semicapX}{RGB}{232,165,165}
\tikzset{
 semicap wire/.style={draw=black,line width=.4pt},
 semicap node/.style={circle,draw=black,text=black,line width=.4pt,
   inner sep=0pt,minimum size=2.1mm},
 semicap phase/.style={semicap node,fill=semicapZ,minimum size=5.2mm,
   text width=4mm,align=center,font=\scriptsize}
}
\DeclareRobustCommand{\semiwebCaptionGate}[1]{%
  \tikz[baseline=-.5ex,x=1mm,y=1mm]{%
    \draw[semicap wire] (-5,0)--(5,0);
    \node[semicap phase] at (0,0) {$#1$};}}
\DeclareRobustCommand{\semiwebCaptionCnot}{%
  \tikz[baseline=-.5ex,x=1mm,y=1mm]{%
    \draw[semicap wire] (-4,2)--(4,2);
    \draw[semicap wire] (-4,-2)--(4,-2);
    \draw[semicap wire] (0,2)--(0,-2);
    \node[semicap node,fill=semicapZ] at (0,2) {};
    \node[semicap node,fill=semicapX] at (0,-2) {};}}
\DeclareRobustCommand{\semiwebCaptionEffect}[1]{%
  \tikz[baseline=-.5ex,x=1mm,y=1mm]{%
    \draw[semicap wire] (-5,0)--(0,0);
    \node[semicap node,fill=semicap#1] at (0,0) {};}}
\DeclareRobustCommand{\semiwebCaptionState}[1]{%
  \tikz[baseline=-.5ex,x=1mm,y=1mm]{%
    \draw[semicap wire] (0,0)--(5,0);
    \node[semicap node,fill=semicap#1] at (0,0) {};}}
\DeclareRobustCommand{\semiwebCaptionPhaseState}[1]{%
  \tikz[baseline=-.5ex,x=1mm,y=1mm]{%
    \draw[semicap wire] (0,0)--(5,0);
    \node[semicap phase] at (0,0) {$#1$};}}

\definecolor{paperpurple}{HTML}{53257F}
\definecolor{newround}{HTML}{277D46}
\providecommand{\ket}[1]{\lvert#1\rangle}

\newcommand{\PL}{P_{\mathrm L}}
\newcommand{\F}{\mathbb F_2}
\newcommand{\Tr}{\operatorname{Tr}}
\newcommand{\rank}{\operatorname{rank}}
\newcommand{\spann}{\operatorname{span}}
\newcommand{\ind}[1]{\mathbf1\!\left[#1\right]}

\newcommand{\arxiv}[1]{[\href{https://arxiv.org/abs/#1}{arXiv:#1}]}

\newcommand{\fig}[3]{%
 \begin{center}
 \includegraphics[width=#2\linewidth]{figures/other_msc/#1.pdf}
 \captionof{figure}{#3}
 \end{center}}
\title{Exact logical error rates for magic state cultivation}
\author{Kwok Ho Wan}
\orcid{0000-0002-1762-1001}
\affiliation{Blackett Laboratory, Imperial College London, South Kensington, London SW7 2AZ, UK}
\affiliation{Mathematical Institute, University of Oxford, Andrew Wiles Building, Woodstock Road, Oxford OX2 6GG, UK}
\email{((initials))1496((at))((9.81))mail.com}
\thanks{current affiliation: PsiQuantum, Palo Alto.}
\author{Ainhoa Zapirain}
\date{\today}
\begin{document}
\maketitle
\begin{abstract}
We compute exactly the acceptance and logical error rates for the distance $d=3$ and $d=5$ magic state cultivation circuits from Clifft \arxiv{2604.27058} and 
SOFT \arxiv{2512.23037} using Pauli propagation and binary tensor contraction. Actual $T$-gates are studied, not the $S$-gate proxy used for sampling. The calculation includes every fault order at several circuit-level noise strengths ($p$). 
We provide a series expansion form to the logical error rates, through order $(p/(1-p))^{10}$. 
The analytical results recover the numerical values from Clifft and SOFT at both $d=3$ and $d=5$ to within their sampling uncertainty. Then, we show that the $d=3$ and $d=5$ circuits actually have fault distances of $d_{\text{fault}}=2$ and $d_{\text{fault}}=3$ respectively, explaining the similar distance degrading effects from the companion code of [\href{https://doi.org/10.22331/q-2026-06-12-2134}{Quantum \textbf{10}, 2134 (2026)}].
\end{abstract}
\section{Introduction and results}\label{sec:results}
Magic state cultivation grows and double-checks a logical magic state, $|T^{\dagger}\rangle = \frac{1}{\sqrt{2}}\Big(|0\rangle + e^{-i\pi/4}|1\rangle\Big)$, injected onto a
small colour code, achieving very low ($10^{-6}$ to $10^{-9}$) logical error rates (LER)~\cite{msc}. Such rates determine the resource cost of a fault-tolerant computation, yet at $10^{-9}$ they are expensive to resolve by direct sampling, and it has been unclear whether the $S$-gate proxy used to estimate them reproduces the actual $T$-gate circuit. Because most attempts are rejected and accepted logical errors are rare, Monte Carlo estimates of the LER are expensive, so we compute the two underlying probabilities instead,
\begin{equation}
 \begin{aligned}
 A(p)&=\Pr(\text{post-selection criteria accepted}),\\
 B(p)&=\Pr\!\left(\text{post-selected and logical error}\right),\\
 \PL(p)&=\frac{B(p)}{A(p)}.
 \end{aligned}
 \label{eq:probabilities}
\end{equation}
Here, $\PL$ is the LER of the post-selected (upon all $+1$ parity detectors~\cite{msc}) state and $p$ is a circuit-level noise parameter~\cite{msc,soft,symft,clifft}. The escape stage is not included.

As a `source of truth' in this manuscript, we use the SOFT circuits~\cite{soft}, also distributed within Clifft~\cite{clifft} and SymFT~\cite{symft} and we use the standard circuit-level noise model of Ref.~\cite{msc}, uniform depolarising noise after every operation with measurement and reset flips, at a common strength $p$ throughout~\cite{msc,soft}. In addition, the SOFT circuits use the unitary injection from the companion code of~\cite{msc} (\texttt{inject[`unitary']}). The full
$d=5$ circuit includes injection, $d=3$ double checking, code growth, $d=5$ double checking, and a noiseless readout for the sake of logical error rate computation. The labels $d=3,5$ refer to the code distances of the underlying seven- and nineteen-qubit triangular colour codes. 
The circuit's fault distances ($d_{\text{fault}}$) under a standard circuit-level noise model are lower: $d_{\text{fault}}=2$ and $d_{\text{fault}}=3$, as shown by the leading order terms in Eqs.~\eqref{eq:d3result} and~\eqref{eq:fullseries}, respectively.

\begin{figure*}[t]
\centering
\includegraphics[width=.325\linewidth]{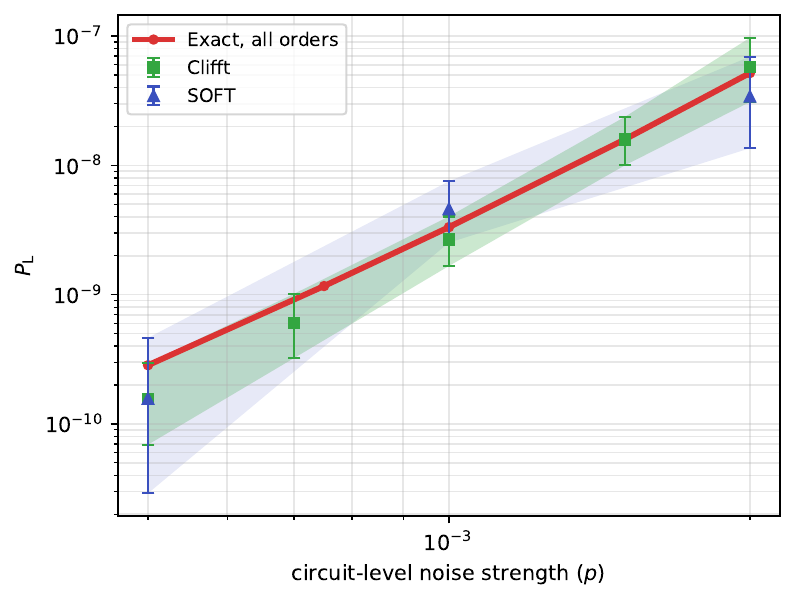}\hfill
\includegraphics[width=.325\linewidth]{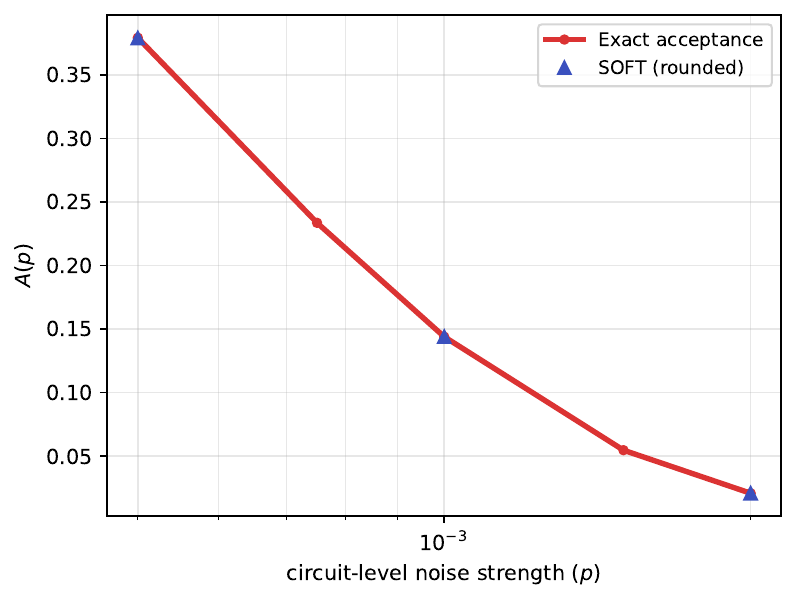}\hfill
\includegraphics[width=.325\linewidth]{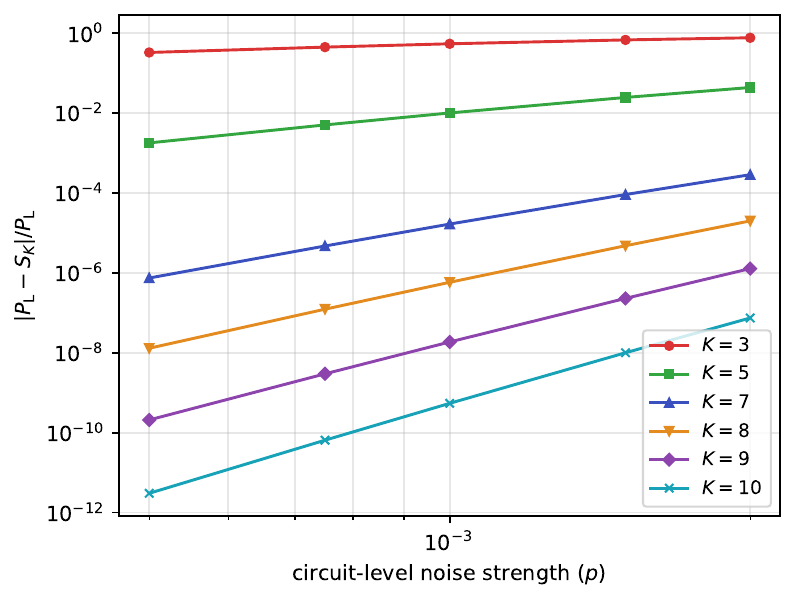}
\caption{$d=5$ cultivation: logical error (left), acceptance (middle), and relative error of the truncated series (right). Shading and error bars show the published LER sampling-error intervals.}
\label{fig:comparison}
\label{fig:series}
\end{figure*}

Writing $x=p/(1-p)$, the $d=5$ result through $x^{10}$, with remainder $O(x^{11})$, is:
\begin{equation}
 \boxed{\scalebox{0.73}{$\displaystyle\begin{aligned}
&P_{\mathrm L}^{(5)}(x)=\frac{574}{375}x^{3}+\frac{972959741}{810000}x^{4}+\frac{2678166036893}{4860000}x^{5}\\[.4ex]
&{}+\frac{46204605652071071}{1458000000}x^{6}+\frac{20165667601860828941}{14580000000}x^{7}\\[.4ex]
&{}+\frac{34708968532201339967}{656100000}x^{8}+\frac{12146976079587122943200741}{6561000000000}x^{9}\\[.4ex]
&{}+\frac{28153927149162391481472671659}{472392000000000}x^{10}+O(x^{11})
\end{aligned}
$}}
 \label{eq:fullseries}
\end{equation}
Each coefficient in Eq.~\eqref{eq:fullseries} is an exact
rational number. The separate all-orders calculation at $p=10^{-3}$ gives
\begin{equation}
 \boxed{\begin{gathered}
 \PL^{(5)}(p=10^{-3})
 =3.32842537\ldots \times10^{-9}
 \end{gathered}} \ .
 \label{eq:anchor}
\end{equation}
Equation~\eqref{eq:anchor} is the decimal expansion of an exact fraction, reconstructed as described in Appendix~\ref{app:arithmetic}. We refer to such values as exact fractions.

The $d=3$ cultivation circuit, which has an independent state-vector benchmark~\cite{msc}, can also be evaluated exactly. At $p=10^{-3}$,
\begin{equation}
 \begin{gathered}
 \PL^{(3)}(p=10^{-3})=8.982790954\ldots\times10^{-7},\\
 \PL^{(3)}(x)=\frac{32}{75}x^2+\frac{2009039}{4500}x^3+O(x^4).
 \end{gathered}
 \label{eq:d3result}
\end{equation}
Its complete coefficients through $x^{10}$ are in
Table~\ref{tab:d3coeff}.

\textbf{The main results}: we compute the LER in two forms: 
\begin{enumerate}
    \item an exact fraction at a fixed noise strength $p$, and 
    \item a series in $x=p/(1-p)$ with exact coefficients for small $p$.
\end{enumerate}
The fraction is the exact value at one $p$. The series can be evaluated at any $p$. Its accuracy is assessed against an exact fraction where one exists, Fig.~\ref{fig:comparison} (right) and Table~\ref{tab:convergence}, and by the remainder bound of Appendix~\ref{app:arithmetic} (Table~\ref{tab:tails}) otherwise.

\subsection{Exact logical error rates across different noise strengths}
Table~\ref{tab:grid} gives a few exact rational evaluations. Following the standard circuit-level noise model, every operation such as preparation, idle, gate and
mid-circuit measurements is noisy, only the final logical readout is noiseless to compute the LER. The escape, grafting and decoding stages~\cite{msc} are outside the scope of this manuscript. Appendix~\ref{app:circuits} specifies the source of the circuits used and the post-selection criteria.

\begin{table}[h]
\centering
\captionsetup{font=small,position=above,skip=4pt}
\caption{Whole-circuit $d=5$ results. Displayed decimals are rounded; the stored values are exact fractions.}
\label{tab:grid}
\footnotesize
\setlength{\tabcolsep}{3pt}
\begin{tabular}{@{}r|l|l@{}}
\toprule
$p$ & $A(p)$ & $P_{\mathrm L}(p)$ \\
\midrule
$0.00050$ & $0.379184137167$ & $2.846165501072\times10^{-10}$ \\
$0.00075$ & $0.233599896243$ & $1.165527889094\times10^{-9}$ \\
$0.00100$ & $0.143955584096$ & $3.328425371418\times10^{-9}$ \\
$0.00150$ & $0.0547199749366$ & $1.591266806530\times10^{-8}$ \\
$0.00200$ & $0.0208263641922$ & $5.175102366142\times10^{-8}$ \\
\bottomrule
\end{tabular}

\end{table}

At $p=10^{-3}$, the acceptance for the $d=5$ cultivation circuits is about
$14.395\ldots\%$ and 
$B=4.7914541\ldots\times10^{-10}$. Direct sampling would therefore require about $2.1\times10^9$
attempts, on average, to observe one accepted outcome with a logical error, whereas the exact calculation in this manuscript sums the noise distribution instead.

\subsection{Contribution of the higher-order terms}
The LERs for $d=3$ and $d=5$ have series
expansions in parameter $x=p/(1-p)$. Take the $d=5$ circuit's series and denote
its expansion to order $K$ by
\begin{equation}
S_K(p)=\sum_{k=3}^{K}L_kx^k.
\end{equation}
Although $\PL$ is small, its leading $x^3$ term is less than half of the LER at $p=10^{-3}$. The higher-order terms in the LER series expansion are not negligible, and their contributions can be compared with the exact LER.

Table~\ref{tab:convergence} compares $S_K(10^{-3})$ with the exact $d=5$ LER in Eq.~\eqref{eq:anchor}. The leading term alone is insufficient, but the truncated sums approach the exact fraction order by order.
\begin{table}[h]
\centering
\captionsetup{font=small,position=above,skip=4pt}
\caption{Truncated whole-circuit series at $p=10^{-3}$, compared with the exact rational value.}
\label{tab:convergence}
\footnotesize
\setlength{\tabcolsep}{3pt}
\begin{tabular}{@{}c|l|l@{}}
\toprule
\shortstack{Highest order\\$K$} & $S_K(10^{-3})$ & $(P_{\mathrm L}-S_K)/P_{\mathrm L}$ \\
\midrule
3 & $1.5352678660\times10^{-9}$ & $5.387\times10^{-1}$ \\
5 & $3.2950960775\times10^{-9}$ & $1.001\times10^{-2}$ \\
7 & $3.3283701142\times10^{-9}$ & $1.660\times10^{-5}$ \\
8 & $3.3284234413\times10^{-9}$ & $5.799\times10^{-7}$ \\
9 & $3.3284253094\times10^{-9}$ & $1.863\times10^{-8}$ \\
10 & $3.3284253696\times10^{-9}$ & $5.427\times10^{-10}$ \\
\bottomrule
\end{tabular}

\end{table}

Figure~\ref{fig:comparison} (right-most) shows the relative error of the various truncated $d=5$ cultivation circuit's LER series expansion compared against the exact fraction. See Appendix~\ref{app:arithmetic} for bounds on the contribution from the omitted terms.
Keeping terms through $x^{10}$ gives good agreement with the exact fraction to a small relative error below $\sim 10^{-7}$.

\subsection{Comparison with numerical simulations}
\begin{figure}[h]
\centering
\includegraphics[width=.98\linewidth]{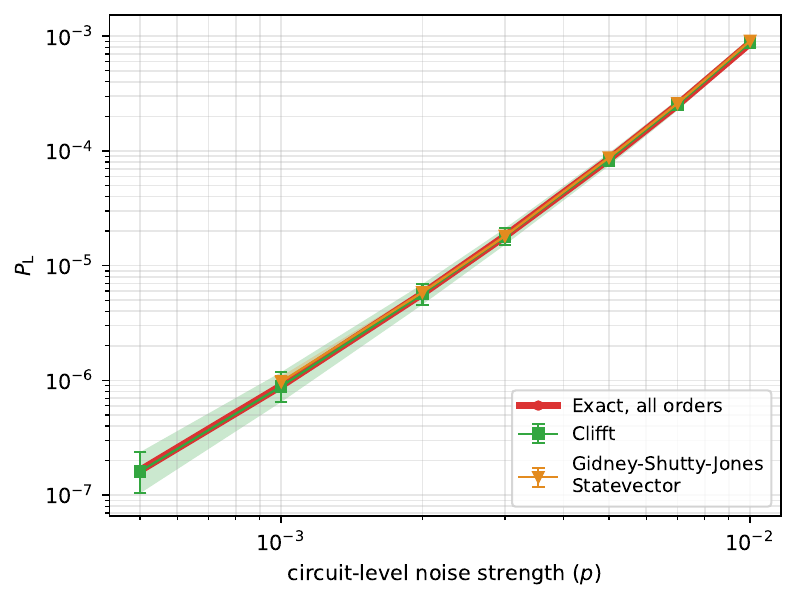}
\caption{$d=3$ cultivation: exact values compared with Clifft and Gidney--Shutty--Jones state-vector results. Interval conventions are given in Appendix~\ref{app:numerics}.}
\label{fig:d3comparison}
\end{figure}

The exact values and the tenth-order series lie inside the published Clifft and SOFT intervals at every shared noise strength (Fig.~\ref{fig:comparison}), and the acceptance agrees with SOFT's reported values~\cite{soft}. At $d=3$ the exact values lie inside the Clifft and Gidney--Shutty--Jones state-vector intervals (Fig.~\ref{fig:d3comparison}).

\section{Method}\label{sec:method}
Pauli propagation~\cite{ppframework} and tensor contraction~\cite{itensor} are central to our exact approach in deriving the LER. 
In short, Pauli propagation expresses the observables as sums of Pauli strings and tracks
their transformations under the operations. 
Firstly, we use Pauli propagation to calculate the effects of the \textit{incoming} error through the heavily non-Clifford gates dominated double-checking circuits. 
Then we encode the gates, noise and post-selection conditions in a tensor network, contract said tensor network to obtain $A(p)$ and $B(p)$ from Eqs.~\eqref{eq:probabilities}, which are then used to compute the LER.

 \begin{figure}[H]\centering
 \begin{adjustbox}{max width=.98\linewidth}\begin{tikzpicture}[x=1cm,y=1cm]
\tikzset{
 stagebox/.style={box,text width=64mm,inner sep=1.6mm},
 leafbox/.style={box,text width=29mm,inner sep=1.4mm,minimum height=9.5mm}}
\node[stagebox] (circ) at (0,0)
 {the noisy cultivation circuit, stage by stage\\ {\scriptsize\color{black!60}(Fig.~\ref{fig:overview})}};
\node[stagebox,below=3.2mm of circ] (pp)
 {Pauli propagation through a double-checking stage:\\
  response $F_A(E;p),F_B(E;p)$ to each incoming error $E$\\ {\scriptsize\color{black!60}(Sec.~\ref{sec:pp})}};
\node[stagebox,below=3.2mm of pp] (grp)
 {group the errors with equal response:\\ one small table per stage\\ {\scriptsize\color{black!60}(Sec.~\ref{sec:grouping})}};
\node[stagebox,below=3.2mm of grp] (tn)
 {tensor network of gates, noise and post-selection:\\ fills each table and joins the stages\\ {\scriptsize\color{black!60}(Sec.~\ref{sec:tensor-contraction})}};
\node[leafbox] (rat) at ($(tn.south)+(-1.75,-1.0)$)
 {rational $p$,\\ modular primes\\ {\scriptsize\color{black!60}(App.~\ref{app:F-fixedp})}};
\node[leafbox] (ser) at ($(tn.south)+(1.75,-1.0)$)
 {polynomials in $x$,\\ degrees $0,\ldots,10$\\ {\scriptsize\color{black!60}(App.~\ref{app:F-series})}};
\node[leafbox,below=3.2mm of rat] (ab)
 {exact $A(p),B(p)$\\ $\PL=B/A$\\ {\scriptsize\color{black!60}(Table~\ref{tab:grid})}};
\node[leafbox,below=3.2mm of ser] (ak)
 {exact $a_k,e_k$\\ series for $\PL$\\ {\scriptsize\color{black!60}(Eq.~\eqref{eq:fullseries})}};
\draw[flow] (circ)--(pp);
\draw[flow] (pp)--(grp);
\draw[flow] (grp)--(tn);
\draw[flow] (tn.south)--(rat.north);
\draw[flow] (tn.south)--(ser.north);
\draw[flow] (rat)--(ab);
\draw[flow] (ser)--(ak);
\end{tikzpicture}\end{adjustbox}
 \caption{Steps of the calculation. Pauli propagation gives the response of each double-checking stage to an incoming error, grouping keeps the tables small, and one tensor network joins the stages. The network is contracted in two arithmetics, giving exact fractions at fixed $p$ and exact series coefficients.}\label{fig:pipeline}\end{figure}
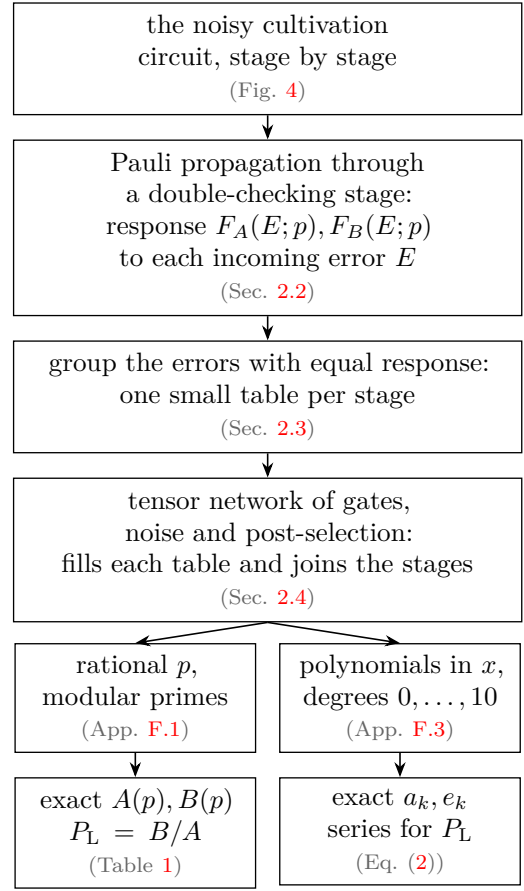
Figure~\ref{fig:pipeline} summarises the steps. Section~\ref{sec:pp} carries the check observables backwards through a double-checking stage, which gives the response of that stage to each incoming Pauli error. Section~\ref{sec:grouping} groups the errors that share a response, so that each stage becomes a small table. Section~\ref{sec:tensor-contraction} writes the gates, noise and post-selection as tensors, contracts them with the incoming labels open to fill each table, and joins the stages into $A(p)$ and $B(p)$. The same network is evaluated twice: with a rational $p$ in modular arithmetic, which gives the exact fractions, and with polynomials in $x=p/(1-p)$ truncated at degree ten, which gives the series coefficients.

\subsection{Method overview}
\begin{figure*}[tp]\centering
 \begin{adjustbox}{max width=.98\linewidth}\input{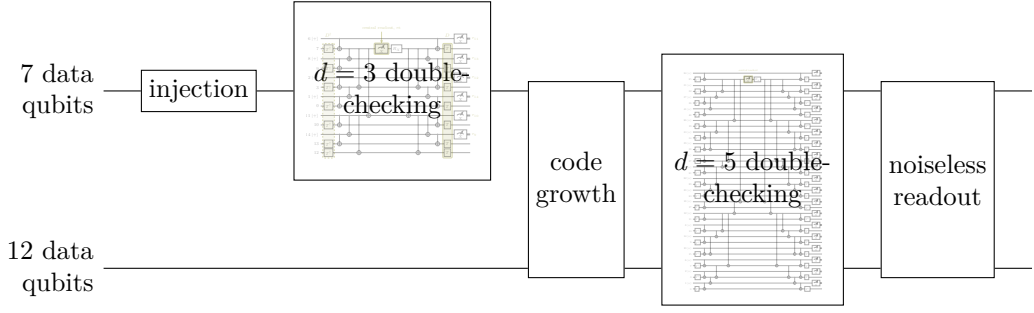}\end{adjustbox}
 \caption{Stages of the full $d=5$ cultivation circuit. Ancillary sub-systems to perform measurements are dropped for simplicity.}\label{fig:overview}\end{figure*}

The $d=5$ cultivation circuit has two non-Clifford checking stages separated by a purely Clifford (code growth) sub-circuit, see Fig.~\ref{fig:overview} for an overview of the $d=5$ cultivation circuit and its stages/sub-circuits. 
Fig.~\ref{fig:d3-incoming} shows one of the smaller double-checking sub-circuit and incoming Pauli error (see Appendix~\ref{app:circuits} for both double-checking circuits). 

 \begin{figure}[H]\centering
 \begin{adjustbox}{max width=.98\linewidth}\begingroup\color{black}
\begin{minipage}{8.4cm}\centering
{\small\textbf{(a)} Earlier stages supply the incoming error}\par\smallskip
\begin{tikzpicture}[font=\footnotesize]
\node[box,minimum width=22mm,minimum height=11mm] (prep) at (0,0)
 {Clifford\\preparation};
\node[box,minimum width=16mm,minimum height=11mm] (inj) at (3.3,0)
 {$T_3^\dagger$\\injection};
\node[box,minimum width=32mm,minimum height=11mm] (enc) at (0,-1.9)
 {Clifford encoding\\and measured parities};
\node[box,minimum width=29mm,minimum height=11mm] (dc) at (5.0,-1.9)
 {$d=3$ double checking\\expanded below};
\draw[flow] (prep)--(inj);
\draw[flow] (inj.south) -- ++(0,-0.4) -| (enc.north);
\draw[flow] (enc)--node[note,above] {incoming\\state}(dc);
\node[note] at (2.6,-2.85) {the earlier noise supplies the weights of $E$};
\end{tikzpicture}\par\medskip
{\small\textbf{(b)} One specified incoming error: $E=X_7$}\par\smallskip
\resizebox{8.4cm}{!}{\begin{tikzpicture}[font=\scriptsize,line width=.45pt,
  response arrow/.style={-{Stealth[length=1.6mm]},line width=.55pt}]
\begin{yquant}[register/minimum height=1.6mm,register/minimum depth=1.6mm,
  register/separation=1.1mm,operator/minimum width=3.6mm,operator/separation=1.7mm]
[name=inrow6] qubit {$6\;\ket{+}$} q6;
[name=inrow7] qubit {$7$} q7;
[name=inrow8] qubit {$8\;\ket{+}$} q8;
[name=inrow9] qubit {$9$} q9;
[name=inrow2] qubit {$2\;\ket{+}$} q2;
[name=inrow3] qubit {$3$} q3;
[name=inrow1] qubit {$1\;\ket{+}$} q1;
[name=inrow0] qubit {$0$} q0;
[name=inrow11] qubit {$11\;\ket{+}$} q11;
[name=inrow10] qubit {$10$} q10;
[name=inrow14] qubit {$14\;\ket{+}$} q14;
[name=inrow13] qubit {$13$} q13;
[name=inrow12] qubit {$12$} q12;
[name=incomingE,style={pauli,minimum size=7mm,font=\scriptsize,text=black}] box {$X$} q7;
align q6,q7,q8,q9,q2,q3,q1,q0,q11,q10,q14,q13,q12;
[name=inentry] box {$T^\dagger$} q0;
box {$T^\dagger$} q3,q7,q9,q10,q12,q13;
align q6,q7,q8,q9,q2,q3,q1,q0,q11,q10,q14,q13,q12;
[name=infoldfirst] cnot q0 | q1;
cnot q3 | q2;
cnot q7 | q6;
cnot q9 | q8;
cnot q10 | q11;
cnot q13 | q14;
align q6,q7,q8,q9,q2,q3,q1,q0,q11,q10,q14,q13,q12;
[name=v2foldtwo] cnot q1 | q3;
cnot q8 | q7;
cnot q14 | q11;
align q6,q7,q8,q9,q2,q3,q1,q0,q11,q10,q14,q13,q12;
[name=v2foldthree] cnot q3 | q7;
cnot q12 | q11;
align q6,q7,q8,q9,q2,q3,q1,q0,q11,q10,q14,q13,q12;
[name=infoldlast] cnot q11 | q7;
align q6,q7,q8,q9,q2,q3,q1,q0,q11,q10,q14,q13,q12;
[name=inhubmeasure,type=qubit] measure {$X$} q7;
box {$R_X$} q7;
align q6,q7,q8,q9,q2,q3,q1,q0,q11,q10,q14,q13,q12;
[name=inunfoldfirst] cnot q11 | q7;
align q6,q7,q8,q9,q2,q3,q1,q0,q11,q10,q14,q13,q12;
cnot q3 | q7;
cnot q12 | q11;
align q6,q7,q8,q9,q2,q3,q1,q0,q11,q10,q14,q13,q12;
cnot q1 | q3;
cnot q8 | q7;
cnot q14 | q11;
align q6,q7,q8,q9,q2,q3,q1,q0,q11,q10,q14,q13,q12;
[name=inunfoldlast] cnot q0 | q1;
cnot q3 | q2;
cnot q7 | q6;
cnot q9 | q8;
cnot q10 | q11;
cnot q13 | q14;
align q6,q7,q8,q9,q2,q3,q1,q0,q11,q10,q14,q13,q12;
[name=inexit] box {$T$} q0;
box {$T$} q3,q7,q9,q10,q12,q13;
align q6,q7,q8,q9,q2,q3,q1,q0,q11,q10,q14,q13,q12;
[name=infinalmeasure] measure {$X$} q14;
measure {$X$} q11,q6,q2,q8,q1;
\end{yquant}

\node[pauli,minimum size=7mm,font=\scriptsize,text=black,
      fill opacity=1,text opacity=1] at (incomingE.center) {$X$};

\coordinate (instage) at ($(incomingE.east)!.5!(inentry.west)$);
\coordinate (intop) at ($(inrow6.north)+(0,8mm)$);
\coordinate (inbottom) at ($(inrow12.south)+(0,-4mm)$);
\coordinate (incentre) at ($(instage)!.5!(infinalmeasure.east)$);
\draw[densely dashed,black!55] (instage |- intop)--(instage |- inbottom);
\node[anchor=south,align=center,font=\scriptsize,text=pppurple!65!black]
  (inputlabel) at ($(incomingE |- intop)+(0,3mm)$)
  {fixed input error\\$E=X_7$};
\draw[response arrow,densely dashed,draw=pppurple!65!black]
  (inputlabel.south)--(incomingE.north);
\node[anchor=south,font=\scriptsize,align=center]
  at ($(incentre |- intop)+(0,8mm)$)
  {$d=3$ double checking\\noise inside this stage is averaged};
\node[anchor=south] at (inentry |- intop) {$D^\dagger$};
\node[anchor=south] at (inexit |- intop) {$D$};
\node[anchor=south] at (infoldfirst |- intop) {$\mathtt{C}_1$};
\node[anchor=south] at (v2foldtwo |- intop) {$\mathtt{C}_2$};
\node[anchor=south] at (v2foldthree |- intop) {$\mathtt{C}_3$};
\node[anchor=south] at (infoldlast |- intop) {$\mathtt{C}_4$};
\draw[black!60] (inunfoldfirst |- intop)--(inunfoldlast |- intop)
  node[midway,above=2pt] {unfold};
\end{tikzpicture}}
\end{minipage}
\endgroup\end{adjustbox}
 \caption{The $d=3$ double-checking circuit and the earlier stages.
The dashed boundary marks this circuit's input and the purple $X_7$ specifies an `incoming error'
present.}\label{fig:d3-incoming}\end{figure}
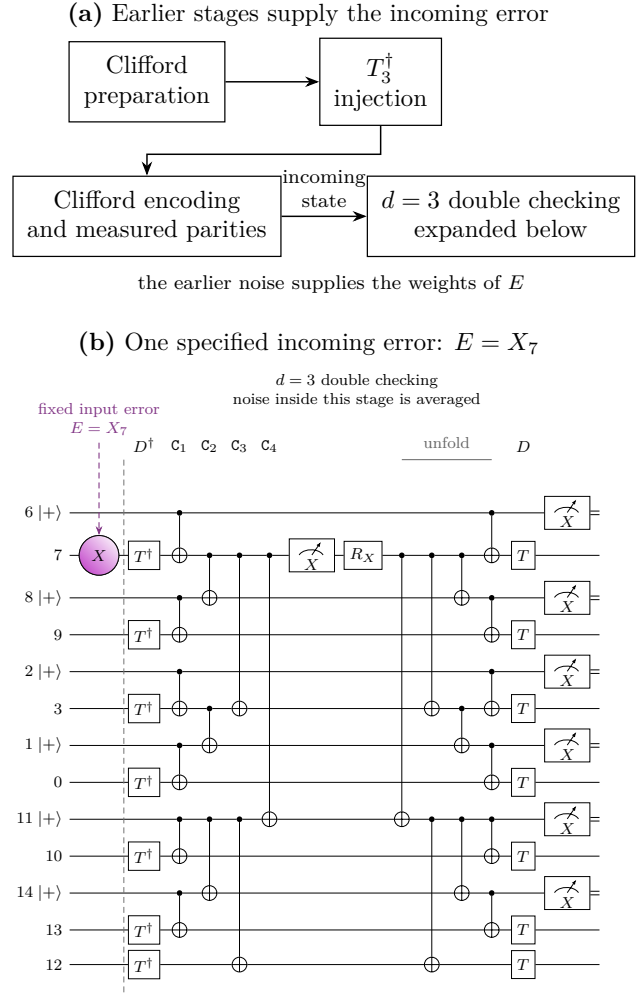

Apart from the single injection gate, the non-Clifford gates occur in the two double-checking stages, and everything between them is Clifford, although the Clifford growth stage still requires the exact average of Appendix~\ref{app:growth}.

Here, `incoming' means present at the entrance to the double-checking stage, before its first $T$/$T^\dagger$ layer, after injection and any earlier stages. For the $d=3$ stage the earlier stages are Clifford apart from the injected state, so their effect on the ideal encoded magic state is a Pauli error $E$ occurring with some probability. The same description holds for the $d=5$ stage, because within each syndrome block the $d=3$ check leaves the ideal state up to a logical Pauli, see Appendix~\ref{app:growth}. We therefore describe the input as a fixed $E$ applied to the ideal state. For example, $E=X_7$ places an $X$ error on data qubit 7, marked in Fig.~\ref{fig:d3-incoming}. The probability of $E$ comes from the earlier stages. The noise inside double checking is averaged separately.

In these double-checking sub-circuits, for each incoming Pauli error $E$, we calculate two numbers (probabilities):
\begin{itemize}
\item $F_A$: the probability of passing post-selection, i.e. all trivial detectors and the specified noiseless final code readout, and
\item $F_B$: the probability of passing post-selection with a logical error in the output state.
\end{itemize}
We call the pair $(F_A,F_B)$ the \emph{response} to that error $E$.
The incoming error is held fixed, while the noise within the double-checking stage is averaged. This separation is what makes the response reusable: one table serves every fault pattern of the earlier stages.

In the full $d=5$ circuit, we retain the probabilities of the
different error labels leaving the earlier $d=3$ double-checking stage.
We then include the noise and post-selection during code growth
to obtain the probabilities of errors entering the final $d=5$ double-checking stage.
We multiply each final $d=5$ double-checking response by the
probability that its incoming error occurs and no earlier
post-selection detector fires. Summing these contributions
gives $A(p)$ and $B(p)$.

We will illustrate Pauli propagation briefly before describing the full calculation next.

\subsection{Pauli propagation}\label{sec:pp}

Pauli propagation~\cite{ppframework} computes an expectation value by carrying the measured observable backwards through the circuit and evaluating it on the input state, so a readout probability becomes a weighted sum of Pauli expectation values. For cultivation, Clifford gates do not increase the number of terms, and the structure of the double-checking stages lets us merge equivalent terms and sum over the noise directly. Rare logical errors therefore never need to be sampled: the LER is computed exactly, or as a series.
{\color{black}For a ZX-calculus description using Pauli semiwebs~\cite{zxflow}, see the discussion and Appendix~\ref{app:semiwebs}.}

As an example, we start by using $\rho$ as the input state to a circuit, $U$ the circuit unitary and
$O$ the observable to be measured. For a unitary circuit,
\begin{equation}
 \begin{gathered}
 \langle O\rangle=\Tr(U\rho U^\dagger O)
 =\Tr(\rho\,U^\dagger O U),\\
 U^\dagger O U=\sum_P c_P P.
 \end{gathered}
 \label{eq:heisenberg}
\end{equation}
The sum runs over tensor products of Pauli operators: $I,X,Y,Z$. A Clifford gate transforms
a Pauli string to another, possibly with a $\pm$ sign. A non-Clifford $T=\text{diag}(1,e^{i\pi/4})$ gate can
split one string into two `paths':
\begin{equation}
 \begin{gathered}
 T^\dagger X T=\frac{X{\color{red}-}Y}{\sqrt2},\quad
 T^\dagger Y T=\frac{X{\color{blue}+}Y}{\sqrt2},\\
 T=\operatorname{diag}(1,e^{i\pi/4}) \ .
 \end{gathered}
 \label{eq:T}
\end{equation}
The {\color{blue}blue ($+$)} and {\color{red}red ($-$)} signs in
Eq.~\eqref{eq:T} belong to the coefficients multiplying the Pauli
operators. Both terms are kept with their signs with nothing sampled. Each propagation path contributes the product
of its coefficients. Contributions to the same Pauli string are added
with their signs, so they can cancel.

Fig.~\ref{fig:pauli} follows an observable backwards through an actual
portion of the $d=3$ double-checking circuit, a single CNOT with circuit-level noise. Starting at $X_{13}$ on the
right to the left. The $T$ gate forks two
terms $X_{13}$ and $Y_{13}$, with coefficients ${\color{blue}+1/\sqrt2}$ and
${\color{red}-1/\sqrt2}$ respecitvely. The CNOT leaves $X_{13}$ unchanged and turns $Y_{13}$
into $Z_{14}Y_{13}$.

 \begin{figure}[h]\centering
 \begin{adjustbox}{max width=.98\linewidth}\begin{minipage}{7.9cm}
\centering
{\small\bfseries (a) A portion of the $d=3$ double-checking circuit}\par\smallskip
\begin{adjustbox}{width=.93\linewidth}
\begin{tikzpicture}[font=\small]
\begin{yquant}[operator/separation=7mm,operator/minimum width=7mm,
  register/separation=4mm]
qubit {$14$} q14;
qubit {$13$} q13;
cnot q13 | q14;
box {$\mathcal N_2$} (q14,q13);
box {$T$} q13;
align q14,q13;
box {$\mathcal N_1$} q14,q13;
output {$X_{13}$} q13;
\end{yquant}
\end{tikzpicture}
\end{adjustbox}\par\smallskip
{\small The circuit runs left to right; the observable starts at the right.}
\par\medskip
{\small\bfseries (b) Follow the observable backwards, from right to left}\par
\begin{tikzpicture}[x=1cm,y=1cm,font=\footnotesize,
  term/.style={pauli,minimum size=11mm,inner sep=0pt,font=\footnotesize},
  start term/.style={initial pauli,minimum size=11mm,inner sep=0pt,font=\footnotesize},
  weight/.style={font=\footnotesize,fill=white,inner sep=.5pt},
  stage/.style={align=center,font=\footnotesize},
  step arrow/.style={-{Stealth[length=1.6mm]},line width=.6pt}]
\node[start term] (o) at (7.35,0) {$X_{13}$};
\node[term] (n1) at (5.65,0) {$X_{13}$};
\node[term] (tx) at (3.95,.8) {$X_{13}$};
\node[term] (ty) at (3.95,-.8) {$Y_{13}$};
\node[term] (nx) at (2.25,.8) {$X_{13}$};
\node[term] (ny) at (2.25,-.8) {$Y_{13}$};
\node[term] (cx) at (.55,.8) {$X_{13}$};
\node[term] (cy) at (.55,-.8) {$Z_{14}Y_{13}$};
\node[stage] at (6.5,1.9) {through\\$\mathcal N_1$};
\node[stage] at (4.8,1.9) {through\\$T_{13}$};
\node[stage] at (3.1,1.9) {through\\$\mathcal N_2$};
\node[stage] at (1.4,1.9) {through\\CNOT};
\draw[step arrow] (o)--node[weight,above=2pt] {$\lambda_1$}(n1);
\draw[step arrow,blue!75!black] (n1)--node[weight,fill=none,above=8pt,sloped] {$+1/\sqrt2$}(tx);
\draw[step arrow,red!80!black] (n1)--node[weight,fill=none,below=8pt,sloped] {$-1/\sqrt2$}(ty);
\draw[step arrow] (tx)--node[weight,above=2pt] {$\lambda_2$}(nx);
\draw[step arrow] (ty)--node[weight,below=2pt] {$\lambda_2$}(ny);
\draw[step arrow] (nx)--node[weight,above=2pt] {$1$}(cx);
\draw[step arrow] (ny)--node[weight,below=2pt] {$1$}(cy);
\node[align=center,text width=7.5cm,font=\small] at (3.95,-2.05)
  {Multiply the weights along each path, then add the resulting Pauli terms.};
\node[draw,inner sep=3mm,font=\normalsize] at (3.95,-4.05)
  {$\displaystyle X_{13}\longmapsto
    \frac{\lambda_1\lambda_2}{\sqrt2}
    \bigl(X_{13}{\color{red}-}Z_{14}Y_{13}\bigr)$};
\end{tikzpicture}
\end{minipage}\end{adjustbox}
 \caption{A $d=3$ double-checking circuit portion and its backward Pauli propagation.
The arrows carry coefficients.}\label{fig:pauli}\end{figure}

\begingroup\color{black}
\begin{subequations}\label{eq:depolarising-channels}
The one-qubit depolarising channel is
\begin{equation}
\begin{aligned}
\mathcal N_1(\rho)
&=(1-p)\rho\\
&\quad+\frac{p}{3}(X\rho X+Y\rho Y+Z\rho Z).
\end{aligned}
\label{eq:depolarising-one}
\end{equation}
The two-qubit depolarising channel is
\begin{equation}
\begin{aligned}
\mathcal N_2(\rho)
&=(1-p)\rho\\
&\quad+\frac{p}{15}
\sum_{\substack{P,Q\in\{I,X,Y,Z\}\\(P,Q)\neq(I,I)}}
(P\otimes Q)\rho(P\otimes Q).
\end{aligned}
\label{eq:depolarising-two}
\end{equation}
\end{subequations}
\par\endgroup
Their physical Pauli faults have positive
probabilities, which we sum exactly. For example, 
\begin{equation}
 \mathcal N_1^\dagger(X)
 =(1-p)X+\frac p3(X-X-X)
 =\lambda_1X,
\end{equation}
where $\lambda_1=1-4p/3$. The dagger here means the adjoint channel, which propagates
the observable backwards through the noise operators~\cite{ppframework}. Similarly,
$\lambda_2=1-16p/15$ for a non-identity two-qubit Pauli operator\footnote{Explicitly, for $P,Q\in\{I,X,Y,Z\}$ and $(P,Q)\neq(I,I)$,
$\mathcal N_2^\dagger(P\otimes Q) =\lambda_2(P\otimes Q)$, where $\lambda_2=1-\frac{16p}{15}$.}. Following all four steps in Fig.~\ref{fig:pauli} gives
\begin{equation}
 X_{13}\longmapsto\frac{\lambda_1\lambda_2}{\sqrt2}
       (X_{13}-Z_{14}Y_{13}).
 \label{eq:localexample}
\end{equation}
To obtain a scalar, one can evaluate the resultant Pauli terms on an input
state and add their weighted contributions. For a Pauli measurement readout with outcome parities $\pm 1$,
the mean outcome determines the two probabilities:
\begin{equation}
 \Pr(O=\pm1)=\frac{1\pm\langle O\rangle}{2}.
 \label{eq:pauli-probability}
\end{equation}
For illustration, one can use input $\ket0_{14}\ket+_{13}$;
$\langle X_{13}\rangle=1$ and $\langle Z_{14}Y_{13}\rangle=0$.
The output mean is therefore $\lambda_1\lambda_2/\sqrt2$.

\subsubsection{Pauli propagation on the double-checking circuit}

We now follow the observable through the whole $d=3$ check of Fig.~\ref{fig:d3-incoming} at $p=0$, to show how the 128 terms produced by the entry layer collapse to a single logical observable. This collapse is what keeps the response tables small. Consider the central $X$-basis measurement on qubit $7$. 
Propagating $X_7$ backwards through the four CNOT layers $\mathtt{C}_4,\mathtt{C}_3,\mathtt{C}_2,\mathtt{C}_1$ gives the strings in
Fig.~\ref{fig:d3-full-propagation}(a). Every step still returns one Pauli string, and the CNOTs spread its support without introducing a sum.

Write $\mathcal D=\{0,3,7,9,10,12,13\}$ for the data qubits.
At the entry layer $D^\dagger=\prod_{q\in\mathcal D}T_q^\dagger$,
backward propagation uses $TXT^\dagger=(X+Y)/\sqrt2$.
The four ancillary $X$ factors have expectation $+1$ on their prepared
$\ket{+}$ states. The operator left on the data is:
\begin{equation}
 R=2^{-7/2}\prod_{q\in\mathcal D}(X_q+Y_q) \ .
 \label{eq:v2-central-operator}
\end{equation}
This product contains 128 Pauli strings. On a codespace input, 112 of the 128 Pauli strings flip one or more code stabiliser
and hence have zero expectation on the ideal encoded state.
Of the remaining 16 Pauli strings, eight act as $X_L$ and eight as $-Y_L$.
Their coefficients combine, forming the logical observable: $M_L$,
\begin{equation}
 \Pi_0R\Pi_0=M_L\Pi_0,\qquad M_L=\frac{X_L-Y_L}{\sqrt2} \ .
 \label{eq:v2-central-logical}
\end{equation}
The ideal state is then:
$\ket{T_L^\dagger}=(\ket{0_L}+e^{-i\pi/4}\ket{1_L})/\sqrt2$,
which has $\langle M_L\rangle=1$\footnote{Pauli operators: $X_L=\prod_{q\in\mathcal D}X_q$, $Z_L=\prod_{q\in\mathcal D}Z_q$,
and $Y_L=iX_LZ_L$.}.

 \begin{figure}[htbp]\centering
 \begin{adjustbox}{max width=.98\linewidth}\begingroup\color{black}
\begin{tikzpicture}[font=\scriptsize,text=black,
  pp term/.style={pauli,minimum size=6.5mm,inner sep=0pt},
  pp first/.style={initial pauli,minimum size=6.5mm,inner sep=0pt},
  pp string/.style={anchor=west,font=\small,align=left,inner sep=0pt}]
\node[note,font=\small] at (3.05,.8)
 {\textbf{(a)} Backwards through the CNOT layers};
\node[pp first] (p0) at (0,0) {};
\node[pp term] (p1) at (0,-1.02) {};
\node[pp term] (p2) at (0,-2.04) {};
\node[pp term] (p3) at (0,-3.06) {};
\node[pp term] (p4) at (0,-4.2) {};
\node[pp string] at (.65,0) {$X_7$};
\node[pp string] at (.65,-1.02) {$X_7X_{11}$};
\node[pp string] at (.65,-2.04) {$X_3X_7X_{11}X_{12}$};
\node[pp string] at (.65,-3.06) {$X_1X_3X_7X_8X_{11}X_{12}X_{14}$};
\node[pp string] at (.65,-4.2)
 {$\begin{aligned}&X_0X_1X_3X_7X_8X_9\\[-1pt]
                  &\qquad X_{10}X_{11}X_{12}X_{13}X_{14}\end{aligned}$};
\foreach \a/\b/\gate in {p0/p1/4,p1/p2/3,p2/p3/2,p3/p4/1}{
 \draw[pauli edge] (\a)--node[note,left=2pt] {$\mathtt{C}_{\gate}$}(\b);
}
\node[note,anchor=north east,align=right] at (6.2,.2)
 {every arrow has\\coefficient $+1$};

\node[note,font=\small] at (3.05,-5.6)
 {\textbf{(b)} Entry layer and code projection};
\node[note,font=\small] at (3.05,-6.4)
 {$2^{-7/2}\displaystyle\prod_{q\in\mathcal D}(X_q+Y_q)$};
\node[pp term] (physical) at (3.05,-7.15) {};
\node[note,anchor=west] at (3.6,-7.15) {128 Pauli strings};
\node[box,minimum width=30mm,minimum height=10mm,font=\scriptsize]
 (project) at (3.05,-8.3) {restrict to the code\\and add signed terms};
\draw[flow] (physical)--(project);
\node[pp term,minimum size=10mm] (xl) at (1.55,-10.15) {$X_L$};
\node[pp term,minimum size=10mm] (yl) at (4.55,-10.15) {$Y_L$};
\draw[pauli edge,blue!75!black] (project.south)--(xl);
\draw[pauli edge,red!80!black] (project.south)--(yl);
\node[note,text=blue!75!black] at (1.1,-9.2) {$+1/\sqrt2$};
\node[note,text=red!80!black] at (5,-9.2) {$-1/\sqrt2$};
\node[note] at (3.05,-11.05)
 {sixteen terms survive:\\eight give $X_L$, eight give $-Y_L$};
\end{tikzpicture}
\endgroup\end{adjustbox}
 \caption{(a) The central $X_7$ propagated backwards, with each string
beside its node. (b) The entry layer and code projection reduce
128 terms to two logical terms.}\label{fig:d3-full-propagation}\end{figure}

The incoming error changes the state on which the propagated operator
is evaluated. For a specified $E$, the central readout probability is
\begin{equation}
 \Pr(m_7=0\mid E)
 =\frac{1+\langle T_L^\dagger|E^\dagger R E|T_L^\dagger\rangle}{2}.
 \label{eq:v2-incoming-mean}
\end{equation}
Substituting $E=I$ gives probability 1, as expected. For $E=X_7$ the conjugation turns $(X_7+Y_7)$ in Eq.~\eqref{eq:v2-central-operator} into $(X_7-Y_7)$ and the probability drops to $1/2$. This is only the readout of the central measurement on qubit 7 in Fig.~\ref{fig:d3-incoming}, however: even when it passes, the six auxiliary measurements and the final code projection detect the syndrome of $X_7$, so with no further faults the check rejects and $F_A(X_7;0)=0$. For $p>0$ a second fault can cancel that syndrome and let $X_7$ pass, which is why every incoming $E$, rejected or not at $p=0$, has an entry $F_A(E;p)$, $F_B(E;p)$ in the table.

With noise, the same backward propagation is averaged over every fault pattern inside the check, keeping only the measurement records with trivial detector parities (which means post-selection is successful). Call the resulting
unnormalised output state $\sigma_E$. Its trace equals the probability of obtaining measurement records
with trivial detector parities, for a fixed incoming error $E$ in either of the two double-checking subcircuits.
The noiseless final readout uses the projector $\Pi_0$ onto the
accepted code space and $\Pi_{\rm bad}$ onto the states within
that code space with the wrong logical outcome. A projector's expectation is the probability
of its event. Thus
\begin{equation}
 \begin{aligned}
 F_A(E;p)&=\Tr(\Pi_0\sigma_E),\\
 F_B(E;p)&=\Tr(\Pi_{\rm bad}\sigma_E).
 \end{aligned}
 \label{eq:response-projectors}
\end{equation}
We evaluate these traces by propagating the projectors backwards. The state $\sigma_E$ need not be stored.
Their Pauli expansions can have negative coefficients, but the final
traces are probabilities. See Appendix~\ref{app:circuits} for these projectors.

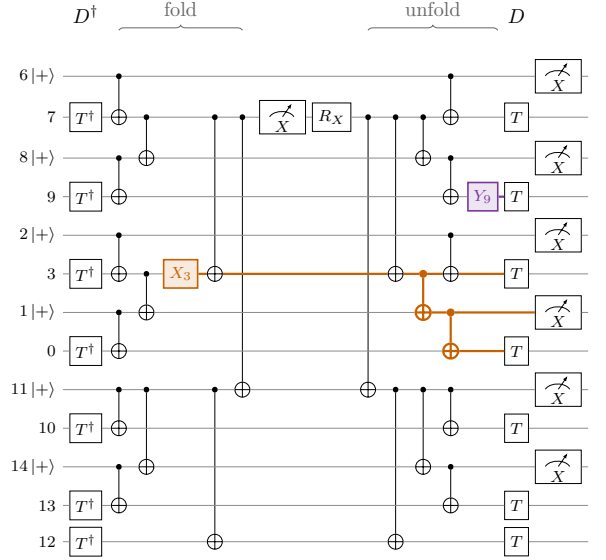
\begin{figure}[htbp]\centering
 \begin{adjustbox}{max width=.98\linewidth}\begingroup
\usetikzlibrary{decorations.pathreplacing}
\definecolor{faultorange}{RGB}{199,97,0}
\definecolor{faultpurple}{RGB}{123,55,156}
\begin{tikzpicture}[font=\scriptsize,line width=.45pt]
\begin{yquant}[register/minimum height=1.7mm,register/minimum depth=1.7mm,register/separation=1.3mm,operator/minimum width=3.6mm,operator/separation=1.2mm,every wire/.style={draw=black!45}]
[name=row6] qubit {$6\,\ket{+}$} q6;
[name=row7] qubit {$7$} q7;
[name=row8] qubit {$8\,\ket{+}$} q8;
[name=row9] qubit {$9$} q9;
[name=row2] qubit {$2\,\ket{+}$} q2;
[name=row3] qubit {$3$} q3;
[name=row1] qubit {$1\,\ket{+}$} q1;
[name=row0] qubit {$0$} q0;
[name=row11] qubit {$11\,\ket{+}$} q11;
[name=row10] qubit {$10$} q10;
[name=row14] qubit {$14\,\ket{+}$} q14;
[name=row13] qubit {$13$} q13;
[name=row12] qubit {$12$} q12;
[name=entry] box {$T^\dagger$} q0;
box {$T^\dagger$} q3;
box {$T^\dagger$} q7;
box {$T^\dagger$} q9;
box {$T^\dagger$} q10;
box {$T^\dagger$} q12;
box {$T^\dagger$} q13;
align q6,q7,q8,q9,q2,q3,q1,q0,q11,q10,q14,q13,q12;
[name=g138n0] cnot q0 | q1;
[name=g138n1] cnot q3 | q2;
[name=g138n2] cnot q7 | q6;
[name=g138n3] cnot q9 | q8;
[name=g138n4] cnot q10 | q11;
[name=g138n5] cnot q13 | q14;
align q6,q7,q8,q9,q2,q3,q1,q0,q11,q10,q14,q13,q12;
[name=g142n0] cnot q1 | q3;
[name=g142n1] cnot q8 | q7;
[name=g142n2] cnot q14 | q11;
align q6,q7,q8,q9,q2,q3,q1,q0,q11,q10,q14,q13,q12;
[name=F1,draw=faultorange,fill=faultorange!13,text=faultorange,line width=.8pt] box {$X_3$} q3;
setstyle {draw=faultorange,line width=1.1pt} q3;
align q6,q7,q8,q9,q2,q3,q1,q0,q11,q10,q14,q13,q12;
align q6,q7,q8,q9,q2,q3,q1,q0,q11,q10,q14,q13,q12;
[name=g146n0] cnot q3 | q7;
[name=g146n1] cnot q12 | q11;
align q6,q7,q8,q9,q2,q3,q1,q0,q11,q10,q14,q13,q12;
[name=g150n0] cnot q11 | q7;
align q6,q7,q8,q9,q2,q3,q1,q0,q11,q10,q14,q13,q12;
[name=hubmeasure,type=qubit] measure {$X$} q7;
[name=hubreset] box {$R_X$} q7;
align q6,q7,q8,q9,q2,q3,q1,q0,q11,q10,q14,q13,q12;
align q6,q7,q8,q9,q2,q3,q1,q0,q11,q10,q14,q13,q12;
[name=g161n0] cnot q11 | q7;
align q6,q7,q8,q9,q2,q3,q1,q0,q11,q10,q14,q13,q12;
[name=g165n0] cnot q3 | q7;
[name=g165n1] cnot q12 | q11;
align q6,q7,q8,q9,q2,q3,q1,q0,q11,q10,q14,q13,q12;
[name=g169n0,style={color=faultorange,draw=faultorange,line width=1.1pt},control style={fill=faultorange}] cnot q1 | q3;
setstyle {draw=faultorange,line width=1.1pt} q1;
[name=g169n1] cnot q8 | q7;
[name=g169n2] cnot q14 | q11;
align q6,q7,q8,q9,q2,q3,q1,q0,q11,q10,q14,q13,q12;
[name=g173n0,style={color=faultorange,draw=faultorange,line width=1.1pt},control style={fill=faultorange}] cnot q0 | q1;
setstyle {draw=faultorange,line width=1.1pt} q0;
[name=g173n1] cnot q3 | q2;
[name=g173n2] cnot q7 | q6;
[name=g173n3] cnot q9 | q8;
[name=g173n4] cnot q10 | q11;
[name=g173n5] cnot q13 | q14;
align q6,q7,q8,q9,q2,q3,q1,q0,q11,q10,q14,q13,q12;
[name=F2,draw=faultpurple,fill=faultpurple!13,text=faultpurple,line width=.8pt] box {$Y_9$} q9;
setstyle {draw=faultpurple,line width=1.1pt} q9;
align q6,q7,q8,q9,q2,q3,q1,q0,q11,q10,q14,q13,q12;
[name=exit] box {$T$} q0;
setstyle {draw=black!45,line width=.45pt} q0;
box {$T$} q3;
setstyle {draw=black!45,line width=.45pt} q3;
box {$T$} q7;
box {$T$} q9;
setstyle {draw=black!45,line width=.45pt} q9;
box {$T$} q10;
box {$T$} q12;
box {$T$} q13;
align q6,q7,q8,q9,q2,q3,q1,q0,q11,q10,q14,q13,q12;
[type=qubit] measure {$X$} q14;
[type=qubit] measure {$X$} q11;
[type=qubit] measure {$X$} q6;
[type=qubit] measure {$X$} q2;
[type=qubit] measure {$X$} q8;
[type=qubit] measure {$X$} q1;
setstyle {draw=black!45,line width=.45pt} q1;
\end{yquant}
\coordinate (heading) at ($(row6)+(0,8mm)$);
\node[anchor=south,font=\small] at (entry |- heading) {$D^\dagger$};
\node[anchor=south,font=\small] at (exit |- heading) {$D$};
\draw[black!60,decorate,decoration={brace,amplitude=2.5pt}] (g138n0 |- heading) -- (g150n0 |- heading) node[midway,above=3pt,font=\small] {fold};
\draw[black!60,decorate,decoration={brace,amplitude=2.5pt}] (g161n0 |- heading) -- (F2.east |- heading) node[midway,above=3pt,font=\small] {unfold};
\end{tikzpicture}

\endgroup\end{adjustbox}
 \caption{A fixed two-fault pattern in the $d=3$ double-checking circuit.
Coloured boxes are $X_3$ and $Y_9$ errors with their propagation traced.}\label{fig:d3-fixed-faults}\end{figure}

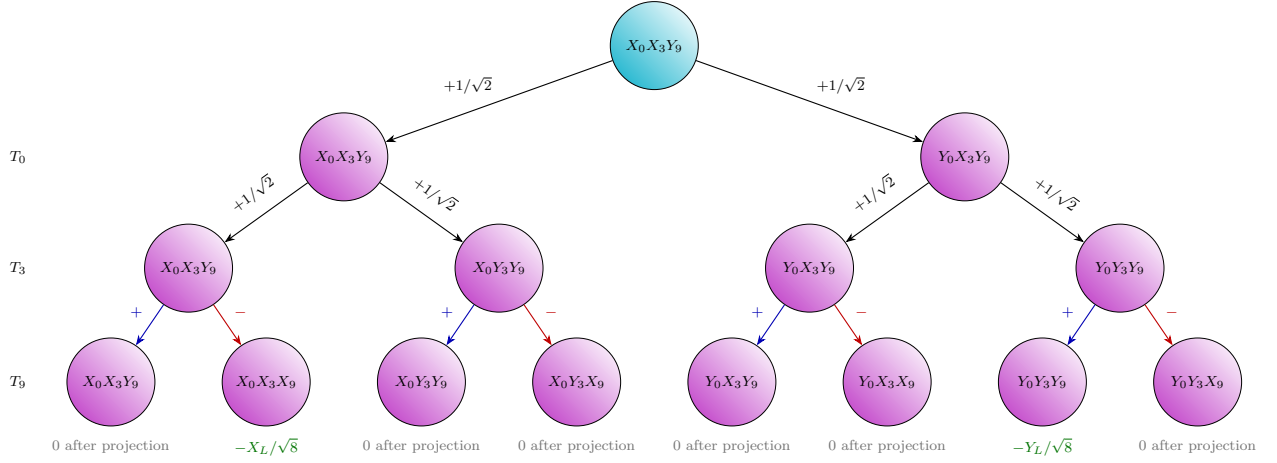
\begin{figure*}[tp]\centering
 \begin{adjustbox}{max width=.98\linewidth}\begin{tikzpicture}[x=1cm,y=1cm,text=black,
pauli/.append style={minimum size=17mm,font=\scriptsize,inner sep=0pt},
label/.style={note,font=\scriptsize}]
\node[initial pauli,minimum size=17mm,font=\scriptsize] (r) at (10.5,0) {$X_0X_3Y_9$};
\node[pauli] (a) at (4.5,-2.15) {$X_0X_3Y_9$};
\node[pauli] (b) at (16.5,-2.15) {$Y_0X_3Y_9$};
\draw[flow] (r)--node[above left,label] {$+1/\sqrt2$}(a);
\draw[flow] (r)--node[above right,label] {$+1/\sqrt2$}(b);
\foreach \name/\x/\txt in {c/1.5/X_0X_3Y_9,d/7.5/X_0Y_3Y_9,e/13.5/Y_0X_3Y_9,f/19.5/Y_0Y_3Y_9}
 {\node[pauli] (\name) at (\x,-4.3) {$\txt$};}
\foreach \from/\to in {a/c,a/d,b/e,b/f}
 {\draw[flow] (\from)--node[midway,sloped,above=4pt,label] {$+1/\sqrt2$}(\to);}
\foreach \name/\x/\txt in {g/0/X_0X_3Y_9,h/3/X_0X_3X_9,i/6/X_0Y_3Y_9,j/9/X_0Y_3X_9,k/12/Y_0X_3Y_9,l/15/Y_0X_3X_9,m/18/Y_0Y_3Y_9,n/21/Y_0Y_3X_9}
 {\node[pauli] (\name) at (\x,-6.5) {$\txt$};}
\foreach \from/\to in {c/g,d/i,e/k,f/m}
 {\draw[flow,blue!70!black] (\from)--node[above left,label] {$+$}(\to);}
\foreach \from/\to in {c/h,d/j,e/l,f/n}
 {\draw[flow,red!75!black] (\from)--node[above right,label] {$-$}(\to);}
\foreach \name in {g,i,j,k,l,n}
 {\node[label,text=black!55] at ($(\name)+(0,-1.25)$) {$0$ after projection};}
\node[label,text=green!45!black] at ($(h)+(0,-1.25)$) {$-X_L/\sqrt8$};
\node[label,text=green!45!black] at ($(m)+(0,-1.25)$) {$-Y_L/\sqrt8$};
\node[label,anchor=east] at (-1.5,-2.15) {$T_0$};
\node[label,anchor=east] at (-1.5,-4.3) {$T_3$};
\node[label,anchor=east] at (-1.5,-6.5) {$T_9$};
\end{tikzpicture}\end{adjustbox}
 \caption{All eight signed terms for a fixed two-fault example.
Every edge contributes $1/\sqrt2$ with its indicated sign. The two surviving terms give acceptance $1/4$ and a logical error.}\label{fig:d3-fault-expansion}\end{figure*}

So far the input carried the error and the check was noiseless. We now reverse the roles: an ideal input, and a fixed pattern of faults inside the check, $X_3$ in the `fold' layers and $Y_9$ in the `unfold' layer (Fig.~\ref{fig:d3-fixed-faults}), chosen because it is the smallest pattern that passes post-selection with a logical error.
The first fault propagates to $X_0X_1X_3$, its ancillary $X_1$ does not
flip the final $X$ readout. The data error before the exit layer is
$Q=X_0X_3Y_9$. Forward propagation through that layer gives
\begin{equation}
 DQD^\dagger=\frac{(X_0+Y_0)(X_3+Y_3)(Y_9-X_9)}{\sqrt8}.
 \label{eq:v2-fixed-fault}
\end{equation}
Fig.~\ref{fig:d3-fault-expansion} shows all eight terms.
The code projection removes six. The surviving operator is
$-(X_L+Y_L)/\sqrt8=-N_L/2$, where
$N_L=(X_L+Y_L)/\sqrt2$ anticommutes with $M_L$.
The accepted state is therefore logically wrong: two faults lead to an accepted logical error, hence the fault distance is $d_\text{fault}=2$. 
A ZX-calculus contraction of the noiseless circuit with these two faults inserted~\cite{wan} confirms it. No detector fires with probability $1/4$, and every accepted output carries the logical error.

\subsection{Grouping equivalent Pauli errors}\label{sec:grouping}
\definecolor{tnblue}{RGB}{0,118,186}
\definecolor{tnorange}{RGB}{181,23,0}
\definecolor{tngreen}{RGB}{1,113,0}
\tikzset{
  tn tensor/.style={rectangle,rounded corners=2pt,draw=black,
    fill=tnblue,text=white,line width=1.1pt,align=center,inner sep=2mm,
    minimum height=9mm,font=\small},
  tn endpoint/.style={tn tensor},
  tn noise/.style={tn tensor,fill=tnorange},
  tn intermediate/.style={tn tensor,fill=tngreen},
  tn constraint/.style={tn tensor,fill=black!16,text=black},
  tn leg/.style={line width=.6pt,draw=black},
  tn bundle/.style={tn leg,double,double distance=.6pt},
  tn copy/.style={circle,fill=black,draw=none,inner sep=0pt,minimum size=2.4pt},
  tn operation/.style={-{Stealth[length=1.8mm]},line width=.6pt,draw=black!65},
  tn label/.style={font=\scriptsize,align=center,inner sep=.8mm},
  tn heading/.style={font=\small,align=center}
}

The injected and then cultivated magic state is encoded in a triangular colour code.
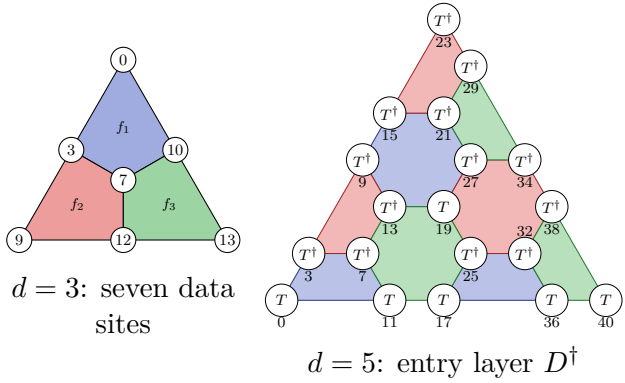
\begin{figure}%
\centering
\begin{minipage}[c]{.38\linewidth}\centering
\begin{adjustbox}{max width=\linewidth}\begin{tikzpicture}[x=1cm,y=1cm,line width=.6pt]
\definecolor{dthreeblue}{RGB}{58,80,190}
\definecolor{dthreered}{RGB}{219,51,51}
\definecolor{dthreegreen}{RGB}{51,166,64}
\coordinate (q0) at (0,3.6);
\coordinate (q3) at (-1.0392,1.8);
\coordinate (q7) at (0,1.2);
\coordinate (q9) at (-2.0784,0);
\coordinate (q10) at (1.0392,1.8);
\coordinate (q12) at (0,0);
\coordinate (q13) at (2.0784,0);
\filldraw[fill=dthreeblue!48] (q0)--(q3)--(q7)--(q10)--cycle;
\filldraw[fill=dthreered!48] (q3)--(q9)--(q12)--(q7)--cycle;
\filldraw[fill=dthreegreen!48] (q7)--(q12)--(q13)--(q10)--cycle;
\node[note] at (0,2.25) {$f_1$};
\node[note] at (-.92,.70) {$f_2$};
\node[note] at (.92,.70) {$f_3$};
\foreach \q in {0,3,7,9,10,12,13}
  \node[circle,draw,fill=white,inner sep=0pt,minimum size=5mm,
        font=\footnotesize] at (q\q) {$\q$};
\end{tikzpicture}\end{adjustbox}
\par\smallskip $d=3$: seven data sites
\end{minipage}\hfill
\begin{minipage}[c]{.58\linewidth}\centering
\begin{adjustbox}{max width=\linewidth}\begin{tikzpicture}[x=.5cm,y=.866025403784cm]
\definecolor{codered}{HTML}{DB3333}
\definecolor{codegreen}{HTML}{33A640}
\definecolor{codeblue}{HTML}{3A50BE}
\coordinate (d5q0) at (0,0);
\coordinate (d5q3) at (1,1);
\coordinate (d5q7) at (3,1);
\coordinate (d5q9) at (3,3);
\coordinate (d5q11) at (4,0);
\coordinate (d5q13) at (4,2);
\coordinate (d5q15) at (4,4);
\coordinate (d5q17) at (6,0);
\coordinate (d5q19) at (6,2);
\coordinate (d5q21) at (6,4);
\coordinate (d5q23) at (6,6);
\coordinate (d5q25) at (7,1);
\coordinate (d5q27) at (7,3);
\coordinate (d5q29) at (7,5);
\coordinate (d5q32) at (9,1);
\coordinate (d5q34) at (9,3);
\coordinate (d5q36) at (10,0);
\coordinate (d5q38) at (10,2);
\coordinate (d5q40) at (12,0);
\filldraw[fill=codeblue!35,draw=codeblue!75!black,line width=.55pt] (d5q0) -- (d5q11) -- (d5q7) -- (d5q3) -- cycle;
\filldraw[fill=codeblue!35,draw=codeblue!75!black,line width=.55pt] (d5q13) -- (d5q19) -- (d5q27) -- (d5q21) -- (d5q15) -- (d5q9) -- cycle;
\filldraw[fill=codeblue!35,draw=codeblue!75!black,line width=.55pt] (d5q17) -- (d5q36) -- (d5q32) -- (d5q25) -- cycle;
\filldraw[fill=codered!35,draw=codered!75!black,line width=.55pt] (d5q3) -- (d5q7) -- (d5q13) -- (d5q9) -- cycle;
\filldraw[fill=codered!35,draw=codered!75!black,line width=.55pt] (d5q15) -- (d5q21) -- (d5q29) -- (d5q23) -- cycle;
\filldraw[fill=codered!35,draw=codered!75!black,line width=.55pt] (d5q25) -- (d5q32) -- (d5q38) -- (d5q34) -- (d5q27) -- (d5q19) -- cycle;
\filldraw[fill=codegreen!35,draw=codegreen!75!black,line width=.55pt] (d5q11) -- (d5q17) -- (d5q25) -- (d5q19) -- (d5q13) -- (d5q7) -- cycle;
\filldraw[fill=codegreen!35,draw=codegreen!75!black,line width=.55pt] (d5q27) -- (d5q34) -- (d5q29) -- (d5q21) -- cycle;
\filldraw[fill=codegreen!35,draw=codegreen!75!black,line width=.55pt] (d5q36) -- (d5q40) -- (d5q38) -- (d5q32) -- cycle;
\node[site,minimum size=6mm,inner sep=0pt,font=\scriptsize] at (d5q0) {$T$};
\node[font=\scriptsize,inner sep=0pt,outer sep=0pt,below=3.2mm] at (d5q0) {0};
\node[site,minimum size=6mm,inner sep=0pt,font=\scriptsize] at (d5q3) {$T^\dagger$};
\node[font=\scriptsize,inner sep=0pt,outer sep=0pt,below=3.2mm] at (d5q3) {3};
\node[site,minimum size=6mm,inner sep=0pt,font=\scriptsize] at (d5q7) {$T^\dagger$};
\node[font=\scriptsize,inner sep=0pt,outer sep=0pt,below=3.2mm] at (d5q7) {7};
\node[site,minimum size=6mm,inner sep=0pt,font=\scriptsize] at (d5q9) {$T^\dagger$};
\node[font=\scriptsize,inner sep=0pt,outer sep=0pt,below=3.2mm] at (d5q9) {9};
\node[site,minimum size=6mm,inner sep=0pt,font=\scriptsize] at (d5q11) {$T$};
\node[font=\scriptsize,inner sep=0pt,outer sep=0pt,below=3.2mm] at (d5q11) {11};
\node[site,minimum size=6mm,inner sep=0pt,font=\scriptsize] at (d5q13) {$T^\dagger$};
\node[font=\scriptsize,inner sep=0pt,outer sep=0pt,below=3.2mm] at (d5q13) {13};
\node[site,minimum size=6mm,inner sep=0pt,font=\scriptsize] at (d5q15) {$T^\dagger$};
\node[font=\scriptsize,inner sep=0pt,outer sep=0pt,below=3.2mm] at (d5q15) {15};
\node[site,minimum size=6mm,inner sep=0pt,font=\scriptsize] at (d5q17) {$T$};
\node[font=\scriptsize,inner sep=0pt,outer sep=0pt,below=3.2mm] at (d5q17) {17};
\node[site,minimum size=6mm,inner sep=0pt,font=\scriptsize] at (d5q19) {$T$};
\node[font=\scriptsize,inner sep=0pt,outer sep=0pt,below=3.2mm] at (d5q19) {19};
\node[site,minimum size=6mm,inner sep=0pt,font=\scriptsize] at (d5q21) {$T^\dagger$};
\node[font=\scriptsize,inner sep=0pt,outer sep=0pt,below=3.2mm] at (d5q21) {21};
\node[site,minimum size=6mm,inner sep=0pt,font=\scriptsize] at (d5q23) {$T^\dagger$};
\node[font=\scriptsize,inner sep=0pt,outer sep=0pt,below=3.2mm] at (d5q23) {23};
\node[site,minimum size=6mm,inner sep=0pt,font=\scriptsize] at (d5q25) {$T^\dagger$};
\node[font=\scriptsize,inner sep=0pt,outer sep=0pt,below=3.2mm] at (d5q25) {25};
\node[site,minimum size=6mm,inner sep=0pt,font=\scriptsize] at (d5q27) {$T^\dagger$};
\node[font=\scriptsize,inner sep=0pt,outer sep=0pt,below=3.2mm] at (d5q27) {27};
\node[site,minimum size=6mm,inner sep=0pt,font=\scriptsize] at (d5q29) {$T^\dagger$};
\node[font=\scriptsize,inner sep=0pt,outer sep=0pt,below=3.2mm] at (d5q29) {29};
\node[site,minimum size=6mm,inner sep=0pt,font=\scriptsize] at (d5q32) {$T^\dagger$};
\node[font=\scriptsize,inner sep=0pt,outer sep=0pt,above=3.2mm] at (d5q32) {32};
\node[site,minimum size=6mm,inner sep=0pt,font=\scriptsize] at (d5q34) {$T^\dagger$};
\node[font=\scriptsize,inner sep=0pt,outer sep=0pt,below=3.2mm] at (d5q34) {34};
\node[site,minimum size=6mm,inner sep=0pt,font=\scriptsize] at (d5q36) {$T$};
\node[font=\scriptsize,inner sep=0pt,outer sep=0pt,below=3.2mm] at (d5q36) {36};
\node[site,minimum size=6mm,inner sep=0pt,font=\scriptsize] at (d5q38) {$T^\dagger$};
\node[font=\scriptsize,inner sep=0pt,outer sep=0pt,below=3.2mm] at (d5q38) {38};
\node[site,minimum size=6mm,inner sep=0pt,font=\scriptsize] at (d5q40) {$T$};
\node[font=\scriptsize,inner sep=0pt,outer sep=0pt,below=3.2mm] at (d5q40) {40};
\end{tikzpicture}\end{adjustbox}
\par\smallskip $d=5$: entry layer $D^\dagger$
\end{minipage}
\caption{Triangular colour codes, number are qubit indices from~\cite{clifft, soft}. On the right, each node shows its entry $T$ or $T^\dagger$ gate in one layer of gates in the $d=5$ double-checking stage.}
\label{fig:patches}
\end{figure}
The $[[7,1,3]]$ ($d=3$) variant code uses seven data qubits, while the $d=5$ cultivation circuit
grows this to a $[[19,1,5]]$ code. In Fig.~\ref{fig:patches}, each coloured face defines an
$X$ and a $Z$ stabiliser, formed by multiplying the corresponding
Pauli operators over the qubits on that face. The code space is the
simultaneous $+1$ eigenspace of these stabilisers~\cite{albert2026handbookerrorcorrectingcodes}. There are six
independent stabiliser generators at $d=3$ and eighteen at $d=5$,
with one logical qubit encoded in either case. Pauli errors that differ by a stabiliser (or product of stabilisers) have the same action
on the encoded magic state and share a response-table entry.

We need one response for each distinguishable incoming error, rather than
one for every Pauli string (or else we need to store exponentially many of them). 
To identify errors that share a response, we first represent each
incoming Pauli string by its $X$ and $Z$ components using two binary
masks, $\boldsymbol a$ and $\boldsymbol b$\footnote{At each qubit $q$,
the tuple $(a_q,b_q)$ maps to the Pauli operators:
$I\leftrightarrow(0,0)$,
$X\leftrightarrow(1,0)$,
$Z\leftrightarrow(0,1)$ and
$Y\leftrightarrow(1,1)$, ignoring the overall phase.}. 
The syndrome records which code stabilisers have been flipped to $-1$ parity (represented by bit $1$)\footnote{Parity $+1$ is represented by bit $0$.}.
For the $d=3$ patch in Fig.~\ref{fig:patches}, the parity-check matrix is
\begin{equation}
 H^{(3)}=
 \begin{pmatrix}
 1&1&1&0&1&0&0\\
 0&1&1&1&0&1&0\\
 0&0&1&0&1&1&1
 \end{pmatrix}.
\end{equation}
The columns follow qubit order $(0,\allowbreak3,\allowbreak7,\allowbreak9,\allowbreak10,\allowbreak12,\allowbreak13)$. The rows correspond to the blue, red and green faces.
Each row marks the qubits belonging to that face.

For either code, we label each incoming error by
\begin{equation}
 \begin{aligned}
 \alpha&=(H\boldsymbol a,\ \boldsymbol1\cdot\boldsymbol a),\\
 \beta&=(H\boldsymbol b,\ \boldsymbol1\cdot\boldsymbol b),
 \end{aligned}
 \label{eq:address}
\end{equation}
all in modulo two. Here $H\boldsymbol a$ and
$H\boldsymbol b$ are the $Z$- and $X$-stabiliser syndromes,
respectively. The final bit in each label is the parity of the
corresponding mask, with $\boldsymbol1$ being the all-ones vector, 
\[
\boldsymbol1 = (1,1,\ldots,1)^{\text{T}} \ .
\]
These two additional bits distinguish logical Pauli classes
with the same syndrome.
Errors with the same $(\alpha,\beta)$ differ by a stabiliser, which acts trivially on the encoded state, so they have the same response.

For example, $X_7$ and $X_0X_3X_{10}$ differ by the blue-face
stabiliser $X_0X_3X_7X_{10}$ and therefore have the same effect
on the encoded state.

 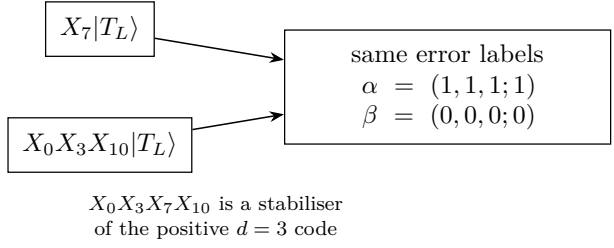
\begin{figure}[htbp]\centering
 \begin{adjustbox}{max width=.98\linewidth}\begin{tikzpicture}[x=1cm,y=1cm]
\node[box] (a) at (0,.8) {$X_7\ket{T_L}$};
\node[box] (b) at (0,-.8) {$X_0X_3X_{10}\ket{T_L}$};
\node[box,text width=41mm] (address) at (4.8,0)
 {same error labels\\$\alpha=(1,1,1;1)$\\$\beta=(0,0,0;0)$};
\draw[flow] (a)--(address);\draw[flow] (b)--(address);
\node[note,align=center] at (1.6,-1.8)
 {$X_0X_3X_7X_{10}$ is a stabiliser\\of the positive $d=3$ code};
\end{tikzpicture}\end{adjustbox}
 \caption{Two Pauli errors share a response-table entry at $d=3$.}\label{fig:frames}\end{figure}

At $d=3$, each label $\alpha$ or $\beta$ has four bits, so each response table has
$16^2=256$ entries. At $d=5$, each label $\alpha$ or $\beta$ has ten bits and a table has
$1024^2$ entries. Grouping stabiliser-equivalent errors reduces the number of
response-table entries from $4^7=16\,384$ to $256$ at $d=3$,
and from $4^{19}\approx2.75\times10^{11}$ to
$1\,048\,576$ at $d=5$. We calculate and store one response
for each equivalence class.
These error labels become tensor indices. Contracting the tensors combines
the double-checking responses with the noise from earlier stages to give
$A(p)$ and $B(p)$.

\subsection{Tensor contraction}\label{sec:tensor-contraction}
The response of a check to an incoming error is a sum over every fault configuration inside the check. Writing the gates, noise and post-selection conditions as small tensors organises that sum: contracting them performs the noise average, and leaving the incoming labels open computes all table entries at once, sharing the internal sums between them.
 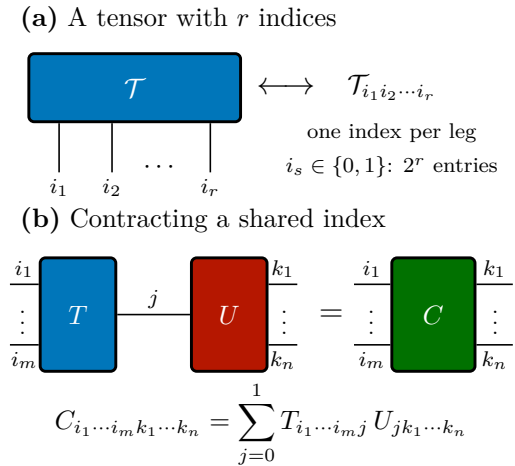
\begin{figure}[htbp]\centering
 \begin{adjustbox}{max width=.98\linewidth}\begin{tikzpicture}[x=1cm,y=1cm]
\node[tn heading,anchor=west] at (0,3.4)
  {\textbf{(a)} A tensor with $r$ indices};
\node[tn tensor,minimum width=28mm] (R) at (1.65,2.5) {$\mathcal T$};
\draw[tn leg] ([xshift=-10mm]R.south)--++(0,-.65)
  node[below,tn label] {$i_1$};
\draw[tn leg] ([xshift=-3mm]R.south)--++(0,-.65)
  node[below,tn label] {$i_2$};
\draw[tn leg] ([xshift=10mm]R.south)--++(0,-.65)
  node[below,tn label] {$i_r$};
\node[font=\small] at (2.0,1.45) {$\cdots$};
\node[font=\large] at (3.65,2.5) {$\longleftrightarrow$};
\node[font=\small] at (5.05,2.5) {$\mathcal T_{i_1i_2\cdots i_r}$};
\node[tn label] at (5.05,1.9) {one index per leg};
\node[tn label] at (5.05,1.45) {$i_s\in\{0,1\}$: $2^r$ entries};

\node[tn heading,anchor=west] at (0,.75)
  {\textbf{(b)} Contracting a shared index};
\node[tn tensor,minimum width=10mm,minimum height=15mm] (T) at (.9,-.5) {$T$};
\node[tn tensor,fill=tnorange,minimum width=10mm,minimum height=15mm]
  (U) at (2.9,-.5) {$U$};
\draw[tn leg] (0,-.1)--node[above,tn label] {$i_1$}
  ([yshift=4mm]T.west);
\draw[tn leg] (0,-.9)--node[below,tn label] {$i_m$}
  ([yshift=-4mm]T.west);
\node[tn label] at (.17,-.5) {$\vdots$};
\draw[tn leg] (T.east)--node[above,tn label] {$j$} (U.west);
\draw[tn leg] ([yshift=4mm]U.east)--node[above,tn label] {$k_1$} (3.8,-.1);
\draw[tn leg] ([yshift=-4mm]U.east)--node[below,tn label] {$k_n$} (3.8,-.9);
\node[tn label] at (3.63,-.5) {$\vdots$};

\node[font=\large] at (4.25,-.5) {$=$};
\node[tn intermediate,minimum width=11mm,minimum height=15mm]
  (C) at (5.6,-.5) {$C$};
\draw[tn leg] (4.55,-.1)--node[above,tn label] {$i_1$}
  ([yshift=4mm]C.west);
\draw[tn leg] (4.55,-.9)--node[below,tn label] {$i_m$}
  ([yshift=-4mm]C.west);
\node[tn label] at (4.78,-.5) {$\vdots$};
\draw[tn leg] ([yshift=4mm]C.east)--node[above,tn label] {$k_1$} (6.65,-.1);
\draw[tn leg] ([yshift=-4mm]C.east)--node[below,tn label] {$k_n$} (6.65,-.9);
\node[tn label] at (6.42,-.5) {$\vdots$};

\node[font=\small] at (3.325,-1.95)
  {$\displaystyle C_{i_1\cdots i_m k_1\cdots k_n}
    =\sum_{j=0}^{1}T_{i_1\cdots i_m j}\,U_{j k_1\cdots k_n}$};
\end{tikzpicture}\end{adjustbox}
 \caption{(a) A tensor with $r$ indices. (b) Contracting $j$ leaves all
$i_1,\ldots,i_m$ and $k_1,\ldots,k_n$ indices open.}\label{fig:tensor-definition}\end{figure}
A tensor is an array whose entries are labelled by indices and defined as follows. For $r$ binary indices, the entries are written
$\mathcal T_{i_1\cdots i_r}$, with $i_s\in\{0,1\}$.
Each leg represents one index. Contracting tensors multiplies
their entries and sums over the connected indices. Open legs
label the indices that remain. A double line bundles several binary indices. Each leg is an index over error labels, not a physical qubit wire. Throughout this manuscript, we draw tensors as rounded boxes with legs as indices, contrasting with square white boxes elsewhere indicating calculation steps.

\subsubsection{Response tables as tensors}
For fixed $p$, the response table $F_A$ is a matrix where the row
index groups the bits of $\alpha$ and column index
groups the bits of $\beta$. Its entry gives the probability (for
that fixed incoming error) that the double-checking stage and noiseless final code readout both pass. The corresponding entry of $F_B$ also requires
a logical error. Both probabilities are averaged over the noise within the
double-checking stage.

\begin{figure}[htbp]\centering
 \begin{adjustbox}{max width=.98\linewidth}\begin{tikzpicture}[x=1cm,y=1cm,
  response label/.style={tn label,font=\footnotesize},
  inner box/.style={rectangle,rounded corners=1pt,draw=black!55,line width=.5pt,inner sep=0pt},
  inner leg/.style={line width=.5pt,draw=black!55}]
\node[tn heading,anchor=west] at (0,5.45)
  {\textbf{(a)} An open $d=3$ response table};
\node[rectangle,rounded corners=2pt,draw=black,line width=1.1pt,
      fill=tnblue,fill opacity=.9,minimum width=26mm,minimum height=30mm]
  (open) at (3.3,3.55) {};
\begin{scope}[shift={(open.center)}]
 \node[inner box,fill=black!16,minimum width=18mm,minimum height=11mm] (L) at (0,.85) {};
 \node[inner box,fill=tngreen,fill opacity=.92,minimum width=18mm,minimum height=2.4mm] (K) at (0,-1.18) {};
 \foreach \i in {-2,-1,0,1,2}{
  \node[inner box,fill=white,fill opacity=.92,minimum width=2.6mm,minimum height=2.6mm] (c\i) at (\i*.33,-.42) {};
  \draw[inner leg] (c\i.north)--(c\i.north |- L.south);
  \draw[inner leg] (c\i.south)--(c\i.south |- K.north);
 }
 \node[fill=tnblue!90!white,text=white,font=\large,inner sep=.5mm] at (0,.03) {$F_O$};
 \foreach \bit/\dy in {1/.45,2/.15,3/-.15,4/-.45}{
  \draw[tn leg] (-2.5,.85+\dy) node[left,response label,inner sep=.4mm] {$\alpha_{\bit}$} -- (L.west |- 0,.85+\dy);
  \draw[tn leg] (L.east |- 0,.85+\dy) -- (2.5,.85+\dy) node[right,response label,inner sep=.4mm] {$\beta_{\bit}$};
 }
\end{scope}
\node[response label] at (3.3,1.75)
  {$\alpha,\beta\in\{0,1\}^4$: $2^8=256$ entries};

\node[tn heading,anchor=west] at (0,1.10)
  {\textbf{(b)} Sum over the incoming errors};
\node[tn tensor,fill=tnorange,minimum width=13mm,minimum height=15mm,font=\large]
  (mu) at (1,-.2) {$\mu$};
\node[tn endpoint,minimum width=15mm,minimum height=15mm,font=\large]
  (closed) at (3.6,-.2) {$F_O$};
\draw[tn bundle] ([yshift=3.5mm]mu.east)
  --node[above,response label] {$\alpha$}
  ([yshift=3.5mm]closed.west);
\draw[tn bundle] ([yshift=-3.5mm]mu.east)
  --node[below,response label] {$\beta$}
  ([yshift=-3.5mm]closed.west);
\node[font=\large] at (4.95,-.2) {$=$};
\node[tn intermediate,minimum width=11mm,minimum height=11mm,font=\large]
  (result) at (6.1,-.2) {$O$};
\node[response label,anchor=north] at (1,-1.3) {earlier-stage\\weights};
\node[response label,anchor=north] at (3.6,-1.3) {double-checking\\response};
\node[response label,anchor=north] at (6.1,-1.3) {probability\\$O=A$ or $B$};
\node[font=\small] at (3.3,-2.55)
  {$\displaystyle O(p)=\sum_{\alpha,\beta}
    \mu(\alpha,\beta;p)\,F_O(\alpha,\beta;p)$};
\end{tikzpicture}\end{adjustbox}
 \caption{(a) Eight binary indices label the $d=3$ response table.
(b) Bundling them into $\alpha,\beta$ and summing with $\mu$ gives $A$ or $B$.
Fig.~\ref{fig:opennet} in Appendix~\ref{app:response}
shows the internal network.}\label{fig:response}\end{figure}
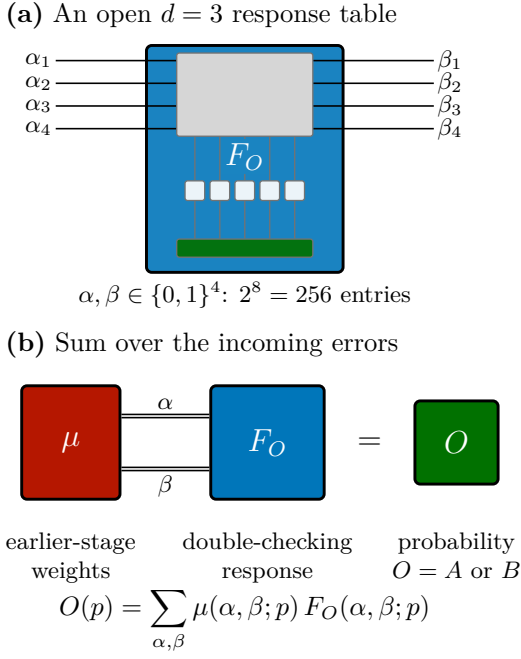

Let $\mu(\alpha,\beta;p)$ be the joint probability that the earlier
stages pass post-selection and supply that incoming error. Then
\begin{equation}
 \begin{aligned}
 A(p)&=\sum_{\alpha,\beta}\mu(\alpha,\beta;p)F_A(\alpha,\beta;p),\\
 B(p)&=\sum_{\alpha,\beta}\mu(\alpha,\beta;p)F_B(\alpha,\beta;p).
 \end{aligned}
 \label{eq:responseaverage}
\end{equation}
For each pair $(\alpha,\beta)$, multiply its weight $\mu$
by the corresponding response $F_A$ or $F_B$, then sum over all pairs.
This is a tensor contraction, shown in Fig.~\ref{fig:response}(b).
Each weight $\mu$ is already the joint probability of passing the earlier post-selection and producing
the specified incoming error. Because the weights sum to the earlier acceptance probability, no further normalisation is needed before taking $\PL=B/A$, giving the LER.

\subsubsection{Tensors for the circuit operations}
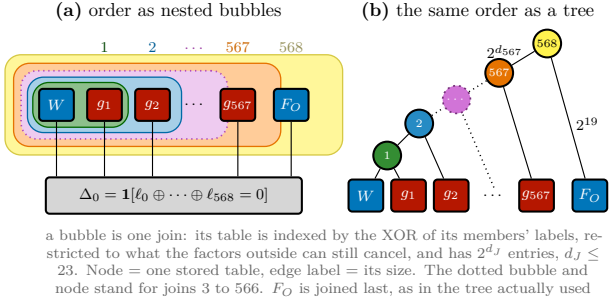
\begin{figure}[H]\centering
 \begin{adjustbox}{max width=.98\linewidth}\begin{tikzpicture}[x=1cm,y=1cm,
  fac/.style={tn tensor,minimum width=6.5mm,minimum height=6mm,inner sep=0.5mm,font=\scriptsize},
  noise/.style={fac,fill=tnorange},
  jn/.style={tn intermediate,circle,minimum size=5.2mm,inner sep=0pt,font=\tiny},
  bub/.style={line width=.7pt,rounded corners=3mm},
  num/.style={font=\scriptsize,anchor=south,inner sep=1pt},
  c1/.style={draw=tngreen!90!black,fill=tngreen!45,text=tngreen!60!black},
  c2/.style={draw=tnblue!90!black,fill=tnblue!35,text=tnblue!80!black},
  cd/.style={draw=pppurple!90!black,fill=pppurple!30,text=pppurple!80!black,dotted,line width=.9pt},
  c3/.style={draw=orange!90!black,fill=orange!40,text=orange!80!black},
  c4/.style={draw=yellow!80!black,fill=yellow!50,text=yellow!50!black},
  elab/.style={tn label,inner sep=0.5pt}]
\node[tn heading] at (2.1,2.25) {\textbf{(a)} order as nested bubbles};
\node[fac] (w) at (0,0.5) {$W$};
\node[noise] (g1) at (0.9,0.5) {$g_1$};
\node[noise] (g2) at (1.8,0.5) {$g_2$};
\node[tn label] (dd) at (2.6,0.5) {$\cdots$};
\node[noise] (g3) at (3.4,0.5) {$g_{567}$};
\node[fac] (f) at (4.4,0.5) {$F_O$};
\node[tn constraint,minimum width=48mm,minimum height=5mm,font=\scriptsize] (d0) at (2.2,-1.2) {$\Delta_0=\ind{\ell_0\oplus\cdots\oplus\ell_{568}=0}$};
\foreach \n in {w,g1,g2,g3,f}{\draw[tn leg] (\n.south)--(\n.south |- d0.north);}
\begin{scope}[on background layer]
\node[bub,c4,fit=(w)(f),inner sep=6.2mm] (b4) {};
\node[bub,c3,fit=(w)(g3),inner sep=4.6mm] (b3) {};
\node[bub,cd,fit=(w)(dd),inner sep=3.2mm] (bd) {};
\node[bub,c2,fit=(w)(g2),inner sep=2.0mm] (b2) {};
\node[bub,c1,fit=(w)(g1),inner sep=1.0mm] (b1) {};
\end{scope}
\node[num,c1,fill=none,draw=none] at (g1 |- b4.north) {1};
\node[num,c2,fill=none,draw=none] at (g2 |- b4.north) {2};
\node[num,cd,fill=none,draw=none,line width=0pt] at (dd |- b4.north) {$\cdots$};
\node[num,c3,fill=none,draw=none] at (g3 |- b4.north) {567};
\node[num,c4,fill=none,draw=none] at (f |- b4.north) {568};
\node[tn heading] at (7.9,2.25) {\textbf{(b)} the same order as a tree};
\node[fac] (tw) at (5.8,-1.2) {$W$};
\node[noise] (t1) at (6.6,-1.2) {$g_1$};
\node[noise] (t2) at (7.4,-1.2) {$g_2$};
\node[tn label] (tdd) at (8.2,-1.2) {$\cdots$};
\node[noise] (t3) at (9.0,-1.2) {$g_{567}$};
\node[fac] (tf) at (10.0,-1.2) {$F_O$};
\node[jn,fill=tngreen!80] (j1) at (6.2,-0.45) {1};
\node[jn,fill=tnblue!85] (j2) at (6.8,0.15) {2};
\node[jn,fill=pppurple!80,draw=pppurple!60!black,dotted,line width=.9pt] (jd) at (7.5,0.6) {$\cdots$};
\node[jn,fill=orange!90!black] (j3) at (8.3,1.1) {567};
\node[jn,fill=yellow!80,text=black] (j4) at (9.2,1.65) {568};
\draw[tn leg] (tw)--(j1); \draw[tn leg] (t1)--(j1);
\draw[tn leg] (j1)--(j2); \draw[tn leg] (t2)--(j2);
\draw[tn leg,dotted,line width=.8pt] (j2)--(jd); \draw[tn leg,dotted,line width=.8pt] (jd)--(j3);
\draw[tn leg,dotted,line width=.8pt] (tdd.north)--(jd);
\draw[tn leg] (t3)--(j3);
\draw[tn leg] (j3)--node[elab,above left=-1pt] {$2^{d_{567}}$}(j4); \draw[tn leg] (tf)--node[elab,right=2.5pt,pos=0.45] {$2^{19}$}(j4);
\node[tn label,text=black!60,align=center,text width=104mm,anchor=north] at (5.0,-1.75) {a bubble is one join: its table is indexed by the XOR of its members' labels, restricted to what the factors outside can still cancel, and has $2^{d_J}$ entries, $d_J\le23$. Node $=$ one stored table, edge label $=$ its size. The dotted bubble and node stand for joins 3 to 566. $F_O$ is joined last, as in the tree actually used};
\end{tikzpicture}\end{adjustbox}
 \caption{A contraction order for the noisy growth stage of Fig.~\ref{fig:overview}, drawn two ways. Its 569 factors, the $d=3$ table $W$, 567 grouped noise factors $g_i$ and the $d=5$ response $F_O$, share one parity constraint on their labels (Appendix~\ref{app:growth}). $\dots$ stand for the omitted factors and joins. (a) Nested bubbles: each bubble is joined into one intermediate table, stored on the legs that cross its boundary. (b) The same order as a tree, one node per join, the dotted node standing for joins 3 to 566, the root being the scalar. The order changes the table sizes, not the contracted result.}\label{fig:orderbubbles}\end{figure}

The eight open legs in the tensor from Fig.~\ref{fig:response}(a) label a completed
response table. To calculate that table, we contract smaller tensors
for the gates, noise and post-selection, keeping the incoming error
labels open, so that the internal sums are shared between table entries. 
We choose the tensor contraction order carefully to keep intermediate tensors small
without changing the results (see Appendix~\ref{app:tensor-examples-v2} for details). Figure~\ref{fig:orderbubbles} shows one such order as nested bubbles and as a tree.

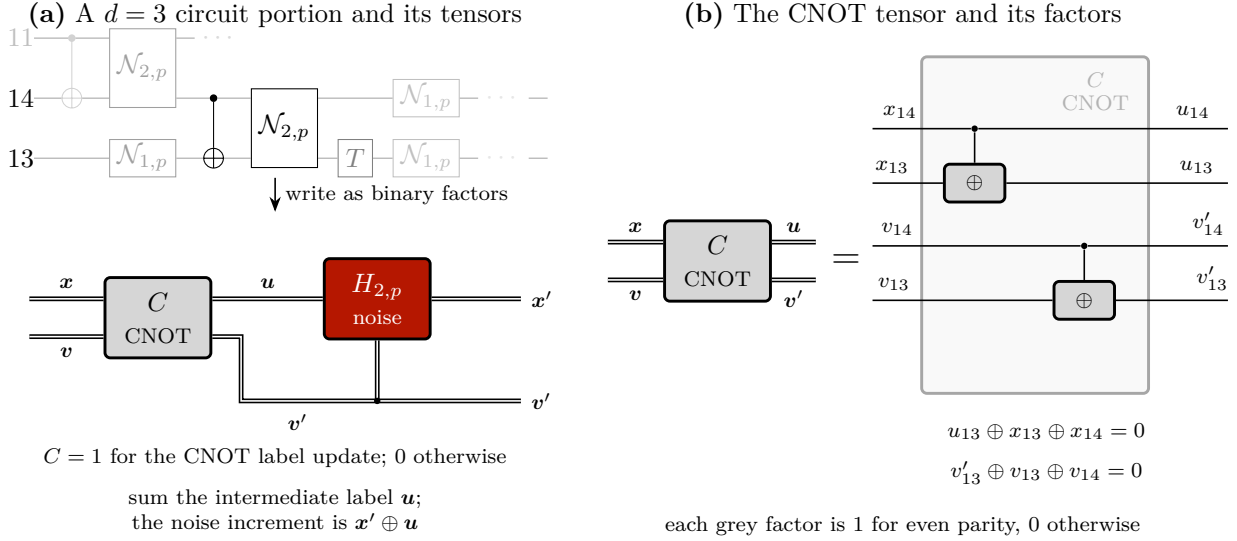
\begin{figure*}[tp]\centering
 \begin{adjustbox}{max width=.98\linewidth}\begin{tikzpicture}[x=1cm,y=1cm,
  tensor/.style={tn tensor,minimum width=14mm,minimum height=10mm,
                 inner sep=1.5mm,font=\small},
  bundle/.style={tn bundle},
  copy bit/.style={tn copy}]

\node[note,font=\small] at (3.4,5.05)
  {\textbf{(a)} A $d=3$ circuit portion and its tensors};
\node[note,font=\small] at (11.7,5.05)
  {\textbf{(b)} The CNOT tensor and its factors};

\node[inner sep=0pt] at (3.4,3.95) {%
  \begin{tikzpicture}[font=\small]
  \begin{yquant}[operator/separation=2.5mm,register/separation=3mm,
    every wire/.append style={draw=black!35}]
  qubit {\color{black!25}$11$} q11;
  qubit {$14$} q14;
  qubit {$13$} q13;
  [style={opacity=.18}] cnot q14 | q11;
  align q11,q14,q13;
  [style={opacity=.38}] box {$\mathcal N_{2,p}$} (q11,q14);
  [style={opacity=.38}] box {$\mathcal N_{1,p}$} q13;
  align q11,q14,q13;
  [style={opacity=.15}] text {$\cdots$} q11;
  discard q11;
  cnot q13 | q14;
  box {$\mathcal N_{2,p}$} (q14,q13);
  [style={opacity=.50}] box {$T$} q13;
  align q14,q13;
  [style={opacity=.25}] box {$\mathcal N_{1,p}$} q14,q13;
  [style={opacity=.10}] text {$\cdots$} q14,q13;
  \end{yquant}
  \end{tikzpicture}};
\draw[flow] (3.4,2.9)--(3.4,2.45)
  node[midway,right,note] {write as binary factors};

\node[tensor,tn constraint] (C) at (1.85,1.05) {$C$\\\scriptsize CNOT};
\node[tensor,tn noise] (N) at (4.75,1.3) {$H_{2,p}$\\\scriptsize noise};
\draw[bundle] (.15,1.3)--node[above,note] {$\boldsymbol x$}
  ([yshift=2.5mm]C.west);
\draw[bundle] (.15,.8)--node[below,note] {$\boldsymbol v$}
  ([yshift=-2.5mm]C.west);
\draw[bundle] ([yshift=2.5mm]C.east)--node[above,note]
  {$\boldsymbol u$}(N.west);
\draw[bundle] (N.east)--(6.65,1.3) node[right,note] {$\boldsymbol x'$};
\draw[bundle] ([yshift=-2.5mm]C.east)--(2.95,.8)--(2.95,-.05)--(6.65,-.05)
  node[right,note] {$\boldsymbol v'$};
\draw[bundle] (4.75,-.05)--(N.south);
\node[copy bit] at (4.75,-.05) {};
\node[note] at (3.7,-.33) {$\boldsymbol v'$};
\node[note] at (3.4,-.8)
  {$C=1$ for the CNOT label update; $0$ otherwise};

\node[note] at (3.4,-1.5)
 {sum the intermediate label $\boldsymbol u$;\\
  the noise increment is $\boldsymbol x'\oplus\boldsymbol u$};

\node[tensor,tn constraint] (Cgroup) at (9.25,1.8) {$C$\\\scriptsize CNOT};
\draw[bundle] (7.8,2.05)--node[above,note] {$\boldsymbol x$}
  ([yshift=2.5mm]Cgroup.west);
\draw[bundle] (7.8,1.55)--node[below,note] {$\boldsymbol v$}
  ([yshift=-2.5mm]Cgroup.west);
\draw[bundle] ([yshift=2.5mm]Cgroup.east)--node[above,note]
  {$\boldsymbol u$}(10.55,2.05);
\draw[bundle] ([yshift=-2.5mm]Cgroup.east)--node[below,note]
  {$\boldsymbol v'$}(10.55,1.55);
\node[font=\Large] at (10.95,1.8) {$=$};

\begin{scope}[on background layer]
 \path[rounded corners=2pt,draw=black!35,line width=1.1pt,
       fill=black!12,fill opacity=.18]
   (11.95,.05) rectangle (14.95,4.5);
 \node[align=center,text=black!30,anchor=north east,
       font=\scriptsize,inner sep=0pt]
   at (14.7,4.30) {$C$\\[-1pt]CNOT};
\end{scope}
\foreach \xx/\yy/\a/\b/\ao/\bo/\name/\labelpos in
 {12.65/3.55/x_{14}/x_{13}/u_{14}/u_{13}/x/.28,
  14.10/2.00/v_{14}/v_{13}/v'_{14}/v'_{13}/v/.12}{
 \begin{scope}[shift={(0,\yy)}]
 \node[tn copy] (dot\name) at (\xx,0) {};
 \draw[tn leg] (11.3,0)--node[pos=\labelpos,above,note] {$\a$}(dot\name);
 \draw[tn leg] (dot\name)--node[pos=.86,above,note] {$\ao$}(16,0);
 \node[tn constraint,minimum width=8mm,minimum height=5mm,
       inner sep=.8mm,font=\scriptsize]
   (xor\name) at (\xx,-.72) {$\oplus$};
 \draw[tn leg] (dot\name)--(xor\name);
 \draw[tn leg] (11.3,-.72)--node[pos=\labelpos,above,note] {$\b$}(xor\name);
 \draw[tn leg] (xor\name)--node[pos=.86,above,note] {$\bo$}(16,-.72);
 \end{scope}
}
\node[note] at (13.6,-.45) {$u_{13}\oplus x_{13}\oplus x_{14}=0$};
\node[note] at (13.6,-1.0) {$v'_{13}\oplus v_{13}\oplus v_{14}=0$};
\node[note] at (11.7,-1.7)
 {each grey factor is $1$ for even parity, $0$ otherwise};
\end{tikzpicture}\end{adjustbox}
 \caption{(a) A circuit portion and its CNOT/noise tensors.
(b) The faint box groups the two parity factors into $C$.
Black dots reuse the same index. Eq.~\eqref{eq:gate-noise-tensor} and Fig.~\ref{fig:opennet} in Appendix~\ref{app:response} place this pair inside the response network.}\label{fig:response-network}\end{figure*}

For example, the CNOT tensor $C$ in Fig.~\ref{fig:response-network}(a)
contains the two parity tensors expanded in panel (b).
An $x$ bit records the $X$ component
of an error. A $v$ bit is a Fourier index for its $Z$ component.
In Fourier form the CNOT acts on the $Z$ labels by the same parity rule as on the $X$ labels, and the noise becomes a multiplicative weight (Appendix~\ref{app:response}).
Each grey tensor is $1$ when its three bits have even parity, and $0$
otherwise. The vector $\boldsymbol u$ labels the error just after the
gate; $\boldsymbol x'$ labels it after the noise. The repeated Fourier
label $\boldsymbol v'$ is shared with the noise tensor $H_{2,p}$.
Noise supplies its weights as defined in Appendix~\ref{app:response}.
The two $T$ layers and final selection are compiled into signed
terminal factors.
Appendix~\ref{app:response} gives their equations and the final transform
from the internal Fourier indices to $F_A,F_B$.

\subsubsection{Exact fractions and series coefficients}
We evaluate the tensor network in two ways to compute the LER. 
\begin{enumerate}
\item At a fixed circuit-level noise strength $p$ expressed as a fraction, we evaluate the whole contraction modulo several certified primes and reconstruct the acceptance and accepted-error probabilities as exact fractions. 
Their ratio is $\PL(p)$ with every fault order included.
Exact rational arithmetic on the whole network is impractical, the fractions run to thousands of digits, whereas modular arithmetic uses fixed-size integers and recovers the exact fraction at the end (Appendix~\ref{app:arithmetic}).
\item As a series expansion, we remove the common no-fault factor and use
polynomials in parameter $x=p/(1-p)$. We keep degrees 0 to 10 in $x$ in every factor, including the earlier-stage weights and final responses. This returns the exact raw acceptance and
accepted-error coefficients through that order. Dividing their series
gives Eq.~\eqref{eq:fullseries}. The contribution of the higher-order terms is bounded (Appendix~\ref{app:arithmetic}).
\end{enumerate}

\section{Discussion}
The full distance $d=5$ magic state cultivation circuit's LER can be calculated exactly on
an everyday device (Apple Macbook M4). The structure exploited is the small syndrome and logical
interface around each non-Clifford double-checking stage (which dominates the non-Clifford cost). 
Pauli propagation reduces each non-Clifford double-checking stage to a response table indexed by the incoming error, 
with stabiliser-equivalent strings merged. Tensor network methods then join these tables to the injection and growth noise, 
whose faults act linearly on the small syndrome-and-logical interface between the stages, so the complete noise average is an exact arithmetic contraction. 
\begin{semiwebaddition}
Pauli propagation also has a graphical interpretation using the Pauli
semiwebs of Ref.~\cite{zxflow}. Fig.~\ref{fig:semiweb-main} shows an
incoming $X_7$ on the $d=3$ double-checking circuit. The red Pauli semiweb in
Fig.~\ref{fig:semiweb-main} shows the Pauli operators on the wires, while the stars mark changes to the $T$-gate phases.
Instead of expanding into signed Pauli terms, this identity keeps a
Pauli string and changes the phases of the circuit it passes through.
Multiplying semiwebs shows where shared Pauli operators and phase changes
cancel. Appendix~\ref{app:semiwebs} defines the notation and gives two
examples.
\end{semiwebaddition}

\begin{figure*}[t]
\centering
  \begingroup\color{black}%
  \input{figures/semiwebs/zx_styles.tex}%
  \pgfsetlayers{background,edgelayer,nodelayer,main}%
  \begin{adjustbox}{max width=.98\linewidth}\input{figures/semiwebs/x7.tikz}\end{adjustbox}%
  \endgroup
\semiwebcaption{A Pauli semiweb for incoming $X_7$ on the $d=3$
double-checking circuit. Red dashed overlays show $X$ Pauli semiwebs;
unhighlighted wires carry $I$. Stars mark phase changes: $+\pi/2$ at the
entry $T^\dagger_7$, and $-\pi/2$ at the six highlighted exit $T$ gates.
\textbf{Legend (up to normalisation):}
\semiwebCaptionGate{\frac{\pi}{4}} ${}=T$ ($-\pi/4$: $T^\dagger$);
\semiwebCaptionGate{\frac{\pi}{2}} ${}=S$ ($-\pi/2$: $S^\dagger$);
\semiwebCaptionCnot ${}={\rm CNOT}$;
measurement outcome $0$: \semiwebCaptionEffect{Z} ${}=\langle+|$ ($X$ basis),
\semiwebCaptionEffect{X} ${}=\langle0|$ ($Z$ basis);
preparations \semiwebCaptionState{X} ${}=|0\rangle$
and \semiwebCaptionState{Z} ${}=|+\rangle$;
resources \semiwebCaptionPhaseState{\frac{\pi}{4}} ${}=|T\rangle$,
\semiwebCaptionPhaseState{\frac{\pi}{2}} ${}=|Y\rangle$.}
\label{fig:semiweb-main}
\end{figure*}

Our method was loosely inspired by early Pauli-propagation papers based on structured Pauli expansions \cite{lowesa,noisylowesa}. SyQMA~\cite{syqma} previously computed symbolic, noise-averaged LER and inferred circuit fault distances from their leading orders in $p$.
Its Steane-code encoding and verification (akin to magic state cultivation but not exactly) example has $p^2$ scaling, under a different circuit and noise conventions.

In addition, the fault-count coefficients $a_k,e_k$ are related to the per-stratum acceptance and accepted-error probabilities that Clifft's stratified importance sampling
estimates~\cite{clifft} and that the Pauli-frame reweighting of~\cite{pfsr} organises by fault number. Here every stratum is summed
exactly rather than sampled. In~\cite{pfsr}, they restricts their $d=5$ sampling
to $k\ge5$ faults on the premise that fewer than $d$ faults cannot yield an
accepted logical error. For the circuits studied here that premise fails,
since $e_3$ and $e_4$ are non-zero and supply about a third of the accepted errors at $p=10^{-3}$ when faults are counted over all 3,564 physical locations, and four fifths when only the 1,996 locations with a non-identity response are counted, Appendix~\ref{app:circuits}.

\begin{figure}[h]\centering
\includegraphics[width=.98\linewidth]{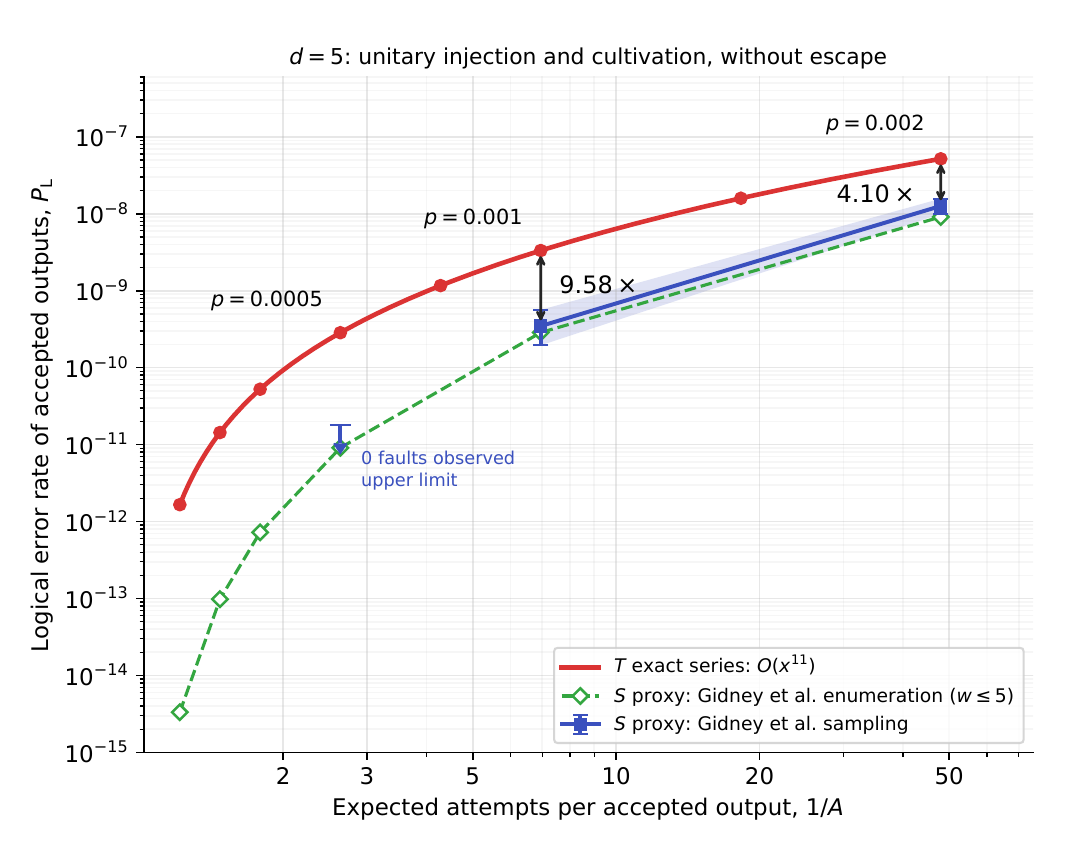}
\caption{$d=5$ cultivation with unitary injection, without escape.
The {\color{red}$T$ curve} uses exact series coefficients through $x^{10}$. The {\color{blue}$S$-proxy data (blue)} and {\color{green}green curve} (that retains detector-error-model terms of at most weight five) are both from~\cite{msc}.}
\label{fig:d5-s-proxy}
\end{figure}

Fig.~\ref{fig:d5-s-proxy} compares the $d=5$ $T$-state series with the
unitary-injection $S$-gate proxy of~\cite{msc}.
At $p=10^{-3}$, the proxy is $\sim 9.6$ times lower (nearly one order of
magnitude). 
Using the proxy to infer a quantum algorithmic estimate will produce an over-optimistic resource count, as the proxy may substantially underestimate the LER of actual $T$-state cultivation.

The leading orders of the series expansion reveal the LER scales as $p^{2}$ at $d=3$ and $p^{3}$ at $d=5$ (see Acknowledgements), not as the intended $p^{3}$ and $p^{5}$ the code distances suggest. 
This explains the apparent distance-degrading behaviour observed in the companion code of~\cite{wan}, albeit with a slightly different injection scheme. 
In a Clifford circuit, detectors are equivalent to closed Pauli webs of its ZX diagram. In the $S$-gate proxy this holds for all 20 declared $d=3$ detectors, but in the $T$-gate circuit only 16 of them are Pauli webs (93 of 107 at $d=5$), and a fault pattern that the $T$-gate circuit accepts can be rejected by the proxy only through the other detectors (Appendix~\ref{app:detector-webs}).
Appendix~\ref{app:other-cultivation} provides exact series coefficients for other flavours of magic state cultivation such as: $\mathbb{RP}^{2}$~\cite{RP2_prxq}, fold-transversal~\cite{fold_msc_prxq} and repaired cultivation~\cite{chan2026diagnosingrestoringdegradedfault}, with comparisons against available numerical results.
Extensions to the escape and grafting stages of cultivation~\cite{msc} are left for future work.

\section{Acknowledgements}
We thank Tim Chan for notifying us of their results~\cite{chan_msc} (paper~\cite{chan2026diagnosingrestoringdegradedfault}) on the reduced fault distance of magic state cultivation circuits. To the best of our knowledge, Chan et al. were the first to identify this reduced fault distance of the cultivation circuits studied in this manuscript. 
We also thank Zhenghao (William) Zhong for the invitation to Oxford as a visitor.
We acknowledge useful discussions with Austin Fowler and Zoë Holmes.
We acknowledge the use of large language models in the refinement of the methods (especially the use of ChatGPT 6 Astra in extending the human-made $d=3$ results to the $d=5$ circuits) and wording of this manuscript. The code generated was checked and tested and we take responsibility for all contents, to the best of our abilities.
The companion code and data are available in~\cite{exact_msc_code}.
\FloatBarrier
\bibliographystyle{quantum}
\bibliography{main}

@misc{msc,
      title={Magic state cultivation: growing {T} states as cheap as {CNOT} gates}, 
      author={Craig Gidney and Noah Shutty and Cody Jones},
      year={2024},
      eprint={2409.17595},
      archivePrefix={arXiv},
      primaryClass={quant-ph},
      url={https://arxiv.org/abs/2409.17595}, 
}

@misc{soft,
      title={{SOFT}: a high-performance simulator for universal fault-tolerant quantum circuits}, 
      author={Riling Li and Keli Zheng and Yiming Zhang and Huazhe Lou and Shenggang Ying and Ke Liu and Xiaoming Sun},
      year={2025},
      eprint={2512.23037},
      archivePrefix={arXiv},
      primaryClass={quant-ph},
      url={https://arxiv.org/abs/2512.23037}, 
}

@misc{clifft,
      title={{Clifft}: {F}ast {E}xact {S}imulation of {N}ear-{Clifford} {Q}uantum {C}ircuits}, 
      author={Bradley A. Chase and Farrokh Labib},
      year={2026},
      eprint={2604.27058},
      archivePrefix={arXiv},
      primaryClass={quant-ph},
      url={https://arxiv.org/abs/2604.27058}, 
}

@misc{symft,
      title={{SymFT}: {U}niversal {F}ault-{T}olerant {Q}uantum {C}ircuit {S}imulation via {S}ymbolic {Clifford}--{Pauli} {F}rames and {S}tabilizer {C}oordinates}, 
      author={Wang Fang and Huazhe Lou and Riling Li},
      year={2026},
      eprint={2607.28600},
      archivePrefix={arXiv},
      primaryClass={quant-ph},
      url={https://arxiv.org/abs/2607.28600}, 
}

@misc{ppframework,
      title={{Pauli} {P}ropagation: {A} {C}omputational {F}ramework for {S}imulating {Q}uantum {S}ystems}, 
      author={Manuel S. Rudolph and Tyson Jones and Yanting Teng and Armando Angrisani and Zo{\"e} Holmes},
      year={2026},
      eprint={2505.21606},
      archivePrefix={arXiv},
      primaryClass={quant-ph},
      url={https://arxiv.org/abs/2505.21606}, 
}

@misc{lowesa,
      title={Classical surrogate simulation of quantum systems with {LOWESA}}, 
      author={Manuel S. Rudolph and Enrico Fontana and Zo{\"e} Holmes and Lukasz Cincio},
      year={2023},
      eprint={2308.09109},
      archivePrefix={arXiv},
      primaryClass={quant-ph},
      url={https://arxiv.org/abs/2308.09109}, 
}

@misc{noisylowesa,
      title={Classical simulations of noisy variational quantum circuits}, 
      author={Enrico Fontana and Manuel S. Rudolph and Ross Duncan and Ivan Rungger and Cristina C{\^i}rstoiu},
      year={2023},
      eprint={2306.05400},
      archivePrefix={arXiv},
      primaryClass={quant-ph},
      url={https://arxiv.org/abs/2306.05400}, 
}

@misc{pfsr,
      title={Computing logical error thresholds with the {Pauli} {F}rame {S}parse {R}epresentation}, 
      author={Thomas Tuloup and Thomas Ayral},
      year={2026},
      eprint={2603.14670},
      archivePrefix={arXiv},
      primaryClass={quant-ph},
      url={https://arxiv.org/abs/2603.14670}, 
}

@misc{syqma,
      title={{SyQMA}: A memory-efficient, symbolic and exact universal simulator for quantum error correction}, 
      author={George Umbrarescu and David Amaro},
      year={2026},
      eprint={2604.15043},
      archivePrefix={arXiv},
      primaryClass={quant-ph},
      url={https://arxiv.org/abs/2604.15043}, 
}

@article{wan,
   title={Simulating magic state cultivation with few {Clifford} terms},
   volume={10},
   ISSN={2521-327X},
   url={http://dx.doi.org/10.22331/q-2026-06-12-2134},
   DOI={10.22331/q-2026-06-12-2134},
   journal={Quantum},
   publisher={Verein zur Forderung des Open Access Publizierens in den Quantenwissenschaften},
   author={Wan, Kwok Ho and Zhong, Zhenghao and Zapirain, Ainhoa},
   year={2026},
   month=jun, pages={2134} }

@article{itensor,
	title={{The {IT}ensor {S}oftware {L}ibrary for {T}ensor {N}etwork {C}alculations}},
	author={Matthew Fishman and Steven R. White and E. Miles Stoudenmire},
	journal={SciPost Phys. Codebases},
	pages={4},
	year={2022},
	publisher={SciPost},
	doi={10.21468/SciPostPhysCodeb.4},
	url={https://scipost.org/10.21468/SciPostPhysCodeb.4}
}

@article{Bravyi_2019,
   title={Simulation of quantum circuits by low-rank stabilizer decompositions},
   volume={3},
   ISSN={2521-327X},
   url={http://dx.doi.org/10.22331/q-2019-09-02-181},
   DOI={10.22331/q-2019-09-02-181},
   journal={Quantum},
   publisher={Verein zur Forderung des Open Access Publizierens in den Quantenwissenschaften},
   author={Bravyi, Sergey and Browne, Dan and Calpin, Padraic and Campbell, Earl and Gosset, David and Howard, Mark},
   year={2019},
   month=sep, pages={181} }

@misc{cultivation_google_expt,
      title={Magic state cultivation on a superconducting quantum processor}, 
      author={Emma Rosenfeld and others},
      year={2025},
      eprint={2512.13908},
      archivePrefix={arXiv},
      primaryClass={quant-ph},
      url={https://arxiv.org/abs/2512.13908}, 
}

@unpublished{chan_msc,
  author = {Chan, Tim},
  title = {Private communications},
  note = {In preparation}
}

@misc{tsim,
      title={Tsim: {F}ast {U}niversal {S}imulator for {Q}uantum {E}rror {C}orrection},
      author={Rafael Haenel and Xiuzhe Luo and Chen Zhao},
      year={2026},
      eprint={2604.01059},
      archivePrefix={arXiv},
      primaryClass={quant-ph},
      url={https://arxiv.org/abs/2604.01059},
}

@misc{pyzx_param,
  author       = {Haenel, Rafael},
  title        = {{PyZX-Param}},
  howpublished = {\url{https://github.com/rafaelha/pyzx}},
  year         = {2026},
  note         = {Fork of PyZX based on {ParamZX:} \url{https://github.com/mjsutcliffe99/ParamZX}},
}

@misc{albert2026handbookerrorcorrectingcodes,
      title={Handbook of Error-Correcting Codes}, 
      author={Victor V. Albert and Philippe Faist},
      year={2026},
      eprint={2606.11484},
      archivePrefix={arXiv},
      primaryClass={quant-ph},
      url={https://arxiv.org/abs/2606.11484}, 
}

@misc{exact_msc_code,
  title = {Exact logical error rates for magic state cultivation:
           companion code and data},
  year = {2026},
  url = {https://github.com/kh428/exact_ler_for_msc},
  howpublished = {\url{https://github.com/kh428/exact_ler_for_msc}},
  note = {Companion code to arXiv:2609.18922}
}

@misc{zxflow,
      title={{ZX-F}low: {A} {F}lexible {C}riterion for {D}eterministic {C}omputation with {ZX-D}iagrams}, 
      author={Aleks Kissinger and John van de Wetering},
      year={2026},
      eprint={2603.09580},
      archivePrefix={arXiv},
      primaryClass={quant-ph},
      url={https://arxiv.org/abs/2603.09580}, 
}

@misc{chan2026diagnosingrestoringdegradedfault,
      title={Diagnosing and Restoring the Degraded Fault Distance of Magic State Cultivation}, 
      author={Tim Chan and Armands Strikis and Zhu Sun and Zhenyu Cai},
      year={2026},
      eprint={2609.17706},
      archivePrefix={arXiv},
      primaryClass={quant-ph},
      url={https://arxiv.org/abs/2609.17706}, 
}

@article{fold_msc_prxq,
  title = {Fold-transversal surface code cultivation},
  author = {Sahay, Kaavya and Tsai, Pei-Kai and Chang, Kathleen (Katie) and Su, Qile and Smith, Thomas B. and Singh, Shraddha and Puri, Shruti},
  journal = {PRX Quantum},
  volume = {7},
  issue = {3},
  pages = {033006},
  numpages = {26},
  year = {2026},
  month = {Jul},
  publisher = {American Physical Society},
  doi = {10.1103/gpvl-lg4c},
  url = {https://link.aps.org/doi/10.1103/gpvl-lg4c}
}

@article{RP2_prxq,
  title = {Efficient {M}agic {S}tate {C}ultivation on {${\mathbb{R}\mathbb{P}}^{2}$}},
  author = {Chen, Zi-Han and Chen, Ming-Cheng and Lu, Chao-Yang and Pan, Jian-Wei},
  journal = {PRX Quantum},
  volume = {7},
  issue = {1},
  pages = {010315},
  numpages = {19},
  year = {2026},
  month = {Jan},
  publisher = {American Physical Society},
  doi = {10.1103/9kys-3whh},
  url = {https://link.aps.org/doi/10.1103/9kys-3whh}
}
\appendix
\section{Circuit, code and post-selection conventions}\label{app:circuits}
This appendix specifies the circuits, code conventions and the post-selection criteria needed to reproduce the numbers in the main text.
The $d=5$ circuit is the corrected cultivation circuit distributed with \href{https://github.com/unitaryfoundation/clifft/blob/main/tests/fixtures/cultivation_d5.stim}{Clifft} and \href{https://github.com/haoliri0/SymFT_Test/blob/main/benchmark/circuit/msc_d5_inject_cultivate_p1e-3.stim}{SymFT}. The standalone $d=3$ circuit is the \href{https://github.com/unitaryfoundation/clifft/blob/main/docs/guide/circuits/circuit_d3_t_gate_p0.001.stim}{Clifft example}, with its \href{https://github.com/haoliri0/SymFT_Test/blob/main/benchmark/circuit/msc_d3_inject_cultivate_p1e-3.stim}{SymFT counterpart}.\footnote{The exact $d=5$ file used has SHA-256 checksum \nolinkurl{859d925c74712efaae8bab1d944f9e5be4e042e226eeafcc6d7bef9d6fbfbae5}. A copy with this checksum is exactly the circuit analysed, byte-for-byte.} The distributed file carries a fixed noise value in its annotations. For each result we replace it, at every location, by the value of $p$ quoted with that result, taken as an exact fraction (for example $p=1/1000$), so that the fixed-$p$ probabilities are rational numbers.

There are 3,564 fault locations in the full $d=5$ cultivation circuit of Refs.~\cite{clifft,soft}. Exactly 1,568 of them have identity response, meaning that none of their outcomes changes $A$ or $B$, so summing over their outcomes gives $(1-p)+p=1$ and they drop out exactly without being set noiseless. The remaining 1,996 consist of
\begin{itemize}
\item injection: 244,
\item $d=3$ double checking: 158,
\item code growth: 908,
\item $d=5$ double checking: 686.
\end{itemize}
The standalone $d=3$ circuit has 518 fault locations, of which 402 have non-identity response.

\subsection{The colour code and the \texorpdfstring{$T$}{T} layers}
For a colour code with face operators $X_f,Z_f$, the terminal code projector and
wrong-outcome projector are
\begin{equation}
 \begin{gathered}
 \Pi_0=\prod_f\frac{I+X_f}{2}\frac{I+Z_f}{2},\\
 \Pi_{\rm bad}=\Pi_0\frac{I-M_L}{2},\qquad
 M_L=\frac{X_L-Y_L}{\sqrt2}.
 \end{gathered}
 \label{eq:projectors}
\end{equation}
In this circuit the target state is $\ket{T_L^\dagger}=(\ket{0_L}+e^{-i\pi/4}\ket{1_L})/\sqrt2$, the $+1$ eigenstate of $M_L$, and the terminal readout that tests it is noiseless.

For $d=3$, order the data sites as $(0,3,7,9,10,12,13)$ and write the parity-check matrix of the $[[7,1,3]]$ colour code as\footnote{The $X$ and $Z$ parity-check matrices are the same.}:
\begin{equation}
 H^{(3)}=\begin{pmatrix}
 1&1&1&0&1&0&0\\
 0&1&1&1&0&1&0\\
 0&0&1&0&1&1&1
 \end{pmatrix}.
 \label{eq:H3}
\end{equation}
Each row gives the support of both an $X$ and a $Z$ face stabiliser.
If $C$ is the row span of $H$, the logical basis is
\begin{equation}
 \ket{0_L}=|C|^{-1/2}\sum_{c\in C}\ket c,\qquad
 \ket{1_L}=X_{\rm all}\ket{0_L}.
 \label{eq:codebasis}
\end{equation}
Here $X_{\rm all}=\prod_{q}X_q$, the product over all data sites, is the transversal $X$, which is $X_L$ for these codes.
Write an $X$ operator on the data qubits as a mask, the binary vector $\boldsymbol a$ of Eq.~\eqref{eq:address} with $a_q=1$ where the operator acts. In the site order $(0,3,7,9,10,12,13)$, for example, $X_7$ has mask $\boldsymbol a=(0,0,1,0,0,0,0)$, the blue face $X_0X_3X_7X_{10}$ has mask $(1,1,1,0,1,0,0)$, the first row of $H^{(3)}$, and $X_L$ has mask $\boldsymbol1=(1,1,1,1,1,1,1)$. The masks of the $X$ stabilisers form the row span $C$ of $H$, and the masks that no $Z$ face detects form its orthogonal complement $C^\perp$. Every face has even weight and even overlap with every other face, so $C$ contains only even-weight masks, while the all-ones mask $\boldsymbol1$, the support of $X_L$, is orthogonal to every face and has odd weight. Hence $C^\perp=C\oplus\spann\{\boldsymbol1\}$: an undetected $X$ mask is either a stabiliser or a stabiliser times $X_L$, and its weight parity, the bit $\boldsymbol1\cdot\boldsymbol a$ of Eq.~\eqref{eq:address}, tells which. The same holds for $Z$ masks with $Z_L$, since the code is self-dual, and Appendix~\ref{app:response} uses this decomposition again.

\begin{table}[htbp]\centering
\captionsetup{position=above,skip=4pt}
\caption{Face supports defining the $d=5$ incidence matrix.}
\label{tab:faces}
\begin{tabular}{rll}\toprule
Face & Colour & $d=5$ source sites\\\midrule
1 & Blue & $0,3,7,11$\\
2 & Blue & $9,13,15,19,21,27$\\
3 & Blue & $17,25,32,36$\\
4 & Red & $3,7,9,13$\\
5 & Red & $15,21,23,29$\\
6 & Red & $19,25,27,32,34,38$\\
7 & Green & $7,11,13,17,19,25$\\
8 & Green & $21,27,29,34$\\
9 & Green & $32,36,38,40$\\\bottomrule
\end{tabular}
\end{table}

\begin{figure}[htbp]\centering
 \begin{adjustbox}{max width=.98\linewidth}\begin{tikzpicture}[font=\small]
\begin{yquant}[operator/separation=6mm,register/separation=5mm]
qubit {$\ket+$} a;
qubits {$\ket{\psi_L}$} d;
box {$D^\dagger$} d;
box {$X_{\rm all}$} d | a;
box {$D$} d;
measure {$X$} a;
output {$+1$} a;
\end{yquant}
\end{tikzpicture}
\qquad $K_+=\dfrac{I+D X_{\rm all}D^\dagger}{2}$\end{adjustbox}
 \caption{The ideal encoded projector associated with double checking.}\label{fig:ideal}\end{figure}
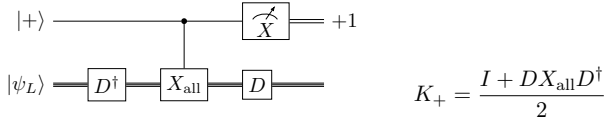

Let $D$ be the exit layer of signed $T$ gates, highlighted in Fig.~\ref{fig:d3circuit}, and $D^\dagger$ the entry layer.
For $d=3$, $D=T^{\otimes7}$. For $d=5$, $D^\dagger$ has $T$ on
sites $0,11,17,19,36,40$ and $T^\dagger$ on the other thirteen data sites,
as drawn in Fig.~\ref{fig:patches}. The encoded check observable is
\begin{equation}
 U=D X_{\rm all}D^\dagger,\qquad K_+=\frac{I+U}{2}.
 \label{eq:idealU}
\end{equation}
Figure~\ref{fig:ideal} is the idealised version of this projector on the code space, with one ancillary qubit and one readout. The double-checking sub-circuits actually used are the CNOT circuits in Figs.~\ref{fig:d3circuit} and~\ref{fig:d5circuit}. The noisy calculation uses those circuits in full, with all their measurement outcomes and fault locations, one fault location at every operation as in the standard circuit-level noise model.

\subsection{Double checking circuits}
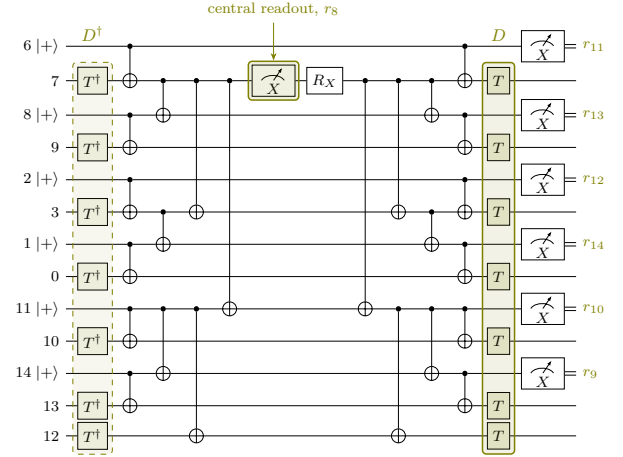
\begin{figure}[htbp]\centering
\begin{adjustbox}{max width=.98\linewidth,max totalheight=.82\textheight}
\begin{tikzpicture}[font=\scriptsize,line width=.45pt]
\begin{yquant}[register/minimum height=1.1mm,register/minimum depth=1.1mm,register/separation=.6mm,operator/minimum width=4mm,operator/separation=2.2mm]
qubit {$6\;\ket+$} q6;
qubit {$7$} q7;
qubit {$8\;\ket+$} q8;
qubit {$9$} q9;
qubit {$2\;\ket+$} q2;
qubit {$3$} q3;
qubit {$1\;\ket+$} q1;
qubit {$0$} q0;
qubit {$11\;\ket+$} q11;
qubit {$10$} q10;
qubit {$14\;\ket+$} q14;
qubit {$13$} q13;
qubit {$12$} q12;
[name=dentry, fill=highlight!8] box {$T^\dagger$} q7,q9,q3,q0,q10,q13,q12;
align q6,q7,q8,q9,q2,q3,q1,q0,q11,q10,q14,q13,q12;
cnot q0 | q1;
cnot q3 | q2;
cnot q7 | q6;
cnot q9 | q8;
cnot q10 | q11;
cnot q13 | q14;
align q6,q7,q8,q9,q2,q3,q1,q0,q11,q10,q14,q13,q12;
cnot q1 | q3;
cnot q8 | q7;
cnot q14 | q11;
align q6,q7,q8,q9,q2,q3,q1,q0,q11,q10,q14,q13,q12;
cnot q3 | q7;
cnot q12 | q11;
align q6,q7,q8,q9,q2,q3,q1,q0,q11,q10,q14,q13,q12;
cnot q11 | q7;
align q6,q7,q8,q9,q2,q3,q1,q0,q11,q10,q14,q13,q12;
[name=m8, type=qubit] measure {$X$} q7;
box {$R_X$} q7;
align q6,q7,q8,q9,q2,q3,q1,q0,q11,q10,q14,q13,q12;
cnot q11 | q7;
align q6,q7,q8,q9,q2,q3,q1,q0,q11,q10,q14,q13,q12;
cnot q3 | q7;
cnot q12 | q11;
align q6,q7,q8,q9,q2,q3,q1,q0,q11,q10,q14,q13,q12;
cnot q1 | q3;
cnot q8 | q7;
cnot q14 | q11;
align q6,q7,q8,q9,q2,q3,q1,q0,q11,q10,q14,q13,q12;
cnot q0 | q1;
cnot q3 | q2;
cnot q7 | q6;
cnot q9 | q8;
cnot q10 | q11;
cnot q13 | q14;
align q6,q7,q8,q9,q2,q3,q1,q0,q11,q10,q14,q13,q12;
[name=dexit, fill=highlight!15] box {$T$} q7,q9,q3,q0,q10,q13,q12;
align q6,q7,q8,q9,q2,q3,q1,q0,q11,q10,q14,q13,q12;
measure {$X$} q14,q11,q6,q2,q8,q1;
output {\textcolor{highlight}{$r_{9}$}} q14;
output {\textcolor{highlight}{$r_{10}$}} q11;
output {\textcolor{highlight}{$r_{11}$}} q6;
output {\textcolor{highlight}{$r_{12}$}} q2;
output {\textcolor{highlight}{$r_{13}$}} q8;
output {\textcolor{highlight}{$r_{14}$}} q1;
\end{yquant}
\node[draw=highlight, thick, fill=highlight, fill opacity=.12, rounded corners=2pt, inner sep=2pt, fit=(m8-0)] (m8box) {};
\node[font=\scriptsize, text=highlight, align=center] (m8lab) at ([yshift=4mm]m8box.north |- current bounding box.north) {central readout, $r_8$};
\draw[-{Stealth[length=1.4mm]}, highlight, line width=.5pt] (m8lab.south) -- (m8box.north);
\node[draw=highlight, thick, fill=highlight, fill opacity=.10, rounded corners=2pt, inner sep=2.6pt, fit=(dexit-0) (dexit-6), label={[font=\footnotesize, text=highlight, yshift=2.5mm]above:$D$}] {};
\node[draw=highlight, dashed, fill=highlight, fill opacity=.06, rounded corners=2pt, inner sep=2.6pt, fit=(dentry-0) (dentry-6), label={[font=\footnotesize, text=highlight, yshift=2.5mm]above:$D^\dagger$}] {};
\end{tikzpicture}
\end{adjustbox}
\caption{The standalone $d=3$ double checking circuit. The entry layer $D^\dagger$ and the exit layer $D$ of $T$ gates are highlighted. Readouts are labelled by binary outcome $r_i$. 
}
\label{fig:d3circuit}
\end{figure}

\begin{figure}[p]\centering
\begin{adjustbox}{max width=.98\linewidth,max totalheight=.88\textheight}
\begin{tikzpicture}[font=\scriptsize,line width=.45pt]
\begin{yquant}[register/minimum height=1.1mm,register/minimum depth=1.1mm,register/separation=.6mm,operator/minimum width=4mm,operator/separation=2.2mm]
qubit {$24\;\ket+$} q24;
qubit {$25$} q25;
qubit {$31\;\ket+$} q31;
qubit {$32$} q32;
qubit {$37\;\ket+$} q37;
qubit {$36$} q36;
qubit {$41\;\ket+$} q41;
qubit {$40$} q40;
qubit {$26\;\ket+$} q26;
qubit {$19$} q19;
qubit {$33\;\ket+$} q33;
qubit {$34$} q34;
qubit {$39\;\ket+$} q39;
qubit {$38$} q38;
qubit {$20\;\ket+$} q20;
qubit {$27$} q27;
qubit {$28\;\ket+$} q28;
qubit {$21$} q21;
qubit {$22\;\ket+$} q22;
qubit {$29$} q29;
qubit {$30\;\ket+$} q30;
qubit {$23$} q23;
qubit {$18\;\ket+$} q18;
qubit {$17$} q17;
qubit {$12\;\ket+$} q12;
qubit {$11$} q11;
qubit {$8\;\ket+$} q8;
qubit {$13$} q13;
qubit {$14\;\ket+$} q14;
qubit {$9$} q9;
qubit {$10\;\ket+$} q10;
qubit {$15$} q15;
qubit {$6\;\ket+$} q6;
qubit {$7$} q7;
qubit {$2\;\ket+$} q2;
qubit {$3$} q3;
qubit {$1\;\ket+$} q1;
qubit {$0$} q0;
box {$T$} q36,q40,q19,q17,q11,q0;
box {$T^\dagger$} q25,q32,q34,q38,q27,q21,q29,q23,q13,q9,q15,q7,q3;
align q24,q25,q31,q32,q37,q36,q41,q40,q26,q19,q33,q34,q39,q38,q20,q27,q28,q21,q22,q29,q30,q23,q18,q17,q12,q11,q8,q13,q14,q9,q10,q15,q6,q7,q2,q3,q1,q0;
cnot q0 | q1;
cnot q3 | q2;
cnot q7 | q6;
cnot q13 | q8;
cnot q15 | q10;
cnot q11 | q12;
cnot q9 | q14;
cnot q17 | q18;
cnot q27 | q20;
cnot q29 | q22;
cnot q25 | q24;
cnot q19 | q26;
cnot q21 | q28;
cnot q23 | q30;
cnot q32 | q31;
cnot q34 | q33;
cnot q36 | q37;
cnot q38 | q39;
cnot q40 | q41;
align q24,q25,q31,q32,q37,q36,q41,q40,q26,q19,q33,q34,q39,q38,q20,q27,q28,q21,q22,q29,q30,q23,q18,q17,q12,q11,q8,q13,q14,q9,q10,q15,q6,q7,q2,q3,q1,q0;
cnot q1 | q3;
cnot q15 | q14;
cnot q30 | q29;
cnot q38 | q33;
cnot q41 | q37;
align q24,q25,q31,q32,q37,q36,q41,q40,q26,q19,q33,q34,q39,q38,q20,q27,q28,q21,q22,q29,q30,q23,q18,q17,q12,q11,q8,q13,q14,q9,q10,q15,q6,q7,q2,q3,q1,q0;
cnot q3 | q7;
cnot q14 | q13;
cnot q33 | q26;
cnot q29 | q28;
cnot q37 | q32;
align q24,q25,q31,q32,q37,q36,q41,q40,q26,q19,q33,q34,q39,q38,q20,q27,q28,q21,q22,q29,q30,q23,q18,q17,q12,q11,q8,q13,q14,q9,q10,q15,q6,q7,q2,q3,q1,q0;
cnot q13 | q12;
cnot q32 | q25;
cnot q28 | q27;
align q24,q25,q31,q32,q37,q36,q41,q40,q26,q19,q33,q34,q39,q38,q20,q27,q28,q21,q22,q29,q30,q23,q18,q17,q12,q11,q8,q13,q14,q9,q10,q15,q6,q7,q2,q3,q1,q0;
cnot q7 | q12;
cnot q27 | q26;
align q24,q25,q31,q32,q37,q36,q41,q40,q26,q19,q33,q34,q39,q38,q20,q27,q28,q21,q22,q29,q30,q23,q18,q17,q12,q11,q8,q13,q14,q9,q10,q15,q6,q7,q2,q3,q1,q0;
cnot q12 | q18;
cnot q26 | q25;
align q24,q25,q31,q32,q37,q36,q41,q40,q26,q19,q33,q34,q39,q38,q20,q27,q28,q21,q22,q29,q30,q23,q18,q17,q12,q11,q8,q13,q14,q9,q10,q15,q6,q7,q2,q3,q1,q0;
cnot q18 | q25;
align q24,q25,q31,q32,q37,q36,q41,q40,q26,q19,q33,q34,q39,q38,q20,q27,q28,q21,q22,q29,q30,q23,q18,q17,q12,q11,q8,q13,q14,q9,q10,q15,q6,q7,q2,q3,q1,q0;
[name=mc, type=qubit] measure {$X$} q25;
box {$R_X$} q25;
align q24,q25,q31,q32,q37,q36,q41,q40,q26,q19,q33,q34,q39,q38,q20,q27,q28,q21,q22,q29,q30,q23,q18,q17,q12,q11,q8,q13,q14,q9,q10,q15,q6,q7,q2,q3,q1,q0;
cnot q18 | q25;
align q24,q25,q31,q32,q37,q36,q41,q40,q26,q19,q33,q34,q39,q38,q20,q27,q28,q21,q22,q29,q30,q23,q18,q17,q12,q11,q8,q13,q14,q9,q10,q15,q6,q7,q2,q3,q1,q0;
cnot q12 | q18;
cnot q26 | q25;
align q24,q25,q31,q32,q37,q36,q41,q40,q26,q19,q33,q34,q39,q38,q20,q27,q28,q21,q22,q29,q30,q23,q18,q17,q12,q11,q8,q13,q14,q9,q10,q15,q6,q7,q2,q3,q1,q0;
cnot q7 | q12;
cnot q27 | q26;
align q24,q25,q31,q32,q37,q36,q41,q40,q26,q19,q33,q34,q39,q38,q20,q27,q28,q21,q22,q29,q30,q23,q18,q17,q12,q11,q8,q13,q14,q9,q10,q15,q6,q7,q2,q3,q1,q0;
cnot q13 | q12;
cnot q32 | q25;
cnot q28 | q27;
align q24,q25,q31,q32,q37,q36,q41,q40,q26,q19,q33,q34,q39,q38,q20,q27,q28,q21,q22,q29,q30,q23,q18,q17,q12,q11,q8,q13,q14,q9,q10,q15,q6,q7,q2,q3,q1,q0;
cnot q3 | q7;
cnot q14 | q13;
cnot q33 | q26;
cnot q29 | q28;
cnot q37 | q32;
align q24,q25,q31,q32,q37,q36,q41,q40,q26,q19,q33,q34,q39,q38,q20,q27,q28,q21,q22,q29,q30,q23,q18,q17,q12,q11,q8,q13,q14,q9,q10,q15,q6,q7,q2,q3,q1,q0;
cnot q1 | q3;
cnot q15 | q14;
cnot q30 | q29;
cnot q38 | q33;
cnot q41 | q37;
align q24,q25,q31,q32,q37,q36,q41,q40,q26,q19,q33,q34,q39,q38,q20,q27,q28,q21,q22,q29,q30,q23,q18,q17,q12,q11,q8,q13,q14,q9,q10,q15,q6,q7,q2,q3,q1,q0;
cnot q0 | q1;
cnot q3 | q2;
cnot q7 | q6;
cnot q13 | q8;
cnot q15 | q10;
cnot q11 | q12;
cnot q9 | q14;
cnot q17 | q18;
cnot q27 | q20;
cnot q29 | q22;
cnot q25 | q24;
cnot q19 | q26;
cnot q21 | q28;
cnot q23 | q30;
cnot q32 | q31;
cnot q34 | q33;
cnot q36 | q37;
cnot q38 | q39;
cnot q40 | q41;
align q24,q25,q31,q32,q37,q36,q41,q40,q26,q19,q33,q34,q39,q38,q20,q27,q28,q21,q22,q29,q30,q23,q18,q17,q12,q11,q8,q13,q14,q9,q10,q15,q6,q7,q2,q3,q1,q0;
box {$T$} q25,q32,q34,q38,q27,q21,q29,q23,q13,q9,q15,q7,q3;
box {$T^\dagger$} q36,q40,q19,q17,q11,q0;
align q24,q25,q31,q32,q37,q36,q41,q40,q26,q19,q33,q34,q39,q38,q20,q27,q28,q21,q22,q29,q30,q23,q18,q17,q12,q11,q8,q13,q14,q9,q10,q15,q6,q7,q2,q3,q1,q0;
measure {$X$} q20,q22,q8,q10,q2,q6,q24,q31,q39,q1,q12,q18,q37,q33,q14,q26,q28,q30,q41;
\end{yquant}
\node[draw=highlight, thick, fill=highlight, fill opacity=.12, rounded corners=2pt, inner sep=2pt, fit=(mc-0), label={[font=\scriptsize, text=highlight, yshift=3.6mm]above:central readout}] {};
\end{tikzpicture}
\end{adjustbox}
\caption{The $d=5$ double checking circuit, with its corrected $T$ pattern. Wires are ordered by the fold-layer components, and four sites idle in this window are not drawn. The central $X$ measurement on site 25 is highlighted.
}
\label{fig:d5circuit}
\end{figure}
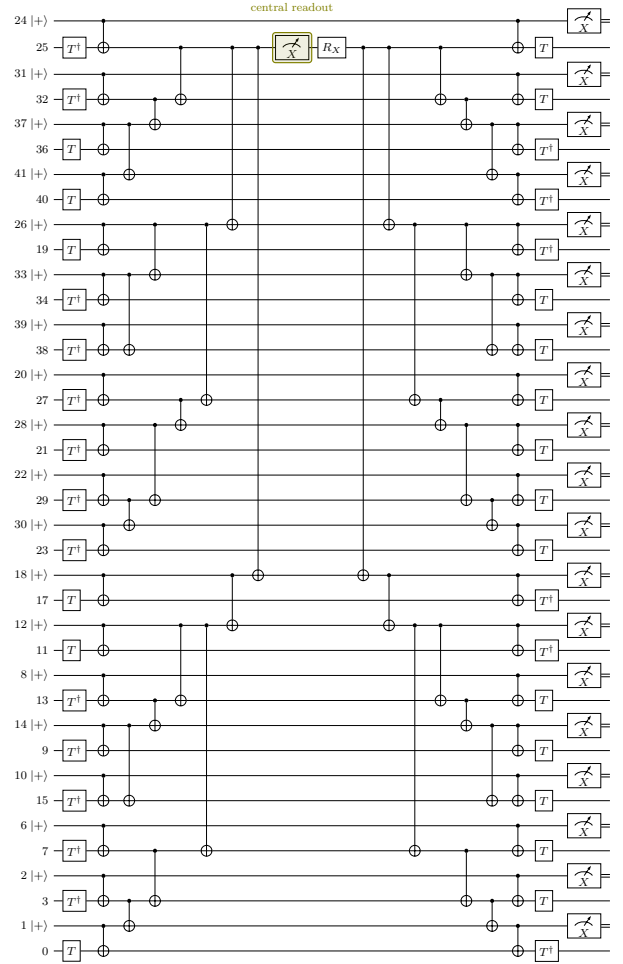

Figures~\ref{fig:d3circuit} and \ref{fig:d5circuit} show the gate sequences of the cultivation double-checking sub-circuits.
In the $d=3$ double-checking circuit, there are four layers of CNOTs, followed by the
central $X$ measurement$+$reset and then reverse unfolding CNOT circuit and final measurements.
Similarly, the $d=5$ double-checking circuit has seven fold layers with varying number of CNOT(s) in each layer followed by similar fold, unfold and measurements.

These double-checking circuits produce many measured bits and post-selection is a set of modulo 2 addition conditions on these measured bits. 
In the $d=3$ sub-circuit, the central readout bit $r_8$ and the auxiliary readout bits enter conditions such as:
\[r_6\oplus r_7\oplus r_8=0 \ , \]
\[r_{11}\oplus r_8=0 \ \text{etc} \ ,\]
these conditions/constraints are known as \textit{detectors}.
Post-selection acceptance occurs when all these detectors return bit 0. 
However, a fault pattern (consisting of many faults) can flips a logical operator but bypasses these detector constraints, as these faults can collectively cancel out in the $\oplus$ addition. 
This leads to a malignant fault pattern that is not caught by post-selection.
This is how the two-fault logical error of Fig.~\ref{fig:d3-fixed-faults} passes post-selection.


\clearpage
\section{An accepted three-fault logical error at $d=5$}
\label{app:d5-fault-v2}
\begin{figure}[tp]\centering
\begin{adjustbox}{max width=.48\textwidth,max totalheight=.87\textheight}
\input{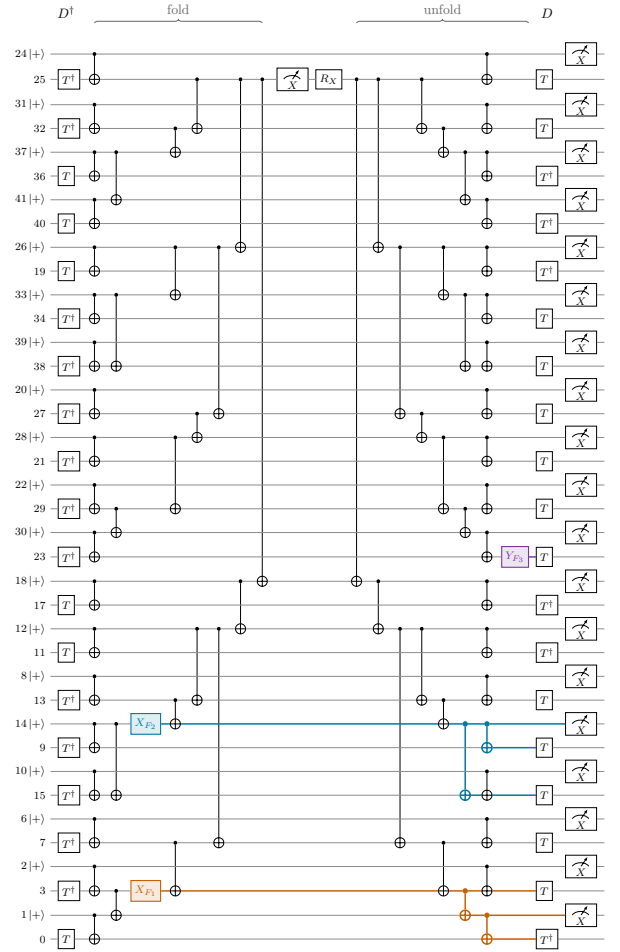}
\end{adjustbox}
\caption{Three faults in the $d=5$ double-checking circuit.
Coloured boxes mark the faults; coloured wires trace their propagation.
Conditioned on this pattern, $A_{\mathcal F}=B_{\mathcal F}=1/16$.}
\label{fig:d5-fixed-faults}
\end{figure}

The $d=5$ double-checking circuit also admits a small fault pattern
that is not caught by post-selection, leading to a logical error in the post-selected state. 
We take the ideal encoded input $\ket{T_L^\dagger}$ and place the three faults shown in
Fig.~\ref{fig:d5-fixed-faults}:
\begin{itemize}
\item $F_1$: $X_3$ after $\text{CNOT}_{3\to1}$ fold,
\item $F_2$: $X_{14}$ after $\text{CNOT}_{14\to15}$ in fold, and
\item $F_3$: $Y_{23}$ after $\text{CNOT}_{30\to23}$ in unfold.
\end{itemize}
No other location is faulty.
Each specified fault is one outcome of a two-qubit depolarising (e.g. $I\otimes X$, $X\otimes Y$ etc)
channel, with probability $p/15$ at its location.

The unfold propagates the first two faults to $X_0X_1X_3$ and
$X_9X_{14}X_{15}$. The ancillary factors $X_1$ and $X_{14}$ commute
with their $X$ readouts. The central and ancillary measurements
all return $+1$. Before the exit layer $D$, the data qubits pick up the error
\begin{equation}
 Q=X_0X_3X_9X_{15}Y_{23}.
 \label{eq:d5-fault-frame-v2}
\end{equation}
The exit gates expand $DQD^\dagger$ into 32 signed Pauli terms.
Only two survive the noiseless code projection $\Pi_0$:
\begin{equation}
 \Pi_0DQD^\dagger\Pi_0 =-\frac{X_L+Y_L}{\sqrt{32}}\Pi_0 =-\frac14N_L\Pi_0,
 \label{eq:d5-fault-projection-v2}
\end{equation}
where the logical operator $N_L=(X_L+Y_L)/\sqrt2$ acts on the nineteen-qubit code.
It anticommutes with $M_L=(X_L-Y_L)/\sqrt2$, so it maps the ideal
$M_L=+1$ magic state to the orthogonal $M_L=-1$ state.
The amplitude factor $1/4$ gives
\begin{equation}
 A_{\mathcal F}=B_{\mathcal F}=\frac1{16},
 \qquad \frac{B_{\mathcal F}}{A_{\mathcal F}}=1.
 \label{eq:d5-fixed-fault-probabilities-v2}
\end{equation}
Thus 1/16 of attempts with this fixed three-fault pattern
pass post-selection, and every accepted output exhibits a logical error.
This is an explicit weight-three failure contributing to the leading
term of Eq.~\eqref{eq:fullseries}. A ZX-calculus contraction of this weight-three fault pattern using techniques from~\cite{wan} further confirms this result.
Hence the $d=5$ cultivation circuit actually has fault distance $d_{\text{fault}} =3$ instead of $5$.

\FloatBarrier

\clearpage
\onecolumn
\newcommand{\matbw}{k=0,\ e_{\rm bit}=0,\ \ell=(0,0,1,0,0,0,0),\qquad M=\begin{pmatrix}1&1&1&1\\1&1&1&1\\1&1&1&1\\1&1&1&1\end{pmatrix},\quad b'=\begin{pmatrix}1\\1\\1\\1\end{pmatrix}}
\newcommand{\solbw}{(0,0,0,1), (0,0,1,0), (0,1,0,0), (0,1,1,1), (1,0,0,0), (1,0,1,1), (1,1,0,1), (1,1,1,0)}
\newcommand{\resAbw}{1/2}\newcommand{\resBbw}{0}
\newcommand{\mattwofault}{k=1,\ e_{\rm bit}=1,\ \ell=(0,0,0,1,0,0,0),\qquad M=\begin{pmatrix}0&1&0&0\\1&0&0&0\\0&0&0&0\\0&0&0&0\end{pmatrix},\quad b'=\begin{pmatrix}0\\1\\0\\0\end{pmatrix}}
\newcommand{\soltwofault}{(1,0,0,0), (1,0,0,1), (1,0,1,0), (1,0,1,1)}
\newcommand{\resAtwofault}{1/4}\newcommand{\resBtwofault}{1/4}
\newcommand{\matxl}{k=1,\ e_{\rm bit}=0,\ \ell=(0,0,0,0,0,0,0),\qquad M=\begin{pmatrix}0&0&0&0\\0&0&0&0\\0&0&0&0\\0&0&0&1\end{pmatrix},\quad b'=\begin{pmatrix}0\\0\\0\\0\end{pmatrix}}
\newcommand{\solxl}{(0,0,0,0), (0,0,1,0), (0,1,0,0), (0,1,1,0), (1,0,0,0), (1,0,1,0), (1,1,0,0), (1,1,1,0)}
\newcommand{\resAxl}{1/2}\newcommand{\resBxl}{0}
\newcommand{\matxlminus}{k=1,\ e_{\rm bit}=1,\ \ell=(0,0,0,0,0,0,0),\qquad M=\begin{pmatrix}0&0&0&0\\0&0&0&0\\0&0&0&0\\0&0&0&1\end{pmatrix},\quad b'=\begin{pmatrix}0\\0\\0\\1\end{pmatrix}}
\newcommand{\solxlminus}{(0,0,0,1), (0,0,1,1), (0,1,0,1), (0,1,1,1), (1,0,0,1), (1,0,1,1), (1,1,0,1), (1,1,1,1)}
\newcommand{\resAxlminus}{1/2}\newcommand{\resBxlminus}{1/2}
\newcommand{\matzs}{k=0,\ e_{\rm bit}=1,\ \ell=(0,0,1,0,0,0,0),\qquad M=\begin{pmatrix}0&0&0&0\\0&0&0&0\\0&0&0&0\\0&0&0&0\end{pmatrix},\quad b'=\begin{pmatrix}1\\1\\1\\0\end{pmatrix}}
\newcommand{\solzs}{}
\newcommand{\resAzs}{0}\newcommand{\resBzs}{0}
\newcommand{\scanexit}{210}\newcommand{\scanentry}{210}\newcommand{\scanincoming}{210}\newcommand{\scancases}{210}
\newcommand{\Hthree}{\begin{pmatrix}1&1&1&0&1&0&0\\0&1&1&1&0&1&0\\0&0&1&0&1&1&1\end{pmatrix}}
\newcommand{\Gfour}{\begin{pmatrix}1&1&1&0&1&0&0\\0&1&1&1&0&1&0\\0&0&1&0&1&1&1\\1&1&1&1&1&1&1\end{pmatrix}}
\newcommand{\masksbw}{\begin{pmatrix}0&0&1&0&0&0&0\\0&0&0&0&0&0&0\\0&0&1&0&0&0&0\\0&0&0&0&0&0&0\end{pmatrix}}
\newcommand{\interbw}{\begin{pmatrix}1&1&1&1\\1&1&1&1\\1&1&1&1\\1&1&1&1\end{pmatrix}}
\newcommand{\elloverbw}{(1,1,1,1)}
\newcommand{\augbw}{\left(\begin{array}{cccc|c}1&1&1&1&1\\1&1&1&1&1\\1&1&1&1&1\\1&1&1&1&1\end{array}\right)}
\newcommand{\rrefbw}{\left(\begin{array}{cccc|c}1&1&1&1&1\\0&0&0&0&0\\0&0&0&0&0\\0&0&0&0&0\end{array}\right)}
\newcommand{\rankbw}{1}\newcommand{\consbw}{consistent}
\newcommand{\phasecountsbw}{8/8/0/0}
\newcommand{\gsumbw}{0.5+0.5i}\newcommand{\gabsbw}{0.5}
\newcommand{\synbw}{(0,0,0)}
\newcommand{\maskstwofault}{\begin{pmatrix}0&0&0&0&0&0&0\\0&0&0&0&0&0&0\\1&1&0&1&0&0&0\\0&0&0&1&0&0&0\end{pmatrix}}
\newcommand{\intertwofault}{\begin{pmatrix}2&1&0&2\\1&2&0&2\\0&0&0&0\\2&2&0&3\end{pmatrix}}
\newcommand{\ellovertwofault}{(0,1,0,1)}
\newcommand{\augtwofault}{\left(\begin{array}{cccc|c}0&1&0&0&0\\1&0&0&0&1\\0&0&0&0&0\\0&0&0&0&0\end{array}\right)}
\newcommand{\rreftwofault}{\left(\begin{array}{cccc|c}1&0&0&0&1\\0&1&0&0&0\\0&0&0&0&0\\0&0&0&0&0\end{array}\right)}
\newcommand{\ranktwofault}{2}\newcommand{\constwofault}{consistent}
\newcommand{\phasecountstwofault}{12/0/4/0}
\newcommand{\gsumtwofault}{0.5}\newcommand{\gabstwofault}{0.25}
\newcommand{\syntwofault}{(0,0,0)}
\newcommand{\masksxl}{\begin{pmatrix}1&1&1&1&1&1&1\\0&0&0&0&0&0&0\\0&0&0&0&0&0&0\\0&0&0&0&0&0&0\end{pmatrix}}
\newcommand{\interxl}{\begin{pmatrix}0&0&0&0\\0&0&0&0\\0&0&0&0\\0&0&0&0\end{pmatrix}}
\newcommand{\elloverxl}{(0,0,0,0)}
\newcommand{\augxl}{\left(\begin{array}{cccc|c}0&0&0&0&0\\0&0&0&0&0\\0&0&0&0&0\\0&0&0&1&0\end{array}\right)}
\newcommand{\rrefxl}{\left(\begin{array}{cccc|c}0&0&0&1&0\\0&0&0&0&0\\0&0&0&0&0\\0&0&0&0&0\end{array}\right)}
\newcommand{\rankxl}{1}\newcommand{\consxl}{consistent}
\newcommand{\phasecountsxl}{8/0/0/8}
\newcommand{\gsumxl}{0.5-0.5i}\newcommand{\gabsxl}{0.5}
\newcommand{\synxl}{(0,0,0)}
\newcommand{\maskszs}{\begin{pmatrix}0&0&0&0&0&0&0\\0&0&0&0&0&0&0\\0&0&0&0&0&0&0\\0&0&1&0&0&0&0\end{pmatrix}}
\newcommand{\interzs}{\begin{pmatrix}0&0&0&0\\0&0&0&0\\0&0&0&0\\0&0&0&0\end{pmatrix}}
\newcommand{\elloverzs}{(1,1,1,1)}
\newcommand{\augzs}{\left(\begin{array}{cccc|c}0&0&0&0&1\\0&0&0&0&1\\0&0&0&0&1\\0&0&0&0&0\end{array}\right)}
\newcommand{\rrefzs}{\left(\begin{array}{cccc|c}0&0&0&0&1\\0&0&0&0&1\\0&0&0&0&1\\0&0&0&0&0\end{array}\right)}
\newcommand{\rankzs}{0}\newcommand{\conszs}{inconsistent}
\newcommand{\phasecountszs}{8/0/8/0}
\newcommand{\gsumzs}{0}\newcommand{\gabszs}{0}
\newcommand{\synzs}{(0,0,0)}
\newcommand{\masksxseven}{\begin{pmatrix}0&0&1&0&0&0&0\\0&0&0&0&0&0&0\\0&0&0&0&0&0&0\\0&0&0&0&0&0&0\end{pmatrix}}
\newcommand{\interxseven}{\begin{pmatrix}0&0&0&0\\0&0&0&0\\0&0&0&0\\0&0&0&0\end{pmatrix}}
\newcommand{\synxseven}{(1,1,1)}
\newcommand{\bmattwofault}{\begin{pmatrix}0&-0.3536+0.3536i\\-0.3536-0.3536i&0\end{pmatrix}}
\newcommand{\binsttwofault}{(0.25,0.25)}
\newcommand{\bmatzs}{\begin{pmatrix}0&0\\0&0\end{pmatrix}}
\newcommand{\binstzs}{(0,0)}
\newcommand{\bmatident}{\begin{pmatrix}1&0\\0&1\end{pmatrix}}
\newcommand{\binstident}{(1,0)}
\newcommand{\dfivephase}{7}
\newcommand{\dfivesites}{0,3,7,9,11,13,15,17,19,21,23,25,27,29,32,34,36,38,40}
\newcommand{\dfivegrid}{{1,1,1,0,1,0,0,0,0,0,0,0,0,0,0,0,0,0,0},{0,0,0,1,0,1,1,0,1,1,0,0,1,0,0,0,0,0,0},{0,0,0,0,0,0,0,1,0,0,0,1,0,0,1,0,1,0,0},{0,1,1,1,0,1,0,0,0,0,0,0,0,0,0,0,0,0,0},{0,0,0,0,0,0,1,0,0,1,1,0,0,1,0,0,0,0,0},{0,0,0,0,0,0,0,0,1,0,0,1,1,0,1,1,0,1,0},{0,0,1,0,1,1,0,1,1,0,0,1,0,0,0,0,0,0,0},{0,0,0,0,0,0,0,0,0,1,0,0,1,1,0,1,0,0,0},{0,0,0,0,0,0,0,0,0,0,0,0,0,0,1,0,1,1,1},{1,1,1,1,1,1,1,1,1,1,1,1,1,1,1,1,1,1,1},{1,0,0,0,1,0,0,1,1,0,0,0,0,0,0,0,1,0,1}}
\newcommand{\dfivesw}{(0,4,0,4,4,4,0,4,0)}
\newcommand{\dfiveuniform}{rejected: The requested check pattern does not preserve the code}
\newcommand{\cperpgrid}{{0,0,0,0,0,0,0},{0,0,1,0,1,1,1},{0,1,1,1,0,1,0},{0,1,0,1,1,0,1},{1,1,1,0,1,0,0},{1,1,0,0,0,1,1},{1,0,0,1,1,1,0},{1,0,1,1,0,0,1},{1,1,1,1,1,1,1},{1,1,0,1,0,0,0},{1,0,0,0,1,0,1},{1,0,1,0,0,1,0},{0,0,0,1,0,1,1},{0,0,1,1,1,0,0},{0,1,1,0,0,0,1},{0,1,0,0,1,1,0}}
\newcommand{\cperplabels}{0000,0010,0100,0110,1000,1010,1100,1110,0001,0011,0101,0111,1001,1011,1101,1111}
\newcommand{\HthreeC}{\begin{pmatrix}\textcolor{tnblue}{1}&\textcolor{tnblue}{1}&\textcolor{tnblue}{1}&0&\textcolor{tnblue}{1}&0&0\\0&\textcolor{red!80!black}{1}&\textcolor{red!80!black}{1}&\textcolor{red!80!black}{1}&0&\textcolor{red!80!black}{1}&0\\0&0&\textcolor{tngreen}{1}&0&\textcolor{tngreen}{1}&\textcolor{tngreen}{1}&\textcolor{tngreen}{1}\end{pmatrix}}
\newcommand{\GfourC}{\begin{pmatrix}\textcolor{tnblue}{1}&\textcolor{tnblue}{1}&\textcolor{tnblue}{1}&0&\textcolor{tnblue}{1}&0&0\\0&\textcolor{red!80!black}{1}&\textcolor{red!80!black}{1}&\textcolor{red!80!black}{1}&0&\textcolor{red!80!black}{1}&0\\0&0&\textcolor{tngreen}{1}&0&\textcolor{tngreen}{1}&\textcolor{tngreen}{1}&\textcolor{tngreen}{1}\\1&1&1&1&1&1&1\end{pmatrix}}
\newcommand{\masksCbw}{\begin{pmatrix}0&0&\colorbox{tnblue!22}{$1$}&0&0&0&0\\0&0&0&0&0&0&0\\0&0&\colorbox{tnblue!22}{$1$}&0&0&0&0\\0&0&0&0&0&0&0\end{pmatrix}}
\newcommand{\interCbw}{\begin{pmatrix}\colorbox{tnblue!22}{$1$}&\colorbox{tnblue!22}{$1$}&\colorbox{tnblue!22}{$1$}&\colorbox{tnblue!22}{$1$}\\\colorbox{tnblue!22}{$1$}&\colorbox{tnblue!22}{$1$}&\colorbox{tnblue!22}{$1$}&\colorbox{tnblue!22}{$1$}\\\colorbox{tnblue!22}{$1$}&\colorbox{tnblue!22}{$1$}&\colorbox{tnblue!22}{$1$}&\colorbox{tnblue!22}{$1$}\\\colorbox{tnblue!22}{$1$}&\colorbox{tnblue!22}{$1$}&\colorbox{tnblue!22}{$1$}&\colorbox{tnblue!22}{$1$}\end{pmatrix}}
\newcommand{\MCbw}{\begin{pmatrix}\colorbox{tnblue!22}{$1$}&\colorbox{tnblue!22}{$1$}&\colorbox{tnblue!22}{$1$}&\colorbox{tnblue!22}{$1$}\\\colorbox{tnblue!22}{$1$}&\colorbox{tnblue!22}{$1$}&\colorbox{tnblue!22}{$1$}&\colorbox{tnblue!22}{$1$}\\\colorbox{tnblue!22}{$1$}&\colorbox{tnblue!22}{$1$}&\colorbox{tnblue!22}{$1$}&\colorbox{tnblue!22}{$1$}\\\colorbox{tnblue!22}{$1$}&\colorbox{tnblue!22}{$1$}&\colorbox{tnblue!22}{$1$}&\fcolorbox{tngreen}{tngreen!18}{$1$}\end{pmatrix}}
\newcommand{\bpCbw}{\begin{pmatrix}\colorbox{tnorange!22}{$1$}\\\colorbox{tnorange!22}{$1$}\\\colorbox{tnorange!22}{$1$}\\\fcolorbox{tngreen}{tngreen!18}{$1$}\end{pmatrix}}
\newcommand{\augCbw}{\left(\begin{array}{cccc|c}\colorbox{black!22}{$1$}&1&1&1&\colorbox{tnorange!22}{$1$}\\\colorbox{black!22}{$1$}&1&1&1&\colorbox{tnorange!22}{$1$}\\\colorbox{black!22}{$1$}&1&1&1&\colorbox{tnorange!22}{$1$}\\\colorbox{black!22}{$1$}&1&1&1&\colorbox{tnorange!22}{$1$}\end{array}\right)}
\newcommand{\rrefCbw}{\left(\begin{array}{cccc|c}\colorbox{black!22}{$1$}&1&1&1&\colorbox{tnorange!22}{$1$}\\0&0&0&0&0\\0&0&0&0&0\\0&0&0&0&0\end{array}\right)}
\newcommand{\pivotsbw}{h_0}
\newcommand{\masksCtwofault}{\begin{pmatrix}0&0&0&0&0&0&0\\0&0&0&0&0&0&0\\\colorbox{tnblue!22}{$1$}&\colorbox{tnblue!22}{$1$}&0&\colorbox{tnblue!22}{$1$}&0&0&0\\0&0&0&\colorbox{tnorange!22}{$1$}&0&0&0\end{pmatrix}}
\newcommand{\interCtwofault}{\begin{pmatrix}2&\colorbox{tnblue!22}{$1$}&0&2\\\colorbox{tnblue!22}{$1$}&2&0&2\\0&0&0&0\\2&2&0&\colorbox{tnblue!22}{$3$}\end{pmatrix}}
\newcommand{\MCtwofault}{\begin{pmatrix}0&\colorbox{tnblue!22}{$1$}&0&0\\\colorbox{tnblue!22}{$1$}&0&0&0\\0&0&0&0\\0&0&0&\fcolorbox{tngreen}{tngreen!18}{$0$}\end{pmatrix}}
\newcommand{\bpCtwofault}{\begin{pmatrix}0\\\colorbox{tnorange!22}{$1$}\\0\\\fcolorbox{tngreen}{tngreen!18}{$0$}\end{pmatrix}}
\newcommand{\augCtwofault}{\left(\begin{array}{cccc|c}0&\colorbox{black!22}{$1$}&0&0&0\\\colorbox{black!22}{$1$}&0&0&0&\colorbox{tnorange!22}{$1$}\\0&0&0&0&0\\0&0&0&0&0\end{array}\right)}
\newcommand{\rrefCtwofault}{\left(\begin{array}{cccc|c}\colorbox{black!22}{$1$}&0&0&0&\colorbox{tnorange!22}{$1$}\\0&\colorbox{black!22}{$1$}&0&0&0\\0&0&0&0&0\\0&0&0&0&0\end{array}\right)}
\newcommand{\pivotstwofault}{h_0, h_1}
\newcommand{\masksCxl}{\begin{pmatrix}\colorbox{tnblue!22}{$1$}&\colorbox{tnblue!22}{$1$}&\colorbox{tnblue!22}{$1$}&\colorbox{tnblue!22}{$1$}&\colorbox{tnblue!22}{$1$}&\colorbox{tnblue!22}{$1$}&\colorbox{tnblue!22}{$1$}\\0&0&0&0&0&0&0\\0&0&0&0&0&0&0\\0&0&0&0&0&0&0\end{pmatrix}}
\newcommand{\interCxl}{\begin{pmatrix}0&0&0&0\\0&0&0&0\\0&0&0&0\\0&0&0&0\end{pmatrix}}
\newcommand{\MCxl}{\begin{pmatrix}0&0&0&0\\0&0&0&0\\0&0&0&0\\0&0&0&\fcolorbox{tngreen}{tngreen!18}{$1$}\end{pmatrix}}
\newcommand{\bpCxl}{\begin{pmatrix}0\\0\\0\\\fcolorbox{tngreen}{tngreen!18}{$0$}\end{pmatrix}}
\newcommand{\augCxl}{\left(\begin{array}{cccc|c}0&0&0&0&0\\0&0&0&0&0\\0&0&0&0&0\\0&0&0&\colorbox{black!22}{$1$}&0\end{array}\right)}
\newcommand{\rrefCxl}{\left(\begin{array}{cccc|c}0&0&0&\colorbox{black!22}{$1$}&0\\0&0&0&0&0\\0&0&0&0&0\\0&0&0&0&0\end{array}\right)}
\newcommand{\pivotsxl}{h_3}
\newcommand{\masksCzs}{\begin{pmatrix}0&0&0&0&0&0&0\\0&0&0&0&0&0&0\\0&0&0&0&0&0&0\\0&0&\colorbox{tnorange!22}{$1$}&0&0&0&0\end{pmatrix}}
\newcommand{\interCzs}{\begin{pmatrix}0&0&0&0\\0&0&0&0\\0&0&0&0\\0&0&0&0\end{pmatrix}}
\newcommand{\MCzs}{\begin{pmatrix}0&0&0&0\\0&0&0&0\\0&0&0&0\\0&0&0&\fcolorbox{tngreen}{tngreen!18}{$0$}\end{pmatrix}}
\newcommand{\bpCzs}{\begin{pmatrix}\colorbox{tnorange!22}{$1$}\\\colorbox{tnorange!22}{$1$}\\\colorbox{tnorange!22}{$1$}\\\fcolorbox{tngreen}{tngreen!18}{$0$}\end{pmatrix}}
\newcommand{\augCzs}{\left(\begin{array}{cccc|c}0&0&0&0&\colorbox{tnorange!22}{$1$}\\0&0&0&0&\colorbox{tnorange!22}{$1$}\\0&0&0&0&\colorbox{tnorange!22}{$1$}\\0&0&0&0&0\end{array}\right)}
\newcommand{\rrefCzs}{\left(\begin{array}{cccc|c}0&0&0&0&\colorbox{tnorange!22}{$1$}\\0&0&0&0&\colorbox{tnorange!22}{$1$}\\0&0&0&0&\colorbox{tnorange!22}{$1$}\\0&0&0&0&0\end{array}\right)}
\newcommand{\pivotszs}{none}
\newcommand{\masksCxseven}{\begin{pmatrix}0&0&\colorbox{tnblue!22}{$1$}&0&0&0&0\\0&0&0&0&0&0&0\\0&0&0&0&0&0&0\\0&0&0&0&0&0&0\end{pmatrix}}
\newcommand{\interCxseven}{\begin{pmatrix}0&0&0&0\\0&0&0&0\\0&0&0&0\\0&0&0&0\end{pmatrix}}

\section{An explicit double-checking response}\label{app:response}

The main text needs two tables per double-checking stage, $F_A(\alpha,\beta;p)$ and $F_B(\alpha,\beta;p)$, indexed by the incoming error $(\alpha,\beta)$. $F_A$ is the probability that the stage passes post-selection, and $F_B$ the probability that it passes with a logical error left uncaught.
The open network of Sec.~\ref{app:B-open} computes the whole table at once. This appendix works through smaller examples explicitly. It gives a closed form for \emph{one} entry, at a fixed incoming error and a fixed pattern of internal faults, and then shows how averaging over the internal faults builds the table.
Every example uses the $d=3$ cultivation circuit in full, with every matrix written out, and Sec.~\ref{app:B-dfive} explains changes needed for the $d=5$ circuit (only size changes).
Ultimately, the use of these tables is to perform the contraction of Fig.~\ref{fig:response}(b), drawn here with its summation written out:
\begin{center}
\begin{tikzpicture}[x=1cm,y=1cm]
\node[font=\Large] at (-1.4,0) {$\displaystyle\sum_{\alpha,\beta}$};
\node[tn noise,minimum width=11mm,minimum height=13mm,font=\large] (mu) at (0,0) {$\mu$};
\node[tn endpoint,minimum width=12mm,minimum height=13mm,font=\large] (F) at (2.4,0) {$F_O$};
\draw[tn bundle] ([yshift=3mm]mu.east)--node[above,tn label] {$\alpha$}([yshift=3mm]F.west);
\draw[tn bundle] ([yshift=-3mm]mu.east)--node[below,tn label] {$\beta$}([yshift=-3mm]F.west);
\node[font=\large] at (3.7,0) {$=$};
\node[tn intermediate,minimum width=9mm,minimum height=9mm,font=\large] (O) at (4.8,0) {$O$};
\node[tn label,scale = 1.08] at (1.6,-1.50) {$\displaystyle\sum_{\alpha,\beta}\mu(\alpha,\beta;p)\,F_O(\alpha,\beta;p)=O(p)$ \ , \\ with $O=A$ or $B$ \ .};
\end{tikzpicture}
\end{center}
At $d=3$ both tensors ($\mu$ and $F_O$) are $16\times16$ tables of probabilities, $\alpha$ and $\beta$ are four bits each, and the tensor contraction above is a sum of 256 products. 
Note that tensors here follow the conventions of Sec.~\ref{sec:tensor-contraction}. 
We will now go through the notation used throughout this appendix section, then derive one entry of the table (Secs.~\ref{app:B-window} to~\ref{app:B-checks}), before averaging over the internal faults to fill the whole table (Secs.~\ref{app:response-average} and~\ref{app:B-open}), and finally, talk about the $d=5$ extension (Sec.~\ref{app:B-dfive}).

\subsection{Notation, all of it}\label{app:B-notation}

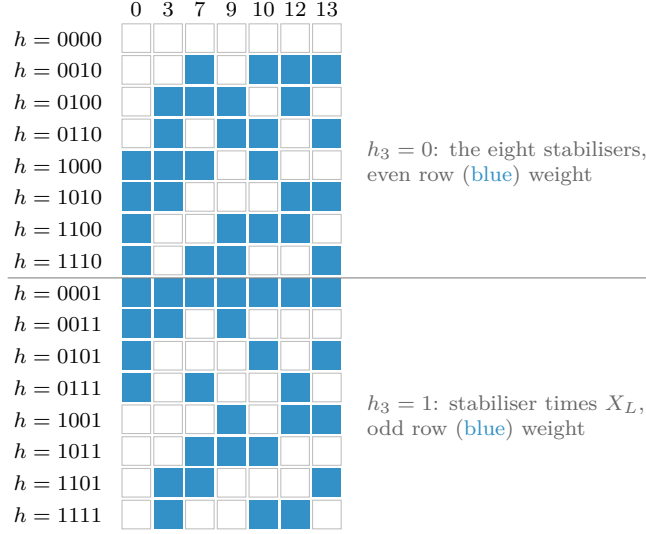
\begin{figure}[H]\centering
\begin{tikzpicture}[x=4.2mm,y=4.2mm]
\foreach \row [count=\r from 0] in \cperpgrid {
  \foreach \bit [count=\c from 0] in \row {
    \ifnum\bit=1 \fill[tnblue!80] (\c,-\r) rectangle ++(0.9,0.9); \else \draw[black!25] (\c,-\r) rectangle ++(0.9,0.9); \fi
  }
}
\foreach \site [count=\c from 0] in {0,3,7,9,10,12,13} {\node[font=\scriptsize] at (\c+0.45,1.35) {$\site$};}
\foreach \h [count=\r from 0] in \cperplabels {\node[font=\scriptsize,anchor=east] at (-0.3,-\r+0.45) {$h=\h$};}
\draw[black!50] (-3.6,-7.1) -- (18.3,-7.1);
\node[font=\scriptsize,anchor=west,align=left,text=black!60] at (7.4,-3.5) {$h_3=0$: the eight stabilisers, \\even row ({\color{tnblue!80}blue}) weight};
\node[font=\scriptsize,anchor=west,align=left,text=black!60] at (7.4,-11.5) {$h_3=1$: stabiliser times $X_L$, \\odd row ({\color{tnblue!80}blue}) weight};
\end{tikzpicture}
\caption{The sixteen masks of $C^\perp$ at $d=3$, one row per coordinate vector $h=h_0h_1h_2h_3$, a filled square where the mask has a 1. The top half is $C$, the bottom half is $C\oplus\boldsymbol1$.}
\label{fig:cperp}
\end{figure}

\paragraph{Sites, masks, faces.} For the $d=3$ colour code, the seven data sites are ordered $(0,3,7,9,10,12,13)$. A Pauli string on the data is written as two \emph{masks}, binary vectors with one bit per site,
\[
P(\boldsymbol a,\boldsymbol b)\ \propto\ \prod_q X_q^{a_q}Z_q^{b_q},\qquad
a_q=1 \text{ where the string has } X \text{ or } Y,\qquad b_q=1 \text{ where it has } Z \text{ or } Y,
\]
so $X_7$ maps to $\boldsymbol a=(0,0,1,0,0,0,0)$, $\boldsymbol b=\boldsymbol0$, and $Y_9$ has $\boldsymbol a=\boldsymbol b=(0,0,0,1,0,0,0)$. 
The overall phase of a Pauli string is dropped. This simplification is benign as each fault pattern is one fixed operator and the probability is invariant to a global phase change. 
The relative signs between terms of the coherent sum below are retained\footnote{Each Pauli of a fault pattern is an operator $e^{i\varphi}X^{\boldsymbol a}Z^{\boldsymbol b}$ inserted at a fixed place, and a record amplitude is linear in each of them, so these phases multiply the whole amplitude and cancel in $|{\cal G}|^2$. The sum ${\cal G}=2^{-m}\sum_h i^{f(h)}$ of Eq.~\eqref{eq:gauss} is different: its terms run over the masks $v(h)\in C^\perp$, and $f(h)$ gives them different phases $\pm1,\pm i$ that interfere.}.
The three faces of the $[[7,1,3]]$ colour code are the rows of $H^{(3)}$ in Eq.~\eqref{eq:H3}:
\[
\begin{aligned}
g_0&=(1,1,1,0,1,0,0)=\{0,3,7,10\}\ \text{(blue)},\\
g_1&=(0,1,1,1,0,1,0)=\{3,7,9,12\}\ \text{(red)},\\
g_2&=(0,0,1,0,1,1,1)=\{7,10,12,13\}\ \text{(green)} .
\end{aligned}
\]
Each face is both an $X$ stabiliser and a $Z$ stabiliser. 
Two sets built from the faces recur everywhere: 
\begin{enumerate}
  \item $C=\spann\{g_0,g_1,g_2\}$, the eight $X$-stabiliser masks, and 
  \item $C^\perp$, the masks with even overlap with every face, that is the $X$ masks no $Z$ face detects. 
\end{enumerate}
For this code $C^\perp=C\oplus\spann\{\boldsymbol1\}$ with $\boldsymbol1=(1,\ldots,1)$ being the support of $X_L$. 
We use the four generators $g_0,g_1,g_2$ and $g_3=\boldsymbol1$ of $C^\perp$, write $r=3$ for the number of faces and $m=r+1=4$. 
As matrices, one column per site,
\[
H^{(3)}=\HthreeC,\qquad
\begin{pmatrix}g_0\\g_1\\g_2\\g_3\end{pmatrix}=\GfourC ,
\]
ones coloured by face, logical row in black. A bit vector $h=(h_0,h_1,h_2,h_3)$ names the mask $v(h)=h_0g_0\oplus h_1g_1\oplus h_2g_2\oplus h_3\boldsymbol1$. The four generators are independent, so the sixteen $h$ are exactly the sixteen masks of $C^\perp$, and $h_3$ says whether the mask lies in the logical coset. Figure~\ref{fig:cperp} draws all sixteen masks.

In short, an error is nothing but a pair of bit masks, and the sixteen masks of $C^\perp$ are the only $X$ errors the code cannot see, which is exactly the set the counting formula below runs over. Before carryin on, we will list out the notation in a table, which will be use throughout this appendix section.

\paragraph{Symbols used below.}
\begin{center}\small
\begin{tabular}{ll}\toprule
symbol & meaning \\\midrule
$\oplus$ & bitwise XOR of two masks, or addition mod 2 of two bits \\
$\wedge$ & bitwise AND of two masks (entrywise product) \\
$\boldsymbol a\cdot\boldsymbol b$ & $\sum_q a_qb_q \bmod 2$, the parity of the overlap \\
$|\boldsymbol a|$ & the number of ones in $\boldsymbol a$, its weight \\
$\ind{\text{cond}}$ & $1$ if the condition holds, $0$ otherwise \\
$\#\{h\in\F^m: Mh=b'\}$ & the number of bit vectors $h$ that solve the system, $\F=\{0,1\}$ \\
$\epsilon_q=\pm1$ & the exponent of $T$ on site $q$ in the exit layer $D=\prod_qT_q^{\epsilon_q}$, all $+1$ at $d=3$ \\
$w_\epsilon(\boldsymbol v)=\sum_q\epsilon_qv_q$ & signed weight: ones of $\boldsymbol v$ on $T$ sites minus ones on $T^\dagger$ sites, $=|\boldsymbol v|$ at $d=3$ \\
$t=\sum_q\epsilon_q\bmod8$ & $7$ for both circuits \\
$s=\pm1$ & the outcome of the central $X$ measurement, bit $r_8$ in Fig.~\ref{fig:d3circuit}, $s=(-1)^{r_8}$ \\
$H\boldsymbol a$ & the $Z$-check syndrome of an $X$ mask, the three parities $g_i\cdot\boldsymbol a$ \\
\bottomrule
\end{tabular}
\end{center}

\paragraph{Coloured boxes in the matrices.} This boxed colour scheme in matrices applies to this appendix only, the later appendices state their own where they start. Every highlighted matrix below is generated from the same computation, and the colours follow one rule. \colorbox{tnblue!22}{Blue} marks anything that comes from an $X$ mask: the masks $\boldsymbol a,\boldsymbol u$, the intersection counts $|\boldsymbol u\wedge g_i\wedge g_j|$, and the ones of $M$. \colorbox{tnorange!22}{Orange} marks anything that comes from a $Z$ mask: the masks $\boldsymbol b,\boldsymbol z$, the overlaps $|\ell\wedge g_i|$, and the ones of $b'$. A \fcolorbox{tngreen}{tngreen!18}{green box} marks the logical slot, index $r=3$, the only place where $k$ and $e_{\rm bit}$ enter. In row reductions, \colorbox{black!22}{grey} marks the pivot columns, the bits of $h$ the system determines. Reading rule: blue decides the shape of the constraints, orange decides whether they can be satisfied, green is where the logical information lives.

\subsection{Where the two Paulis sit}\label{app:B-window}

\begin{figure}[H]\centering
\begin{tikzpicture}[font=\small]
\begin{yquant}[operator/separation=3.2mm,register/separation=6mm]
qubits {data} d;
qubits {anc.} c;
[name=dentry] box {$D^\dagger$} d;
[name=fold] box {fold CX} (d,c);
[name=mx] box {$M_X$} c;
[name=unfold] box {unfold CX} (d,c);
[name=dexit] box {$D$} d;
align -;
hspace {2.5mm} -;
[name=meas] measure c;
\end{yquant}
\node[draw=black!40,fill=black!10,fill opacity=.35,rounded corners=3pt,inner sep=2mm,fit=(fold-0) (unfold-0)] (cliff) {};
\node[note,text=black!60,above=0.5mm of cliff.north] {Clifford part};
\foreach \n/\t in {dentry/135, fold/138--150, mx/{154, $r_8$}, unfold/161--173, dexit/177}{
  \node[note,text=black!60] at ($(\n-0 |- meas-0.south)+(0,-4.5mm)$) {\t};
}
\node[note,text=black!60,anchor=west] at ($(meas-0.west |- meas-0.south)+(0,-4.5mm)$) {180, $r_9..r_{14}$};
\draw[dashed,black!60] ([xshift=-2mm,yshift=-16mm]dentry-0.west) -- ([xshift=-2mm,yshift=8mm]dentry-0.west)
  node[above=1mm,note,align=center] {entrance\\$P(\boldsymbol a,\boldsymbol b)$};
\draw[dashed,black!60] ([xshift=-1.6mm,yshift=-16mm]dexit-0.west) -- ([xshift=-1.6mm,yshift=8mm]dexit-0.west)
  node[above=1mm,note,align=center] {exit side of the Clifford part\\$P(\boldsymbol u,\boldsymbol z)$};
\end{tikzpicture}
\caption{The $d=3$ window after Eq.~\eqref{eq:covariance} has been applied to the channels beside the $T$ layers. The numbers under the boxes are the line numbers in the standalone $d=3$ source file of Appendix~\ref{app:circuits}. The central measurement on site 7 sits between the fold and unfold CNOT layers, with an implicit reset (not drawn). Figure~\ref{fig:move} shows the same step on one wire.}
\label{fig:window}
\end{figure}
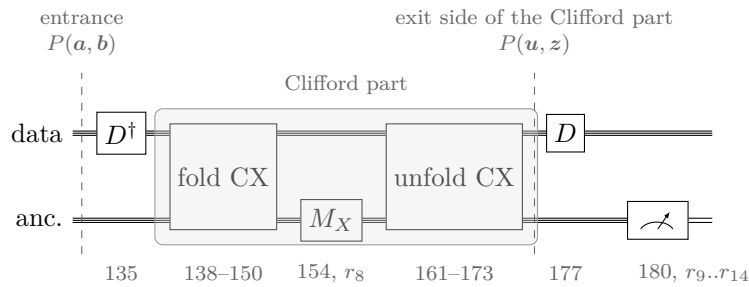

Figure~\ref{fig:window} is the $d=3$ double-checking circuit of Fig.~\ref{fig:d3circuit}, drawn with the line numbers of the source file, which we call the \textit{window}.
The incoming error $P(\boldsymbol a,\boldsymbol b)$ acts on the ideal encoded state $\ket{T_L^\dagger}$ at the entrance, before the entry layer $D^\dagger$. Inside, every operation is followed by a depolarising channel. Depolarising covariance lets the channels beside the $T$ layers be commuted through them,
\begin{equation}
 {\cal N}_{1,p}(V\rho V^\dagger)=V{\cal N}_{1,p}(\rho)V^\dagger ,
 \label{eq:covariance}
\end{equation}
for any one-qubit unitary $V$, here $T$ or $T^\dagger$. It holds because $X\rho X+Y\rho Y+Z\rho Z=2I-\rho$ for unit-trace $\rho$, so
\begin{equation}
 {\cal N}_{1,p}(\rho)=(1-p)\rho+\tfrac p3\big(2I-\rho\big)=\Big(1-\tfrac{4p}{3}\Big)\rho+\tfrac{4p}{3}\,\tfrac I2 ,
 \label{eq:nonidentity}
\end{equation}
and a unitary commutes with both terms. Three consequences of applying Eq.~\eqref{eq:covariance} to the channels beside the $T$ layers matter below.
\begin{enumerate}
\item The nonidentity part $\mathcal E(\rho)=\tfrac13\sum_{P\ne I}P\rho P=\tfrac13(2I-\rho)$ obeys the same identity on its own, so in the expansion over fault sets
\begin{equation}
 \prod_{j}\mathcal N^{(j)}_{1,p}=\sum_{S}(1-p)^{N-|S|}{\color{magenta}p^{|S|}}\prod_{j\in S}\mathcal E^{(j)}
 \label{eq:faultcount}
\end{equation}
every term keeps its factor {\color{magenta}$p^{|S|}$}, which is why the series coefficients are unaffected.
\item Only the averaged channel is commuted through, not its individual Pauli terms. With $\rho'=T^\dagger\rho T$,
\begin{equation}
 T^\dagger(X\rho X)T=\tfrac12\,(X-Y)\,\rho'\,(X-Y),
 \label{eq:singleterm}
\end{equation}
which has cross terms $X\rho'Y$ and is not one Pauli conjugation, because $T^\dagger XT=(X-Y)/\sqrt2$ is not a Pauli.
\item Once Eq.~\eqref{eq:covariance} has been applied, all internal noise lives in the Clifford part, where every fault is a Pauli and Paulis propagate to Paulis.
\end{enumerate}
The formula below therefore summarises all internal faults by one data Pauli $P(\boldsymbol u,\boldsymbol z)$, the net error just before the exit layer $D$, with the ancilla effects handled by the detectors. The covariance identity fails for any channel that is not a mixture of identity and full depolarisation, for instance pure dephasing.

\begin{figure}[H]\centering
\begin{adjustbox}{max width=\linewidth}\begin{tabular}{@{}c@{\hspace{4mm}}c@{\hspace{4mm}}c@{}}
\begin{tabular}[b]{@{}c@{}}actual faults: $X_3$ in the fold, $Y_9$ after the unfold\\propagate to $Q=X_0X_3Y_9$ at the exit side, $A=B=1/4$\end{tabular} & &
\begin{tabular}[b]{@{}c@{}}$Q=X_0X_3Y_9$ inserted just before $D$\\same numbers, $A=B=1/4$\end{tabular}\\[2mm]
\begin{tikzpicture}[font=\small,baseline=(current bounding box.center)]
\begin{yquant}[operator/separation=2.2mm,register/separation=5mm]
qubits {data} d;
qubits {anc.} c;
box {$D^\dagger$} d;
box {fold CX} (d,c);
[fill=tnorange!30,draw=tnorange!70!black] box {$X_3$} d;
box {fold CX} (d,c);
box {$M_X$} c;
box {unfold CX} (d,c);
[fill=tnorange!30,draw=tnorange!70!black] box {$Y_9$} d;
box {$D$} d;
align -;
measure c;
\end{yquant}
\end{tikzpicture} & {\Large$=$} &
\begin{tikzpicture}[font=\small,baseline=(current bounding box.center)]
\begin{yquant}[operator/separation=2.2mm,register/separation=5mm]
qubits {data} d;
qubits {anc.} c;
box {$D^\dagger$} d;
box {fold CX} (d,c);
box {$M_X$} c;
box {unfold CX} (d,c);
[fill=tnorange!30,draw=tnorange!70!black] box {$Q$} d;
box {$D$} d;
align -;
measure c;
\end{yquant}
\end{tikzpicture}\\[13mm]
 & {\Large$\neq$} & \\[6mm]
\multicolumn{3}{c}{\begin{tabular}[b]{@{}c@{}}$Q$ inserted just after $D^\dagger$: its $Z_9$ part propagates onto ancilla 8\\a detector fires, $A=0$\end{tabular}}\\[2mm]
\multicolumn{3}{c}{%
\begin{tikzpicture}[font=\small,baseline=(current bounding box.center)]
\begin{yquant}[operator/separation=2.2mm,register/separation=5mm]
qubits {data} d;
qubits {anc.} c;
box {$D^\dagger$} d;
[fill=tnorange!30,draw=tnorange!70!black] box {$Q$} d;
box {fold CX} (d,c);
box {$M_X$} c;
box {unfold CX} (d,c);
box {$D$} d;
align -;
measure c;
\end{yquant}
\end{tikzpicture}}
\end{tabular}\end{adjustbox}
\caption{Where the internal Pauli is inserted. Top: the actual two-fault pattern and $Q=X_0X_3Y_9$ at the exit side of the Clifford part give the same numbers. Bottom: $Q$ at the entry side does not, its $Z_9$ part reaches ancilla 8 and a detector fires.}
\label{fig:qposition}
\end{figure}
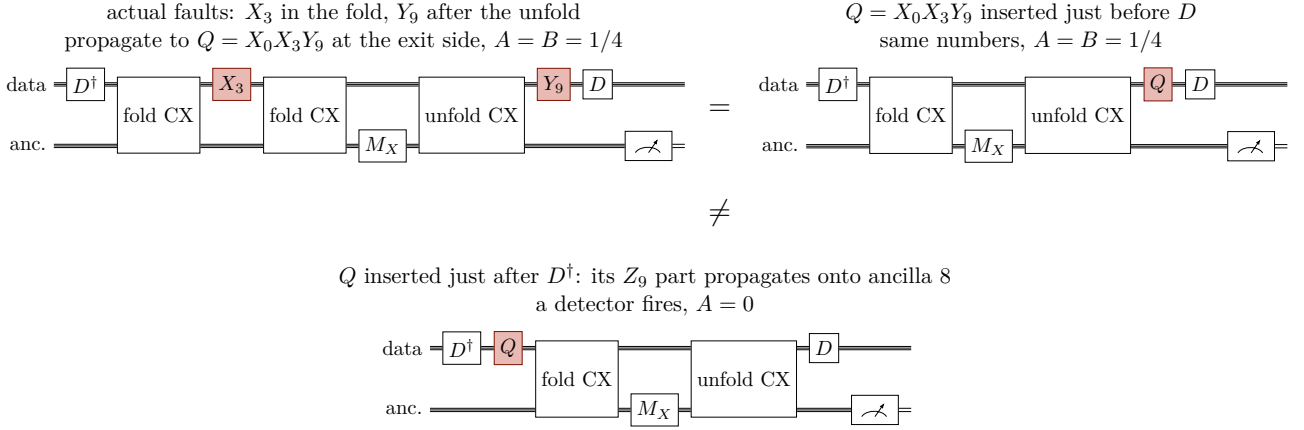

Two positions are natural for $P(\boldsymbol u,\boldsymbol z)$, the entry side or the exit side of the Clifford part, and a simulation decides it (Sec.~\ref{app:B-checks}). The main text's two-fault pattern of Eq.~\eqref{eq:v2-fixed-fault} and Fig.~\ref{fig:d3-fault-expansion}, $X_3$ in the fold and $Y_9$ in the unfold, has net data error $Q=X_0X_3Y_9$ before the exit layer. Inserting $Q$ at the exit side reproduces the actual pattern exactly, $A=B=1/4$. Inserting it at the entry side does not: its $Z_9$ part would propagate onto ancilla 8 and fire a detector. 
Figure~\ref{fig:qposition} draws the three cases.

 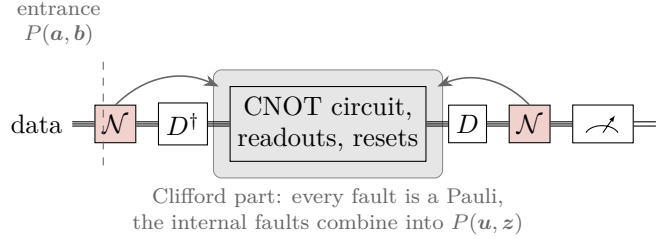
\begin{figure}[htbp]\centering
 \begin{adjustbox}{max width=.98\linewidth}\begin{tikzpicture}[font=\small]
\begin{yquant}[operator/separation=3mm,register/separation=5mm]
qubits {data} d;
[name=nin,fill=tnorange!20] box {$\mathcal N$} d;
[name=dentry] box {$D^\dagger$} d;
[name=cl] box {CNOT circuit,\\readouts, resets} d;
[name=dexit] box {$D$} d;
[name=nout,fill=tnorange!20] box {$\mathcal N$} d;
measure d;
\end{yquant}
\begin{scope}[on background layer]
\node[draw=black!45,fill=black!10,rounded corners=3pt,inner sep=2.2mm,fit=(cl-0),
      label={[font=\scriptsize,text=black!60,align=center]below:Clifford part: every fault is a Pauli,\\
      the internal faults combine into $P(\boldsymbol u,\boldsymbol z)$}] (bubble) {};
\end{scope}
\draw[dashed,black!60] ([xshift=-1.6mm,yshift=-3mm]nin-0.south) -- ([xshift=-1.6mm,yshift=6.5mm]nin-0.north)
  node[anchor=south east,xshift=1mm,font=\scriptsize,align=center] {entrance\\$P(\boldsymbol a,\boldsymbol b)$};
\draw[flow,black!60] (nin-0.north) to[bend left=35] ([xshift=-1.5mm]cl-0.north west);
\draw[flow,black!60] (nout-0.north) to[bend right=35] ([xshift=1.5mm]cl-0.north east);
\end{tikzpicture}\end{adjustbox}
 \caption{Equation~\eqref{eq:covariance} compressed. The noise beside the $T$ layers is commuted through into the Clifford part, where every fault is a Pauli.}\label{fig:move}\end{figure}
With both Paulis placed, one entry of the table is a single coherent sum, which the next section evaluates.

\subsection{One entry: a coherent sum, then a bit count}\label{app:B-entry}

\paragraph{Where the sum comes from.} Fix $P(\boldsymbol a,\boldsymbol b)$, $P(\boldsymbol u,\boldsymbol z)$ and $s$. The exit layer expands every $X$ into $(X\pm Y)/\sqrt2$, and the entry layer does the same backwards, so a single Pauli fault pattern becomes a sum of Pauli strings with amplitudes $2^{-|\mathrm{support}|/2}$ and eighth-root-of-unity phases. The final code readout projects onto the code space. On the code space an $X$ string acts as $X_L$ or as the identity depending on its mask, and it is killed unless its mask is in $C^\perp$. What is left is a sum over the sixteen masks $v(h)$ with a phase $i^{f(h)}$ each,
\begin{equation}
 \begin{aligned}
 {\cal G}&=2^{-m}\sum_{h\in\F^m}i^{f(h)},\\
 f(h)&=\underbrace{2\,\ell\cdot v(h)}_{\text{$Z$ part vs the mask}}\ \underbrace{-\,w_\epsilon(\boldsymbol u\wedge v(h))}_{\text{$T$ phases on the $X$ overlap}}\ +\ \underbrace{(tk+2e_{\rm bit})\,h_r}_{\text{logical slot}}\pmod 4 ,
 \end{aligned}
 \label{eq:gauss}
\end{equation}
with $k$, $e_{\rm bit}$ and $\ell$ defined in Eq.~\eqref{eq:binaryparameters} below and $v(h)=\bigoplus_ih_ig_i$. The three pieces are: a sign $(-1)^{\ell\cdot v}$ from commuting the total $Z$ mask $\ell$ past the string, a power of $i$ from the $T$ gates acting where the internal $X$ mask overlaps the string, and a term that only the logical coset ($h_r=1$) sees. The acceptance probability of the pattern is $|\mathcal G|^2$.

\paragraph{The same sum, concretely, for the two-fault pattern.} Take $Q=X_0X_3Y_9$, so $\boldsymbol u=\{0,3,9\}$, $\boldsymbol z=\{9\}$, no incoming error, $s=+1$. The exit layer conjugates each factor, $TXT^\dagger=(X+Y)/\sqrt2$ and $TYT^\dagger=(Y-X)/\sqrt2$, so
\[
DQD^\dagger=\frac{(X_0+Y_0)(X_3+Y_3)(Y_9-X_9)}{\sqrt8},
\]
eight signed strings, Fig.~\ref{fig:d3-fault-expansion} of the main text. The final code readout keeps a string only if it commutes with every stabiliser, that is if both its $X$ mask and its $Z$ mask have even overlap with every face. The $X$ mask is $\{0,3,9\}$ for all eight terms and passes. The $Z$ mask is the set of sites carrying $Y$, and only two of the eight pass:
\begin{center}\small\begin{tabular}{lcccc}\toprule
term & $X$ mask & $Z$ mask & $X$-check syndrome of the $Z$ mask & survives $\Pi_0$? \\\midrule
$+X_0X_3Y_9$ & $(1,1,0,1,0,0,0)$ & $(0,0,0,1,0,0,0)$ & $(0,1,0)$ & no \\
$-X_0X_3X_9$ & $(1,1,0,1,0,0,0)$ & $(0,0,0,0,0,0,0)$ & $(0,0,0)$ & yes \\
$+X_0Y_3Y_9$ & $(1,1,0,1,0,0,0)$ & $(0,1,0,1,0,0,0)$ & $(1,0,0)$ & no \\
$-X_0Y_3X_9$ & $(1,1,0,1,0,0,0)$ & $(0,1,0,0,0,0,0)$ & $(1,1,0)$ & no \\
$+Y_0X_3Y_9$ & $(1,1,0,1,0,0,0)$ & $(1,0,0,1,0,0,0)$ & $(1,1,0)$ & no \\
$-Y_0X_3X_9$ & $(1,1,0,1,0,0,0)$ & $(1,0,0,0,0,0,0)$ & $(1,0,0)$ & no \\
$+Y_0Y_3Y_9$ & $(1,1,0,1,0,0,0)$ & $(1,1,0,1,0,0,0)$ & $(0,0,0)$ & yes \\
$-Y_0Y_3X_9$ & $(1,1,0,1,0,0,0)$ & $(1,1,0,0,0,0,0)$ & $(0,1,0)$ & no \\
\bottomrule\end{tabular}
\end{center}
\begin{center}
\begin{tikzpicture}[x=1cm,y=1cm]
\node[box,text width=26mm,font=\small] (q) at (0,0) {one Pauli\\$Q=X_0X_3Y_9$};
\node[box,text width=30mm,font=\small] (e) at (4.2,0) {exit layer $D$\\$2^{3}=8$ signed strings};
\node[box,text width=30mm,font=\small] (p) at (8.4,0) {code readout $\Pi_0$\\2 strings survive};
\node[box,text width=26mm,font=\small] (m) at (12.4,0) {$2\times2$ logical matrix\\$-(X_L+Y_L)/\sqrt8$};
\draw[flow] (q)--(e); \draw[flow] (e)--(p); \draw[flow] (p)--(m);
\node[note,below=1mm of e] {amplitudes $\pm2^{-3/2}$};
\node[note,below=1mm of p] {masks in $C^\perp$ only};
\node[note,below=1mm of m] {$A=\|\cdot\|^2=\tfrac14$};
\end{tikzpicture}
\end{center}
\textbf{Why only two of the eight strings survive.} $Y_9$ alone has $Z$ mask $\{9\}$, and site 9 lies in the red face, so the red $X$ check flips. Adding $Y_0Y_3$ puts $\{0,3\}$ into the blue face as well, and now every face is overlapped an even number of times. 
This is the even-overlap condition of B.1 again, applied to a $Z$ mask against the $X$ checks, which have the same supports: a string survives the code readout exactly when its $Z$ mask lies in $C^\perp$.
The surviving pair $-X_0X_3X_9+Y_0Y_3Y_9$ acts on the code space as $-(X_L+Y_L)/\sqrt8$. Our implementation computes exactly this restriction, keeping every interfering phase as an exact eighth root of unity, and works in the basis $\ket{0_L},\ket{1_L}$. The restricted operator is
\[
\bmattwofault\;=\;-\frac{1}{\sqrt8}\begin{pmatrix}0&1-i\\1+i&0\end{pmatrix}=-\frac{X_L+Y_L}{\sqrt8},
\]
which applied to the ideal input and projected on the central readout outcome $s=+1$ gives $(A,B)=\binsttwofault$. With no fault at all the same computation gives the identity and $(A,B)=\binstident$. This is the coherent sum evaluated directly. The bit count below is the shortcut that gives the same numbers without listing strings.

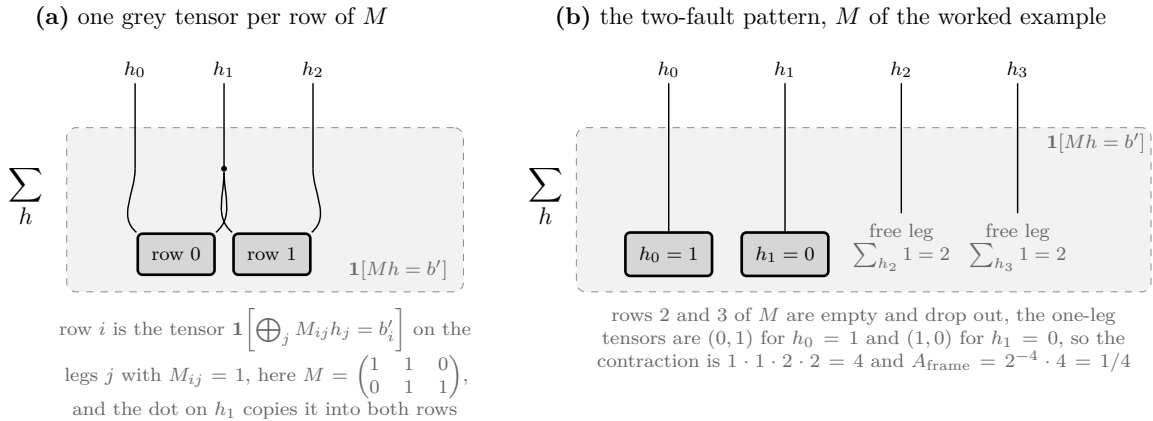
\begin{figure}[H]\centering
\begin{adjustbox}{max width=.9\linewidth}\begin{tikzpicture}[x=1cm,y=1cm,
  bigbox/.style={draw=black!45,dashed,fill=black!16,fill opacity=.30,rounded corners=4pt},
  row/.style={tn constraint,minimum width=11mm,minimum height=6mm,font=\scriptsize}]
\node[tn heading,anchor=west] at (-1.3,3.9) {\textbf{(a)} one grey tensor per row of $M$};
\node[font=\Large] at (-1.3,1.3) {$\displaystyle\sum_{h}$};
\node[bigbox,minimum width=58mm,minimum height=24mm] (A) at (2.2,1.1) {};
\node[tn label,anchor=south east,text=black!60] at ([xshift=-1.5mm,yshift=1mm]A.south east) {$\ind{Mh=b'}$};
\foreach \j/\x in {0/0.3,1/1.6,2/2.9}{
  \draw[tn leg] (\x,2.95)--(\x,1.7);
  \node[tn label,anchor=south] at (\x,2.95) {$h_\j$};
}
\node[row] (r0) at (0.9,0.45) {row 0};
\node[row] (r1) at (2.3,0.45) {row 1};
\coordinate (c0) at (0.3,1.7); \node[tn copy] (c1) at (1.6,1.7) {}; \coordinate (c2) at (2.9,1.7);
\draw[tn leg] (c0) to[out=-90,in=140] (r0.north west);
\draw[tn leg] (c1) to[out=-90,in=40] (r0.north east);
\draw[tn leg] (c1) to[out=-90,in=140] (r1.north west);
\draw[tn leg] (c2) to[out=-90,in=40] (r1.north east);
\node[tn label,anchor=north,text=black!60,align=center,text width=62mm] at (2.2,-0.25)
  {row $i$ is the tensor $\ind{\textstyle\bigoplus_j M_{ij}h_j=b'_i}$ on the legs $j$ with $M_{ij}=1$, here $M=\begin{pmatrix}1&1&0\\0&1&1\end{pmatrix}$, and the dot on $h_1$ copies it into both rows};
\node[tn heading,anchor=west] at (6.3,3.9) {\textbf{(b)} the two-fault pattern, $M$ of the worked example};
\node[font=\Large] at (6.3,1.3) {$\displaystyle\sum_{h}$};
\node[bigbox,minimum width=84mm,minimum height=24mm] (B) at (10.95,1.1) {};
\node[tn label,anchor=north east,text=black!60] at (B.north east) {$\ind{Mh=b'}$};
\foreach \j/\x in {0/8.1,1/9.8,2/11.5,3/13.2}{
  \node[tn label,anchor=south] at (\x,2.95) {$h_\j$};
}
\draw[tn leg] (8.1,2.95)--(8.1,0.75);
\draw[tn leg] (9.8,2.95)--(9.8,0.75);
\draw[tn leg] (11.5,2.95)--(11.5,1.05);
\draw[tn leg] (13.2,2.95)--(13.2,1.05);
\node[row] (s1) at (8.1,0.45) {$h_0=1$};
\node[row] (s0) at (9.8,0.45) {$h_1=0$};
\node[tn label,text=black!60,align=center] at (11.5,0.55) {free leg\\$\sum_{h_2}1=2$};
\node[tn label,text=black!60,align=center] at (13.2,0.55) {free leg\\$\sum_{h_3}1=2$};
\node[tn label,anchor=north,text=black!60,align=center,text width=84mm] at (10.95,-0.25)
  {rows 2 and 3 of $M$ are empty and drop out, the one-leg tensors are $(0,1)$ for $h_0=1$ and $(1,0)$ for $h_1=0$, so the contraction is $1\cdot1\cdot2\cdot2=4$ and $A_{\rm frame}=2^{-4}\cdot4=1/4$};
\end{tikzpicture}\end{adjustbox}
\caption{The indicator $\ind{Mh=b'}$ as one grey tensor, drawn transparent so that its factorisation into one grey tensor per row of $M$ is visible. (a) The rule on a generic example, with the copy tensor of Eq.~\eqref{eq:copytensor} on the shared leg. (b) The two-fault pattern.}
\label{fig:countfactor}
\end{figure}

\paragraph{The bit count.} The probability is $|{\cal G}|^2$ with ${\cal G}$ the sum of $2^m$ fourth roots of unity in Eq.~\eqref{eq:gauss}\footnote{The string amplitudes and the magic input carry eighth roots of unity, $e^{\pm i\pi/4}$. Each term of ${\cal G}$ is a product of two of them, so ${\cal G}$ is a sum of fourth roots, $i^{f(h)}\in\{\pm1,\pm i\}$.}, and squaring it term by term would mean $4^m$ products. But $f$ of Eq.~\eqref{eq:gauss} is a \emph{quadratic} function of $h$ over the integers mod 4 (it has the form of $q(x)$, like Eq.~(69) from~\cite{Bravyi_2019}), and its polarisation $f(h\oplus d)-f(h)-f(d)=2h^{\!\top}Md \pmod 4$ is linear with a symmetric binary matrix $M$.
Summing over $h$ first in $|\mathcal G|^2=2^{-2m}\sum_{h,d}i^{f(h\oplus d)-f(h)}$ makes every $d\notin\ker M$ cancel, since $\sum_{h}(-1)^{h\cdot Md}=2^m\ind{Md=0}$,
\begin{equation}
 |{\cal G}|^2=2^{-m}\sum_{d\in\ker M}i^{f(d)} .
 \label{eq:gausssquare}
\end{equation}
For $d\in\ker M$ the mask $\boldsymbol u\wedge v(d)$ lies in $C^\perp$ and has parity $kd_r$, and the corrected code obeys $w_\epsilon(w)=t(|w|\bmod2)\pmod4$ for every $w\in C^\perp$, so $f(d)=2b'\cdot d\pmod 4$ with $b'$ as below. The same orthogonality then leaves $|{\cal G}|^2=0$ or $|{\cal G}|^2=2^{-m}|\ker M|=2^{-\rank M}$, and since $M$ is symmetric this is exactly the consistency condition for $Mh=b'$. The ingredients are
\begin{equation}
 \begin{aligned}
 k&=|\boldsymbol u\oplus\boldsymbol a|\bmod2,\\
 e_{\rm bit}&=\ind{s=-1}\oplus(|\boldsymbol z|\bmod2),\\
 \ell&=\boldsymbol b\oplus\boldsymbol z\oplus(\boldsymbol u\wedge\boldsymbol a),
 \end{aligned}
 \label{eq:binaryparameters}
\end{equation}
\begin{equation}
 \begin{aligned}
 M_{ij}&=|\boldsymbol u\wedge g_i\wedge g_j|\bmod2\ \oplus\ k\,\ind{i=j=r},\\
 b'_i&=\ell\cdot g_i\oplus e_{\rm bit}\ind{i=r},
 \end{aligned}
 \label{eq:countmatrix}
\end{equation}
and the result is: if $H(\boldsymbol u\oplus\boldsymbol a)\ne0$ the final code readout rejects the fault pattern, otherwise
\begin{equation}
 A_{\rm frame}=2^{-m}\,\#\{h\in\F^m: Mh=b'\},\qquad B_{\rm frame}=e_{\rm bit}\,A_{\rm frame}.
 \label{eq:positivecount}
\end{equation}
A failed ancillary detector also sets both responses to zero, $A_{\rm frame}=B_{\rm frame}=0$. A consistent system has $A_{\rm frame}=2^{-\rank M}$, an inconsistent one has zero response. The matrix is $4\times4$ at $d=3$ and $10\times10$ at $d=5$, so one entry costs practically nothing. As a tensor, the count is one grey indicator with $m$ legs, every leg summed:
\begin{center}
\begin{tikzpicture}[x=1cm,y=1cm]
\node[font=\large,anchor=east] at (-1.6,0) {$A_{\rm frame}=\dfrac{1}{2^{m}}$};
\node[font=\Large] at (-0.7,0) {$\displaystyle\sum_{h\in\F^m}$};
\node[tn constraint,minimum width=34mm,minimum height=10mm] (M) at (2.2,0) {$\ind{Mh=b'}$};
\foreach \i/\xx in {0/1.0,1/1.8,2/2.6,3/3.4}{
  \draw[tn leg] (\xx,1.0)--(\xx,0.5);
  \node[tn label,anchor=south] at (\xx,1.0) {$h_\i$};
}
\node[font=\large,anchor=west] at (4.2,0) {$,\qquad B_{\rm frame}=e_{\rm bit}\,A_{\rm frame}.$};
\end{tikzpicture}
\end{center}
The grey box has $2^m$ binary numbers, and the contraction is their sum. Each row of $M$ is itself a parity constraint, so $\ind{Mh=b'}$ factorises into $m$ smaller grey tensors, one per row, joined through the legs they share.
Where two rows use the same bit the leg carries a black dot, the copy tensor
\begin{equation}
 \vcenter{\hbox{\begin{tikzpicture}[x=1cm,y=1cm]
  \node[tn copy] (d) at (0,0) {};
  \draw[tn leg] (d)--(-0.45,0.4) node[tn label,anchor=south east,inner sep=1pt] {$a$};
  \draw[tn leg] (d)--(0.45,0.4) node[tn label,anchor=south west,inner sep=1pt] {$b$};
  \draw[tn leg] (d)--(0,-0.45) node[tn label,anchor=north,inner sep=1pt] {$c$};
 \end{tikzpicture}}}\;=\;\delta_{abc}=\ind{a=b=c},
 \label{eq:copytensor}
\end{equation}
which hands one bit to every row that needs it. Figure~\ref{fig:countfactor} draws the factorised form, and the examples below use it.

\paragraph{Every symbol of the count, with its reason.}
\begin{description}
\item[$k$.] Parity of the total $X$ mask. On the code space an $X$ string of even weight in $C^\perp$ is a stabiliser and of odd weight is a stabiliser times $X_L$, so $k=1$ means the combined $X$ error is a logical $X$ up to a stabiliser. It enters only the logical slot $(r,r)$ of $M$.
\item[$e_{\rm bit}$.] The logical-error flag. Each internal $Z$ on a data site flips the sign of $M_L$ once and a $-1$ central outcome flips it again, so the surviving state has $M_L=s(-1)^{|\boldsymbol z|}$. The accepted part is entirely good or entirely bad, which is why $B=e_{\rm bit}A$. An \emph{incoming} $Z$ does not enter $e_{\rm bit}$: it enters $\ell$ and is caught by the $X$ checks, or, if logical, by the linear system, see the examples. This all-or-nothing rule holds once Eq.~\eqref{eq:covariance} has been applied. An individual fault at its original position can carry a fractional conditional error.
\item[$\ell$.] The total $Z$ mask, incoming plus internal, plus a correction on the sites where the two $X$ masks overlap: there $X_qX_q=I$ but the $T$ phases between the two errors leave a $Z_q$ behind. It supplies the right-hand side $b'$.
\item[$M$.] Off the diagonal, the quadratic part of $f$: the $T$ phases on the intersection of two generators with the internal $X$ mask. On the diagonal, the parity of the internal $X$ mask on each generator, with the logical slot corrected by $k$. Without an internal $X$ mask, $M$ is zero apart from the logical slot.
\item[$b'$.] The right-hand side of the linear system, with $f(d)=2\,b'\cdot d$ on $\ker M$: for a face, the parity of the total $Z$ mask on that face, which is exactly the $X$-check syndrome bit. For the logical row, the parity on $\boldsymbol1$ corrected by the flag.
\item[Rejection.] It has two faces. A $Z$-check syndrome, $H(\boldsymbol u\oplus\boldsymbol a)\ne0$, is an early exit. An $X$-check syndrome shows up as an inconsistent system, $b'\notin\operatorname{im}M$.
\end{description}

\subsection{Worked examples with all the matrices}\label{app:B-examples}
The formula is compact enough that its content is best seen by running it. This section does so on five cases, every matrix written out. The first, an incoming $X_7$ cancelled by an internal $X_7$, is a pattern that is trivial as a Pauli yet accepted only half the time, because the $T$ layer between the two errors turns their product into a phase. The second is the main text's two-fault pattern, accepted with probability $1/4$ and always wrong, the smallest failure of the $d=3$ check. The three short ones exercise the remaining branches of the formula: the early exit on a $Z$ syndrome, an internal $Z$ that sets the logical flag, and the logical slot with both central signs.

Legend for the matrices: \colorbox{tnblue!22}{blue} from an $X$ mask, \colorbox{tnorange!22}{orange} from a $Z$ mask, \fcolorbox{tngreen}{tngreen!18}{green} the logical slot, \colorbox{black!22}{grey} a pivot column.

\paragraph{Incoming $X_7$, internal $X_7$, $s=+1$.} The incoming $X_7$ is the example of Fig.~\ref{fig:d3-incoming} in the main text, and the second $X_7$ sits at the exit side of the Clifford part of Fig.~\ref{fig:window}. The four masks, rows $\boldsymbol a,\boldsymbol b,\boldsymbol u,\boldsymbol z$:
\[
\masksCbw .
\]
Site 7 lies in all three faces. $\boldsymbol u\oplus\boldsymbol a=\boldsymbol0$, so the $Z$-check syndrome is $\synbw$ and nothing is rejected early. The intersection counts that build $M$ and the overlaps that build $b'$ are
\[
\big(|\boldsymbol u\wedge g_i\wedge g_j|\big)_{ij}=\interCbw,\qquad \big(|\ell\wedge g_i|\big)_i=\elloverbw ,
\]
all odd (blue) because site 7 is in every generator, so every entry of $M$ will be 1. Reducing mod 2 and adding the logical corrections, here none since $k=e_{\rm bit}=0$,
\begin{equation}
 k=0,\ e_{\rm bit}=0,\ \ell=(0,0,1,0,0,0,0),\qquad M=\MCbw,\quad b'=\bpCbw .
 \label{eq:workedmatrix}
\end{equation}
Blue marks the ones of $M$, orange the ones of $b'$, the green boxes are the logical slot. The augmented system and its row reduction over $\F$:
\[
\augCbw\ \longrightarrow\ \rrefCbw .
\]
Pivot $\pivotsbw$ (grey), $\rank M=\rankbw$, \consbw, so $2^{4-\rankbw}=8$ solutions: all four rows are the same constraint $h_0\oplus h_1\oplus h_2\oplus h_3=1$, and the reduction makes that visible by wiping three rows. The solutions are
\[ h\in\{\solbw\}, \]
eight of sixteen, so $A_{\rm frame}=\resAbw$ and $B_{\rm frame}=\resBbw$. The Gauss sum tells the same story with phases, $f(h)=|h|^2\bmod4$, eight terms equal to $1$, eight equal to $i$, and $|(1+i)/2|^2=1/2$:
\begin{center}\small\begin{tabular}{ccccc}\toprule
$h$ & $|v(h)|$ & $f(h)\bmod4$ & $i^{f(h)}$ & solves $Mh=b'$? \\\midrule
$(0,0,0,0)$ & 0 & 0 & $1$ & no \\
$(0,0,0,1)$ & 7 & 1 & $i$ & yes \\
$(0,0,1,0)$ & 4 & 1 & $i$ & yes \\
$(0,0,1,1)$ & 3 & 0 & $1$ & no \\
$(0,1,0,0)$ & 4 & 1 & $i$ & yes \\
$(0,1,0,1)$ & 3 & 0 & $1$ & no \\
$(0,1,1,0)$ & 4 & 0 & $1$ & no \\
$(0,1,1,1)$ & 3 & 1 & $i$ & yes \\
$(1,0,0,0)$ & 4 & 1 & $i$ & yes \\
$(1,0,0,1)$ & 3 & 0 & $1$ & no \\
$(1,0,1,0)$ & 4 & 0 & $1$ & no \\
$(1,0,1,1)$ & 3 & 1 & $i$ & yes \\
$(1,1,0,0)$ & 4 & 0 & $1$ & no \\
$(1,1,0,1)$ & 3 & 1 & $i$ & yes \\
$(1,1,1,0)$ & 4 & 1 & $i$ & yes \\
$(1,1,1,1)$ & 3 & 0 & $1$ & no \\
\bottomrule\end{tabular}
\end{center}
\begin{center}
\begin{tikzpicture}[x=1cm,y=1cm]
\draw[black!40] (0,0) circle (1.2);
\draw[black!30] (-1.4,0)--(1.4,0); \draw[black!30] (0,-1.4)--(0,1.4);
\node[font=\scriptsize,anchor=east] at (-1.25,0) {$-1$};
\node[font=\scriptsize,anchor=north] at (0,-1.25) {$-i$};
\fill[tnblue] (1.2,0) circle (2.5pt) node[right=4pt,font=\small] {$1$: 8 terms};
\fill[tnblue] (0,1.2) circle (2.5pt) node[above=4pt,font=\small] {$i$: 8 terms};
\draw[flow,tnorange,thick] (0,0)--(0.6,0.6);
\node[font=\small] at (0,-1.9) {$\mathcal G=\tfrac{8+8i}{16}=\gsumbw$};
\node[font=\small] at (0,-2.4) {phase counts $1/i/-1/-i$: $\phasecountsbw$, so $|\mathcal G|^2=\gabsbw$};
\end{tikzpicture}
\qquad
\begin{tikzpicture}[x=1cm,y=1cm]
\node[tn constraint,minimum width=58mm,minimum height=11mm] (odd) at (3,0) {$\Delta_{\rm odd}(h_0,h_1,h_2,h_3)=\ind{h_0\oplus h_1\oplus h_2\oplus h_3=1}$};
\foreach \i in {0,1,2,3}{
  \draw[tn leg] (1+1.33*\i,1.15)--(1+1.33*\i,0.57);
  \node[tn label,anchor=south] at (1+1.33*\i,1.15) {$h_{\i}$};
}
\node[tn label,font=\small] at (3,-1.1) {$\displaystyle A_{\rm frame}=\frac1{16}\sum_{h}\Delta_{\rm odd}(h)=\frac8{16}$};
\end{tikzpicture}
\end{center}
The left-hand picture is the complex plane. The unit circle carries the four values $1,i,-1,-i$ that $i^{f(h)}$ can take, each blue dot is labelled with how many of the sixteen terms of Eq.~\eqref{eq:gauss} land there, and the orange arrow is their average, ${\cal G}$. The right-hand picture is the count as a tensor: one grey box whose sixteen entries are the last column of the table, eight ones, summed over all four legs and divided by sixteen. It is the factorised form of Fig.~\ref{fig:countfactor} with the four identical rows of $M$ merged into one four-leg tensor. The four legs are the bits $h_i$, not qubits. Physically, the incoming $X_7$ and the internal $X_7$ cancel as $X$ errors but not as phases: the $T^\dagger$ layer sits between them and $X_7T_7^\dagger X_7\propto T_7$, so the pattern behaves like a $Z$-type phase on site 7 that leaves the state half inside and half outside the target. The check accepts with probability $1/2$, and every accepted output is the correct state, so $(A,B)=(1/2,0)$.

\paragraph{The main text's two-fault pattern.} The pattern of Fig.~\ref{fig:d3-fault-expansion}, with its net Pauli moved to the exit side as in Fig.~\ref{fig:qposition}. $\boldsymbol u=\{0,3,9\}$, $\boldsymbol z=\{9\}$, $\boldsymbol a=\boldsymbol b=\boldsymbol0$, $s=+1$:
\[
\masksCtwofault .
\]
The $Z$ checks: $g_0\cap\boldsymbol u=\{0,3\}$, $g_1\cap\boldsymbol u=\{3,9\}$, $g_2\cap\boldsymbol u=\varnothing$, all even, syndrome $\syntwofault$, no early rejection. Intersection counts and overlaps:
\[
\big(|\boldsymbol u\wedge g_i\wedge g_j|\big)_{ij}=\interCtwofault,\qquad \big(|\ell\wedge g_i|\big)_i=\ellovertwofault .
\]
Mod 2, with the logical corrections $k=1$ in $M_{33}$ and $e_{\rm bit}=1$ in $b'_3$:
\[ k=1,\ e_{\rm bit}=1,\ \ell=(0,0,0,1,0,0,0),\qquad M=\MCtwofault,\quad b'=\bpCtwofault . \]
Blue: the odd intersection counts $(0,1)$ and $(1,0)$ became the two ones of $M$. Orange: the overlap of $\ell=\{9\}$ with the red face became $b'_1=1$. Green: $M_{33}=|\boldsymbol u\wedge\boldsymbol1\wedge\boldsymbol1|\bmod2\oplus k=1\oplus1=0$ and $b'_3=|\ell\wedge\boldsymbol1|\bmod2\oplus e_{\rm bit}=1\oplus1=0$, the two logical corrections cancel the raw values there.
\[
\augCtwofault\ \longrightarrow\ \rrefCtwofault .
\]
Pivots $\pivotstwofault$ (grey), $\rank M=\ranktwofault$, \constwofault, so $2^{4-\ranktwofault}=4$ solutions. Row 0 says $h_1=0$, row 1 says $h_0=1$, rows 2 and 3 are empty. The solutions are
\[ h\in\{\soltwofault\}, \]
four of sixteen, so $A_{\rm frame}=\resAtwofault$ and, since $e_{\rm bit}=1$, $B_{\rm frame}=\resBtwofault$. This is the acceptance $1/4$ with a logical error of Fig.~\ref{fig:d3-fault-expansion} in the main text.
\begin{center}\small\begin{tabular}{ccccc}\toprule
$h$ & $|v(h)|$ & $f(h)\bmod4$ & $i^{f(h)}$ & solves $Mh=b'$? \\\midrule
$(0,0,0,0)$ & 0 & 0 & $1$ & no \\
$(0,0,0,1)$ & 7 & 0 & $1$ & no \\
$(0,0,1,0)$ & 4 & 0 & $1$ & no \\
$(0,0,1,1)$ & 3 & 0 & $1$ & no \\
$(0,1,0,0)$ & 4 & 0 & $1$ & no \\
$(0,1,0,1)$ & 3 & 0 & $1$ & no \\
$(0,1,1,0)$ & 4 & 0 & $1$ & no \\
$(0,1,1,1)$ & 3 & 0 & $1$ & no \\
$(1,0,0,0)$ & 4 & 2 & $-1$ & yes \\
$(1,0,0,1)$ & 3 & 2 & $-1$ & yes \\
$(1,0,1,0)$ & 4 & 2 & $-1$ & yes \\
$(1,0,1,1)$ & 3 & 2 & $-1$ & yes \\
$(1,1,0,0)$ & 4 & 0 & $1$ & no \\
$(1,1,0,1)$ & 3 & 0 & $1$ & no \\
$(1,1,1,0)$ & 4 & 0 & $1$ & no \\
$(1,1,1,1)$ & 3 & 0 & $1$ & no \\
\bottomrule\end{tabular}
\end{center}
\begin{center}
\begin{tikzpicture}[x=1cm,y=1cm]
\node[tn label,anchor=south] at (0.6,1.2) {$h_0$}; \node[tn label,anchor=south] at (2.0,1.2) {$h_1$};
\node[tn label,anchor=south] at (3.4,1.2) {$h_2$}; \node[tn label,anchor=south] at (4.8,1.2) {$h_3$};
\node[tn constraint,minimum width=11mm,minimum height=7mm,font=\scriptsize] (r1) at (0.6,0) {$h_0=1$};
\node[tn constraint,minimum width=11mm,minimum height=7mm,font=\scriptsize] (r0) at (2.0,0) {$h_1=0$};
\draw[tn leg] (0.6,1.2)--(r1.north); \draw[tn leg] (2.0,1.2)--(r0.north);
\draw[tn leg] (3.4,1.2)--(3.4,0.2); \draw[tn leg] (4.8,1.2)--(4.8,0.2);
\node[tn label,align=center] at (4.1,-0.55) {free legs\\(rows 2 and 3 are empty)};
\node[tn label,font=\small] at (2.7,-1.3) {$A_{\rm frame}=\tfrac1{16}\sum_h \ind{h_0=1}\ind{h_1=0}=\tfrac{4}{16}$};
\end{tikzpicture}
\qquad
\begin{tikzpicture}[x=1cm,y=1cm]
\draw[black!40] (0,0) circle (1.2);
\draw[black!30] (-1.4,0)--(1.4,0); \draw[black!30] (0,-1.4)--(0,1.4);
\node[font=\scriptsize,anchor=south] at (0,1.25) {$i$}; \node[font=\scriptsize,anchor=north] at (0,-1.25) {$-i$};
\fill[tnblue] (1.2,0) circle (2.5pt) node[right=4pt,font=\small] {$1$: 12 terms};
\fill[tnblue] (-1.2,0) circle (2.5pt) node[left=4pt,font=\small] {$-1$: 4 terms};
\draw[flow,tnorange,thick] (0,0)--(0.6,0);
\node[font=\small,anchor=south west] at (0.1,0.08) {$\mathcal G=\tfrac{12-4}{16}$};
\node[font=\small] at (0,-1.9) {phase counts $1/i/-1/-i$: $\phasecountstwofault$};
\node[font=\small] at (0,-2.4) {$\mathcal G=\gsumtwofault$, $|\mathcal G|^2=\gabstwofault$};
\end{tikzpicture}
\end{center}
On the left, the linear system as a tensor network: each nonzero row of $M$ is one grey constraint on the legs it touches. It is Fig.~\ref{fig:countfactor}(b). Here each grey box is a one-leg tensor, $(0,1)$ for $h_0=1$ and $(1,0)$ for $h_1=0$, and a free open leg summed contributes a factor $2$, so the contraction is $1\cdot1\cdot2\cdot2=4$ and $A_{\rm frame}=4/16$. On the right, the same answer from the phases, in the complex-plane picture of the previous example: the sixteen terms of Eq.~\eqref{eq:gauss} are spread over $1$ and $-1$ and cancel down to $|\mathcal G|^2=1/4$.

\paragraph{Three short ones.} Three single errors, each worked the same way: write the masks, apply the formula, read off $(A,B)$. They show what the two long examples did not: an error the code readout rejects, an internal $Z$ that makes the accepted output wrong, and a logical $X$ read out with both values of the central sign $s$.
\begin{itemize}
\item \emph{Incoming $X_7$ alone.} Masks $\masksCxseven$, syndrome $H(\boldsymbol a)=\synxseven\ne0$: rejected before anything else, $A=B=0$. This is the main text's $F_A(X_7;0)=0$. $M$ and $b'$ are never formed. The mask is not one of the sixteen of Fig.~\ref{fig:cperp}.
\item \emph{Internal $Z_7$.} Placed at the exit side of the Clifford part as in Fig.~\ref{fig:window}. Masks $\masksCzs$. $\boldsymbol u=\boldsymbol0$, so the intersection table is $\interCzs$ and $M=0$, while $\ell=\{7\}$ has overlaps $\elloverzs$, giving $b'=(1,1,1,0)$ after the $e_{\rm bit}=1$ correction in the last slot:
\[ M=\MCzs,\quad b'=\bpCzs, \qquad \augCzs\ \longrightarrow\ \rrefCzs, \]
\conszs, no solution, $A=0$. The three ones in $b'$ are the $X$-check syndrome of $Z_7$: rejection by a syndrome shows up as inconsistency, not as an early exit. The coherent-sum computation agrees, its $2\times2$ matrix is $\bmatzs$ and the result $\binstzs$.
\item \emph{Incoming $X_L$, both outcomes.} $\boldsymbol1$ is the last generator row of Fig.~\ref{fig:cperp}, and the two outcomes are those of the central readout $r_8$ highlighted in Fig.~\ref{fig:d3circuit}. $\boldsymbol a=\boldsymbol1$, $k=1$, only the logical slot of $M$ is set:
\[ k=1,\ e_{\rm bit}=0,\ \ell=\boldsymbol0,\qquad M=\MCxl,\quad b'=\bpCxl \]
for $s=+1$, eight solutions, $A=\resAxl$, $B=\resBxl$. For $s=-1$ the flag flips, $b'=(0,0,0,1)$, again eight solutions, $A=\resAxlminus$, $B=\resBxlminus$. Physically $X_L\ket{T_L^\dagger}\propto\ket{T_L}$ is an equal superposition of the $M_L=\pm1$ states: half the time the check projects onto the good state, half the time onto the bad one, and the central bit records which.
\end{itemize}

The next subsection provides a few independent computations that are sanity tests for the $(A,B)$ computed in this subsection's examples.

\subsection{Sanity tests of the worked examples}\label{app:B-checks}

Every probability of the previous subsection (for the $d=3$ circuit) was evaluated in three independent ways, which agree in every case:
\begin{enumerate}
\item the formula, Eqs.~\eqref{eq:binaryparameters} to \eqref{eq:positivecount}, written out directly,
\item the coherent sum, which builds the exact $2\times2$ code-restricted matrix of $DQD^\dagger$ over the eighth roots of unity and applies it to the ideal input, and
\item a 13-qubit state-vector simulation of the actual $d=3$ window from the source file, with the ideal encoded input, the incoming Pauli at the entrance and the internal Pauli at the exit side of the Clifford part, acceptance being every measured bit equal to the fault-free reference.
\end{enumerate}
The Gauss sum of Eq.~\eqref{eq:gauss} was evaluated term by term as well.
\begin{center}\adjustbox{max width=\linewidth}{\begin{tabular}{lcccccccc}\toprule
case & $\boldsymbol a$ & $\boldsymbol b$ & $\boldsymbol u$ & $\boldsymbol z$ & $s$ & formula $(A,B)$ & code $(A,B)$ & simulation $(A,B)$ \\\midrule
fault-free & $\varnothing$ & $\varnothing$ & $\varnothing$ & $\varnothing$ & $+1$ & $(1,0)$ & $(1,0)$ & $(1,0)$ \\
incoming X7 only & $7$ & $\varnothing$ & $\varnothing$ & $\varnothing$ & $+1$ & $(0,0)$ & $(0,0)$ & $(0,0)$ \\
incoming X7, internal X7 & $7$ & $\varnothing$ & $7$ & $\varnothing$ & $+1$ & $(1/2,0)$ & $(1/2,0)$ & $(0.5,0)$ \\
incoming X7, internal X7, $s=-1$ & $7$ & $\varnothing$ & $7$ & $\varnothing$ & $-1$ & $(0,0)$ & $(0,0)$ & $(0,0)$ \\
incoming $X_L$ & $\text{all}$ & $\varnothing$ & $\varnothing$ & $\varnothing$ & $+1$ & $(1/2,0)$ & $(1/2,0)$ & $(0.5,0)$ \\
incoming $X_L$, $s=-1$ & $\text{all}$ & $\varnothing$ & $\varnothing$ & $\varnothing$ & $-1$ & $(1/2,1/2)$ & $(1/2,1/2)$ & $(0.5,0.5)$ \\
internal Z7 & $\varnothing$ & $\varnothing$ & $\varnothing$ & $7$ & $+1$ & $(0,0)$ & $(0,0)$ & $(0,0)$ \\
incoming $Z_L$ & $\varnothing$ & $\text{all}$ & $\varnothing$ & $\varnothing$ & $+1$ & $(0,0)$ & $(0,0)$ & $(0,0)$ \\
incoming $Z_L$, $s=-1$ & $\varnothing$ & $\text{all}$ & $\varnothing$ & $\varnothing$ & $-1$ & $(1,1)$ & $(1,1)$ & $(1,1)$ \\
moved X0 X3 Y9 (Fig. 7 pattern) & $\varnothing$ & $\varnothing$ & $0,3,9$ & $9$ & $+1$ & $(1/4,1/4)$ & $(1/4,1/4)$ & $(0.25,0.25)$ \\
literal X3 fold, Y9 unfold & $\varnothing$ & $\varnothing$ & $\varnothing$ & $\varnothing$ & $+1$ & $(1/4,1/4)$ & $(1/4,1/4)$ & $(0.25,0.25)$ \\
\bottomrule\end{tabular}
}\end{center}
Beyond these cases, 40 random $(\boldsymbol a,\boldsymbol b,\boldsymbol u,\boldsymbol z)$ and all \scancases\ one- and two-site Paulis, internal at either side of the Clifford part and incoming at the entrance, agree case by case. 
The row ``literal $X_3$ fold, $Y_9$ unfold'' is the two-fault pattern simulated at its actual positions, compared with Eqs.~\eqref{eq:binaryparameters} to \eqref{eq:positivecount} evaluated on its net Pauli $Q=X_0X_3Y_9$ at the exit side.
These two conventions have to match, and they do. 
Inserting $Q$ at the exit side of the Clifford part reproduces the pattern at its actual positions. For $s=-1$ the source's detectors require the $X$ faces to read $-1$ and one ancilla, compared with $r_8$, to read $1$, with the logical readout frame following $r_8$. 
In the standalone $d=3$ circuit the detector $r_6\oplus r_7\oplus r_8=0$ with $r_6=r_7=0$ forces $s=+1$, so only that branch is ever accepted there.

So far one entry at a time, for one fixed pattern of internal faults. The table needs the average over all internal fault patterns, and the next subsection shows how local tensors perform that average without enumerating patterns: each depolarising channel becomes a small kernel, each CNOT a rule for combining kernels, and the whole check an open network whose contraction is the table.

\subsection{Averaging the channels with local tensors}\label{app:response-average}
The tensor pictures follow the conventions of Sec.~\ref{sec:tensor-contraction}. A rounded box is a function of the bits on its legs, grey an indicator, orange-red a noise weight, blue an input, green an intermediate sum. A leg joining two boxes is summed, an open leg is an output index, a black dot is the copy tensor of Eq.~\eqref{eq:copytensor}, and a double line carries several bits.

\paragraph{The mixed kernel.} A Pauli channel on one qubit has four probabilities $P(e_x,e_z)$, for depolarising noise $P(0,0)=1-p$ and $P(1,0)=P(0,1)=P(1,1)=p/3$. We keep the $X$ increment $e_x$ as it is and Fourier-transform the $Z$ increment,
\begin{equation}
 H(e_x,v)=\sum_{e_z}P(e_x,e_z)(-1)^{v\cdot e_z},\qquad
 H_{1,p}=\begin{pmatrix}1-\tfrac{2p}{3}&1-\tfrac{4p}{3}\\[.3ex]\tfrac{2p}{3}&0\end{pmatrix}
 \label{eq:mixed}
\end{equation}
with rows $e_x=0,1$ and columns $v=0,1$. Check one entry: $e_x=1$, $v=1$ collects the $X$ and $Y$ events with signs $+$ and $-$, $\tfrac p3-\tfrac p3=0$. As a matrix product, with $P$ indexed by $(e_x,e_z)$ and $W$ the one-bit Walsh matrix,
\begin{equation}
 H_{1,p}=P\,W=\begin{pmatrix}1-p&\tfrac p3\\[.2ex]\tfrac p3&\tfrac p3\end{pmatrix}\begin{pmatrix}1&1\\1&-1\end{pmatrix} .
 \label{eq:depol1}
\end{equation}
The picture below is Eq.~\eqref{eq:depol1} as a contraction: the orange box is the channel $P$, the blue box the Walsh matrix $W$, and summing their shared leg $e_z$ gives the kernel $H_{1,p}$, the object every later picture is built from.
\begin{center}
\begin{tikzpicture}[x=1cm,y=1cm]
\node[tn noise,minimum width=12mm,minimum height=9mm] (P) at (0,0) {$P$};
\node[tn tensor,minimum width=9mm,minimum height=9mm] (W) at (2.2,0) {$W$};
\node[font=\large] at (1.1,0.75) {$\sum_{e_z}$};
\draw[tn leg] (-1.3,0)--node[above,tn label] {$e_x$}(P.west);
\draw[tn leg] (P.east)--node[above,tn label] {$e_z$}(W.west);
\draw[tn leg] (W.east)--node[above,tn label] {$v$}(3.5,0);
\node[font=\large] at (4.3,0) {$=$};
\node[tn noise,minimum width=14mm,minimum height=9mm] (H) at (6.0,0) {$H_{1,p}$};
\draw[tn leg] (4.9,0)--node[above,tn label] {$e_x$}(H.west);
\draw[tn leg] (H.east)--node[above,tn label] {$v$}(7.2,0);
\end{tikzpicture}
\end{center}
In the picture the leg $e_z$ is the summed index of the matrix product. 
For two qubits, $e_x,v\in\F^2$ and the joint kernel is
\begin{equation}
 H_{2,p}(e_x,v)=(1-16p/15)\ind{e_x=0}+\frac{4p}{15}\ind{v=0} .
 \label{eq:depol2}
\end{equation}
Specified $X,Y,Z$ flips give, respectively,
\begin{equation}
 H_X=\begin{pmatrix}1-p&1-p\\p&p\end{pmatrix},\qquad
 H_Y=\begin{pmatrix}1-p&1-p\\p&-p\end{pmatrix},\qquad
 H_Z=\begin{pmatrix}1&1-2p\\0&0\end{pmatrix}.
 \label{eq:flipkernels}
\end{equation}
The minus signs in $H$ are not interference: $H$ is the Walsh transform of a classical distribution, unlike the amplitude sum of Eq.~\eqref{eq:gauss}. The reason for the transform is that a CNOT maps $X$ masks forward and $Z$ masks backward, and in the Fourier variable $v$ dual to $z$ the backward rule becomes the same forward rule as for $x$, the control bit added to the target, Eq.~\eqref{eq:mixedCNOT}. One grey parity tensor per bit then serves both components, and the channel weight is a plain factor in $(e_x,v)$. The next paragraph pushes one CNOT with its noise through this rule.

\paragraph{One gate with its noise.} A CNOT with control $c$ and target $t$ updates the mixed coordinates as
\begin{equation}
 (x_c,x_t)\mapsto(x_c,x_t\oplus x_c),\qquad
 (v_c,v_t)\mapsto(v_c,v_t\oplus v_c),
 \label{eq:mixedCNOT}
\end{equation}
that is, in matrix form over $\F$,
\[
\begin{pmatrix}x_c\\x_t\end{pmatrix}\mapsto\begin{pmatrix}1&0\\1&1\end{pmatrix}\begin{pmatrix}x_c\\x_t\end{pmatrix},\qquad
\begin{pmatrix}v_c\\v_t\end{pmatrix}\mapsto\begin{pmatrix}1&0\\1&1\end{pmatrix}\begin{pmatrix}v_c\\v_t\end{pmatrix},\qquad
\text{whereas}\quad\begin{pmatrix}z_c\\z_t\end{pmatrix}\mapsto\begin{pmatrix}1&1\\0&1\end{pmatrix}\begin{pmatrix}z_c\\z_t\end{pmatrix}.
\]
The $v$ matrix is the transpose of the $z$ matrix, which is what ``dual character'' means, and it happens to equal the $x$ matrix. The $v$ bits are dual characters, not forward $Z$-error masks. The rule preserves the pairing $v\cdot e_z$. The tensor $C$ in Fig.~\ref{fig:response-network}(a) is the indicator that both rules hold. Joining it to the following depolarising channel gives
\begin{equation}
 \sum_{\boldsymbol u}
 C(\boldsymbol u,\boldsymbol v';\boldsymbol x,\boldsymbol v)\,
 H_{2,p}(\boldsymbol x'\oplus\boldsymbol u,\boldsymbol v').
 \label{eq:gate-noise-tensor}
\end{equation}
\begin{center}
\begin{tikzpicture}[x=1cm,y=1cm]
\node[tn constraint,minimum width=14mm,minimum height=13mm] (C) at (2,1.05) {$C$\\\scriptsize CNOT};
\node[tn noise,minimum width=16mm,minimum height=10mm] (N) at (5.15,1.35) {$H_{2,p}$\\\scriptsize noise};
\draw[tn bundle] (.1,1.35)--node[above,tn label] {$\boldsymbol x$} ([yshift=3mm]C.west);
\draw[tn bundle] (.1,.75)--node[below,tn label] {$\boldsymbol v$} ([yshift=-3mm]C.west);
\draw[tn bundle] ([yshift=3mm]C.east)--node[above,tn label] {$\boldsymbol u$}(N.west);
\node[font=\large] at (3.55,2.2) {$\sum_{\boldsymbol u}$};
\draw[tn bundle] (N.east)--node[above,tn label] {$\boldsymbol x'$}(7.5,1.35);
\draw[tn bundle] ([yshift=-3mm]C.east)--(3.15,.75)--(3.15,-.15)--(7.5,-.15);
\draw[tn bundle] (5.15,-.15)--(N.south);
\node[tn copy] at (5.15,-.15) {};
\node[tn label,anchor=north east,inner sep=1pt] at (5.05,-.25) {$\delta$, Eq.~\eqref{eq:copytensor}};
\node[tn label,anchor=north] at (6.35,-.2) {$\boldsymbol v'$};
\end{tikzpicture}
\end{center}
Here $\boldsymbol u$ is the $X$ label just after the gate and $\boldsymbol x'$ the label after the noise; their XOR is the error increment supplied to the noise kernel, and $\boldsymbol v'$ appears in $C$, in $H_{2,p}$ and on the output leg, so the three are joined by the copy tensor $\delta$ of Eq.~\eqref{eq:copytensor}. Take an example at $p=10^{-3}$: $\boldsymbol x=(1,0)$ and $\boldsymbol v=(0,0)$ on the pair (control, target). $C$ forces $\boldsymbol u=(1,1)$ and $\boldsymbol v'=(0,0)$, so the block has one term: its value is $H_{2,p}(\boldsymbol x'\oplus(1,1),(0,0))$, which is $1-4p/5=0.9992$ at $\boldsymbol x'=(1,1)$ (the $X$ mask is unchanged, while all possible $Z$ errors are included in the noise average) and $4p/15\approx0.00027$ at $\boldsymbol x'=(1,0)$ (an $X$-mask flip on the target). Any $\boldsymbol v'\ne(0,0)$ gives $0$. Summing $\boldsymbol u$ is the noise average for this gate.

One gate with its noise is thus one small tensor, and chaining them along the CNOT circuit averages over every internal fault pattern at once. After all this work the mission has not changed: the tables $F_A$ and $F_B$, which give $A$ and $B$ and hence $\PL$. We are now one step from them.

\subsection{The open network that fills the table}\label{app:B-open}

Everything so far produced one entry of the table for one incoming error. Here the incoming labels are left open and everything else is contracted, so a single contraction returns the whole of $F_A$ and $F_B$, the objects that Eq.~\eqref{eq:responseaverage} turns into $A$, $B$ and $\PL$. We do it in a Fourier-transformed basis for the incoming $Z$ label. With $y=(y_0,\ldots,y_{r-1},c)\in\F^m$ define
\begin{equation}
 G^{(j)}(\alpha,y)=\sum_\beta(-1)^{\beta\cdot y}\big[F_A(\alpha,\beta)-2jF_B(\alpha,\beta)\big],\qquad j=0,1,
 \label{eq:G}
\end{equation}
so $G^{(0)}$ is the transformed acceptance $P_{\rm good}+P_{\rm bad}$ and $G^{(1)}$ the transformed difference $P_{\rm good}-P_{\rm bad}$. The bit $c$ is a Fourier coordinate, not a measured outcome. 
The Fourier transform is used because the tensor network is a sum over binary variables of products of small factors, and the $Z$ side of the incoming error enters the count only through parities such as $\ell\cdot g_i$, that is through characters, so it is cheaper to compute the characters $(-1)^{y\cdot h}$ directly, indexed by $y$, and transform once at the end than to keep the incoming $Z$ label $\beta$ as an open index on every intermediate tensor. The same applies to the central sign $s$, it enters only through a character, so the network is built with its Fourier partner $c$ and transformed at the end.

\paragraph{The site factor, term by term.} Choose a binary right inverse $R$ of the label map $\boldsymbol a\mapsto\alpha$ in Eq.~\eqref{eq:address} and fix the representative $\boldsymbol a=R\alpha$. At site $q$ let $s_q=\bigoplus_{i:q\in g_i}y_i$, and introduce summed face bits $u_i,w_i$ and a logical-coset bit $\lambda$. The terminal site factor is
\begin{equation}\small
 {\cal C}_q=\underbrace{\ind{v_q\oplus s_q=j}}_{\text{sector }j}\,
 \underbrace{\ind{x_q\oplus a_q\oplus\lambda\oplus\textstyle\bigoplus_{i:q\in g_i}\!u_i=0}}_{\text{code projection}}\,
 \underbrace{\ind{x_qs_q\oplus a_qc\oplus\textstyle\bigoplus_{i:q\in g_i}\!w_i=0}}_{\text{linear system}}\,
 \underbrace{(-1)^{a_qx_q(s_q\oplus c)}}_{\text{sign}} .
 \label{eq:sitefactor}
\end{equation}
In Eq.\eqref{eq:sitefactor}, the first indicator selects the character sector $j$, the same $j$ as in $G^{(j)}$. The second is $H(\boldsymbol u\oplus\boldsymbol a)=0$ in disguise: the total $X$ mask must be a sum of faces ($u_i$) plus optionally the logical mask ($\lambda$), site by site. The third is the linear system $Mh=b'$ site by site, with $w_i$ playing the role of the row multipliers. The sign is the leftover of the $\boldsymbol u\wedge\boldsymbol a$ correction in $\ell$. Its products are binary AND operations and the sign is retained in the contraction.
Figure~\ref{fig:terminalsplit} separates the three indicators from the sign.
{\color{black}
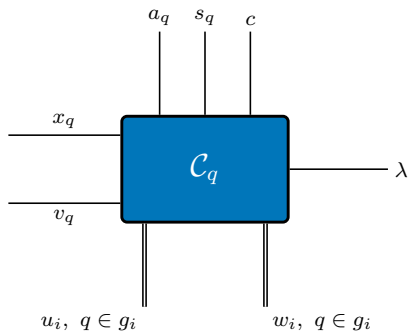
\begin{figure}[H]\centering
\begin{adjustbox}{max width=.7\linewidth}\begin{tikzpicture}[x=1cm,y=1cm,
  tensor/.style={tn tensor,minimum width=25mm,minimum height=10mm,
                 inner sep=1.5mm,font=\small},
  bundle/.style={tn bundle},
  copy bit/.style={tn copy}]
\node[note] at (3.7,4.43)
 {example: incoming bit $a_q=1$, central character $c=0$};
\node[tensor,tn constraint] (R) at (3.7,2.65)
 {$R^{(j)}$\\\scriptsize selection and support};
\node[tensor] (P) at (3.7,.25)
 {$\Phi$\\\scriptsize phase sign};
\draw[bundle] (3.7,3.9)--(R.north);
\node[note,anchor=west] at (3.88,3.65) {remaining code\\and character bits};
\coordinate (xdot) at (1.35,2.05);
\coordinate (sdot) at (6.05,2.05);
\draw[wire] (.65,2.05)--(xdot) node[midway,above,note] {$x_q$};
\draw[wire] (xdot)|-([yshift=1.8mm]R.west);
\draw[wire] (xdot)|-(P.west);
\draw[wire] (6.75,2.05)--(sdot) node[midway,above,note] {$s_q$};
\draw[wire] (sdot)|-([yshift=1.8mm]R.east);
\draw[wire] (sdot)|-(P.east);
\node[copy bit] at (xdot) {};
\node[copy bit] at (sdot) {};
\node[note] at (3.7,1.43)
 {$R^{(j)}=0$ for forbidden labels\\$R^{(j)}=1$ for permitted labels};
\node[note,font=\small] at (3.7,-1.15)
 {$\Phi(x_q,s_q)=(-1)^{x_qs_q}
   =\begin{pmatrix}+1&+1\\+1&-1\end{pmatrix}_{x_q,s_q}$};
\node[note] at (3.7,-2.05)
 {the same $x_q,s_q$ enter both factors;\\
  keep the minus sign when summing them};
\end{tikzpicture}\end{adjustbox}
\caption{The site factor of Eq.~\eqref{eq:sitefactor} for $a_q=1$, $c=0$, split into the three indicators $R^{(j)}$ (grey) and the sign $\Phi=(-1)^{x_qs_q}$ (blue). The dots copy $x_q$ and $s_q$ into both factors, so the product is taken entry by entry and the negative entry survives.}
\label{fig:terminalsplit}
\end{figure}}
\begin{center}
\begin{tikzpicture}[x=1cm,y=1cm]
\node[tn tensor,minimum width=22mm,minimum height=14mm,font=\large] (Cq) at (0,0) {$\mathcal C_q$};
\draw[tn leg] (-2.6,0.45)--node[above,tn label] {$x_q$}([yshift=4.5mm]Cq.west);
\draw[tn leg] (-2.6,-0.45)--node[below,tn label] {$v_q$}([yshift=-4.5mm]Cq.west);
\draw[tn leg] ([xshift=-6mm]Cq.north)--++(0,1.1) node[above,tn label] {$a_q$};
\draw[tn leg] (Cq.north)--++(0,1.1) node[above,tn label] {$s_q$};
\draw[tn leg] ([xshift=6mm]Cq.north)--++(0,1.1) node[above,tn label] {$c$};
\draw[tn bundle] ([xshift=-8mm]Cq.south)--++(0,-1.1) node[below left,tn label] {$u_i,\ q\in g_i$};
\draw[tn bundle] ([xshift=8mm]Cq.south)--++(0,-1.1) node[below right,tn label] {$w_i,\ q\in g_i$};
\draw[tn leg] (Cq.east)--++(1.3,0) node[right,tn label] {$\lambda$};
\node[tn label,text=black!60] at (0,-2.9) {one such tensor per data site, all seven share $\lambda$, $c$ and the face bits};
\end{tikzpicture}
\end{center}
For a site in three faces this tensor has $2^{12}$ entries. The three parity indicators each halve the support, leaving $512$ nonzero entries, and the sign turns an entry into $-1$ exactly when $a_qx_q(s_q\oplus c)=1$. At the central measurement the insertion is $v_{\rm old}\oplus v_{\rm new}\oplus c=j$. Fourier expansion of the ancillary conditions then gives
\begin{equation}
 G_c^{(j)}(\alpha,y_0,\ldots,y_{r-1};p)=2^{-n_{\rm aux}}\sum_{\text{internal bits}}\prod_\nu{\cal T}_\nu,\qquad
 n_{\rm aux}=6\ (d=3),\quad19\ (d=5),
 \label{eq:opennetwork}
\end{equation}
where the factors ${\cal T}_\nu$ are the channel kernels, the CNOT constraints, the central insertion and Eq.~\eqref{eq:sitefactor}. Several $X$ masks share one label $\alpha$, but the condition $\boldsymbol a=R\alpha$ picks exactly one of them per $\alpha$, so each open value of $\alpha$ is counted once and no entry is multiplied by the size of the coset. For fixed $c,j$ the output has $2^{4+3}$ entries at $d=3$ and $2^{10+9}$ at $d=5$.
\begin{figure}[H]\centering
\begin{adjustbox}{max width=\linewidth}
\begin{tikzpicture}[x=1cm,y=1cm]
\begin{scope}[scale=0.82,transform shape,shift={(0.4,0.7)}]
\node[tn constraint,minimum width=95mm,minimum height=10mm,font=\small] (L) at (6,4.2) {label maps and shared code indices ($\boldsymbol a=R\alpha$, $s_q$, $\lambda$, $u_i$, $w_i$)};
\foreach \i/\dy in {0/.3,1/.1,2/-.1,3/-.3}{
 \draw[tn leg] (0.4,4.2+\dy)--([yshift=\dy cm]L.west);
 \draw[tn leg] ([yshift=\dy cm]L.east)--(11.6,4.2+\dy);
}
\node[tn label] at (0.25,4.85) {$\alpha$};
\node[tn label] at (11.75,4.85) {$y$};
\foreach \q/\xx in {0/2.2,3/3.45,7/4.7,9/5.95,10/7.2,12/8.45,13/9.7}{
 \node[tn tensor,minimum width=9mm,minimum height=8mm,font=\small] (C\q) at (\xx,2.3) {$\mathcal C_{\q}$};
 \draw[tn bundle] (\xx,3.7)--(C\q.north);
 \draw[tn bundle] (C\q.south)--(\xx,1.35);
}
\node[tn intermediate,minimum width=95mm,minimum height=10mm,font=\small] (K) at (6,0.85) {CNOT constraints, noise kernels, central insertion};
\node[font=\Large,anchor=east] at (-0.1,2.5) {$G^{(j)}(\alpha,y;p)=\dfrac{1}{2^{6}}\mathop{\scalebox{1.2}{$\displaystyle\sum$}}\limits_{\substack{\text{\scriptsize internal}\\\text{\scriptsize bits}}}$};
\node[tn label] at (6,-0.1) {one number per open label pair $(\alpha,y)$};
\end{scope}
\end{tikzpicture}
\end{adjustbox}
\caption{The open network of Eq.~\eqref{eq:opennetwork} at $d=3$, the incoming labels $\alpha$ and $y$ left open. Its contraction is the table $G^{(j)}(\alpha,y;p)$, one number per open label pair, which Eq.~\eqref{eq:inverseF} turns into $F_A$ and $F_B$.}
\label{fig:opennet}
\end{figure}
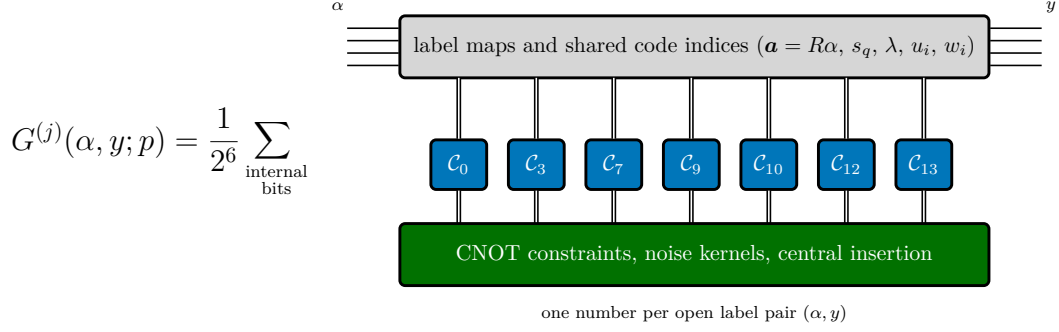
Figure~\ref{fig:opennet} is the network of Eq.~\eqref{eq:opennetwork} at $d=3$, its three layers described from the top down. Top bar: the incoming labels and the bits shared between sites. Middle: one site factor per data site. Bottom bar: the Clifford part with its noise, exactly the gate-plus-noise blocks of the previous subsection. At $d=3$ the network has $2^{4}$ open values of $\alpha$ and $2^{4}$ of $y$ for each $j$, so one contraction returns $512$ numbers, and the factor $1/2^{6}$ is the six ancilla projections. Finally invert the incoming $Z$ transform,
\begin{align}
 F_A(\alpha,\beta)&=2^{-m}\sum_y(-1)^{\beta\cdot y}G^{(0)}(\alpha,y),
 \nonumber\\
 F_B(\alpha,\beta)&=2^{-m-1}\sum_y(-1)^{\beta\cdot y}\big[G^{(0)}(\alpha,y)-G^{(1)}(\alpha,y)\big],
 \label{eq:inverseF}
\end{align}
one more contraction with $W^{\otimes m}$, the Walsh matrix of Eq.~\eqref{eq:depol1} on every $y$ leg, a $16\times16$ matrix of $\pm1$ entries at $d=3$ applied to each row $G^{(0)}(\alpha,\cdot)$. The factor $\tfrac12$ turns $(P_{\rm good}+P_{\rm bad})-(P_{\rm good}-P_{\rm bad})$ into the wrong probability $P_{\rm bad}$.
\begin{center}
\begin{tikzpicture}[x=1cm,y=1cm]
\node[font=\large] at (-1.3,0) {$F_A(\alpha,\beta)=\dfrac{1}{2^{m}}$};
\node[font=\Large] at (0.5,0) {$\displaystyle\sum_{y}$};
\node[tn intermediate,minimum width=15mm,minimum height=13mm,font=\large] (G) at (2.4,0) {$G^{(0)}$};
\draw[tn bundle] (1.2,0)--node[above,tn label] {$\alpha$}(G.west);
\node[tn tensor,minimum width=8mm,minimum height=8mm] (W) at (4.6,0) {$W^{\otimes m}$};
\draw[tn bundle] (G.east)--node[above,tn label] {$y$}(W.west);
\draw[tn bundle] (W.east)--node[above,tn label] {$\beta$}(6.0,0);
\node[tn label] at (2.6,-1.2) {$W_{\beta_iy_i}=(-1)^{\beta_iy_i}$ on each of the $m$ bits, and $F_B$ uses $(G^{(0)}-G^{(1)})/2$ instead};
\end{tikzpicture}
\end{center}
The resulting $F_A,F_B$ are nonnegative ordinary probabilities for each incoming label pair $(\alpha,\beta)$ and are the tables used in Eq.~\eqref{eq:physicalgrowth}.

\subsection{The same objects at \texorpdfstring{$d=5$}{d=5}}\label{app:B-dfive}

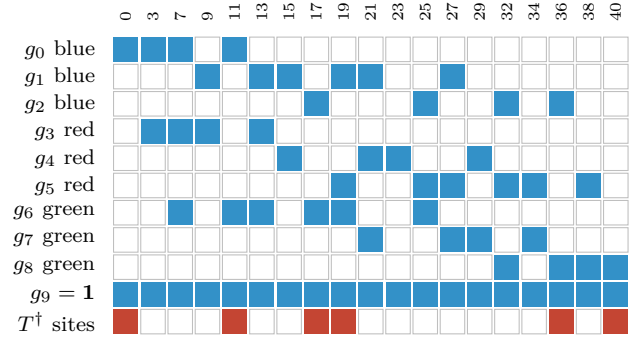
\begin{figure}[H]\centering
\begin{tikzpicture}[x=3.6mm,y=3.6mm]
\foreach \row [count=\r from 0] in \dfivegrid {
  \foreach \bit [count=\c from 0] in \row {
    \ifnum\bit=1
      \ifnum\r=10 \fill[tnorange!80] (\c,-\r) rectangle ++(0.9,0.9); \else \fill[tnblue!80] (\c,-\r) rectangle ++(0.9,0.9); \fi
    \else \draw[black!25] (\c,-\r) rectangle ++(0.9,0.9); \fi
  }
}
\foreach \site [count=\c from 0] in \dfivesites {\node[font=\tiny,rotate=90,anchor=west] at (\c+0.45,1.1) {$\site$};}
\foreach \lab [count=\r from 0] in {$g_0$ blue,$g_1$ blue,$g_2$ blue,$g_3$ red,$g_4$ red,$g_5$ red,$g_6$ green,$g_7$ green,$g_8$ green,$g_9=\boldsymbol1$,$T^\dagger$ sites} {\node[font=\scriptsize,anchor=east] at (-0.3,-\r+0.45) {\lab};}
\end{tikzpicture}
\caption{The $d=5$ incidence matrix $H^{(5)}$ (nine faces), the logical row $\boldsymbol1$, and the sites where the exit layer applies $T^\dagger$ (orange). Face sizes are $(4,6,4,4,4,6,6,4,4)$.}
\label{fig:dfive}
\end{figure}

Nothing changes in the formula except the sizes: nineteen data sites, the nine faces of Table~\ref{tab:faces}, $r=9$, $m=10$, and $\epsilon_q=-1$ on the six sites $0,11,17,19,36,40$ where the exit layer applies $T^\dagger$, Fig.~\ref{fig:dfive}. The signed face weights $\dfivesw$ are all divisible by four and $t=13-6=\dfivephase$, which is the condition for $DX_{\rm all}D^\dagger$ to be a logical operator. The implementation checks it and rejects the all-$T$ pattern.

\clearpage
\newcommand{\Tone}{\begin{pmatrix}1&2\\0&1\end{pmatrix}}
\newcommand{\Ttwo}{\begin{pmatrix}1&1\\2&0\end{pmatrix}}
\newcommand{\Tthree}{\begin{pmatrix}0&1\\1&1\end{pmatrix}}
\newcommand{\Tfour}{\begin{pmatrix}2&1\\1&0\end{pmatrix}}
\newcommand{\Mtwelve}{\begin{pmatrix}\colorbox{tngreen!22}{$5$}&\colorbox{tngreen!22}{$1$}\\\colorbox{tngreen!22}{$2$}&\colorbox{tngreen!22}{$0$}\end{pmatrix}}
\newcommand{\Mthirtyfour}{\begin{pmatrix}\colorbox{tngreen!22}{$1$}&\colorbox{tngreen!22}{$0$}\\\colorbox{tngreen!22}{$3$}&\colorbox{tngreen!22}{$1$}\end{pmatrix}}
\newcommand{\Zvalue}{8}
\newcommand{\Kthirteen}{\begin{pmatrix}0&\colorbox{tnorange!22}{$1$}&\colorbox{tnorange!22}{$1$}&\colorbox{tnorange!22}{$1$}\\0&\colorbox{tnorange!22}{$2$}&\colorbox{tnorange!22}{$2$}&\colorbox{tnorange!22}{$2$}\\0&0&0&0\\0&\colorbox{tnorange!22}{$1$}&\colorbox{tnorange!22}{$1$}&\colorbox{tnorange!22}{$1$}\end{pmatrix}}
\newcommand{\Ktwentyfour}{\begin{pmatrix}\colorbox{tnorange!22}{$2$}&\colorbox{tnorange!22}{$1$}&\colorbox{tnorange!22}{$2$}&\colorbox{tnorange!22}{$1$}\\\colorbox{tnorange!22}{$4$}&\colorbox{tnorange!22}{$2$}&0&0\\\colorbox{tnorange!22}{$1$}&0&\colorbox{tnorange!22}{$1$}&0\\\colorbox{tnorange!22}{$2$}&0&0&0\end{pmatrix}}
\newcommand{\Kentries}{9/9}
\newcommand{\Mtwelveexample}{1\cdot1 + 2\cdot2 = 5}
\newcommand{\Cqzero}{\begin{pmatrix}0&0&0&0\\\colorbox{tnblue!22}{$1$}&0&0&0\end{pmatrix}}
\newcommand{\Cqlam}{\begin{pmatrix}\colorbox{tnblue!22}{$1$}&0&0&\colorbox{tnblue!22}{$1$}\\0&0&0&0\end{pmatrix}}
\newcommand{\CqW}{\begin{pmatrix}0&0&0&0\\0&0&0&\colorbox{tnorange!22}{$-1$}\end{pmatrix}}
\newcommand{\Cqnonzero}{8}
\newcommand{\Cqneg}{2}
\newcommand{\Phimatrix}{\begin{pmatrix}1&1\\1&\colorbox{tnorange!22}{$-1$}\end{pmatrix}}
\newcommand{\toyOA}{725571/800000}
\newcommand{\toyOB}{19701/800000}
\newcommand{\toyPL}{199/7329}
\newcommand{\toyW}{\begin{pmatrix}9/10&1/20\\1/40&1/40\end{pmatrix}}
\newcommand{\toyFA}{\begin{pmatrix}1&1/2\\0&0\end{pmatrix}}
\newcommand{\toyFB}{\begin{pmatrix}0&1/2\\0&0\end{pmatrix}}
\newcommand{\joinentries}{4}

\providecommand{\hl}[2]{\colorbox{#1!22}{#2}}
\section{Contraction orders and the growth average}\label{app:tensor-examples-v2}

This appendix describes two things the later appendices require. First, a contraction order and its tree, shown on a toy example of four contracted tensors. Then the growth average written as a sum over fault configurations, which Appendix~\ref{app:growth} evaluates as one contraction.
\subsection{Two orders for the same contraction}
A contraction order specifies which tensors to combine first. Draw a bubble around a group, multiply their entries, sum the indices used only inside that group, and retain the collective tensor legs. 
This results in one tensor that can then be combined with the next group. 
In Fig.~\ref{fig:contraction-tree} the four indices are binary and the complete scalar is
\begin{equation}
 Z=\sum_{a,b,c,d}T_1(a,b)T_2(b,c)T_3(c,d)T_4(d,a).
 \label{eq:treeexample}
\end{equation}
{\color{black}
 \begin{figure}[H]\centering
 \begin{adjustbox}{max width=.98\linewidth}\begin{tikzpicture}[x=1cm,y=1cm,
  original/.style={tn tensor,fill=tnorange,minimum width=8mm,minimum height=8mm,inner sep=1mm},
  merged/.style={tn intermediate,minimum width=14mm,inner sep=1.5mm},
  index/.style={tn label,fill=white,inner sep=.5mm}]
\node[tn heading] at (2.8,5.65) {\textbf{(a)} Group neighbouring tensors};
\node[tn heading] at (10.3,5.65) {\textbf{(b)} Group opposite tensors};
\node[original] (t1) at (1.1,4.4) {$T_1$};
\node[original] (t2) at (4.5,4.4) {$T_2$};
\node[original] (t4) at (1.1,2.6) {$T_4$};
\node[original] (t3) at (4.5,2.6) {$T_3$};
\begin{scope}[on background layer]
\node[draw=tngreen!70,fill=tngreen!5,dashed,rounded corners=5mm,
  fit=(t1)(t2),inner sep=4mm] {};
\node[draw=tngreen!70,fill=tngreen!5,dashed,rounded corners=5mm,
  fit=(t4)(t3),inner sep=4mm] {};
\end{scope}
\draw[tn leg] (t1)--node[index,above] {$b$}(t2);
\draw[tn leg] (t4)--node[index,below] {$d$}(t3);
\draw[tn leg] (t1)--node[index,left] {$a$}(t4);
\draw[tn leg] (t2)--node[index,right] {$c$}(t3);
\draw[tn operation] (2.8,1.75)--node[tn label,right] {sum the internal $b,d$}(2.8,1.0);
\node[merged,minimum height=12mm] (m12) at (1.1,.15) {$M_{12}$};
\node[merged,minimum height=12mm] (m34) at (4.5,.15) {$M_{34}$};
\draw[tn leg] ([yshift=2.5mm]m12.east)--node[index,above] {$a$}
  ([yshift=2.5mm]m34.west);
\draw[tn leg] ([yshift=-2.5mm]m12.east)--node[index,below] {$c$}
  ([yshift=-2.5mm]m34.west);
\node[tn label] at (2.8,-.85) {two binary indices per tensor: \textbf{4 entries}};
\draw[tn operation] (2.8,-1.2)--node[tn label,right] {sum $a,c$}(2.8,-1.8);
\node[tn label,font=\small] at (2.8,-2.45) {$Z=\displaystyle\sum_{a,c}\underbrace{\textstyle\sum_b T_1(a,b)T_2(b,c)}_{M_{12}(a,c)}\;\underbrace{\textstyle\sum_d T_3(c,d)T_4(d,a)}_{M_{34}(c,a)}$};

\node[original] (s1) at (8.6,4.4) {$T_1$};
\node[original] (s2) at (12,4.4) {$T_2$};
\node[original] (s4) at (8.6,2.6) {$T_4$};
\node[original] (s3) at (12,2.6) {$T_3$};
\begin{scope}[on background layer]
\foreach \angle in {-27.897,27.897}{
  \begin{scope}[shift={(10.3,3.5)},rotate=\angle]
  \path[draw=tnorange!75,fill=tnorange!5,dashed,line width=.6pt]
    (-2.62,0)
    .. controls (-2.62,.48) and (-2.31,.74) .. (-1.92,.74)
    .. controls (-1.25,.74) and (-.98,.16) .. (0,.16)
    .. controls (.98,.16) and (1.25,.74) .. (1.92,.74)
    .. controls (2.31,.74) and (2.62,.48) .. (2.62,0)
    .. controls (2.62,-.48) and (2.31,-.74) .. (1.92,-.74)
    .. controls (1.25,-.74) and (.98,-.16) .. (0,-.16)
    .. controls (-.98,-.16) and (-1.25,-.74) .. (-1.92,-.74)
    .. controls (-2.31,-.74) and (-2.62,-.48) .. cycle;
  \end{scope}
}
\end{scope}
\draw[tn leg] (s1)--node[index,above] {$b$}(s2);
\draw[tn leg] (s4)--node[index,below] {$d$}(s3);
\draw[tn leg] (s1)--node[index,left] {$a$}(s4);
\draw[tn leg] (s2)--node[index,right] {$c$}(s3);
\draw[tn operation] (10.3,1.75)--node[tn label,right] {no internal index to sum}(10.3,1.0);
\node[merged,minimum height=17mm] (k13) at (8.6,.05) {$K_{13}$};
\node[merged,minimum height=17mm] (k24) at (12,.05) {$K_{24}$};
\foreach \y/\lab in {6/a,2/b,-2/c,-6/d}{
  \draw[tn leg] ([yshift=\y mm]k13.east)--node[index] {$\lab$}
    ([yshift=\y mm]k24.west);
}
\node[tn label] at (10.3,-1.15) {four binary indices per tensor: \textbf{16 entries}};
\draw[tn operation] (10.3,-1.5)--node[tn label,right] {sum $a,b,c,d$}(10.3,-1.9);
\node[tn label,font=\small] at (10.3,-2.55) {$Z=\displaystyle\sum_{a,b,c,d}\underbrace{T_1(a,b)T_3(c,d)}_{K_{13}(a,b,c,d)}\;\underbrace{T_2(b,c)T_4(d,a)}_{K_{24}(a,b,c,d)}$};
\end{tikzpicture}\end{adjustbox}
 \caption{Two orders for the same network. Each bubble becomes one tensor, and its exposed indices determine its size.}\label{fig:contraction-tree}\end{figure}
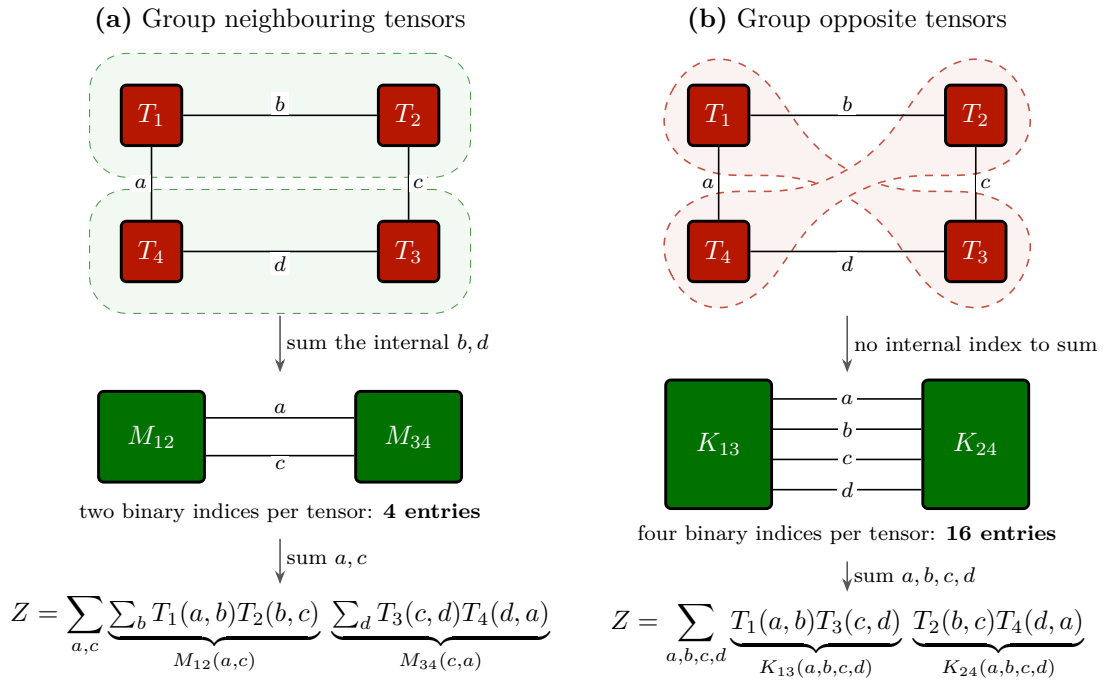}
\FloatBarrier

For example, take the tensors, with (rows, columns) being the (first, second) indices,
\[
T_1=\Tone,\quad T_2=\Ttwo,\quad T_3=\Tthree,\quad T_4=\Tfour .
\]
Joining neighbours first sums $b$ and $d$,
\begin{equation}
 \begin{aligned}
 M_{12}(a,c)&=\sum_bT_1(a,b)T_2(b,c)=\Mtwelve,\\
 M_{34}(c,a)&=\sum_dT_3(c,d)T_4(d,a)=\Mthirtyfour,\\
 Z&=\sum_{a,c}M_{12}(a,c)M_{34}(c,a)=\Zvalue ,
 \end{aligned}
 \label{eq:treejoins}
\end{equation}
for instance $M_{12}(0,0)=\Mtwelveexample$. Each intermediate has four entries. Joining opposite tensors first sums nothing, because $T_1$ and $T_3$ share no index, so the intermediates keep all four legs open. Each of $K_{13}(a,b,c,d)=T_1(a,b)T_3(c,d)$ and $K_{24}(a,b,c,d)=T_2(b,c)T_4(d,a)$ then has sixteen entries, and $Z=\sum_{a,b,c,d}K_{13}K_{24}=\Zvalue$ again. The order changes the storage and the arithmetic but both calculations sum to the same $Z$.
{\color{black}
\begin{center}
\begin{tikzpicture}[x=1cm,y=1cm]
\node[tn tensor,fill=tnorange,minimum width=8mm,minimum height=8mm] (t1) at (0,1.6) {$T_1$};
\node[tn tensor,fill=tnorange,minimum width=8mm,minimum height=8mm] (t2) at (2.2,1.6) {$T_2$};
\node[tn tensor,fill=tnorange,minimum width=8mm,minimum height=8mm] (t4) at (0,0) {$T_4$};
\node[tn tensor,fill=tnorange,minimum width=8mm,minimum height=8mm] (t3) at (2.2,0) {$T_3$};
\draw[tn leg] (t1)--node[above,tn label] {$b$}(t2); \draw[tn leg] (t4)--node[below,tn label] {$d$}(t3);
\draw[tn leg] (t1)--node[left,tn label] {$a$}(t4); \draw[tn leg] (t2)--node[right,tn label] {$c$}(t3);
\node[font=\Large] at (-1.6,0.8) {$\displaystyle\sum_{a,b,c,d}$};
\node[font=\large] at (3.4,0.8) {$=$};
\node[font=\Large] at (4.4,0.8) {$\displaystyle\sum_{a,c}$};
\node[tn intermediate,minimum width=12mm,minimum height=11mm] (m12) at (5.6,0.8) {$M_{12}$};
\node[tn intermediate,minimum width=12mm,minimum height=11mm] (m34) at (7.6,0.8) {$M_{34}$};
\draw[tn leg] ([yshift=2.5mm]m12.east)--node[above,tn label] {$a$}([yshift=2.5mm]m34.west);
\draw[tn leg] ([yshift=-2.5mm]m12.east)--node[below,tn label] {$c$}([yshift=-2.5mm]m34.west);
\node[font=\large] at (8.7,0.8) {$=\ \Zvalue$};
\end{tikzpicture}
\end{center}}

The order in which tensors are joined does not change the answer but greatly affects the size of the intermediate tables. For a large network, contraction order decides whether the contraction fits in memory at all. 
The growth stage (see Fig.~\ref{fig:overview}) has 569 factors, so the contraction order decides how large the intermediate arrays blow up to. The plan used for the growth sum keeps the largest stored table at $2^{23}$ entries while the largest single join multiplies $2^{27}$ pairs of entries before summing them.

\subsection{Including the noisy growth stage}\label{app:growth-overview-v2}
The earlier $d=3$ stage supplies weights for its outgoing syndrome and logical Pauli label. Clifford growth propagates these labels and adds the labels of its faults. They combine by XOR: two faults may cancel a detector mismatch, so they must not be rejected separately. For a growth-fault configuration $g$, let $w_g$ be its probability, $h(g)$ its detector-compatibility label, $s(g)$ its required input syndrome, and $v(g)$ its output syndrome and logical frame. Write $W_{s,t}$ for the earlier-stage weight with syndrome $s$ and logical Pauli label $t$. Then
\begin{equation}
 \begin{gathered}
 O=\sum_{g,t}w_g\,\ind{h(g)=0}\,
       W_{s(g),t}\,F_O\!\left(\Gamma(v(g),t)\right),\\
 O\in\{A,B\}.
 \end{gathered}
 \label{eq:physicalgrowth}
\end{equation}
The map $\Gamma$ retains the outgoing syndrome and XORs $t$ into the logical frame. This is Eq.~\eqref{eq:responseaverage} with the earlier weights written explicitly. 
Reading Eq.~\eqref{eq:physicalgrowth} from right to left, $F_O$ is the final $d=5$ response table, indexed by the label that arrives at the last check. $W_{s,t}$ is the first-stage table, first-stage post-selection included. 
In between, a configuration $g$ happens with probability $w_g$, is kept only if its detector label is zero, needs the input syndrome $s(g)$, and emits $v(g)$. 

As a toy example, take two syndromes, two tags and four growth configurations at $p=10^{-2}$. Rows are indexed by the syndrome, columns by the tag of $W$ or by the frame of $F_O$, so that
\[
W=\toyW,\qquad F_A=\toyFA,\qquad F_B=\toyFB .
\]
Each configuration contributes one row per tag:
{\color{black}
\begin{center}\small\begin{tabular}{ccccccc}\toprule
$g$ & $t$ & $w_g$ & $h(g)$ & $s(g)$ & $\Gamma(v(g),t)$ & contribution to $A$, to $B$ \\\midrule
1 & 0 & $9801/10000$ & 0 & 0 & $(0,0)$ & $88209/100000$, $0$ \\
1 & 1 & $9801/10000$ & 0 & 0 & $(0,1)$ & $9801/400000$, $9801/400000$ \\
2 & 0 & $99/10000$ & 0 & 1 & $(0,1)$ & $99/800000$, $99/800000$ \\
2 & 1 & $99/10000$ & 0 & 1 & $(0,0)$ & $99/400000$, $0$ \\
3 & 0 & $99/10000$ & 1 & 0 & $(1,0)$ & $0$, $0$ \\
3 & 1 & $99/10000$ & 1 & 0 & $(1,1)$ & $0$, $0$ \\
4 & 0 & $1/10000$ & 0 & 0 & $(1,0)$ & $0$, $0$ \\
4 & 1 & $1/10000$ & 0 & 0 & $(1,1)$ & $0$, $0$ \\
\bottomrule\end{tabular}
\end{center}}
Summing the last column gives $A=\toyOA$ and $B=\toyOB$. Configuration 3 is dropped by $h(g)=1$ whatever its weight, configuration 2 only ever meets the $s=1$ row of $W$, its frame flip moves the tag before it reaches $F_O$, and configuration 4 is compatible but lands on output syndrome 1, where both $F_A$ and $F_B$ vanish in this toy example. 

Two things the formula insists on.
\begin{enumerate}
\item All 64 input syndromes remain in $W$: a growth fault can cancel an input syndrome, as configuration 2 does, so the syndrome-0 row alone would miss accepted events. 
\item Growth faults combine by XOR with each other and with the earlier label before anything is rejected. 
\end{enumerate}
Growth removes cross-syndrome coherence in the way proved in Appendix~\ref{app:growth}, which is what makes a table of probabilities $W_{s,t}$ sufficient. 
Nothing is post-selected between the two stages. Every first-stage syndrome enters growth, and the noiseless final code projection of the standalone $d=3$ calculation is not applied at this interface.
 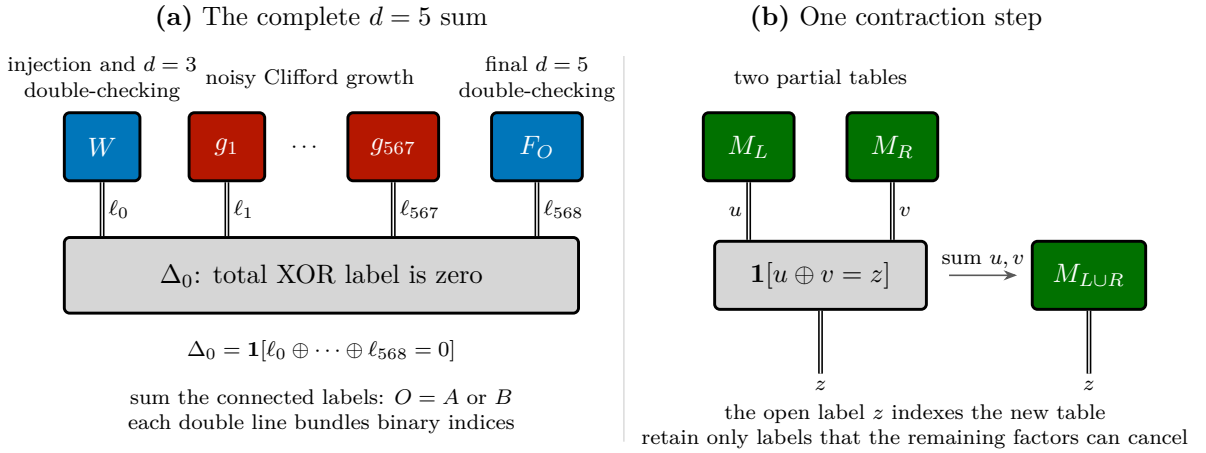
\begin{figure}[H]\centering
 \begin{adjustbox}{max width=.98\linewidth}\begin{tikzpicture}[x=1cm,y=1cm]
\node[tn heading] at (3.45,4.75) {\textbf{(a)} The complete $d=5$ sum};
\node[tn heading] at (11.05,4.75) {\textbf{(b)} One contraction step};

\node[tn label] at (.55,3.95) {injection and $d=3$\\double-checking};
\node[tn label] at (3.3,3.95) {noisy Clifford growth};
\node[tn label] at (6.3,3.95) {final $d=5$\\double-checking};
\node[tn endpoint,minimum width=10mm] (w) at (.55,3.05) {$W$};
\node[tn noise,minimum width=10mm] (g1) at (2.2,3.05) {$g_1$};
\node[tn label] at (3.25,3.05) {$\cdots$};
\node[tn noise,minimum width=12mm] (gn) at (4.4,3.05) {$g_{567}$};
\node[tn endpoint,minimum width=12mm] (f) at (6.3,3.05) {$F_O$};
\node[tn constraint,minimum width=68mm,minimum height=10mm] (delta) at (3.45,1.35)
  {$\Delta_0$: total XOR label is zero};
\foreach \name/\lab in {w/\ell_0,g1/\ell_1,gn/\ell_{567},f/\ell_{568}}{
  \draw[tn bundle] (\name.south)--node[right,tn label] {$\lab$}
    (\name.south |- delta.north);
}
\node[tn label] at (3.45,.35)
  {$\Delta_0=\mathbf1[\ell_0\oplus\cdots\oplus\ell_{568}=0]$};
\node[tn label] at (3.45,-.45)
  {sum the connected labels: $O=A$ or $B$\\
   each double line bundles binary indices};

\draw[black!20] (7.45,-.85)--(7.45,4.25);
\node[tn label] at (10.05,3.95) {two partial tables};
\node[tn intermediate,minimum width=12mm] (ml) at (9.1,3.05) {$M_L$};
\node[tn intermediate,minimum width=12mm] (mr) at (11,3.05) {$M_R$};
\node[tn constraint,minimum width=28mm] (xor) at (10.05,1.35)
  {$\mathbf1[u\oplus v=z]$};
\draw[tn bundle] (ml.south)--node[left,tn label] {$u$}(ml.south |- xor.north);
\draw[tn bundle] (mr.south)--node[right,tn label] {$v$}(mr.south |- xor.north);
\draw[tn bundle] (xor.south)--(10.05,.05) node[below,tn label] {$z$};
\draw[tn operation] (11.75,1.35)--node[above,tn label] {sum $u,v$}(12.65,1.35);
\node[tn intermediate,minimum width=15mm] (parent) at (13.6,1.35) {$M_{L\cup R}$};
\draw[tn bundle] (parent.south)--(13.6,.05) node[below,tn label] {$z$};
\node[tn label] at (11.3,-.65)
  {the open label $z$ indexes the new table\\
   retain only labels that the remaining factors can cancel};
\end{tikzpicture}\end{adjustbox}
 \caption{The $d=5$ tensor network and one XOR contraction. Rounded boxes are tensors and the arrow denotes a calculation step.}\label{fig:growth-network}\end{figure}

Fig.~\ref{fig:growth-network}(a) groups the growth noise into 567 distributions $g_i$ between $W$ and the final response $F_O$. The grey constraint matches their labels by requiring a zero total XOR. Panel (b) combines two partial tables $M_L(u),M_R(v)$ by summing over $u\oplus v=z$ and keeping the output label $z$. With two-bit labels and three entries each,
{\color{black}
\begin{center}\small\begin{tabular}{cccc}\toprule
$z$ & pairs $(u,v)$ with $u\oplus v=z$ & $\sum M_L(u)M_R(v)$ & value \\\midrule
$(0,0)$ & $(00,00)$, $(10,10)$ & $9/10\cdot4/5 + 1/20\cdot1/10$ & $29/40$ \\
$(0,1)$ & $(10,11)$, $(01,00)$ & $1/20\cdot1/10 + 1/20\cdot4/5$ & $9/200$ \\
$(1,0)$ & $(00,10)$, $(10,00)$, $(01,11)$ & $9/10\cdot1/10 + 1/20\cdot4/5 + 1/20\cdot1/10$ & $27/200$ \\
$(1,1)$ & $(00,11)$, $(01,10)$ & $9/10\cdot1/10 + 1/20\cdot1/10$ & $19/200$ \\
\bottomrule\end{tabular}
\end{center}}
the parent table has \joinentries\ entries, one per reachable $z$. Only labels that can be cancelled by the remaining factors are stored. Appendix~\ref{app:growth} gives the binary constraints, contraction and exact Fourier slicing over all characters

\clearpage
\newcommand{\rhotoy}{\begin{pmatrix}0.5&0.1&\colorbox{tnorange!22}{$0.2$}&0\\0.1&0.2&0&\colorbox{tnorange!22}{$0.1$}\\\colorbox{tnorange!22}{$0.2$}&0&0.2&0.05\\0&\colorbox{tnorange!22}{$0.1$}&0.05&0.1\end{pmatrix}}
\newcommand{\rhodeph}{\begin{pmatrix}\colorbox{tnblue!22}{$0.5$}&\colorbox{tnblue!22}{$0.1$}&0&0\\\colorbox{tnblue!22}{$0.1$}&\colorbox{tnblue!22}{$0.2$}&0&0\\0&0&\colorbox{tnblue!22}{$0.2$}&\colorbox{tnblue!22}{$0.05$}\\0&0&\colorbox{tnblue!22}{$0.05$}&\colorbox{tnblue!22}{$0.1$}\end{pmatrix}}
\newcommand{\Kout}{\begin{pmatrix}0.03125&0.00625&0&0\\0.00625&0.0125&0&0\\0&0&0&0\\0&0&0&0\end{pmatrix}}
\newcommand{\lamval}{700035623/703125000}
\newcommand{\gzero}{1403160623/1406250000}
\newcommand{\gone}{3089377/1406250000}
\newcommand{\lamnum}{0.995606}
\newcommand{\gzeronum}{0.997803}
\newcommand{\gonenum}{0.002197}
\newcommand{\Pgx}{44802279872/44865134955}
\newcommand{\gxzero}{44901139936/44865134955}
\newcommand{\gxone}{98860064/44865134955}
\newcommand{\Atoy}{997003/1000000}
\newcommand{\Atoynum}{0.997003000}
\newcommand{\survivors}{000, 111}
\newcommand{\Czero}{499001/500000}
\newcommand{\Cone}{249001/250000}
\newcommand{\Czeronum}{0.998002000}
\newcommand{\Conenum}{0.996004000}

\providecommand{\hl}[2]{\colorbox{#1!22}{#2}}
\section{The complete growth average}\label{app:growth}

The growth stage of Fig.~\ref{fig:overview} is where the $d=5$ averaging becomes hard as its 908 noise channels sit between the two non-Clifford double-checking stages (Figs.~\ref{fig:d3circuit} and~\ref{fig:d5circuit}) and their faults combine by XOR with the first-stage labels, so no fault can be rejected on its own and enumerating all fault patterns is out of reach.
This appendix section shows how to fix this issue and sum the tensors exactly as one tensor contraction between the $d=3$ table $W$ and the $d=5$ response $F_O$.
Figure~\ref{fig:growth-network} shows the solution, one tensor network whose three ingredients are:
\begin{enumerate}
\item the first-stage table, 
\item the growth-noise distributions, and 
\item  the final double-checking response. 
\end{enumerate}
This appendix derives their interface labels and the contraction used to sum them. 
Summing this network exactly, with no fault pattern ever enumerated, takes the steps of the flow chart below.

\begin{center}
\adjustbox{max width=\linewidth}{%
\begin{tikzpicture}[x=1cm,y=1cm]
\tikzset{st/.style={box,text width=30mm,inner sep=1.5mm,font=\small}}
\node[st] (a) at (0,0) {\ref{app:E-blocks}: why a table of probabilities per syndrome block is enough};
\node[st,right=3mm of a] (b) {\ref{app:E-table}: the first-stage table $W_{s,t}$, $64\times4$};
\node[st,right=3mm of b] (c) {\ref{app:E-labels}: every factor as a distribution over binary labels, one shared XOR constraint};
\node[st,right=3mm of c] (d) {\ref{app:E-boundary}, \ref{app:E-tree}: contract along a tree, keeping only labels the rest can cancel};
\node[st,right=3mm of d] (e) {\ref{app:E-fourier}: eight Fourier characters cut the label space, averaged exactly};
\draw[flow] (a)--(b); \draw[flow] (b)--(c); \draw[flow] (c)--(d); \draw[flow] (d)--(e);
\end{tikzpicture}}
\end{center}

\noindent\begin{minipage}{\linewidth}
\medskip\noindent\textbf{Notation.}\par\nopagebreak
{\color{black}
\begin{center}\small
\begin{tabular}{lp{11.5cm}}\toprule
symbol & meaning \\\midrule
$W$, $F_O$ & the two endpoint tables, the first-stage table and the final-response table at the ends of the growth network, Fig.~\ref{fig:growthfactors} \\
$K_r$, $K_{g,r}$ & the growth instrument for accepted record $r$, without and with the growth-fault configuration $g$ \\
$\Pi_s$ & the projector onto the $d=3$ code block with syndrome $s$ (64 of them, $\Pi_0$ is the code space) \\
$V$ & the ideal map from the positive $d=3$ logical basis to the positive $d=5$ logical basis \\
$E_{\rm in}$, $E_{\rm out}$ & Paulis that move a block to syndrome $s$ on the way in, and carry the output syndrome and logical signs \\
$W_{s,t}$ & first-stage table: weight of input syndrome $s$ ($6$ bits) and logical Pauli tag $t$ ($2$ bits), post-selection included \\
$\ell_i$, $f_i(\ell_i)$ & the 75-bit label of factor $i$ and its weight at that label, $f_0=W$, $f_{568}=F_O$ \\
$\Delta_0$ & the single constraint $\ind{\ell_0\oplus\cdots\oplus\ell_{568}=0}$ \\
$S$, $r_g$, $m_g$ & the image subspace of a group of channels, its rank, the number of channels in the group \\
$\lambda(p)=\prod_\ell(1-c_\ell p)$ & the group's damping factor, $c_\ell=2,\ 4/3,\ 16/15$ for flip, one-qubit, two-qubit depolarising \\
$g_p(v)$, $g_x(v)$ & the group's distribution over labels, at fixed $p$ and as a polynomial in $x$ \\
$\mathcal B_J=S_J\cap S_{\bar J}$ & the labels a subset $J$ of factors can produce that the rest can still cancel \\
$M_J(z)$ & the table a tree node stores: total weight of subset $J$ at label $z\in\mathcal B_J$ \\
$U\oplus K$, $\chi$, $f_{i,\chi}$ & a split of the 75-bit label space with $\dim K=3$, a character of $K$, the factor transformed on $K$ \\
\bottomrule
\end{tabular}
\end{center}}
\end{minipage}
\paragraph{Colours, for this appendix only.} In tensor pictures \colorbox{black!22}{grey} is an indicator, \colorbox{tnorange!22}{orange-red} a noise distribution, \colorbox{tnblue!22}{blue} a table given at either end of the network, \colorbox{tngreen!22}{green} an intermediate table.

\subsection{Why syndrome blocks can be treated separately}\label{app:E-blocks}
The state leaving the first double-checking stage is not confined to the $d=3$ code space, since nothing measures its stabilisers before growth. As a side note, this explains the need to measure the code stabilisers in the cultivation experiment of~\cite{cultivation_google_expt}.\footnote{The experiment makes the same point in hardware. Cultivation measures no code stabilisers itself, and adding syndrome extraction cycles after cultivation improved the logical error by about two orders of magnitude.}
It is a superposition and mixture over all 64 syndrome blocks, with coherence between blocks.
Written in those blocks, the state is
\begin{equation}
 \rho=\sum_{s,s'}\Pi_s\,\rho\,\Pi_{s'} .
 \label{eq:blocks}
\end{equation}
The diagonal terms $s=s'$ are what the table $W_{s,t}$ records, one weight per block $s$ and logical Pauli label $t$. The off-diagonal terms $s\neq s'$ are the coherence between blocks. If the growth instrument could see them, the table would not be enough. What makes it enough is a property of growth itself. Every Kraus branch $K$ of the growth stage must satisfy
\begin{equation}
 K\rho K^\dagger=K\Big(\sum_s\Pi_s\,\rho\,\Pi_s\Big)K^\dagger ,
 \label{eq:blockreq}
\end{equation}
so that the off-diagonal terms of Eq.~\eqref{eq:blocks} never contribute. This cannot be assumed. For the SOFT growth circuit it follows from a stabiliser calculation on its Choi state, Eq.~\eqref{eq:dephasing} in Section~\ref{subsubsec:off-diag} below. The supplement of Ref.~\cite{cultivation_google_expt} gives a related caution about post-selection. In its example, a small coherent rotation of the ideal $d=3$ state produces a tomographically reconstructed Bloch vector with $\langle X\rangle^2+\langle Y\rangle^2>1$: the different measurement bases post-select different ensembles. This is an inconsistency in the reconstruction, not a non-physical output state. They sidestep the issue experimentally, with kickback tomography and added syndrome extraction cycles, and note that Pauli twirling would only approximate it away. Here nothing is approximated, since Eq.~\eqref{eq:dephasing} shows that the growth instrument itself never sees the off-diagonal blocks.

\subsubsection{Lack of off-diagonal contributions}
\label{subsubsec:off-diag}
The claim is Eq.~\eqref{eq:dephasing}. To derive it, entangle the seven input data sites with seven untouched reference sites in Bell pairs and propagate the actual growth circuit for a fixed accepted record.
Its Choi stabilisers include six reference-only input-code generators, eighteen output-code generators, and logical $X/X$ and $Z/Z$ correlations. The remaining sites are product states at their resets. Ten random measurements give Choi norm squared $2^{-10}$. Multiplying by the input dimension $2^7$ gives $\Tr(K_r^\dagger K_r)=1/8$. The input code projector has rank two, so
\begin{equation}
 K_r=e^{i\theta_r}\frac14 V\Pi_0.
 \label{eq:choi}
\end{equation}
Here $V$ maps the positive $d=3$ logical basis to the positive $d=5$ logical basis.

Pauli faults change the signs of these stabilisers, but not their unsigned measurement algebra. Every nonzero noisy branch consequently has the form
\begin{equation}
 K_{g,r}=e^{i\theta_{g,r}}\,\frac14\,
 \underbrace{E_{\rm out}}_{\text{output Pauli}}\;
 \underbrace{V}_{\text{encode}}\;
 \underbrace{E_{\rm in}^\dagger}_{\text{block $s$ to code space}}\;
 \underbrace{\Pi_s}_{\text{select block $s$}} ,
 \label{eq:noisychoi}
\end{equation}
read from right to left. Here $E_{\rm in}$ is the Pauli with $E_{\rm in}\Pi_0E_{\rm in}^\dagger=\Pi_s$, so its adjoint returns block $s$ to the domain of $V$, and $E_{\rm out}$ is a Pauli on the output qubits carrying the output syndrome and the logical signs. A branch whose deterministic measurements are incompatible with the record is zero. Since $\Pi_s\Pi_{s'}=\delta_{ss'}\Pi_s$, the sandwich sees a single diagonal block of Eq.~\eqref{eq:blocks},
\begin{equation}
 K_{g,r}\,\rho\,K_{g,r}^\dagger
 =\frac1{16}\,E_{\rm out}VE_{\rm in}^\dagger\,\big(\Pi_s\rho\,\Pi_s\big)\,E_{\rm in}V^\dagger E_{\rm out}^\dagger ,
 \label{eq:oneblock}
\end{equation}
and the phase $e^{i\theta_{g,r}}$ drops out. Hence, for arbitrary $\rho$,
\begin{equation}
 K_{g,r}\rho K_{g,r}^\dagger
 =K_{g,r}\left(\sum_s\Pi_s\rho\Pi_s\right)K_{g,r}^\dagger.
 \label{eq:dephasing}
\end{equation}
This establishes the required syndrome dephasing. It preserves all 64 input blocks, rather than retaining only $\Pi_0\rho\Pi_0$.

Of the 59 measurement bits recorded during growth, four are random and fixed by no detector, so there are $2^4=16$ accepted records, each carrying the factor $1/4$ of Eq.~\eqref{eq:noisychoi}. In the binary compilation of the instrument these four bits have zero columns in both the compatibility constraints and the Choi signs, so they neither reject a fault configuration nor change the Pauli frames, and $K_{g,r}$ is the same operator for all sixteen records. Summing Eq.~\eqref{eq:oneblock} over the records therefore cancels the $1/16$,
\begin{equation}
 \sum_r K_{g,r}\,\rho\,K_{g,r}^\dagger=E_{\rm out}VE_{\rm in}^\dagger\,\big(\Pi_s\rho\,\Pi_s\big)\,E_{\rm in}V^\dagger E_{\rm out}^\dagger ,
\end{equation}
and no further factor of sixteen multiplies $W$.

\subsection{The first-stage table}\label{app:E-table}
Within each retained syndrome block, the first-stage calculation finds the logical output of every branch to be exactly one of the four states $P_L\ket{T_L^\dagger}$ with $P_L\in\{I_L,X_L,Y_L,Z_L\}$, that is, the ideal magic state or the ideal magic state with one logical Pauli error applied. The two-bit tag $t$ records which. 
The table $W_{s,t}$ has $64\times4$ entries, six syndrome bits and two logical-Pauli bits, and retains the branch weights including first-stage post-selection. 
This classification is performed with exact arithmetic on all 8,388,608 compiled frame queries, in $\mathbb Z[\zeta_8]$, the ring of integer combinations $a+b\zeta_8+c\zeta_8^2+d\zeta_8^3$ of powers of the eighth root of unity $\zeta_8=e^{i\pi/4}$, so $\zeta_8^2=i$ and $\zeta_8^4=-1$. Every amplitude of a Clifford$+T$ circuit is an element of this ring divided by a power of $\sqrt2$, so no rounding ever occurs. The four-state property is specific to this circuit and need not hold for an arbitrary non-Clifford circuit. 
\subsection{Binary labels and factorisation}\label{app:E-labels}
Of the 1,207 noise channels between the two non-Clifford double-checking stages, 299 have an identity response and drop out. The 908 relevant ones are 130 flips, 456 one-qubit depolarisers and 322 two-qubit depolarisers. Propagating their Pauli increments through the Clifford instrument gives 49 independent compatibility bits and 26 interface bits: six incoming syndrome bits, eighteen outgoing syndrome bits and two logical bits. The combined label space has dimension 75.

Each of $W$, the growth-channel distributions and $F_O$ becomes a table over this 75-bit space, nonzero only on its own coordinates. $W$ uses the six input-syndrome coordinates and the two logical coordinates. $F_O$ uses the eighteen output-syndrome and two logical coordinates. 
Setting the total label to zero does three things at once:
\begin{enumerate}
\item matches the incoming syndrome on the $F_O$ side to the outgoing one on the $W$ side, 
\item XORs the logical tags as in Eq.~\eqref{eq:physicalgrowth}, and 
\item enforces every detector, since a compatibility bit left at 1 means a detector fired. 
\end{enumerate}
If $\ell_i$ is the label of factor $f_i$, the scalar is
\begin{equation}
 \begin{gathered}
 O=\sum_{\ell_0\oplus\cdots\oplus\ell_{568}=0}
       \prod_{i=0}^{568}f_i(\ell_i), \ \ \text{ where }
 f_0=W,\quad f_{568}=F_O.
 \end{gathered}
 \label{eq:convolution}
\end{equation}
The 567 intermediate factors are exact groups of channels whose faults land in the same label subspace. 
In Fig.~\ref{fig:growthfactors}, the grey tensor is $\Delta_0=\ind{\ell_0\oplus\cdots\oplus\ell_{568}=0}$. 
Contracting it with the 569 factors is precisely Eq.~\eqref{eq:convolution}. 
The labels share one 75-bit space, but each factor is stored only on the few coordinates it uses, and $\Delta_0$ is never stored at all. 
It enters only as the 75 parity equations under the sum in Eq.~\eqref{eq:convolution}, which the contraction enforces.
{\color{black}
 \begin{figure}[H]\centering
 \begin{adjustbox}{max width=.98\linewidth}\begin{tikzpicture}[x=1cm,y=1cm]
\node[tn endpoint] (w) at (.45,1.5) {$W$};
\node[tn noise] (g1) at (2.2,1.5) {$g_1$};
\node[tn label] at (3.5,1.5) {$\cdots$};
\node[tn noise] (gm) at (4.8,1.5) {$g_{567}$};
\node[tn endpoint] (f) at (6.85,1.5) {$F_O$};
\node[tn constraint,minimum width=75mm] (zero) at (3.65,-.25) {$\Delta_0$};
\foreach \name/\lab in {w/\ell_0,g1/\ell_1,gm/\ell_{567},f/\ell_{568}}{
  \draw[tn bundle] (\name.south)--node[right,tn label] {$\lab$}
    (\name.south |- zero.north);
}
\node[tn label] at (3.65,-1.2)
  {$\Delta_0=\mathbf1[\ell_0\oplus\ell_1\oplus\cdots\oplus\ell_{568}=0]$};
\node[tn label] at (3.65,-2.05)
  {each label has 49 compatibility bits and 26 interface bits\\
   multiply the tensor entries and sum the connected labels};
\end{tikzpicture}\end{adjustbox}
 \caption{The 569 factors share one total-label constraint.}\label{fig:growthfactors}\end{figure}
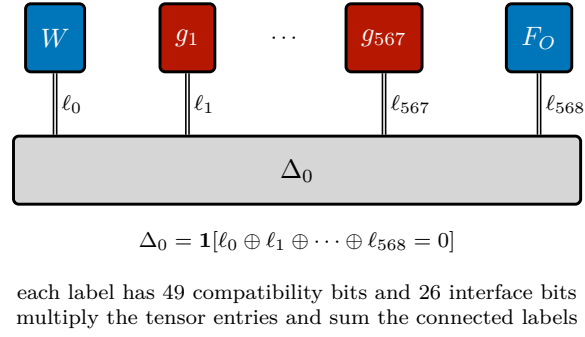}

\paragraph{One group of channels.} 
Three observations turn the 908 noise channels of the growth stage, each a bit flip or a one- or two-qubit depolariser, into the 567 grouped tensors $g_i$ of Fig.~\ref{fig:growthfactors}, each with a closed-form table over its labels.
\begin{enumerate}
\item Every $f_i$ in Eq.~\eqref{eq:convolution} is a table of probabilities indexed by a label. For one channel it is tiny, one entry per outcome.
\item A group of channels whose outcomes land in the same label subspace $S$ is merged into one factor. Its label is the XOR of the individual labels, so its table is the convolution of the small tables.
\item A character transform turns that convolution into a product, and each channel's transform on a nontrivial character is a single number, $1-c_\ell p$. Multiplying these numbers and transforming back gives the table with no enumeration of outcome combinations.
\end{enumerate}
Take a group of $m_g$ channels sharing the subspace $S$ of rank $r_g$. On the trivial character of $S$ every channel contributes $1$. A nontrivial character is $-1$ on exactly half of each channel's Pauli outcomes, so channel $\ell$ contributes $1-c_\ell p$, with $c_\ell=2$ for a bit flip, $4/3$ for a one-qubit and $16/15$ for a two-qubit depolariser. Their product is the group's damping factor,
\begin{equation}
 \lambda(p)=\prod_{\ell=1}^{m_g}(1-c_\ell p),\qquad
 c_\ell\in\{2,4/3,16/15\}.
 \label{eq:damping}
\end{equation}
Transforming back gives the grouped distribution
\begin{equation}
 g_p(v)=\lambda(p)\ind{v=0}
        +\frac{1-\lambda(p)}{2^{r_g}}\ind{v\in S}.
 \label{eq:group}
\end{equation}
It includes all outcomes of all channels in the group, with no approximation. The number $1-c_\ell p$ is the factor by which one noise channel of strength $p$ shrinks a Pauli operator, as in Section~\ref{sec:pp}. The table $g_p(v)$ is what each orange box $g_i$ of Fig.~\ref{fig:growthfactors} holds, namely the factor $f_i(\ell_i)$ of Eq.~\eqref{eq:convolution} for $i=1,\dots,567$.
Note that the input $v$ (to $g_p$) is the label $\ell_i$ summed over in Eq.~\eqref{eq:convolution}.

\paragraph{Why independent weights cannot be averaged one at a time.} With every factor known, one might hope to average each one on its own and multiply the results. Post-selection forbids this. The constraint $\Delta_0$ couples the factors, so whether one fault is accepted depends on the others. Figure~\ref{fig:paritytoy} shows the smallest case. For three independent bits with labels $(1,0),(1,1),(0,1)$ and one constraint, total label zero, the eight assignments are
{\color{black}
\begin{center}\small\begin{tabular}{cccc}\toprule
$f_1f_2f_3$ & total label & weight & kept? \\\midrule
$000$ & $(0,0)$ & $(1-p)(1-p)(1-p)$ & yes \\
$001$ & $(0,1)$ & $(1-p)(1-p)p$ & no \\
$010$ & $(1,1)$ & $(1-p)p(1-p)$ & no \\
$011$ & $(1,0)$ & $(1-p)pp$ & no \\
$100$ & $(1,0)$ & $p(1-p)(1-p)$ & no \\
$101$ & $(1,1)$ & $p(1-p)p$ & no \\
$110$ & $(0,1)$ & $pp(1-p)$ & no \\
$111$ & $(0,0)$ & $ppp$ & yes \\
\bottomrule\end{tabular}
\end{center}}
Only $\survivors$ have zero total label, so the acceptance probability of the toy example in Fig.~\ref{fig:paritytoy} is
\begin{equation}
 A_{\rm toy}=(1-p)^3+p^3,
 \label{eq:toy}
\end{equation}
which is $\Atoy=\Atoynum$ at $p=10^{-3}$. A single non-zero label fails, yet the three non-zero labels pass together. Rejecting faults one at a time would lose the $p^3$ term. The orange-red tensors in Fig.~\ref{fig:paritytoy} give the independent weights $g_i(0)=1-p$ and $g_i(1)=p$. The two grey tensors impose $f_1=f_2$ and $f_2=f_3$. The central dot copies $f_2$ into both conditions. This makes the complete accepted sum visible: only the assignments $000$ and $111$ survive the contraction. The growth sum is this picture with 569 factors and a 75-bit constraint. The next two sections show how to contract it without ever forming the $2^{75}$ label space.
{\color{black}
 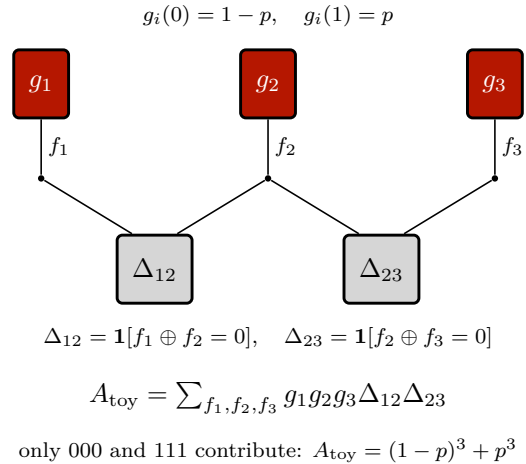
\begin{figure}[H]\centering
 \begin{adjustbox}{max width=.98\linewidth}\begin{tikzpicture}[x=1cm,y=1cm]
\foreach \i/\x in {1/0,2/3,3/6}{
  \node[tn noise] (g\i) at (\x,1.5) {$g_{\i}$};
  \node[tn copy] (f\i) at (\x,.25) {};
  \draw[tn leg] (g\i.south)--node[right,tn label] {$f_{\i}$}(f\i);
}
\node[tn constraint] (c1) at (1.5,-.95) {$\Delta_{12}$};
\node[tn constraint] (c2) at (4.5,-.95) {$\Delta_{23}$};
\draw[tn leg] (f1)--([xshift=-3mm]c1.north);
\draw[tn leg] (f2)--([xshift=3mm]c1.north);
\draw[tn leg] (f2)--([xshift=-3mm]c2.north);
\draw[tn leg] (f3)--([xshift=3mm]c2.north);
\node[tn label] at (3,2.4) {$g_i(0)=1-p,\quad g_i(1)=p$};
\node[tn label] at (3,-1.85)
  {$\Delta_{12}=\mathbf1[f_1\oplus f_2=0],\quad
    \Delta_{23}=\mathbf1[f_2\oplus f_3=0]$};
\node[tn label,font=\small] at (3,-2.65)
  {$A_{\rm toy}=\sum_{f_1,f_2,f_3}g_1g_2g_3\Delta_{12}\Delta_{23}$};
\node[tn label] at (3,-3.35)
  {only $000$ and $111$ contribute: $A_{\rm toy}=(1-p)^3+p^3$};
\end{tikzpicture}\end{adjustbox}
 \caption{Independent noise tensors coupled by post-selection. The sum is written out below the network.}\label{fig:paritytoy}\end{figure}}

\subsection{Which partial labels must be stored?}\label{app:E-boundary}
For a subset $J$ of the 569 factors $f_i$ of Eq.~\eqref{eq:convolution}, the boxes of Fig.~\ref{fig:growthfactors}, let $S_J$ be the span of its possible labels and let $S_{\bar J}$ be the span of the remaining factors' labels.
Only a partial label in
\begin{equation}
 {\cal B}_J=S_J\cap S_{\bar J}
 \label{eq:boundary}
\end{equation}
can be matched by an equal label from the remaining factors, so that the total XOR label $\ell_0\oplus\cdots\oplus\ell_{568}$ vanishes. 
A component outside the intersection has exactly zero contribution to Eq.~\eqref{eq:convolution}, whatever its numerical size. 
Whenever a subset $J$ of the factors has been contracted, only the entries of its table with labels in $\mathcal B_J$ are kept before it is joined with another subset of the 569 factors.
In Fig.~\ref{fig:boundary}, the left table $M_L$ belongs to one contracted subset and the right table $M_R$ to the remaining factors. The left labels span $e_1,e_2$ and the right labels $e_2,e_3$, so only the $e_2$ coordinate can cross the cut.
Writing the labels as $z_1e_1+z_2e_2$ and $z'_2e_2+z_3e_3$, their sum vanishes only if $z_1=z_3=0$ and $z_2=z'_2$. 
This is the simplest example of restricting a table before the next contraction.
{\color{black}
 \begin{figure}[H]\centering
 \begin{adjustbox}{max width=.98\linewidth}\begin{tikzpicture}[x=1cm,y=1cm]
\node[tn constraint,minimum width=8mm] (p1) at (0,0) {$\delta_0$};
\node[tn intermediate,minimum width=11mm] (l) at (2.05,0) {$M_L$};
\node[tn intermediate,minimum width=11mm] (r) at (5.25,0) {$M_R$};
\node[tn constraint,minimum width=8mm] (p3) at (7.3,0) {$\delta_0$};
\draw[tn leg] (p1)--node[above,tn label] {$z_1$}(l);
\draw[tn leg] (l)--node[above,tn label,fill=white] {$z_2$}(r);
\draw[tn leg] (r)--node[above,tn label] {$z_3$}(p3);
\draw[dashed,black!45] (3.65,-.85)--(3.65,1.0);
\node[tn label] at (2.05,-1.05) {left labels\\$z_1e_1+z_2e_2$};
\node[tn label] at (5.25,-1.05) {right labels\\$z_2e_2+z_3e_3$};
\node[tn label] at (3.65,-2.1)
  {$\delta_0(z)=\mathbf1[z=0]$: $z_1=z_3=0$\\
   only the $e_2$ coordinate can cross the cut};
\end{tikzpicture}\end{adjustbox}
 \caption{Only the shared label coordinate can connect the two partial tables.}\label{fig:boundary}\end{figure}
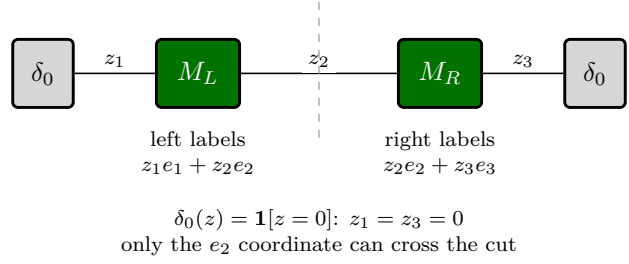}

\subsection{What a tree node stores and computes}\label{app:E-tree}
Each leaf of the growth contraction tree contains one of the factors $f_i$ in Eq.~\eqref{eq:convolution}. A node containing the subset $J$ of leaves stores a table indexed by their combined label $z$,
\begin{equation}
 M_J(z)=\sum_{\substack{\{\ell_i:i\in J\}\\
                      \bigoplus_{i\in J}\ell_i=z}}
                  \prod_{i\in J}f_i(\ell_i),
 \qquad z\in{\cal B}_J.
 \label{eq:treemessage}
\end{equation}
One entry therefore combines many fault configurations. Only the label needed by the remaining factors survives. The individual configurations do not have to be stored.
When a node joins disjoint sets of leaves $L$ and $R$, the node's table is
\begin{equation}
 \begin{gathered}
 M_{L\cup R}(z)=
 \sum_{\substack{u\in{\cal B}_L,\ v\in{\cal B}_R\\u\oplus v=z}}
            M_L(u)M_R(v),\qquad z\in{\cal B}_{L\cup R}.
 \end{gathered}
 \label{eq:treeconvolution}
\end{equation}
{\color{black}
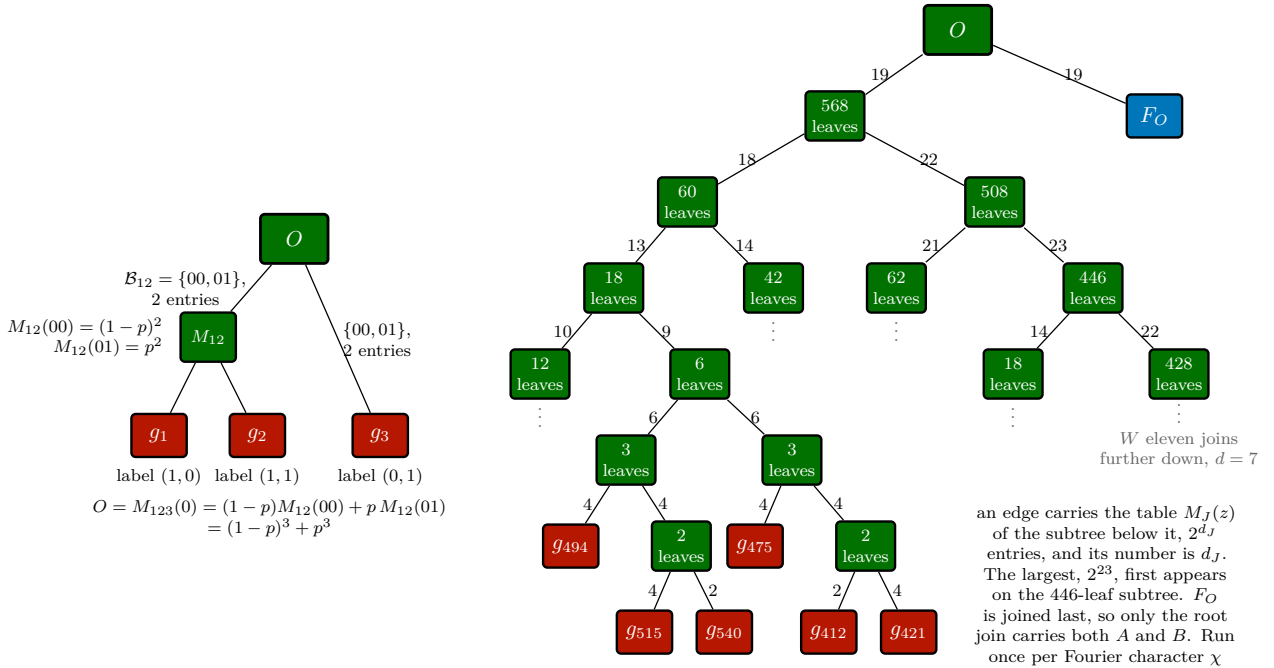
\begin{figure}[H]\centering
\begin{adjustbox}{max width=\linewidth}\begin{tikzpicture}[x=1cm,y=1cm,
  leaf/.style={tn noise,minimum width=9mm,minimum height=7mm,font=\small,inner sep=1mm},
  endl/.style={tn endpoint,minimum width=9mm,minimum height=7mm,font=\small},
  node/.style={tn intermediate,minimum width=9mm,minimum height=8mm,font=\scriptsize,align=center,inner sep=1mm},
  root/.style={tn intermediate,minimum width=11mm,minimum height=8mm,font=\small,draw=black,line width=1.3pt},
  elab/.style={tn label,align=center,inner sep=1pt}]
\node[tn heading,anchor=west] at (-1.2,9.9) {\textbf{(a)} the three-fault toy example, Eq.~\eqref{eq:toy}};
\node[leaf] (f1) at (0,2.40) {$g_1$};
\node[leaf] (f2) at (1.6,2.40) {$g_2$};
\node[leaf] (f3) at (3.6,2.40) {$g_3$};
\node[tn label,anchor=north] at (0,1.95) {label $(1,0)$};
\node[tn label,anchor=north] at (1.6,1.95) {label $(1,1)$};
\node[tn label,anchor=north] at (3.6,1.95) {label $(0,1)$};
\node[node] (m12) at (0.8,4.00) {$M_{12}$};
\node[root] (r) at (2.2,5.60) {$O$};
\draw[tn leg] (f1)--(m12); \draw[tn leg] (f2)--(m12);
\draw[tn leg] (m12)--node[elab,left=1pt] {$\mathcal B_{12}=\{00,01\}$,\\2 entries}(r);
\draw[tn leg] (f3)--node[elab,right=1pt] {$\{00,01\}$,\\2 entries}(r);
\node[tn label,anchor=north,align=center] at (1.8,1.45) {$O=M_{123}(0)=(1-p)M_{12}(00)+p\,M_{12}(01)$\\$=(1-p)^3+p^3$};
\node[tn label,anchor=east,align=right] at (0.15,4.00) {$M_{12}(00)=(1-p)^2$\\$M_{12}(01)=p^2$};
\node[tn heading,anchor=west] at (5.9,9.9) {\textbf{(b)} the selected tree of the growth sum, one branch followed down to its leaves};
\node[root] (R) at (13.0,9.0) {$O$};
\node[endl] (FO) at (16.2,7.6) {$F_O$};
\node[node] (n568) at (11.0,7.6) {568\\leaves};
\node[node] (n60) at (8.6,6.2) {60\\leaves};
\node[node] (n508) at (13.6,6.2) {508\\leaves};
\node[node] (n18) at (7.4,4.8) {18\\leaves};
\node[node] (n42) at (10.0,4.8) {42\\leaves};
\node[node] (n62) at (12.0,4.8) {62\\leaves};
\node[node] (n446) at (15.2,4.8) {446\\leaves};
\node[node] (n12) at (6.2,3.4) {12\\leaves};
\node[node] (n6) at (8.8,3.4) {6\\leaves};
\node[node] (n18b) at (13.9,3.4) {18\\leaves};
\node[node] (n428) at (16.6,3.4) {428\\leaves};
\node[node] (n3a) at (7.6,2.0) {3\\leaves};
\node[node] (n3b) at (10.3,2.0) {3\\leaves};
\node[leaf] (g494) at (6.7,0.6) {$g_{494}$};
\node[node] (n2a) at (8.5,0.6) {2\\leaves};
\node[leaf] (g475) at (9.7,0.6) {$g_{475}$};
\node[node] (n2b) at (11.5,0.6) {2\\leaves};
\node[leaf] (g515) at (7.9,-0.8) {$g_{515}$};
\node[leaf] (g540) at (9.2,-0.8) {$g_{540}$};
\node[leaf] (g412) at (10.9,-0.8) {$g_{412}$};
\node[leaf] (g421) at (12.2,-0.8) {$g_{421}$};
\draw[tn leg] (R)--node[elab,right=1pt,pos=0.5] {19}(FO);
\draw[tn leg] (R)--node[elab,left=1pt,pos=0.5] {19}(n568);
\draw[tn leg] (n568)--node[elab,left=1pt,pos=0.5] {18}(n60);
\draw[tn leg] (n568)--node[elab,right=1pt,pos=0.5] {22}(n508);
\draw[tn leg] (n60)--node[elab,left=1pt,pos=0.5] {13}(n18);
\draw[tn leg] (n60)--node[elab,right=1pt,pos=0.5] {14}(n42);
\draw[tn leg] (n508)--node[elab,left=1pt,pos=0.5] {21}(n62);
\draw[tn leg] (n508)--node[elab,right=1pt,pos=0.5] {23}(n446);
\draw[tn leg] (n18)--node[elab,left=1pt,pos=0.5] {10}(n12);
\draw[tn leg] (n18)--node[elab,right=1pt,pos=0.5] {9}(n6);
\draw[tn leg] (n446)--node[elab,left=1pt,pos=0.5] {14}(n18b);
\draw[tn leg] (n446)--node[elab,right=1pt,pos=0.5] {22}(n428);
\draw[tn leg] (n6)--node[elab,left=1pt,pos=0.5] {6}(n3a);
\draw[tn leg] (n6)--node[elab,right=1pt,pos=0.5] {6}(n3b);
\draw[tn leg] (n3a)--node[elab,left=1pt,pos=0.5] {4}(g494);
\draw[tn leg] (n3a)--node[elab,right=1pt,pos=0.5] {4}(n2a);
\draw[tn leg] (n3b)--node[elab,left=1pt,pos=0.5] {4}(g475);
\draw[tn leg] (n3b)--node[elab,right=1pt,pos=0.5] {4}(n2b);
\draw[tn leg] (n2a)--node[elab,left=1pt,pos=0.5] {4}(g515);
\draw[tn leg] (n2a)--node[elab,right=1pt,pos=0.5] {2}(g540);
\draw[tn leg] (n2b)--node[elab,left=1pt,pos=0.5] {2}(g412);
\draw[tn leg] (n2b)--node[elab,right=1pt,pos=0.5] {4}(g421);
\node[tn label,text=black!60] at (6.2,2.8) {$\vdots$};
\node[tn label,text=black!60] at (10.0,4.2) {$\vdots$};
\node[tn label,text=black!60] at (12.0,4.2) {$\vdots$};
\node[tn label,text=black!60] at (13.9,2.8) {$\vdots$};
\node[tn label,text=black!60,align=center] at (16.6,2.5) {$\vdots$\\$W$ eleven joins\\further down, $d=7$};
\node[tn label,anchor=north,align=center,text width=44mm] at (15.4,1.35) {an edge carries the table $M_J(z)$ of the subtree below it, $2^{d_J}$ entries, and its number is $d_J$. The largest, $2^{23}$, first appears on the 446-leaf subtree. $F_O$ is joined last, so only the root join carries both $A$ and $B$. Run once per Fourier character $\chi$};
\end{tikzpicture}\end{adjustbox}
\caption{The contraction tree as a tree. (a) The three-fault toy example of Eq.~\eqref{eq:toy}: two leaves join into a node that stores two entries on its boundary, and the root gives $A_{\rm toy}$. (b) The tree actually used for the growth sum, 569 leaves and 1,137 nodes found by the search: its top levels, and the 60-leaf branch followed down to real leaves $g_i$. Boxes give subtree leaf counts, edge numbers the boundary dimension $d_J$ of the table stored on them, dots elided subtrees. Tables of the largest size, $2^{23}$ entries, sit on the 446-leaf subtree and on two subtrees inside it, $W$ is a leaf fifteen joins deep with a 7-bit boundary, and $F_O$ is joined last.}
\label{fig:treenodes}
\end{figure}}
Figure~\ref{fig:treenodes} draws the tree, for the toy example and for the top of the tree actually used.
Restricting the child tables loses no compatible term. If the parent label $z$ can be cancelled by the factors outside $L\cup R$, then $u=z\oplus v$ can be cancelled by $R$ and those outside factors, so $u\in{\cal B}_L$, and likewise $v\in{\cal B}_R$. At the root no factors remain, so $\Delta_0$ leaves only $z=0$, and $M_{\rm all}(0)$ is $O$. Panel (a) of Fig.~\ref{fig:treenodes} does this for the toy example of Eq.~\eqref{eq:toy}, where fault 3 contributes only the labels $0$ or $(0,1)$, so only those two entries of $M_{12}$ can be matched and are stored, and the $p^2$ entry must survive to the last join, where that fault cancels it.

An edge carries its table and a binary basis of ${\cal B}_J$, $2^{d_J}$ entries for a boundary of dimension $d_J$. Each join is evaluated in coordinates adapted to its two children, with Walsh transforms on their shared coordinates, so Eq.~\eqref{eq:treeconvolution} never lists every pair of entries. The order and the bases are chosen together by a heuristic search that prices products, Walsh additions, coordinate changes and storage. The chosen order stores at most $2^{23}$ entries per table, and its largest join multiplies $2^{27}$ pairs of entries before summing them. The same tree is run once per Fourier character below and the eight root values are averaged. Every subtree is computed once and reused for $A$ and $B$, which differ only in $F_O$, so only the joins that contain $F_O$ are done twice.

\subsection{Eight exact Fourier characters}\label{app:E-fourier}
The boundaries of Section~\ref{app:E-boundary} bound the tables but do not make them small enough on their own. Projecting three label coordinates out of every factor, and restoring them exactly by summing over the characters of the projected space, shortens every boundary that carries those coordinates, at the cost of running the tree eight times. Choose one common decomposition $\F^{75}=U\oplus K$ of the 75-bit label space, XOR being its addition, with $\dim K=3$ and $\dim U=72$. Every label is then uniquely $\ell=(u,k)$ with $u\in U$ and $k\in K$, the three coordinates in $K$ being the ones projected out. For each of the eight linear functionals $\chi$ of $K$, with $\chi\cdot k\in\{0,1\}$, define
\begin{equation}
 f_{i,\chi}(u)=\sum_{k\in K}(-1)^{\chi\cdot k}f_i(u,k),
 \qquad \chi\in K^*.
 \label{eq:partialfourier}
\end{equation}
Character orthogonality gives the exact identity
\begin{equation}
 O=\frac18\sum_{\chi\in K^*}
       \sum_{u_0\oplus\cdots\oplus u_{568}=0}
                       \prod_i f_{i,\chi}(u_i).
 \label{eq:eight}
\end{equation}
Every factor uses the same partition and all eight characters are included. The transformed factors can be signed. The complete average restores the three projected conditions. Nothing is dropped. The 72-dimensional quotient is never stored as a dense $2^{72}$ array, since each join first uses Eq.~\eqref{eq:boundary}.

For the toy example of Eq.~\eqref{eq:toy}, keep the first condition $f_1\oplus f_2=0$ and project the second, $f_2\oplus f_3=0$. Weighting each assignment by $(-1)^{\chi(f_2\oplus f_3)}$, the two character contractions are
\[
\begin{gathered}
C_0=(1-p)^2+p^2=\Czero,\qquad C_1=(1-2p)^2=\Cone,\\
\tfrac12(C_0+C_1)=\Atoy ,
\end{gathered}
\]
at $p=10^{-3}$. Their half-sum $\tfrac12(C_0+C_1)$ is exactly $(1-p)^3+p^3$, as in Fig.~\ref{fig:fouriertoy}. This is how slicing saves storage without discarding a condition. Signed intermediate values are harmless here because everything is exact, first as rationals and then, in Appendix~\ref{app:arithmetic}, modulo primes. What must not be done is to drop a character or to treat a signed intermediate as a probability.
{\color{black}
 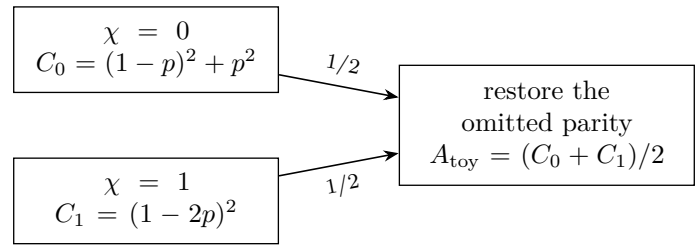
\begin{figure}[H]\centering
 \begin{adjustbox}{max width=.98\linewidth}\begin{tikzpicture}[x=1cm,y=1cm]
\node[box,text width=31mm] (zero) at (0,1)
 {$\chi=0$\\$C_0=(1-p)^2+p^2$};
\node[box,text width=31mm] (one) at (0,-1)
 {$\chi=1$\\$C_1=(1-2p)^2$};
\node[box,text width=35mm] (mean) at (5.3,0)
 {restore the omitted parity\\$A_{\rm toy}=(C_0+C_1)/2$};
\draw[flow] (zero)--node[note,above,sloped] {$1/2$}(mean);
\draw[flow] (one)--node[note,below,sloped] {$1/2$}(mean);
\end{tikzpicture}\end{adjustbox}
 \caption{Both characters restore the missing parity in the three-fault example.}\label{fig:fouriertoy}\end{figure}}

\clearpage
\newcommand{\dfivelead}{3}
\newcommand{\dfiveleadval}{\frac{574}{375}}
\newcommand{\dfiveaone}{\frac{166}{3}}
\newcommand{\dfiverestored}{\frac{4870}{3}}
\newcommand{\dthreelead}{2}
\newcommand{\dthreeleadval}{\frac{32}{75}}
\newcommand{\dthreeaone}{\frac{77}{3}}
\newcommand{\dthreerestored}{\frac{425}{3}}
\newcommand{\cauchytoy}{(1+2x^1+3x^2)(1+1x^1+1x^2)\equiv 1+3x^1+6x^2}
\newcommand{\cauchyw}{1,3,6}
\newcommand{\heightfive}{157.9}
\newcommand{\heightthree}{114.6}
\newcommand{\boundtenfive}{334871366019039939384856251352775393280000000000}
\newcommand{\boundtenbits}{158}
\newcommand{\aten}{\frac{7771271684969588874728818721617}{147622500000000}}
\newcommand{\Uaten}{65190151826709372991261166558706139136}
\newcommand{\denflip}{500}
\newcommand{\dendepolone}{750}
\newcommand{\dendepoltwo}{1875}
\newcommand{\growthbits}{9022}
\newcommand{\primebits}{23912}
\newcommand{\reduceddigits}{5748}
\newcommand{\crtrA}{0, 7}
\newcommand{\crtrB}{2, 2}
\newcommand{\prothk}{3722901}
\newcommand{\prothm}{40}
\newcommand{\protha}{5}
\newcommand{\prothq}{4093372938558898177}
\newcommand{\prothsmall}{13=3\cdot2^2+1,\quad 2^{6}=64\equiv-1\pmod{13}}
\newcommand{\deltaval}{2.68960\times10^{-16}}
\newcommand{\deltapub}{2.68959\times10^{-16}}
\newcommand{\taillo}{3.328425294\times10^{-9}}
\newcommand{\tailhi}{3.328427163\times10^{-9}}
\newcommand{\tailrel}{5.613\times10^{-7}}
\newcommand{\tailrelpub}{5.61\times10^{-7}}
\newcommand{\Sten}{3.328425370\times10^{-9}}
\newcommand{\LERdec}{3.328425371\times10^{-9}}
\newcommand{\taylorresid}{5.427\times10^{-10}}
\newcommand{\Adec}{0.143955584096}
\newcommand{\Aten}{0.143955584096}

\providecommand{\hl}[2]{\colorbox{#1!22}{#2}}
\section{Exact arithmetic, coefficients and remainders}\label{app:arithmetic}
Every number in this main text concerns an exact rational. This appendix describes how these numbers are computed and certified. The TLDR route is:
{\color{black}
\begin{center}
\adjustbox{max width=\linewidth}{%
\begin{tikzpicture}[x=1cm,y=1cm]
\tikzset{st/.style={box,text width=30mm,inner sep=1.5mm,font=\small}}
\node[st] (a) at (0,0) {\ref{app:F-fixedp}: fixed $p$, compute modulo primes, reconstruct the two integers $N_A,N_B$};
\node[st,right=3mm of a] (b) {\ref{app:F-reuse}: reuse the exact final-response table across primes};
\node[st,right=3mm of b] (c) {\ref{app:F-series}: series, raw coefficients $a_k,e_k$ in $x=p/(1-p)$, then the quotient $L_k$};
\node[st,right=3mm of c] (e) {\ref{app:F-tail}: what the omitted orders can do to $\PL$, an interval that contains it};
\draw[flow] (a)--(b); \draw[flow] (b)--(c); \draw[flow] (c)--(e); %
\end{tikzpicture}}
\end{center}}

\noindent\begin{minipage}{\linewidth}
\medskip\noindent\textbf{Notation.}\par\nopagebreak
{\color{black}
\begin{center}\small
\begin{tabular}{lp{11.5cm}}\toprule
symbol & meaning \\\midrule
$W$, $F_O$ & the two endpoint tables, the first-stage table and the final-response table at the ends of the growth network, Fig.~\ref{fig:growthfactors} \\
$D$, $D_W$, $D_F$ & a positive integer that clears every denominator of $A$ and $B$ at the chosen $p$, and the same for the first-stage and final-response tables \\
$N_A=DA$, $N_B=DB$ & the two integers actually reconstructed, $0\le N_B\le N_A\le D$ \\
$q$, $k2^m+1$, $a$ & a prime field, the Proth form of the primes used, and the witness with $a^{(q-1)/2}\equiv-1\pmod q$ \\
$x=p/(1-p)$ & the series variable: a $k$-fault pattern has weight $(1-p)^N x^k$ times its Pauli factors \\
$a_k$, $e_k$ & raw coefficients: accepted weight and accepted-wrong weight of all $k$-fault patterns, relevant-location convention, $N=1996$ ($402$ at $d=3$) \\
$L_k$ & the coefficients of $\PL$ itself, from the quotient $e(x)/a(x)$ \\
$\nu\in\{1,3,15\}$ & the number of Pauli outcomes of a location: flip, one-qubit, two-qubit depolarising \\
$\mathbb F_q[x]/(x^{11})$ & polynomials with coefficients mod $q$, truncated after $x^{10}$ \\
$A_K$, $E_K$, $\delta=A-A_K$ & the sums through order $K$, and the omitted acceptance weight \\
$S_K=\sum_{k\le K}L_kx^k$ & the truncated series for $\PL$ \\
\bottomrule
\end{tabular}
\end{center}}
\end{minipage}
\paragraph{Colours, for this appendix only.} \hl{tnblue}{Blue} marks an exact quantity that is computed or reconstructed, \hl{tnorange}{orange} marks a bound or a certificate, something only guaranteed to be at least as large as the truth, and \hl{tngreen}{green} marks a value confirmed by an independent recomputation.

\subsection{Rational probabilities at fixed \texorpdfstring{$p$}{p}}\label{app:F-fixedp}
\paragraph{Why not floating point.} 
$B$ at $p=10^{-3}$ is $4.79\times10^{-10}$ and comes out of a signed sum, where the terms inside sum can have either $\pm$ sign and their magnitudes are far larger than the sum itself.
The signs have two sources, the eight Fourier characters of Appendix~\ref{app:growth}, whose transformed factors are signed, and the interfering phases of the response sums of Appendix~\ref{app:response}.
A double-precision floating-point number stores only 53 significant bits, about sixteen decimal digits. Cancellation amplifies its rounding error, and the accuracy lost depends on the magnitudes of the intermediate contributions relative to their sum, not on the small value of $B$ alone. Modular arithmetic avoids this rounding error and reconstructs integers with tens of thousands of bits exactly.
So every contraction is done in prime fields $\mathbb F_q$\footnote{All primes used are Proth primes, $q=k2^m+1$ with odd $k<2^m$, in the range $2^{61}<q<2^{62}$. Each certificate records a witness $a$ with $a^{(q-1)/2}\equiv-1\pmod q$, which proves primality by Proth's theorem, \url{https://en.wikipedia.org/wiki/Proth\%27s_theorem}. A tiny instance is $\prothsmall$, and one in the range used is $q=\prothq=\prothk\cdot2^{\prothm}+1$ with witness $\protha$.}, where arithmetic is exact and cheap, and the answer is put back together at the end by the Chinese remainder theorem (see next).

\paragraph{The Chinese remainder theorem (CRT)/Sunzi's theorem\footnote{Let's not coarse grain out Sunzi.}.} An integer is fixed by its remainders modulo several primes, provided it is smaller than the product of those primes. The remainders are cheap, since every contraction can be done modulo a prime with fixed-size integers, and the integer is assembled from them at the end. This is how $N_A$ and $N_B$ are found. Once the product of the primes used exceeds $D\ge N_A\ge N_B$, both integers are determined, and any further primes serve as independent verifications.

For a small example take the primes $3$ and $5$ and arrange the integers $0,\dots,14$ by their remainders, Fig.~\ref{fig:crt}. Every cell of the grid is filled exactly once, so the remainders $2$ and $3$ can only belong to $N=8$. 
 \begin{figure}[H]\centering
 \begin{adjustbox}{max width=.98\linewidth}\begin{tikzpicture}[x=1cm,y=1cm,
  cell/.style={draw=black!60,minimum width=9mm,minimum height=8mm,font=\small,inner sep=0pt},
  hit/.style={cell,fill=black!15,draw=black,line width=1.1pt},
  hd/.style={font=\scriptsize,text=black!70}]
\foreach \r in {0,1,2}{\node[hd,anchor=east] at (-0.6,-\r) {$\r$};}
\node[hd,rotate=90] at (-1.4,-1) {$N\bmod3$};
\foreach \c in {0,...,4}{\node[hd] at (\c,0.65) {$\c$};}
\node[hd] at (2,1.1) {$N\bmod5$};
\foreach \N in {0,...,14}{
  \pgfmathtruncatemacro{\r}{mod(\N,3)}\pgfmathtruncatemacro{\c}{mod(\N,5)}
  \ifnum\N=8 \node[hit] at (\c,-\r) {$\N$};\else \node[cell] at (\c,-\r) {$\N$};\fi}
\node[hd,anchor=north,align=center,text width=62mm] at (2,-2.55) {remainders $(2,3)$ occur once, at $N=8$\\with primes $q_1,\dots,q_n$ the grid has $\prod_jq_j$ cells, one per integer below the product};
\end{tikzpicture}\end{adjustbox}
 \caption{The Chinese remainder theorem for the primes $3$ and $5$. The fifteen integers below $15$ fill the grid of remainder pairs exactly once, so the remainders $(2,3)$ identify $N=8$.}\label{fig:crt}\end{figure}
In the full calculation the reconstruction primes are chosen so that their product exceeds the sufficient denominator $D$ at each $p$. Table~\ref{tab:cost} gives the prime counts and the denominator bounds, the largest of which has 21,137 bits.

\paragraph{The denominator.} 
Every number in the main text is an exact rational. At a fixed $p$ the two probabilities are $A=N_A/D$ and $B=N_B/D$, with a denominator $D$ that can be written down in advance from the structure of the network, so it suffices to compute the integers $N_A,N_B$ modulo enough primes and to reassemble them by the Chinese remainder theorem. The series coefficients are obtained the same way, order by order in $x=p/(1-p)$. The route is
Let $D_W,D_F$ be sufficient positive denominators of the first-stage and final-response tables at the chosen rational $p$. One sufficient denominator for both full probabilities is
\begin{equation}
 D=2^{75}D_WD_F\prod_{\ell\in{\rm growth}}\operatorname{den}(1-c_\ell p) \ ,
 \label{eq:denominator}
\end{equation}
where $\operatorname{den}(r)$ is the denominator of the rational $r$ in lowest terms\footnote{Write $r$ as a reduced fraction $a/b$ with $\gcd(a,b)=1$ and $b>0$. Then $\operatorname{den}(r)=b$.} and $c_\ell\in\{2,4/3,16/15\}$ is the damping constant of growth channel $\ell$ from Eq.~\eqref{eq:damping}.
Equation~\eqref{eq:denominator} lists where every denominator can come from: the $2^{-75}$ of one global Fourier normalisation, the two endpoint tables, and one damping factor $1-c_\ell p$ per growth channel. 
To see this, use Eq.~\eqref{eq:eight} with $K$ the whole 75-bit space, $O=2^{-75}\sum_\chi\prod_i\hat f_i(\chi)$, before any slicing.
The unnormalised transforms of $W,F$ are integer sums and retain their denominators. 
Each physical growth channel contributes either one or its damping factor $1-c_\ell p$ to any character, see Appendix~\ref{app:growth}. 
The only global Fourier normalisation is $2^{-75}$. 
So $N_A = D\times A$ and $N_B = D\times B$ are integers whatever contraction tree is used, since the argument never looks at the normalisations inside a particular tree.
At $p=10^{-3}$, the three damping factors have denominators \hl{tnblue}{$\denflip$}, \hl{tnblue}{$\dendepolone$} and \hl{tnblue}{$\dendepoltwo$}, and the 908 growth channels alone contribute about \growthbits\ bits to $D$. 

This $D$ is chosen to be deliberately generous (e.g. it keeps every damping-factor denominator separately, 9,022 bits from the growth channels alone at $p=10^{-3}$, although most of them cancel in the reduced fraction)
It does not matter that $D$ is far larger than the reduced denominator of $\PL$, which comes from $N_A$ after cancellation. 
It only matters that $D$ is a true common denominator, because the Chinese remainder theorem needs the product of the primes to exceed the integers being reconstructed.

\paragraph{Reconstruct integers, not fractions.} 
Evaluate Eq.~\eqref{eq:eight} modulo each prime $q$, with every rational entry taken modulo $q$, which is possible because $q$ divides none of the denominators. 
Multiplying the result by $D\bmod q$ gives the residues of the integers $N_A=DA$ and $N_B=DB$. 
Since $B\le A\le1$ are probabilities, then:
\begin{equation}
 0\le N_B\le N_A\le D,
 \label{eq:height}
\end{equation}
so both integers lie in a known range and are recovered exactly once the product of the primes exceeds $D$.
Once the product of distinct primes exceeds $D$, the Chinese remainder theorem identifies each integer uniquely in this range. Reconstruct $N_A,N_B$ separately and then form $\PL=N_B/N_A$ over the rationals. Dividing by $A$ inside a prime field is unnecessary and can fail when its residue is zero. The toy example in Fig.~\ref{fig:crt-toy}: $A=7/15$, $B=2/15$, $D=15$, so $N_A=7$, $N_B=2$. Modulo $7$ and $11$ the residues are $N_A\equiv(\crtrA)$ and $N_B\equiv(\crtrB)$, and since $7\cdot11>15$ the reconstruction returns \hl{tngreen}{$N_A=7$, $N_B=2$}, $\PL=2/7$. In $\mathbb F_7$ the residue of $A$ is $0$ and $B/A$ does not exist. In Fig.~\ref{fig:primes}, rows are primes and columns are the eight characters of Appendix~\ref{app:growth}. Along a row the characters are averaged first, exactly, inside $\mathbb F_q$. Down the columns the residues are combined by the Chinese remainder theorem. 
Both directions are sums with a normalisation, and neither drops a character or a residue.
\color{black}
 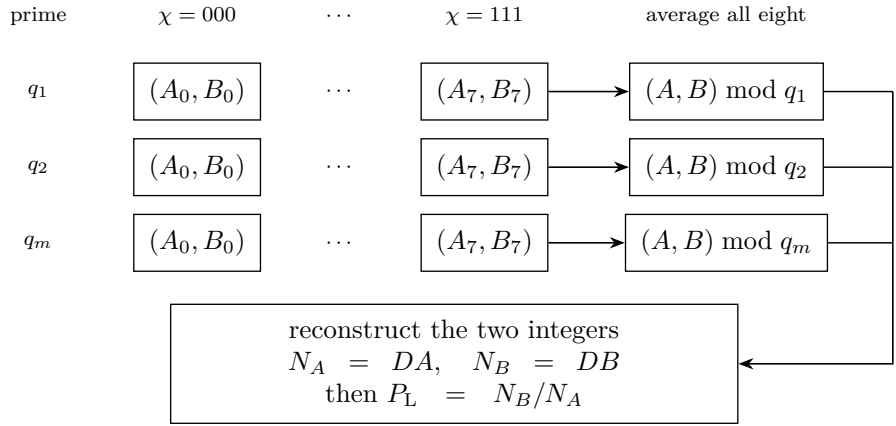
\begin{figure}[H]\centering
 \begin{adjustbox}{max width=.98\linewidth}\begin{tikzpicture}[x=1cm,y=1cm]
\node[note] at (0,2) {prime};
\node[note] at (2.1,2) {$\chi=000$};
\node[note] at (4,2) {$\cdots$};
\node[note] at (5.9,2) {$\chi=111$};
\node[note] at (9.1,2) {average all eight};
\foreach \y/\label in {1/{q_1},0/{q_2},-1/{q_m}}{
 \node[note] at (0,\y) {$\label$};
 \node[box] (a\y) at (2.1,\y) {$(A_0,B_0)$};
 \node[note] at (4,\y) {$\cdots$};
 \node[box] (b\y) at (5.9,\y) {$(A_7,B_7)$};
 \node[box] (r\y) at (9.1,\y) {$(A,B)\bmod\label$};
 \draw[flow] (b\y)--(r\y);
}
\node[box,text width=71mm] (crt) at (5.5,-2.6)
 {reconstruct the two integers\\$N_A=D A,\quad N_B=D B$\\then $P_{\rm L}=N_B/N_A$};
\foreach \y in {1,0,-1}{\draw[wire] (r\y.east)--(11.3,\y);}
\draw[flow] (11.3,1)--(11.3,-2.6)--(crt.east);
\end{tikzpicture}\end{adjustbox}
 \caption{Characters impose the projected conditions and primes reconstruct exact integers. All character columns are averaged before reconstruction down the rows.}\label{fig:primes}\end{figure}
 \begin{figure}[H]\centering
 \begin{adjustbox}{max width=.98\linewidth}\begin{tikzpicture}[x=1cm,y=1cm]
\node[box,text width=28mm] (p) at (0,0)
 {$A=7/15$\\$B=2/15$\\$D=15$};
\node[box,text width=36mm] (r) at (4.2,0)
 {$N_A: 0\pmod7,\ 7\pmod{11}$\\$N_B: 2\pmod7,\ 2\pmod{11}$};
\node[box,text width=31mm] (n) at (9,0)
 {$7\times11>D$\\$N_A=7,\quad N_B=2$\\$P_{\rm L}=2/7$};
\draw[flow] (p)--(r);\draw[flow] (r)--(n);
\end{tikzpicture}\end{adjustbox}
 \caption{A small reconstruction example. $A$ vanishes modulo seven, but separate integer reconstruction still gives $B/A=2/7$.}\label{fig:crt-toy}\end{figure}
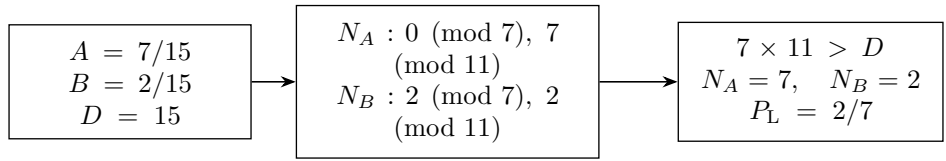

\subsection{One response table for all primes}\label{app:F-reuse}
The final-response table $F_O$ is the expensive part of every contraction, but its entries have a far smaller denominator bound $D_F$ than the full answer. So it is computed exactly once and reused for every prime.
\begin{enumerate}
\item Contract the table modulo the first few primes, until the product of those primes exceeds $D_F$. The Chinese remainder theorem then gives every entry as an exact integer.
\item For each later prime, reduce those exact entries modulo the prime. No new endpoint contraction is needed.
\item Contract the table afresh at two further primes and compare with the reduced entries, as a check.
\end{enumerate}
The grouping of entries with equal residues at the first prime, re-checked at each reconstruction prime, only speeds up step 1. The cache is exact and specific to $p$. The $p=10^{-3}$ repeat reuses it and the first-stage table but recomputes the whole growth sum with a new contraction order, and reproduces $A$, $B$ and $\PL$ exactly, so it is a repeat that shares the two endpoint tables $W$ and $F_O$ with the first run, rather than an independent simulation.

\subsection{Physical fault coefficients}\label{app:F-series}
Finally, the series. The main text reports $A$, $B$ and $\PL$ not only at fixed $p$ but as power series in $x=p/(1-p)$, Eq.~\eqref{eq:fullseries}. Their coefficients are exact rationals too, obtained by the same modular contraction with one change: every number in the network is replaced by a polynomial in $x$, truncated after $x^{10}$. This section says what the coefficients count, how $\PL$'s coefficients follow from them, and how the truncated polynomials are handled.

\paragraph{What the raw coefficients count.} Each of the $N=1996$ fault locations at $d=5$ contributes a weight
\begin{equation}
 \begin{cases} 1-p & \text{no fault},\\[2pt] p/\nu & \text{one of its $\nu$ Pauli outcomes, } \nu\in\{1,3,15\},\end{cases}
 \label{eq:locationweight}
\end{equation}
so a pattern of $k$ faulted locations has weight $(1-p)^{N-k}p^k=(1-p)^Nx^k$ times $1/\nu$ per faulted location. Pulling out the common factor $(1-p)^N$,
\begin{equation}
 \begin{aligned}
 A(p)&=(1-p)^N a(x),& a(x)&=\sum_{k=0}^N a_kx^k,\\
 B(p)&=(1-p)^N e(x),& e(x)&=\sum_{k=0}^N e_kx^k.
 \end{aligned}
 \label{eq:raww}
\end{equation}
Here $a_k$ is the sum over all patterns of $k$ faulted locations of their probability of being accepted, and $e_k$ the same sum with the probability of being accepted with a logical error, as in Eq.~\eqref{eq:probabilities}. For instance $a_1=\dfiveaone$ at $d=5$ is the sum, over every location and each of its $\nu$ Pauli outcomes, of the acceptance probability of that single fault divided by $\nu$. Since every term is a probability and there are $\binom Nk$ patterns of size $k$,
\begin{equation}
 \begin{gathered}
 a_0=1,\qquad 0\le e_k\le a_k\le\binom Nk,\\
 L_k=e_k-\sum_{j=1}^k a_j L_{k-j}.
 \end{gathered}
 \label{eq:recursion}
\end{equation}

\paragraph{From $a_k,e_k$ to the coefficients of $\PL$.} The logical error rate is the quotient $\PL=B/A=e(x)/a(x)=\sum_kL_kx^k$, and multiplying out $a(x)\sum_kL_kx^k=e(x)$ and comparing coefficients gives the last line of Eq.~\eqref{eq:recursion}. Thus $L_k$ includes the acceptance correction and is not itself a positive fault-set count. It can be negative. The leading nonzero $L_k$ equals $e_k$. Running this recursion on the raw $a_k,e_k$ of Tables~\ref{tab:d3raw} and~\ref{tab:d5raw} recovers every $L_k$ of Table~\ref{tab:d3coeff} and of Eq.~\eqref{eq:fullseries}, which were obtained by polynomial division instead, at $d=5$ from $L_{\dfivelead}=\dfiveleadval$ and at $d=3$ from $L_{\dthreelead}=\dthreeleadval$. See Appendix~\ref{app:numerics} for the coefficient tables. The first rows at $d=5$:
{\color{black}
\begin{center}\small\begin{tabular}{clll}\toprule
$k$ & $e_k$ & $\sum_{j=1}^{k}a_jL_{k-j}$ & $L_k$ from the recursion (equal to Eq.~\eqref{eq:fullseries}) \\\midrule
3 & $\frac{574}{375}$ & $0$ & $\frac{574}{375}$ \\
4 & $\frac{1041564221}{810000}$ & $\frac{95284}{1125}$ & $\frac{972959741}{810000}$ \\
5 & $\frac{1012441365667}{1620000}$ & $\frac{89789515027}{1215000}$ & $\frac{2678166036893}{4860000}$ \\
6 & $\frac{6646695516492431}{97200000}$ & $\frac{26747913547657697}{729000000}$ & $\frac{46204605652071071}{1458000000}$ \\
\bottomrule\end{tabular}
\end{center}}
and at $d=3$:
{\color{black}
\begin{center}\small\begin{tabular}{clll}\toprule
$k$ & $e_k$ & $\sum_{j=1}^{k}a_jL_{k-j}$ & $L_k$ from the recursion (equal to Table~\ref{tab:d3coeff}) \\\midrule
2 & $\frac{32}{75}$ & $0$ & $\frac{32}{75}$ \\
3 & $\frac{2058319}{4500}$ & $\frac{2464}{225}$ & $\frac{2009039}{4500}$ \\
4 & $\frac{3046313977}{90000}$ & $\frac{804996719}{67500}$ & $\frac{1183791011}{54000}$ \\
5 & $\frac{3010251653251}{1518750}$ & $\frac{2154828089887}{2025000}$ & $\frac{5576522343343}{6075000}$ \\
\bottomrule\end{tabular}
\end{center}}
The 1,568 locations with an identity response were left out of $N$. Restoring them multiplies both $a$ and $e$ by $(1+x)^{1568}$, which turns $a_1=\dfiveaone$ into $\dfiverestored$ but leaves every $L_k$ unchanged, since the factor cancels in the quotient.

\paragraph{How the coefficients are computed.} Every scalar in the contraction becomes a polynomial in $x$ with coefficients modulo $q$, truncated after $x^{10}$, that is, an element of the ring $\mathbb F_q[x]/(x^{11})$. A product coefficient is a Cauchy sum,
\begin{equation}
 [uv]_k=\sum_{j=0}^k u_jv_{k-j}.
 \label{eq:cauchy}
\end{equation}
As a tensor, the product of two coefficient vectors is the contraction of the two vectors with the three-leg indicator $\ind{j+\ell=k}$, the legs $j$ and $\ell$ summed and $k$ left open:
{\color{black}
\begin{center}
\begin{tikzpicture}[x=1cm,y=1cm]
\node[tn tensor,minimum width=10mm,minimum height=9mm] (u) at (0,0.7) {$u$};
\node[tn tensor,minimum width=10mm,minimum height=9mm] (v) at (0,-0.7) {$v$};
\node[tn constraint,minimum width=22mm,minimum height=24mm,font=\small] (c) at (2.6,0) {$\ind{j+\ell=k}$};
\draw[tn leg] (u.east)--node[above,tn label] {$j$}(c.west|-u.east);
\draw[tn leg] (v.east)--node[below,tn label] {$\ell$}(c.west|-v.east);
\draw[tn leg] (c.east)--node[above,tn label] {$k$}(4.3,0);
\node[font=\large] at (5.0,0) {$=$};
\node[tn intermediate,minimum width=12mm,minimum height=9mm] (w) at (6.2,0) {$uv$};
\draw[tn leg] (w.east)--node[above,tn label] {$k$}(7.5,0);
\node[tn label] at (3.2,-1.5) {toy example: $\cauchytoy\pmod{x^3}$, coefficients $(\cauchyw)$};
\end{tikzpicture}
\end{center}}
Truncation is exact for the degrees kept, because a term of degree above ten can never contribute to a coefficient of degree ten or less. The endpoint tables must be prepared in the same ring. Fixed-$p$ tables do not supply their coefficients.

\paragraph{Grouped factors as polynomials.} In Appendix~\ref{app:growth} a group of $m_g$ channels became one table $g_p(v)$, Eq.~\eqref{eq:group}, built from the damping factor $\lambda(p)=\prod_\ell(1-c_\ell p)$. For the series the same table has to be written in $x$, with the no-fault weight $(1-p)^{m_g}$ of the group removed, because that weight is already part of the global factor $(1-p)^N$ in Eq.~\eqref{eq:raww}. Dividing Eq.~\eqref{eq:group} by $(1-p)^{m_g}$ and using $1-c_\ell p=(1-p)\bigl(1+(1-c_\ell)x\bigr)$ and $1/(1-p)=1+x$ gives
\begin{equation}
 \begin{aligned}
 P_g(x)&=\prod_{\ell=1}^{m_g}\bigl(1+(1-c_\ell)x\bigr),\\
 g_x(v)&=P_g(x)\ind{v=0}\\
       &\quad+\frac{(1+x)^{m_g}-P_g(x)}{2^{r_g}}\ind{v\in S}.
 \end{aligned}
 \label{eq:polygroup}
\end{equation}
Here $P_g(x)$ is the damping factor in the new variable, and $(1+x)^{m_g}$ is the $1$ of Eq.~\eqref{eq:group} after the division. That term is what makes the coefficient of $x^k$ count $k$-fault patterns correctly: the group stands for $m_g$ locations, each of which may fault, not for one effective fault. Replacing $(1+x)^{m_g}$ by $1+x$, as if the group were a single location, would undercount every pattern with two or more faults inside the group.

For example, for a group of one flip, one one-qubit and one two-qubit depolariser at $p=10^{-3}$, $P_g(x)=\Pgx$, and multiplying $g_x(0)=\gxzero$ and $g_x(e)=\gxone$ by $(1-p)^3$ returns exactly $g_p(0)$ and $g_p(e)$ of Appendix~\ref{app:growth}. The series is therefore not an approximation of the fixed-$p$ numbers but an exact computation in a different ring.

To reconstruct $a_k,e_k$ by the Chinese remainder theorem, clear their denominators: $U_{A,k}=2^{31}15^ka_k$ and $U_{B,k}=2^{31}15^ke_k$, the coefficients of $A$ and $B$ scaled to integers, the $2^{31}$ clearing the denominators of the response probabilities ($2^{11}$ at $d=3$)\footnote{The denominators $2^{11}$ and $2^{31}$ arise as follows. Summing the codewords of Eq.~\eqref{eq:codebasis}, a check with $r$ faces has a logical $2\times2$ matrix whose entries are an eighth root of unity times a Gaussian integer $a+bi$, divided by $2^{r+1}$, the root being fixed by the two logical basis bits and a frame parity. Composing the $r=3$ and $r=9$ checks on the normalised magic state keeps this form and gives amplitudes with denominator $2^{14}\sqrt2$, hence probabilities with denominator dividing $2^{29}$. The wrong-outcome projector $(I-M_L)/2$ adds at most a factor four, giving $2^{31}$. One $r=3$ check gives the analogous $2^{11}$. The accepted growth-record sum adds no denominator, by Appendix~\ref{app:growth}.} and the $15^k$ the Pauli outcome weights of $k$ faulted locations, and that $0\le U_{B,k}\le U_{A,k}\le2^{31}15^k\binom Nk$, since every term of $a_k$ is a probability. At $k=10$ this bound has $\boundtenbits$ bits ($115$ at $d=3$), so three primes above $2^{61}$ reconstruct every coefficient at $d=5$, and two suffice at $d=3$ with a third held out as a check, after which the recursion of Eq.~\eqref{eq:recursion} is run over the rationals.

\subsection{A remainder bound that retains post-selection}\label{app:F-tail}
Let
\begin{equation}
 \begin{aligned}
 A_K&=(1-p)^N\sum_{k=0}^K a_kx^k, & \lim_{K\to\infty}A_K&=A,\\
 E_K&=(1-p)^N\sum_{k=0}^K e_kx^k, & \lim_{K\to\infty}E_K&=B,\\
 \delta&=A-A_K\ \ \text{(omitted acceptance)}.
 \end{aligned}
 \label{eq:partialAE}
\end{equation}
Since $0\le e_k\le a_k$ for every $k$, Eq.~\eqref{eq:recursion}, and $x\ge0$, summing the omitted orders $k>K$ gives $0\le B-E_K\le\delta$. If the all-orders acceptance is known, then
\begin{equation}
 \frac{E_K}{A}\le\PL\le\frac{E_K+\delta}{A}.
 \label{eq:tail}
\end{equation}
If only an upper bound $A\le A_+$ is available, the coupled form is
\begin{equation}
 \frac{E_K}{A_+}\le\PL\le
                  \frac{E_K+A_+-A_K}{A_+}.
 \label{eq:tailupper}
\end{equation}
To see this, use $A\le A_+$ twice. The lower end of Eq.~\eqref{eq:tail} satisfies $E_K/A\ge E_K/A_+$. For the upper end write
\[
 \frac{E_K+\delta}{A}=\frac{E_K+\overbrace{A-A_K}^{\delta}}{A}=1-\frac{A_K-E_K}{A}\ \underbrace{\le}_{A\le A_+}\ 1-\frac{A_K-E_K}{A_+}=\frac{E_K+A_+-A_K}{A_+},
\]
where the inequality uses $A_K-E_K\ge0$ and $A\le A_+$. The same expression shows that the upper end increases with $A_+$, so a looser bound only widens the interval. No series convergence assumption is used in either inequality.

These intervals bound $\PL$ itself, not its Taylor polynomial $S_K=\sum_{k\le K}L_kx^k$, whose remainder follows by subtracting $S_K$ from both ends. Table~\ref{tab:tails} lists the intervals $[E_K/A,\ (E_K+\delta)/A]$ of Eq.~\eqref{eq:tail} at $K=10$ for a few values of $p$. At $p=10^{-3}$ the interval has relative width $\tailrel$, its length divided by $\PL$, whereas the truncated series misses the exact value by only $\taylorresid$. The interval is a thousand times wider than the true error because it lets every omitted accepted pattern be a logical error, while in fact most of them are not.
{\color{black}
\begin{table}[H]\centering\small
\captionsetup{position=above,skip=4pt}
\caption{Tenth-order intervals of Eq.~\eqref{eq:tail} for $\PL$, endpoints rounded outward.}
\label{tab:tails}
\begin{adjustbox}{max width=\linewidth}\begin{tabular}{rlll}
\toprule
$p$ & $S_{10}$ & Certified interval for $P_{\mathrm L}$ & Relative width \\
\midrule
$0.0005$ & $2.84616550\times10^{-10}$ & $[2.84616550,2.84616551]\times10^{-10}$ & $3.23\times10^{-9}$ \\
$0.00075$ & $1.16552789\times10^{-9}$ & $[1.16552788,1.16552797]\times10^{-9}$ & $6.79\times10^{-8}$ \\
$0.001$ & $3.32842537\times10^{-9}$ & $[3.32842529,3.32842717]\times10^{-9}$ & $5.61\times10^{-7}$ \\
$0.0015$ & $1.59126679\times10^{-8}$ & $[1.59126614,1.59128218]\times10^{-8}$ & $1.01\times10^{-5}$ \\
$0.002$ & $5.17510198\times10^{-8}$ & $[5.17508665,5.17546273]\times10^{-8}$ & $7.27\times10^{-5}$ \\
\bottomrule
\end{tabular}
\end{adjustbox}
\end{table}}

\clearpage
\twocolumn

\section{Validation and numerical details}\label{app:numerics}
\subsection{Validations}
\begin{enumerate}
\item The response tables of Appendix~\ref{app:response} were computed two ways, by the Gauss sums of Eq.~\eqref{eq:gauss} and by direct codeword amplitudes, on thousands of frames at both distances, and the $d=3$ window was simulated as a 13-qubit state vector with the faults at their literal positions. All three agree on every case.
\item The two-fault $d=3$ pattern of Fig.~\ref{fig:d3-fault-expansion}
was checked with both doubled probability diagrams in bloqade-tsim~\cite{tsim} and
amplitude diagrams in pyzx\_param~\cite{pyzx_param}. The three-fault $d=5$ pattern of
Appendix~\ref{app:d5-fault-v2} was checked with pyzx\_param, using the spider cutting with like-term collection of Ref.~\cite{wan} followed by stabiliser decomposition. The results agree with $(A,B)=(1/4,1/4)$ at $d=3$ and $(1/16,1/16)$ at $d=5$.
These are floating-point scalar checks of fixed fault patterns, not
independent rational certificates for the full noisy circuits.
\item Every fixed-$p$ reconstruction passes the prime-product bound, an unused-prime check and reverse-order CRT, and the repeat at $p=10^{-3}$ with a new contraction order reproduces $A$, $B$ and $\PL$ exactly.
\item The degree-ten coefficients agree with the earlier calculation through degree seven, and multiplying the quotient back into $a(x)$ reproduces $e(x)$ through degree ten, Eq.~\eqref{eq:recursion}.
\end{enumerate}

\subsection{Published comparisons}
Clifft values are read from Fig.~6 of Ref.~\cite{clifft} and SOFT values from Fig.~2 of Ref.~\cite{soft}. At every shared $p$ the exact value lies within their sampling uncertainty. The $d=3$ series expansion also matches with state-vector data from~\cite{msc} in Fig.~\ref{fig:d3comparison}.

\subsection{Measured cost}
Table~\ref{tab:cost} lists the worker time per point, including final-response preparation, cache construction, growth contraction and assembly, on single-threaded C++ code executed on an Apple Macbook (M4)\footnote{In today's value: $<1000\text{USD}$.}. The $p=10^{-3}$ repeat, which reused the cached endpoint tables of Appendix~\ref{app:F-reuse}, took 1.710 worker core-hours and the complete degree-ten calculation 9.857 worker core-hours.

{\color{black}
\begin{table*}[b]\centering\small
\captionsetup{position=above,skip=4pt}
\caption{Prime counts are reconstruction plus held-out checks; bound bits refer to the sufficient full denominator.}
\label{tab:cost}
\begin{adjustbox}{max width=\linewidth}\begin{tabular}{rllll}
\toprule
$p$ & Bound bits & Endpoint primes & Full primes & Worker core-hours \\
\midrule
$0.0005$ & 21,137 & 124+2 & 347+1 & 11.090 \\
$0.00075$ & 20,477 & 118+2 & 336+1 & 10.434 \\
$0.001$ & 19,737 & \emph{reused} & 324+1 & 1.710 \\
$0.0015$ & 18,186 & 107+2 & 299+1 & 9.297 \\
$0.002$ & 18,080 & 101+2 & 297+1 & 8.642 \\
\bottomrule
\end{tabular}
\end{adjustbox}
\end{table*}}

\subsection{Exact coefficients}
Table~\ref{tab:d3coeff} and ~\ref{tab:d3raw} gives the $d=3$ series expanded coefficients, while Table~\ref{tab:d5raw} gives the $d=5$ numbers.
\begin{table}[H]\centering\small
\captionsetup{position=above,skip=4pt}
\caption{The $d=3$ coefficients in $P^{(3)}_\text{L}(x)=\sum_kL_kx^k$.}
\label{tab:d3coeff}
\begin{tabular}{r@{\qquad}l}
\toprule
$k$ & $L_k$ \\
\midrule
2 & $\frac{32}{75}$ \\[1.2ex]
3 & $\frac{2009039}{4500}$ \\[1.2ex]
4 & $\frac{1183791011}{54000}$ \\[1.2ex]
5 & $\frac{5576522343343}{6075000}$ \\[1.2ex]
6 & $\frac{7072572702107929}{243000000}$ \\[1.2ex]
7 & $\frac{3056416241583144971}{3645000000}$ \\[1.2ex]
8 & $\frac{14000422301261560415557}{656100000000}$ \\[1.2ex]
9 & $\frac{352460798986209480998423}{820125000000}$ \\[1.2ex]
10 & $\frac{2800398004859242449492508511}{590490000000000}$ \\[1.2ex]
\bottomrule
\end{tabular}

\end{table}

\begin{table*}[p]\centering\small
\captionsetup{position=above,skip=4pt}
\caption{The $d=3$ raw acceptance and accepted-error coefficients.}
\label{tab:d3raw}
\begin{minipage}[t]{.48\linewidth}\centering
\begin{tabular}{r@{\qquad}l}
\toprule
$k$ & $a_k$ \\
\midrule
0 & $1$ \\[1.2ex]
1 & $\frac{77}{3}$ \\[1.2ex]
2 & $\frac{109433}{100}$ \\[1.2ex]
3 & $\frac{203757811}{6750}$ \\[1.2ex]
4 & $\frac{7640477983}{8100}$ \\[1.2ex]
5 & $\frac{3345046003099}{121500}$ \\[1.2ex]
6 & $\frac{745483248873863}{911250}$ \\[1.2ex]
7 & $\frac{16316109725855291383}{683437500}$ \\[1.2ex]
8 & $\frac{2814706392354594541187}{4100625000}$ \\[1.2ex]
9 & $\frac{740868888180195867334063}{38443359375}$ \\[1.2ex]
10 & $\frac{2424452455579323062802516617}{4613203125000}$ \\[1.2ex]
\bottomrule
\end{tabular}

\end{minipage}\hfill
\begin{minipage}[t]{.48\linewidth}\centering
\begin{tabular}{r@{\qquad}l}
\toprule
$k$ & $e_k$ \\
\midrule
0 & $0$ \\[1.2ex]
1 & $0$ \\[1.2ex]
2 & $\frac{32}{75}$ \\[1.2ex]
3 & $\frac{2058319}{4500}$ \\[1.2ex]
4 & $\frac{3046313977}{90000}$ \\[1.2ex]
5 & $\frac{3010251653251}{1518750}$ \\[1.2ex]
6 & $\frac{8250007730158441}{91125000}$ \\[1.2ex]
7 & $\frac{223846253563431581}{60750000}$ \\[1.2ex]
8 & $\frac{695773372331253998653}{5125781250}$ \\[1.2ex]
9 & $\frac{473468329549333222967513}{102515625000}$ \\[1.2ex]
10 & $\frac{1347503441280814153010426947}{9226406250000}$ \\[1.2ex]
\bottomrule
\end{tabular}

\end{minipage}
\end{table*}

\begin{table*}[p]\centering\small
\captionsetup{position=above,skip=4pt}
\caption{Whole-circuit $d=5$ raw acceptance and accepted-error coefficients.}
\label{tab:d5raw}
\begin{minipage}[t]{.48\linewidth}\centering
\begin{tabular}{r@{\qquad}l}
\toprule
$k$ & $a_k$ \\
\midrule
0 & $1$ \\[1.2ex]
1 & $\frac{166}{3}$ \\[1.2ex]
2 & $\frac{874357}{180}$ \\[1.2ex]
3 & $\frac{803222513}{3375}$ \\[1.2ex]
4 & $\frac{461785319813}{36000}$ \\[1.2ex]
5 & $\frac{6813675679442473}{12150000}$ \\[1.2ex]
6 & $\frac{35673102687723752419}{1458000000}$ \\[1.2ex]
7 & $\frac{7127303166983391609601}{7290000000}$ \\[1.2ex]
8 & $\frac{25074527061623692681497017}{656100000000}$ \\[1.2ex]
9 & $\frac{7051054714711008821570505193}{4920750000000}$ \\[1.2ex]
10 & $\frac{7771271684969588874728818721617}{147622500000000}$ \\[1.2ex]
\bottomrule
\end{tabular}

\end{minipage}\hfill
\begin{minipage}[t]{.48\linewidth}\centering
\begin{tabular}{r@{\qquad}l}
\toprule
$k$ & $e_k$ \\
\midrule
0 & $0$ \\[1.2ex]
1 & $0$ \\[1.2ex]
2 & $0$ \\[1.2ex]
3 & $\frac{574}{375}$ \\[1.2ex]
4 & $\frac{1041564221}{810000}$ \\[1.2ex]
5 & $\frac{1012441365667}{1620000}$ \\[1.2ex]
6 & $\frac{6646695516492431}{97200000}$ \\[1.2ex]
7 & $\frac{267643169821800115741}{43740000000}$ \\[1.2ex]
8 & $\frac{70659679997414182779587}{164025000000}$ \\[1.2ex]
9 & $\frac{527877089790266980903761157}{19683000000000}$ \\[1.2ex]
10 & $\frac{98061225429543733697140528883}{65610000000000}$ \\[1.2ex]
\bottomrule
\end{tabular}

\end{minipage}
\end{table*}

\clearpage
\onecolumn
\section{Other cultivation circuits}
\label{app:other-cultivation}
\captionsetup{hypcap=false}

We compare other cultivation circuits, ending with a
noiseless code projection and logical readout. Escape and following
stages are omitted for consistency with the main text. We adapt the methods
of the main text to these circuits --- using tensor contraction and compare the resultant LER series with the 
available numerical results.

Recall, $A$ is the probability of acceptance and $B$ is the probability of
acceptance with a logical error, and $\PL=B/A$. Throughout we use this series expansion
\[
 x=\frac{p}{1-p},\qquad S_K(p)=\sum_{k=0}^{K}L_kx^k.
\]
The coefficients below are exact for the specified variants of cultivation. Each protocol's bare distance labels $d=3,5$ must not be confused with the circuit fault distance - $d_\text{fault}$.

\subsection*{1. \texorpdfstring{$\mathbb{RP}^{2}$}{RP2}}

For the actual $T$ converted cultivation circuits from~\cite{RP2_prxq}, we obtain
\begin{equation}
P_{L,\mathbb{RP}^2}^{(3)}(x)=\frac{131}{450}x^{2}+\frac{9874093}{10800}x^{3}+\frac{17172616651}{540000}x^{4}+\frac{27243842363149}{24300000}x^{5}+O(x^{6}).
\end{equation}

\begin{equation}
P_{L,\mathbb{RP}^2}^{(5)}(x)=\frac{10439}{13500}x^{3}+\frac{20300775541}{51840000}x^{4}+\frac{616219992737}{1728000}x^{5}+\frac{428120035103643799}{20736000000}x^{6}+O(x^{7}).
\end{equation}

The leading powers of $x$ establish fault distances $d_{\text{fault}}=2,3$
for $\mathbb{RP}^{2}$ at $d=3,5$, respectively.

\fig{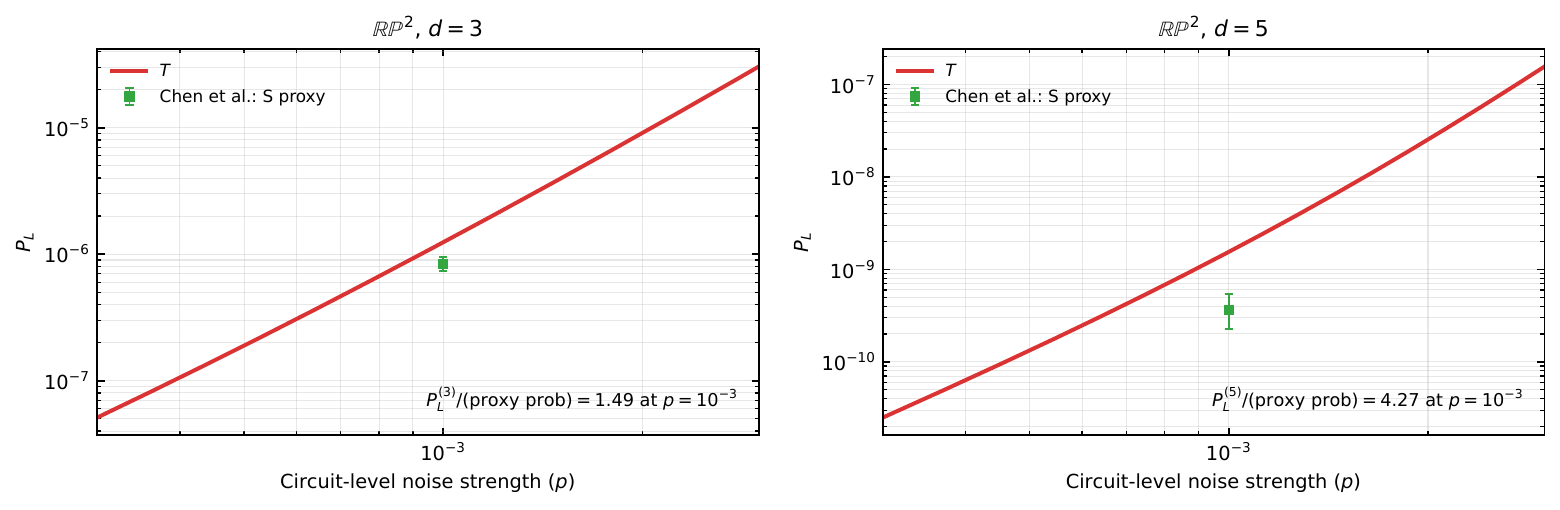}{1}{Actual $T$ cultivation series expanded LER and supplied the $S$-gate proxy samples for
$\mathbb{RP}^{2}$~\cite{RP2_prxq}.}

\subsection*{2. Fold-transversal}

For the fold-transversal circuits of~\cite{fold_msc_prxq}, we obtain
\begin{equation}
P_{L,\mathrm{Fold}}^{(3)}(x)=\frac{3909023513}{41674500}x^{3}+\frac{23378177959691}{12602368800}x^{4}+\frac{21954457843901229433}{396974617200000}x^{5}+O(x^{6}).
\end{equation}

\begin{equation}
P_{L,\mathrm{Fold}}^{(5)}(x)=\frac{3493956797}{17781120000}x^{4}+\frac{140969009479148306917}{1411465305600000}x^{5}+\frac{241056568979596536874210373}{32012033131008000000}x^{6}+O(x^{7}).
\end{equation}

The leading powers of $x$ establish fault distances $d_{\text{fault}}=3,4$
for fold-transversal cultivation at $d=3,5$, respectively.

\fig{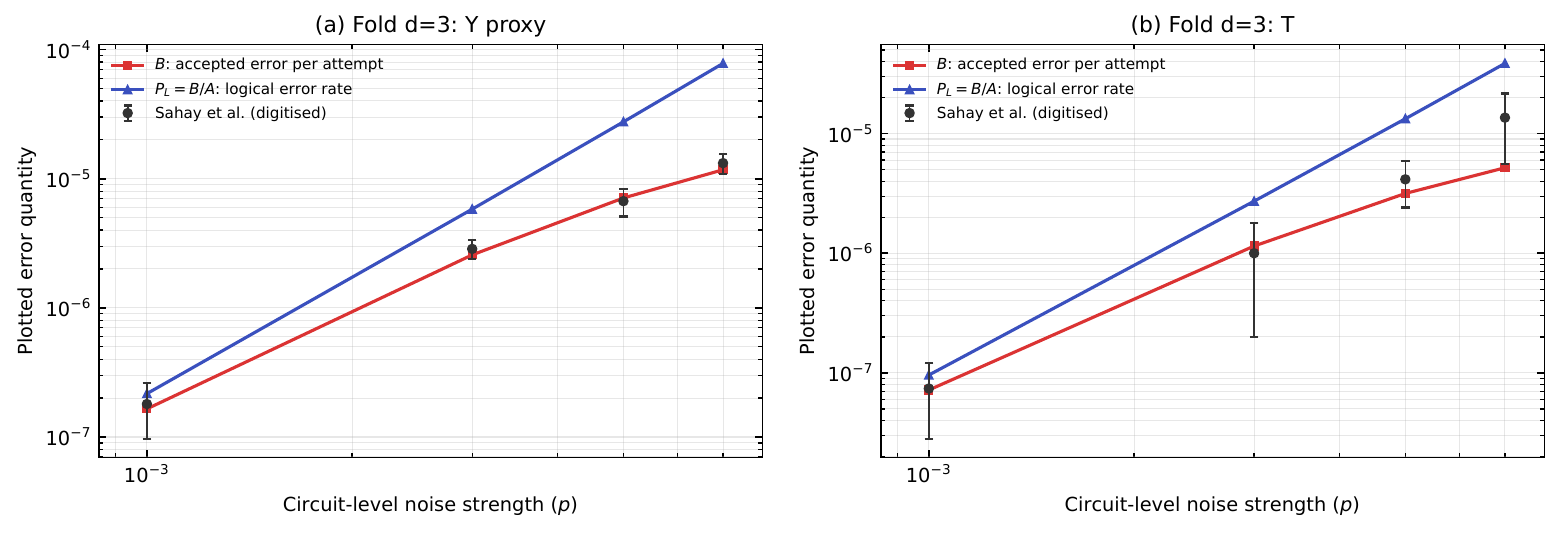}{1}{Published Fold $d=3$ data compared with $B$ per
initial attempt and $\PL=B/A$. The $Y$-proxy curves use independent samples. Black points and
error bars are digitised from the original figures.}

The $Y$-proxy and $T$ data are consistent with $B$ rather than $B/A$. The original no-escape plotting code and counts are unavailable and we cannot confirm the source of the discrepancy. 
It could be possible that the authors were plotting $B$ instead of $B/A$? Or we made a mistake somewhere here.

\subsection*{3. Chan et al.: repaired circuits accompanying arXiv version 1}

For the supplied three-round repaired circuits associated in~\cite{chan2026diagnosingrestoringdegradedfault}, the full-circuit series are
{\small
\begin{equation}
P_{L,\mathrm{Chan},\,v1}^{(3)}(x)=\frac{8514181}{13500}x^{3}+\frac{3637750171}{101250}x^{4}+\frac{9638727601309}{6075000}x^{5}+O(x^{6}).
\end{equation}
\begin{equation}
P_{L,\mathrm{Chan},\,v1}^{(5)}(x)=\frac{53656}{16875}x^{4}+\frac{15796640737639}{24300000}x^{5}+\frac{5280067584675593}{121500000}x^{6}+\frac{22550691826538722573}{10935000000}x^{7}+O(x^{8}).
\end{equation}

}

The leading powers of $x$ establish fault distances $d_{\text{fault}}=3,4$
for Chan et al.'s repaired v1 circuits at $d=3,5$, respectively.

\fig{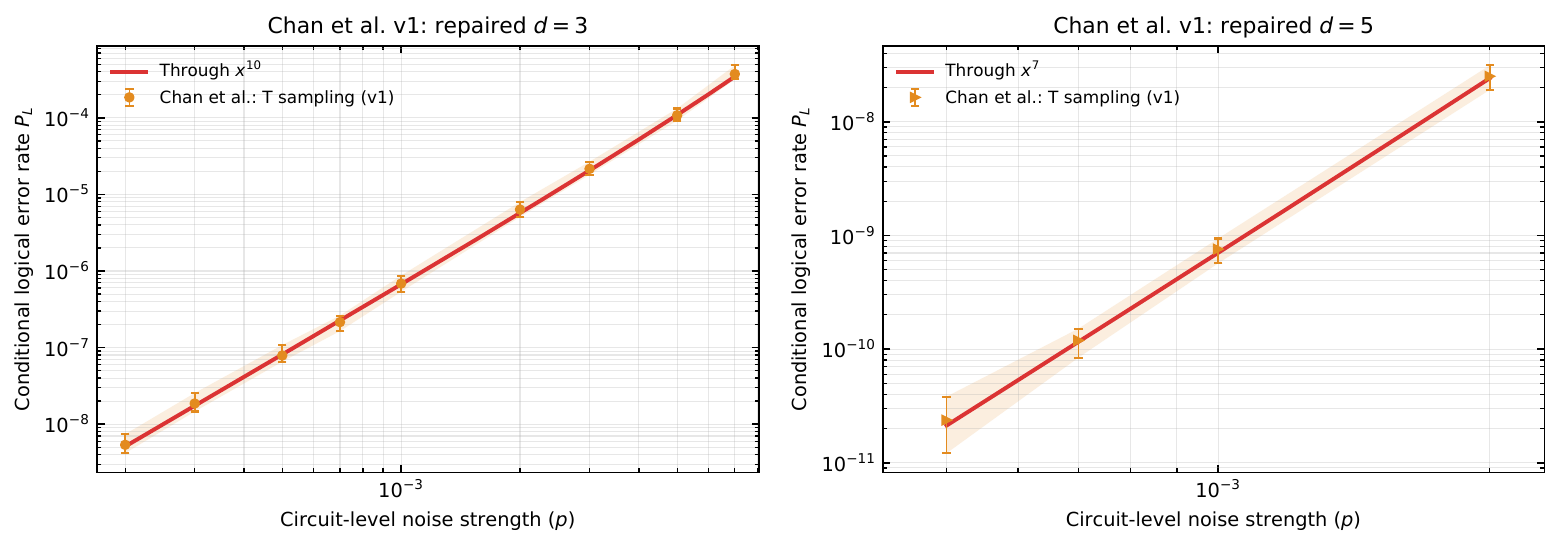}{1}{Repaired arXiv v1 circuits: exact-coefficient series
and digitised actual-$T$ sampling. The $d=3$ line retains terms through $x^{10}$ and the
$d=5$ line through $x^7$. These sampling points belong to arXiv v1, not the further
distance-restored four-round circuits in their arXiv v2.}

The four displayed $d=5$ series values lie inside the sampling
uncertainty intervals. For $d=5$, we found an explicit weight-four fault pattern that
passes post-selection and produces a logical error: $X_{10},X_4,X_2$
during the three growth rounds and $Y_3$ during final double checking.
Conditioned on this pattern and no other faults, $A_F=B_F=1/4$.
Separate gate-level ZX calculus contraction reproduce the uncaught logical error.

The following subsection shows results for their arXiv v2 circuits, which add a
fourth stabiliser round.

\subsection*{4. Chan et al.: repaired circuits accompanying arXiv version 2}

The new repaired circuit adds a fourth noisy stabiliser round before final
$d=5$ double checking. The complete series, including injection, both
double-checking stages and all growth noise, is now:
\begin{equation}
P_{L,\mathrm{new}}^{(5)}(x)=\frac{350644373171}{540000}x^5+\frac{15818820811063247}{364500000}x^6+O(x^7).
\label{eq:new-scaling}
\end{equation}

All lower-order coefficients vanish. Thus the circuit fault distance is finally restored to 5.

\begin{center}
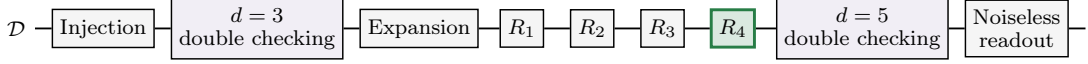

\begin{tikzpicture}[font=\scriptsize,line width=.55pt]
\begin{yquant}[register/minimum height=4mm,register/minimum depth=4mm,
 operator/minimum width=7mm,operator/separation=2.3mm,
 every wire/.append style={draw=black,line width=.7pt}]
qubit {$\mathcal D$} q;
[fill=black!4] box {Injection} q;
[fill=paperpurple!8] box {\shortstack{$d=3$\\double checking}} q;
[fill=black!4] box {Expansion} q;
[fill=black!4] box {$R_1$} q;
[fill=black!4] box {$R_2$} q;
[fill=black!4] box {$R_3$} q;
[fill=newround!17,draw=newround,line width=1.1pt] box {$R_4$} q;
[fill=paperpurple!8] box {\shortstack{$d=5$\\double checking}} q;
[fill=black!4] box {\shortstack{Noiseless\\readout}} q;
\end{yquant}
\end{tikzpicture}

\captionof{figure}{The supplied repair. The additional noisy stabiliser
round is green. Boxes group physical gates; the wire represents the data register.}
\end{center}

\fig{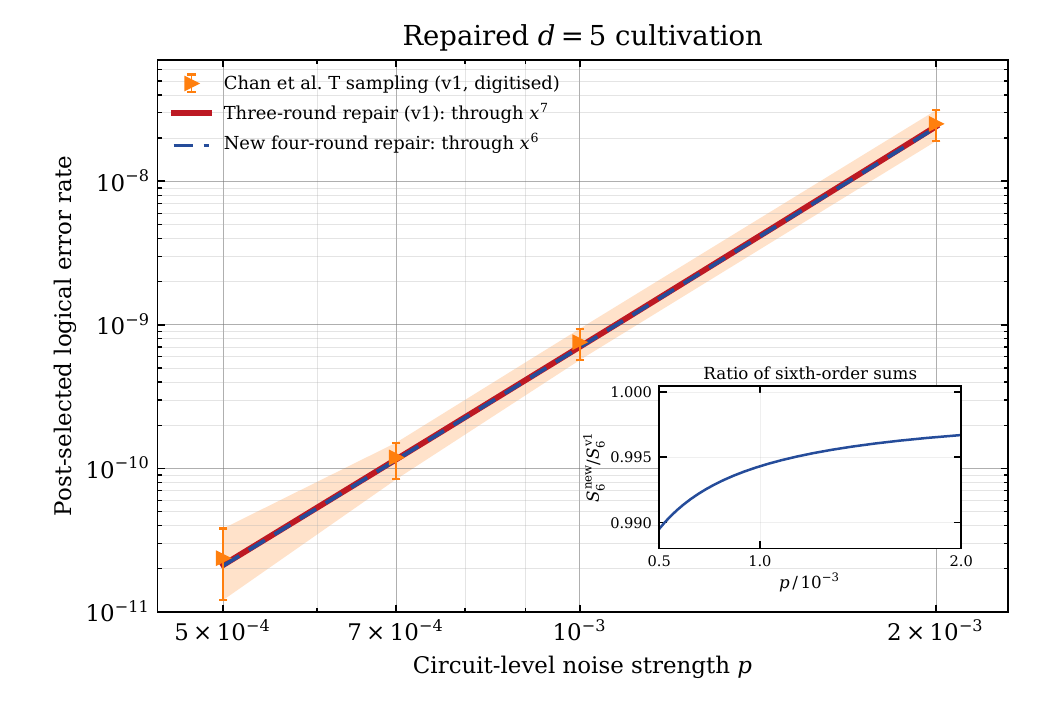}{.93}{The new four-round series plotted over the arXiv v1
comparison. Sampling points belong to arXiv v1 only. The inset compares both
circuits truncated at sixth order.}

An independent ZX calculus contraction confirmed rejection of the earlier weight-four fault.
At $p=10^{-3}$, the series through $x^6$ differ by only $0.57\%$, so the
shorter three-round circuit gives nearly the same LER here, although the
fourth round raises the fault distance from $4 \rightarrow 5$, but is somewhat not impactful in the only fruitful figure of merit - LER.

\clearpage
\begin{semiwebaddition}
\section{Pauli semiwebs in double checking}
\label{app:semiwebs}

\subsection{Definition and relation to Pauli propagation}
A Pauli semiweb assigns $I,X,Y$ or $Z$ to every wire of a ZX diagram.
At each spider, the tensor product $w_v$ of the Pauli operators on its legs acts on its tensor as
\begin{equation}
 w_v|v(\theta_v)\rangle=\lambda_v|v(\theta_v+\delta_v)\rangle.
 \label{eq:semiweb-local}
\end{equation}
Here $|v(\theta_v)\rangle$ is the spider tensor written as a state,
$\lambda_v$ is a nonzero scalar, and a phase shift $\delta_v\ne0$
modulo $2\pi$ is a \emph{defect}. The Pauli operators on each Hadamard tensor's legs must leave it unchanged
up to a scalar. With no defects, the semiweb is a \emph{Pauli web}~\cite{zxflow}.
A non-Clifford diagram can have both Pauli webs and semiwebs with defects.
The accompanying \texttt{pauli\_semiwebs.pdf} gives a basis of the full
semiweb space for each $d=3,5$ double-checking diagram, with editable TikZ sources.

The relation to Pauli propagation is already apparent with only a single $T$ gate:
\begin{equation}
 T^\dagger X=e^{-i\pi/4}XT,
 \qquad T^\dagger XT=\frac{X-Y}{\sqrt2}.
 \label{eq:semiweb-T}
\end{equation}
The first identity keeps the Pauli label and changes the gate's phase;
the second expands the conjugated observable into signed Pauli terms.
Both retain the coherent phases.

Let $K^{(3)}$ represent the noiseless $d=3$ double-checking stage with every $X$
measurement selected as $+1$, before the terminal code projection.
The local identities in Eq.~\eqref{eq:semiweb-local} give
\begin{equation}
 K^{(3)} P_{\rm in}=\lambda P_{\rm out}K^{(3)}_{\boldsymbol\delta},
 \label{eq:semiweb-global}
\end{equation}
where $K^{(3)}_{\boldsymbol\delta}$ has the starred phases changed and $\lambda$
includes the scalar phases from the local identities and contractions.
For Fig.~\ref{fig:semiweb-main}, this is
\begin{equation}
 K^{(3)}X_7=e^{5i\pi/4}X_0X_3X_9X_{10}X_{12}X_{13}K^{(3)}_{\boldsymbol\delta}.
 \label{eq:semiweb-X7}
\end{equation}
The diagram describes the selected branch. It shows changes to phases and measurement outcomes, including cancellations;
full post-selection must still be checked.

\subsection{Multiplication and cancellation}
Products of semiwebs on the same diagram are semiwebs~\cite{zxflow}.
To multiply two semiwebs, multiply their Pauli operators on each wire, including scalar phases such as
$XZ=-iY$. Defect shifts must be composed, rather than added.

\paragraph{A non-Clifford defect that cancels.}
Both $W_{X_7}$ and $W_{X_0}$ put $X$ on both legs of the
$\pi/4$ spider at the exit of qubit $0$. Their product gives $XX=I$,
as shown in Fig.~\ref{fig:semiweb-red-local}. Its phase changes as
\begin{equation}
 \frac{\pi}{4}\xrightarrow{X\otimes X}-\frac{\pi}{4}
 \xrightarrow{X\otimes X}\frac{\pi}{4}.
 \label{eq:semiweb-phase-cancellation}
\end{equation}
The second change is $+\pi/2$ after the first has acted. Other defects remain.

\begin{figure}[p]
\centering
{\color{black}\textbf{(a)} $W_{X_7}$: incoming $X_7$.}\par\vspace{1mm}
  \begingroup\color{black}%
  \input{figures/semiwebs/zx_styles.tex}%
  \pgfsetlayers{background,edgelayer,nodelayer,main}%
  \begin{adjustbox}{max width=.91\linewidth}\input{figures/semiwebs/x7_cancellation.tikz}\end{adjustbox}%
  \endgroup
\par\vspace{1mm}
{\color{black}\textbf{(b)} $W_{X_0}$: incoming $X_0$.}\par\vspace{1mm}
  \begingroup\color{black}%
  \input{figures/semiwebs/zx_styles.tex}%
  \pgfsetlayers{background,edgelayer,nodelayer,main}%
  \begin{adjustbox}{max width=.91\linewidth}\input{figures/semiwebs/x0_cancellation.tikz}\end{adjustbox}%
  \endgroup
\par\vspace{1mm}
{\color{black}\textbf{(c)} Product $W_{X_7}W_{X_0}$: incoming $X_0X_7$.}\par\vspace{1mm}
  \begingroup\color{black}%
  \input{figures/semiwebs/zx_styles.tex}%
  \pgfsetlayers{background,edgelayer,nodelayer,main}%
  \begin{adjustbox}{max width=.91\linewidth}\input{figures/semiwebs/x0x7_cancellation.tikz}\end{adjustbox}%
  \endgroup
\semiwebcaption{Cancellation of a shared non-Clifford defect on the complete
$d=3$ double-checking circuit. Red overlays show $X$ Pauli semiwebs; stars mark
defects. The dashed circle picks out the same $T$-gate at qubit $0$ in all three
panels. In (c), $XX=I$ on each leg and the shared defect cancels, while other defects
remain.}
\label{fig:semiweb-red-local}
\end{figure}

\paragraph{Two measurement defects that cancel.}
The $Z_0$ and $Z_3$ semiwebs each flip the central $X$ measurement on
qubit $7$ and the final ancilla $X$ measurement on qubit $6$.
Their product cancels both measurement defects,
Fig.~\ref{fig:semiweb-cancellation}:
\begin{equation}
 W_{Z_0}W_{Z_3}=W_{Z_0Z_3},\qquad K^{(3)}Z_0Z_3=Z_0Z_3K^{(3)}.
 \label{eq:semiweb-Z-product}
\end{equation}
This product is a defect-free Pauli web.

\begin{figure}[p]
\centering
{\color{black}\textbf{(a)} Incoming $Z_0$: two measurement defects, $\delta=\pi$.}\par\vspace{1mm}
  \begingroup\color{black}%
  \input{figures/semiwebs/zx_styles.tex}%
  \pgfsetlayers{background,edgelayer,nodelayer,main}%
  \begin{adjustbox}{max width=.98\linewidth}\input{figures/semiwebs/z0.tikz}\end{adjustbox}%
  \endgroup
\par\vspace{1mm}
{\color{black}\textbf{(b)} Incoming $Z_3$: the same two measurement defects.}\par\vspace{1mm}
  \begingroup\color{black}%
  \input{figures/semiwebs/zx_styles.tex}%
  \pgfsetlayers{background,edgelayer,nodelayer,main}%
  \begin{adjustbox}{max width=.98\linewidth}\input{figures/semiwebs/z3.tikz}\end{adjustbox}%
  \endgroup
\par\vspace{1mm}
{\color{black}\textbf{(c)} Product $W_{Z_0}W_{Z_3}$: the shared $Z$ operators and measurement defects cancel.}\par\vspace{1mm}
  \begingroup\color{black}%
  \input{figures/semiwebs/zx_styles.tex}%
  \pgfsetlayers{background,edgelayer,nodelayer,main}%
  \begin{adjustbox}{max width=.98\linewidth}% Semiweb: Z0Z3
% All phase metadata use units of pi/4; unchanged nodes have delta=0.
\begin{tikzpicture}[x=1.03cm,y=.495cm]
\tikzset{Z dot/.append style={minimum size=1.8mm},X dot/.append style={minimum size=1.8mm},
Z phase dot/.append style={shape=circle,rounded corners=0pt,scale=1,font={\fontsize{6}{7}\selectfont},minimum size=6mm,text width=4.5mm,align=center,inner sep=.05mm,outer sep=0pt},
Z Web/.append style={preaction={line width=.85mm,draw={rgb,255:red,120;green,200;blue,120}}}}
\begin{pgfonlayer}{nodelayer}
\node [style=none] (n0) at (0,0) {};
\node [style=none] (n1) at (0,-3) {};
\node [style=none] (n2) at (0,-5) {};
\node [style=none] (n3) at (0,-7) {};
\node [style=none] (n4) at (0,-8) {};
\node [style=none] (n5) at (0,-10) {};
\node [style=none] (n6) at (0,-11) {};
\node [style=Z dot] (n7) at (0,-12) {};
\node [style=Z dot] (n8) at (0,-9) {};
\node [style=Z dot] (n9) at (0,-4) {};
\node [style=Z dot] (n10) at (0,-2) {};
\node [style=Z dot] (n11) at (0,-6) {};
\node [style=Z dot] (n12) at (0,-1) {};
\node [style=Z phase dot] (n13) at (1,0) {$-\frac{\pi}{4}$};
\node [style=Z phase dot] (n14) at (1,-3) {$-\frac{\pi}{4}$};
\node [style=Z phase dot] (n15) at (1,-5) {$-\frac{\pi}{4}$};
\node [style=Z phase dot] (n16) at (1,-7) {$-\frac{\pi}{4}$};
\node [style=Z phase dot] (n17) at (1,-8) {$-\frac{\pi}{4}$};
\node [style=Z phase dot] (n18) at (1,-10) {$-\frac{\pi}{4}$};
\node [style=Z phase dot] (n19) at (1,-11) {$-\frac{\pi}{4}$};
\node [style=Z dot] (n20) at (2,-1) {};
\node [style=X dot] (n21) at (2,0) {};
\node [style=Z dot] (n22) at (2,-2) {};
\node [style=X dot] (n23) at (2,-3) {};
\node [style=Z dot] (n24) at (2,-4) {};
\node [style=X dot] (n25) at (2,-5) {};
\node [style=Z dot] (n26) at (2,-6) {};
\node [style=X dot] (n27) at (2,-7) {};
\node [style=Z dot] (n28) at (2,-9) {};
\node [style=X dot] (n29) at (2,-8) {};
\node [style=Z dot] (n30) at (2,-12) {};
\node [style=X dot] (n31) at (2,-11) {};
\node [style=Z dot] (n32) at (3,-3) {};
\node [style=X dot] (n33) at (3,-1) {};
\node [style=Z dot] (n34) at (3,-5) {};
\node [style=X dot] (n35) at (3,-6) {};
\node [style=Z dot] (n36) at (3,-9) {};
\node [style=X dot] (n37) at (3,-12) {};
\node [style=Z dot] (n38) at (4,-5) {};
\node [style=X dot] (n39) at (4,-3) {};
\node [style=Z dot] (n40) at (4,-9) {};
\node [style=X dot] (n41) at (4,-10) {};
\node [style=Z dot] (n42) at (5,-5) {};
\node [style=X dot] (n43) at (5,-9) {};
\node [style=Z dot] (n44) at (6,-5) {};
\node [style=Z dot] (n45) at (7,-5) {};
\node [style=Z dot] (n46) at (8,-5) {};
\node [style=X dot] (n47) at (8,-9) {};
\node [style=Z dot] (n48) at (9,-5) {};
\node [style=X dot] (n49) at (9,-3) {};
\node [style=Z dot] (n50) at (9,-9) {};
\node [style=X dot] (n51) at (9,-10) {};
\node [style=Z dot] (n52) at (10,-3) {};
\node [style=X dot] (n53) at (10,-1) {};
\node [style=Z dot] (n54) at (10,-5) {};
\node [style=X dot] (n55) at (10,-6) {};
\node [style=Z dot] (n56) at (10,-9) {};
\node [style=X dot] (n57) at (10,-12) {};
\node [style=Z dot] (n58) at (11,-1) {};
\node [style=X dot] (n59) at (11,0) {};
\node [style=Z dot] (n60) at (11,-2) {};
\node [style=X dot] (n61) at (11,-3) {};
\node [style=Z dot] (n62) at (11,-4) {};
\node [style=X dot] (n63) at (11,-5) {};
\node [style=Z dot] (n64) at (11,-6) {};
\node [style=X dot] (n65) at (11,-7) {};
\node [style=Z dot] (n66) at (11,-9) {};
\node [style=X dot] (n67) at (11,-8) {};
\node [style=Z dot] (n68) at (11,-12) {};
\node [style=X dot] (n69) at (11,-11) {};
\node [style=Z phase dot] (n70) at (12,0) {$\frac{\pi}{4}$};
\node [style=Z phase dot] (n71) at (12,-3) {$\frac{\pi}{4}$};
\node [style=Z phase dot] (n72) at (12,-5) {$\frac{\pi}{4}$};
\node [style=Z phase dot] (n73) at (12,-7) {$\frac{\pi}{4}$};
\node [style=Z phase dot] (n74) at (12,-8) {$\frac{\pi}{4}$};
\node [style=Z phase dot] (n75) at (12,-10) {$\frac{\pi}{4}$};
\node [style=Z phase dot] (n76) at (12,-11) {$\frac{\pi}{4}$};
\node [style=Z dot] (n77) at (13,-12) {};
\node [style=Z dot] (n78) at (13,-9) {};
\node [style=Z dot] (n79) at (13,-4) {};
\node [style=Z dot] (n80) at (13,-2) {};
\node [style=Z dot] (n81) at (13,-6) {};
\node [style=Z dot] (n82) at (13,-1) {};
\node [style=none] (n83) at (13,0) {};
\node [style=none] (n84) at (13,-3) {};
\node [style=none] (n85) at (13,-5) {};
\node [style=none] (n86) at (13,-7) {};
\node [style=none] (n87) at (13,-8) {};
\node [style=none] (n88) at (13,-10) {};
\node [style=none] (n89) at (13,-11) {};
\node [style=none,font=\scriptsize,anchor=east,text=black] at (-.17,0) {$q_{0}$ $\mathbf Z$};
\node [style=none,font=\scriptsize,anchor=west] at (13.18,0) {$Z$};
\node [style=none,font=\scriptsize,anchor=east,text=black!55] at (-.17,-1) {$q_{1}$};
\node [style=none,font=\scriptsize,anchor=east,text=black!55] at (-.17,-2) {$q_{2}$};
\node [style=none,font=\scriptsize,anchor=east,text=black] at (-.17,-3) {$q_{3}$ $\mathbf Z$};
\node [style=none,font=\scriptsize,anchor=west] at (13.18,-3) {$Z$};
\node [style=none,font=\scriptsize,anchor=east,text=black!55] at (-.17,-4) {$q_{6}$};
\node [style=none,font=\scriptsize,anchor=east,text=black] at (-.17,-5) {$q_{7}$};
\node [style=none,font=\scriptsize,anchor=east,text=black!55] at (-.17,-6) {$q_{8}$};
\node [style=none,font=\scriptsize,anchor=east,text=black] at (-.17,-7) {$q_{9}$};
\node [style=none,font=\scriptsize,anchor=east,text=black] at (-.17,-8) {$q_{10}$};
\node [style=none,font=\scriptsize,anchor=east,text=black!55] at (-.17,-9) {$q_{11}$};
\node [style=none,font=\scriptsize,anchor=east,text=black] at (-.17,-10) {$q_{12}$};
\node [style=none,font=\scriptsize,anchor=east,text=black] at (-.17,-11) {$q_{13}$};
\node [style=none,font=\scriptsize,anchor=east,text=black!55] at (-.17,-12) {$q_{14}$};
\node [style=none,font=\scriptsize] at (1,.85) {Entry};
\node [style=none,font=\scriptsize] at (3.5,.85) {Fold};
\node [style=none,font=\scriptsize] at (6.5,.85) {Measure / reset};
\node [style=none,font=\scriptsize] at (9.5,.85) {Unfold};
\node [style=none,font=\scriptsize] at (12,.85) {Exit};
\end{pgfonlayer}
\begin{pgfonlayer}{edgelayer}
\draw [style=Z Web] (n0.center) to (n13.center);
\draw [style=Z Web] (n1.center) to (n14.center);
\draw (n2.center) to (n15.center);
\draw (n3.center) to (n16.center);
\draw (n4.center) to (n17.center);
\draw (n5.center) to (n18.center);
\draw (n6.center) to (n19.center);
\draw (n12.center) to (n20.center);
\draw [style=Z Web] (n13.center) to (n21.center);
\draw (n10.center) to (n22.center);
\draw [style=Z Web] (n14.center) to (n23.center);
\draw (n9.center) to (n24.center);
\draw (n15.center) to (n25.center);
\draw (n11.center) to (n26.center);
\draw (n16.center) to (n27.center);
\draw (n8.center) to (n28.center);
\draw (n17.center) to (n29.center);
\draw (n7.center) to (n30.center);
\draw (n19.center) to (n31.center);
\draw [style=Z Web] (n23.center) to (n32.center);
\draw [style=Z Web] (n20.center) to (n33.center);
\draw (n25.center) to (n34.center);
\draw (n26.center) to (n35.center);
\draw (n28.center) to (n36.center);
\draw (n30.center) to (n37.center);
\draw (n34.center) to (n38.center);
\draw (n32.center) to (n39.center);
\draw (n36.center) to (n40.center);
\draw (n18.center) to (n41.center);
\draw (n38.center) to (n42.center);
\draw (n40.center) to (n43.center);
\draw (n42.center) to (n44.center);
\draw (n45.center) to (n46.center);
\draw (n43.center) to (n47.center);
\draw (n46.center) to (n48.center);
\draw (n39.center) to (n49.center);
\draw (n47.center) to (n50.center);
\draw (n41.center) to (n51.center);
\draw (n49.center) to (n52.center);
\draw [style=Z Web] (n33.center) to (n53.center);
\draw (n48.center) to (n54.center);
\draw (n35.center) to (n55.center);
\draw (n50.center) to (n56.center);
\draw (n37.center) to (n57.center);
\draw [style=Z Web] (n53.center) to (n58.center);
\draw [style=Z Web] (n21.center) to (n59.center);
\draw [style=Z Web] (n22.center) to (n60.center);
\draw [style=Z Web] (n52.center) to (n61.center);
\draw (n24.center) to (n62.center);
\draw (n54.center) to (n63.center);
\draw (n55.center) to (n64.center);
\draw (n27.center) to (n65.center);
\draw (n56.center) to (n66.center);
\draw (n29.center) to (n67.center);
\draw (n57.center) to (n68.center);
\draw (n31.center) to (n69.center);
\draw [style=Z Web] (n59.center) to (n70.center);
\draw [style=Z Web] (n61.center) to (n71.center);
\draw (n63.center) to (n72.center);
\draw (n65.center) to (n73.center);
\draw (n67.center) to (n74.center);
\draw (n51.center) to (n75.center);
\draw (n69.center) to (n76.center);
\draw (n68.center) to (n77.center);
\draw (n66.center) to (n78.center);
\draw (n62.center) to (n79.center);
\draw (n60.center) to (n80.center);
\draw (n64.center) to (n81.center);
\draw (n58.center) to (n82.center);
\draw [style=Z Web] (n70.center) to (n83.center);
\draw [style=Z Web] (n71.center) to (n84.center);
\draw (n72.center) to (n85.center);
\draw (n73.center) to (n86.center);
\draw (n74.center) to (n87.center);
\draw (n75.center) to (n88.center);
\draw (n76.center) to (n89.center);
\draw [style=Z Web] (n20.center) to (n21.center);
\draw [style=Z Web] (n22.center) to (n23.center);
\draw (n24.center) to (n25.center);
\draw (n26.center) to (n27.center);
\draw (n28.center) to (n29.center);
\draw (n30.center) to (n31.center);
\draw [style=Z Web] (n32.center) to (n33.center);
\draw (n34.center) to (n35.center);
\draw (n36.center) to (n37.center);
\draw (n38.center) to (n39.center);
\draw (n40.center) to (n41.center);
\draw (n42.center) to (n43.center);
\draw (n46.center) to (n47.center);
\draw (n48.center) to (n49.center);
\draw (n50.center) to (n51.center);
\draw [style=Z Web] (n52.center) to (n53.center);
\draw (n54.center) to (n55.center);
\draw (n56.center) to (n57.center);
\draw [style=Z Web] (n58.center) to (n59.center);
\draw [style=Z Web] (n60.center) to (n61.center);
\draw (n62.center) to (n63.center);
\draw (n64.center) to (n65.center);
\draw (n66.center) to (n67.center);
\draw (n68.center) to (n69.center);
\end{pgfonlayer}
\end{tikzpicture}\end{adjustbox}%
  \endgroup
\semiwebcaption{Cancellation of the two measurement defects.
Green overlays show $Z$ Pauli semiwebs; violet stars mark $\pi$ defects in
selected measurement outcomes.}
\label{fig:semiweb-cancellation}
\end{figure}

\FloatBarrier
\end{semiwebaddition}

\clearpage
\section{Pauli webs of the declared detectors}
\label{app:detector-webs}
We draw all the Pauli semiwebs that correspond to checks, one for each declared detector, in
the full $d=3$ and $d=5$ cultivation circuits of Gidney et al.~\cite{msc}, as provided by
SOFT~\cite{soft}, in the accompanying \texttt{detector\_webs.pdf}.

\begin{figure}[h]
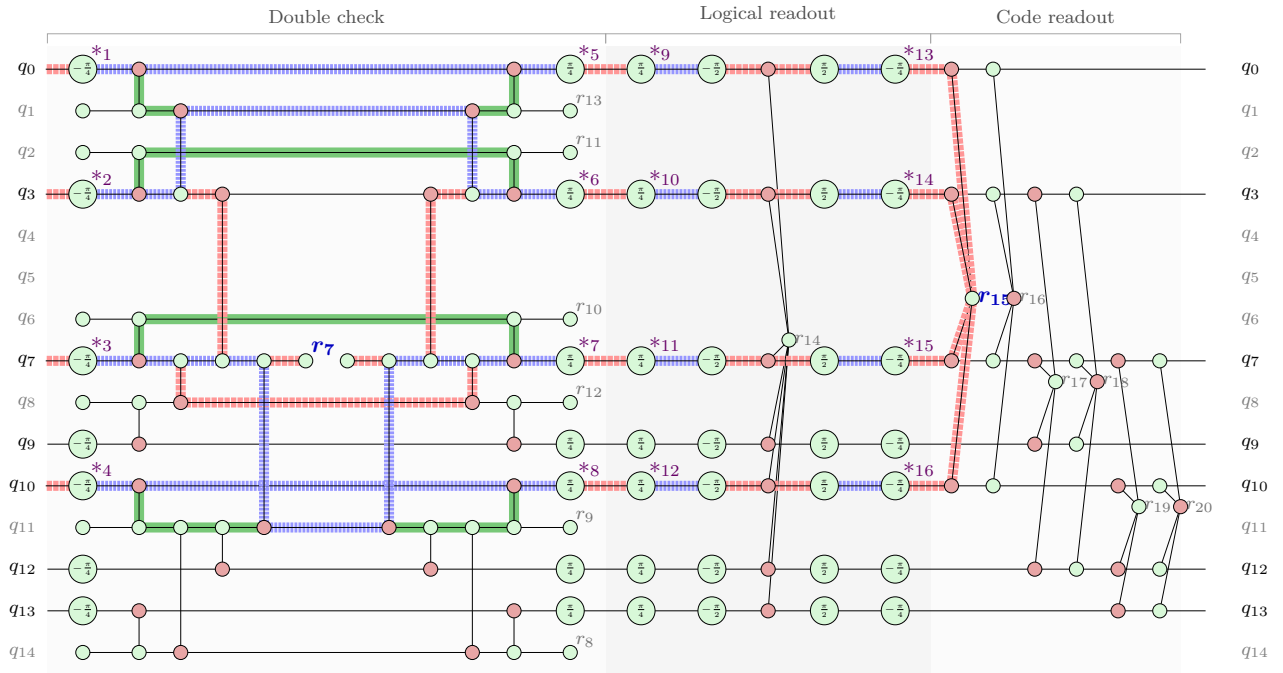

\centering
  \begingroup\color{black}%
  \input{figures/semiwebs/zx_styles.tex}%
  \pgfsetlayers{background,edgelayer,nodelayer,main}%
  \begin{adjustbox}{max width=.98\linewidth}\input{figures/detector_webs/d3_det014_check.tikz}\end{adjustbox}%
  \endgroup
\caption{Closed semiweb for the $d=3$ detector $r_7\oplus r_{15}=0$, the central
readout and the stabilizer $X_0X_3X_7X_{10}$, from the double check to the end of the circuit.
Red, green and blue overlays are $X$, $Z$ and $Y$ labels, and wires cut at the left edge
continue into the injection. The 16 violet stars are defects of $-\pi/2$ at the entry and
readout $T^\dagger$ gates and $+\pi/2$ at the exit and readout $T$ gates. With every $T$
replaced by $S$ the same labels form a Pauli web.}
\label{fig:detector-web}
\end{figure}

Some of the declared detectors are not Pauli webs of the $T$ circuit,
Table~\ref{tab:detector-webs}. Their closed semiwebs have defects on $T$ spiders,
Fig.~\ref{fig:detector-web}, so they depend on the phases of individual $T$ gates. An $X$ or $Y$
fault next to a $T$ gate turns it into $T^\dagger$ and can make them random. The other
detectors respond to every Pauli fault in the same deterministic way in the $T$ circuit and the
$S$ proxy. In the proxy every declared detector is a Pauli web, so it flips deterministically
and the proxy always rejects a pattern that flips one of them, while the $T$ circuit may accept
the same pattern. The detectors in Table~\ref{tab:detector-webs} are therefore the only ones that
separate the two, perhaps explaining the lower fault distances.

\begin{table}[h]
\centering
\caption{Declared detectors of the full circuits, numbered from 0 in source order.
Pauli-web detectors hold whatever the phase of each $T$ gate. The others depend on those
phases.}
\label{tab:detector-webs}
\begin{tabular}{lrrrl}
\toprule
 & Declared & Pauli webs & Not Pauli webs & Detectors that are not Pauli webs\\
\midrule
$d=3$ & 20 & 16 & 4 & 7, 14, 16, 18\\
$d=5$ & 107 & 93 & 14 & 7, 19, 20, 23, 69, 89, 91, 93, 95, 97, 99, 101, 103, 105\\
\bottomrule
\end{tabular}
\end{table}

\end{document}